\documentclass[11pt,a4paper,twoside,openright,notitlepage]{book}
\pdfoutput=1

\usepackage[utf8]{inputenc}
\usepackage[T1]{fontenc} 
\usepackage{lmodern}
\usepackage[czech,english]{babel}
\usepackage[a4paper,twoside,bindingoffset=7.5mm]{geometry}

\usepackage[unicode]{hyperref}
\hypersetup{pdftitle=Heat processes in non-equilibrium stochastic systems}
\hypersetup{pdfauthor=Jiří Pešek}

\usepackage[font={small}]{caption}

\usepackage{amsmath}
\usepackage{amssymb}
\usepackage{amsthm}

\usepackage{mathrsfs}

\usepackage[usenames,dvipsnames]{xcolor}
\usepackage{graphicx}
\usepackage{rotating}
\usepackage{float}
\usepackage{rotfloat}
\usepackage{subcaption}
\usepackage{multirow}
\usepackage{makeidx}

\usepackage{xparse}

\usepackage{fancyhdr}

\hypersetup{
	colorlinks, linkcolor={Red},
	citecolor={Bittersweet}, urlcolor={Emerald}
}

\author{Jiří Pešek}
\title{Heat processes in non-equilibrium stochastic systems}

\makeindex
\numberwithin{equation}{chapter}

\DeclareFontFamily{OMX}{MnSymbolE}{}
\DeclareSymbolFont{largesymbolsMn}{OMX}{MnSymbolE}{m}{n}
\DeclareFontShape{OMX}{MnSymbolE}{m}{n}{
<-6> MnSymbolE5
<6-7> MnSymbolE6
<7-8> MnSymbolE7
<8-9> MnSymbolE8
<9-10> MnSymbolE9
<10-12> MnSymbolE10
<12-> MnSymbolE12}{}
\DeclareMathDelimiter{\llangle}{\mathopen}{largesymbolsMn}{'164}{largesymbolsMn}{'164}
\DeclareMathDelimiter{\rrangle}{\mathclose}{largesymbolsMn}{'171}{largesymbolsMn}{'171}
\newcommand{\dbar}{{\mathchar'26\mkern-11mu \rmd}}

\newcommand{\id}{{\mathrm {id}}}
\newcommand{\rmd}{{\mathrm d}}
\newcommand{\rme}{{\mathrm e}}
\newcommand{\rmi}{{\mathrm i}}
\newcommand{\gen}{{\mathcal L}}
\newcommand{\proj}{{\mathcal P}}
\newcommand{\gensc}{{\hat{\mathscr L}}}
\newcommand{\projsc}{{\hat{\mathscr P}}}
\newcommand{\reals}{{\mathbb R}}
\newcommand{\integers}{{\mathbb Z}}
\newcommand{\eff}{{\mathrm{eff}}}

\DeclareMathOperator*{\supp}{supp}
\DeclareMathOperator*{\tr}{Tr}
\DeclareMathOperator*{\sgn}{sgn}

\newcommand{\err}[2][1]{{\mathrm O\left({\ifnum \pdfstrcmp{#1}{1}=0 #2 \else { #2 }^{ #1 } \fi}\right)}}
\newcommand{\rate}[3][]{{ \ifx{ #1 }{} {k\left( #2 \rightarrow #3 \right)} \else {k_{ #1 }\!\left( #2 \rightarrow #3 \right)} \fi }}
\newcommand{\curr}[3][]{{ \ifx{ #1 }{} {j\left( #2 \rightarrow #3)} \else {j_{ #1 }\!\left( #2 \rightarrow #3 \right)} \fi }}

\DeclareDocumentCommand{\prob}{O{}O{}m}{{ \ifx{ #2 }{} \ifx{ #1 }{} \mathbb P \left( #3 \right) \else \mathbb P^{ #1 } \left( #3 \right) \fi \else \ifx{ #1 }{} \mathbb P_{ #2 } \left( #3 \right) \else \mathbb P^{ #1 }_{ #2 } \left( #3 \right) \fi \fi }}
\DeclareDocumentCommand{\cprob}{O{}O{}mm}{{ \ifx{ #2 }{} \ifx{ #1 }{} \mathbb P \left( #3 \middle| #4 \right) \else \mathbb P^{ #1 } \left( #3 \middle| #4 \right) \fi \else \ifx{ #1 }{} \mathbb P_{ #2 } \left( #3 \middle| #4 \right) \else \mathbb P^{ #1 }_{ #2 } \left( #3 \middle| #4 \right) \fi \fi }}
\DeclareDocumentCommand{\dprob}{O{}O{}m}{{ \rmd \ifx{ #2 }{} \ifx{ #1 }{} \mathbb P \left( #3 \right) \else \mathbb P^{ #1 } \left( #3 \right) \fi \else \ifx{ #1 }{} \mathbb P_{ #2 } \left( #3 \right) \else \mathbb P^{ #1 }_{ #2 } \left( #3 \right) \fi \fi }}
\DeclareDocumentCommand{\dcprob}{O{}O{}mm}{{ \rmd \ifx{ #2 }{} \ifx{ #1 }{} \mathbb P \left( #3 \middle| #4 \right) \else \mathbb P^{ #1 } \left( #3 \middle| #4 \right) \fi \else \ifx{ #1 }{} \mathbb P_{ #2 } \left( #3 \middle| #4 \right) \else \mathbb P^{ #1 }_{ #2 } \left( #3 \middle| #4 \right) \fi \fi }}

\renewcommand{\vec}[1]{{\boldsymbol #1}}

\begin{document}
\pagestyle{empty}
\thispagestyle{empty}

%Titulni strana
\begin{center}

\large

Charles University in Prague

\medskip

Faculty of Mathematics and Physics

\vfill

{\bf\Large DOCTORAL THESIS}

\vfill

\centerline{\mbox{\includegraphics[width=60mm]{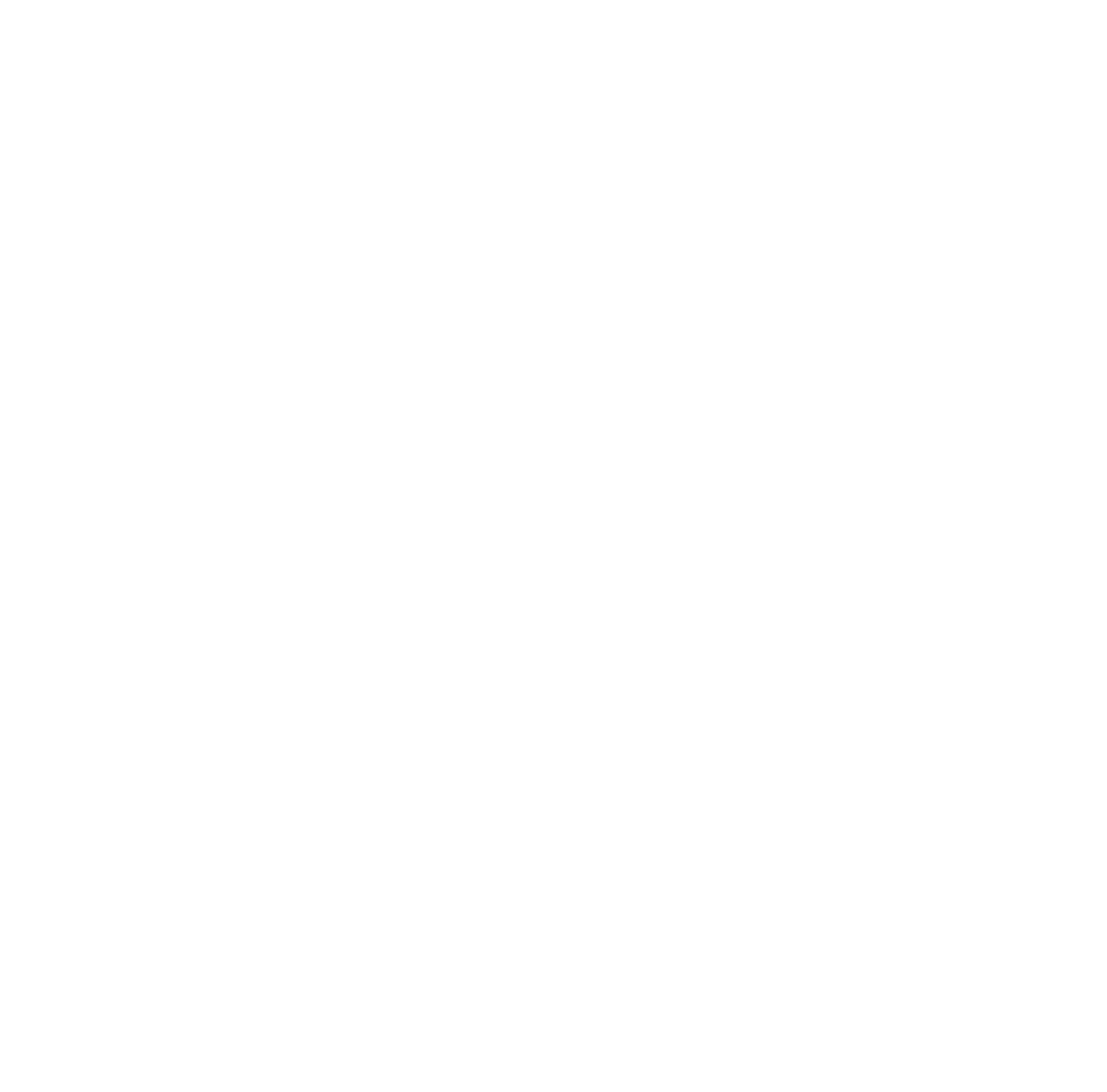}}}

\vfill
\vspace{5mm}

{\LARGE Jiří Pešek}

\vspace{15mm}

% Název práce přesně podle zadání
{\LARGE\bfseries Heat processes in non-equilibrium stochastic systems}

\vfill

% Název katedry nebo ústavu, kde byla práce oficiálně zadána
% (dle Organizační struktury MFF UK)
Institute of Theoretical Physics

\vfill

\begin{tabular}{rl}

Supervisor of the doctoral thesis: & Mgr. Karel Netočný, Ph.D. \\
\noalign{\vspace{2mm}}
Study programme: & Physics \\
\noalign{\vspace{2mm}}
Specialization: & Theoretical physics, \\
& Astronomy and Astrophysics \\
\end{tabular}

\vfill

% Zde doplňte rok
Prague 2013

\end{center}

\newpage

%%% Na tomto místě mohou být napsána případná poděkování (vedoucímu práce,
%%% konzultantovi, tomu, kdo zapůjčil software, literaturu apod.)

\cleardoublepage

\begin{otherlanguage}{czech}
\noindent
\subsubsection{Poděkování}
Tato práce vznikala pod záštitou Fyzikálního ústavu Akademie věd ČR, v.v.i.,
které patří velký dík za příjemné a motivující pracovní prostředí. 
V první řadě bych chtěl poděkovat svému vedoucímu Karlu Netočnému za velmi podnětné diskuze a za to, 
že jsem právě díky němu měl možnost proniknout alespoň trochu do tajů statistické fyziky a termodynamiky. 
Nesmím však opomenout jeho velkou trpělivost, obětavost a ochotu pomoci, za což mu také patří velký dík. 

Chci také poděkovat kolegům z kanceláře Pavlovi Augustinskému a Vladovi Pokornému za podnětné diskuze a příjemně strávený čas v kanceláři. 

Určitě velký dík patří skupině okolo profesora Maese v Belgické Lovani za jejich přijetí, mnoho podnětných dotazů a diskuzí a v neposlední řadě za plodnou spolupráci. 
Jmenovitě pak především Christianu Maesovi, Eliranovi, Simi a Winnymu.
\end{otherlanguage}

\newpage
\cleardoublepage

%%% Strana s čestným prohlášením k disertační práci

\vglue 0pt plus 1fill

\noindent
I declare that I carried out this doctoral thesis independently, and only with the cited sources, literature and other professional sources.

\medskip\noindent
I understand that my work relates to the rights and obligations under the Act No.~121/2000 Coll., the Copyright Act, as amended, in particular the fact that the Charles University in Prague has the right to conclude a license agreement on the use of this work as a school work pursuant to Section~60 paragraph~1 of the Copyright Act.

\vspace{10mm}

\hbox{\hbox to 0.5\hsize{%
In Prague 5$^\text{th}$ September 2013
\hss}\hbox to 0.5\hsize{%
Jiří Pešek
\hss}}

\vspace{20mm}
\newpage
\cleardoublepage
%%% Povinná informační strana disertační práce

\vbox to 0.5\vsize{
\setlength\parindent{0mm}
\setlength\parskip{5mm}

\begin{otherlanguage}{czech}
\noindent{\bf Název práce:}
Tepelné procesy v nerovnovážných stochastických systémech \\
% přesně dle zadání
{\bf Autor:}
Jiří Pešek \\
{\bf Katedra:}  % Případně Ústav: 
Ústav teoretické fyziky MFF UK \\
% dle Organizační struktury MFF UK
{\bf Vedoucí disertační práce:} 
Mgr. Karel Netočný, Ph.D., Fyzikální ústav AV ČR, v.v.i. \\
% dle Organizační struktury MFF UK, případně plný název pracoviště mimo MFF UK
{\bf Abstrakt:} \\ 
% abstrakt v rozsahu 80-200 slov; nejedná se však o opis zadání disertační práce
Tato disertační práce je věnována teoretickému studiu pomalých ter\-mo\-dy\-na\-mi\-ckých procesů v nerovnovážných stochastických systémech. 
Jejím hlavním výsledkem je fyzikálně a matematicky konzistentní konstrukce relevantních termodynamických veličin v kvazistatické limitě pro širokou třídu nerovnovážných modelů. 
Jako aplikaci obecných metod zavádí přirozené nerovnovážně zobecnění tepelné kapacity a detailně analyzuje jeho vlastnosti, včetně anomálního chování daleko od rovnováhy. 
Vyvinuté metody jsou dále použity na příbuzný problém oddělování časových škál, kde u\-mo\-žňu\-jí přesněji popisovat efektivní dynamiku pomalých i rychlých stupňů volnosti. \\ \\ 
{\bf Klíčová slova:}
% 3 až 5 klíčových slov
stochastická termodynamika, kvazistatické procesy, stacionární stavy
\end{otherlanguage}

\vss}\nobreak\vbox to 0.49\vsize{
\setlength\parindent{0mm}
\setlength\parskip{5mm}

\noindent{\bf Title:} 
Heat processes in non-equilibrium stochastic systems \\
{\bf Author:}
Jiří Pešek \\ 
{\bf Department:}
Institute of Theoretical Physics \\ 
% dle Organizační struktury MFF UK v angličtině
{\bf Supervisor:}
Mgr. Karel Netočný, Ph.D., Institute of Physics ASCR, v.v.i. \\ 
% dle Organizační struktury MFF UK, případně plný název pracoviště
% mimo MFF UK v angličtině
{\bf Abstract:} \\
This thesis is devoted to the theoretical study of slow thermodynamic processes in non-equilibrium stochastic systems. Its main result is a physically and mathematically consistent construction of relevant thermodynamic quantities in the quasistatic limit for a large class of non-equilibrium models. As an application of general methods a natural non-equilibrium generalization of heat capacity is introduced and its properties are analyzed in detail, including an anomalous far-from-equilibrium behavior. The developed methods are further applied to the related problem of time-scale separation where they enable to describe the effective dynamics of both slow and fast degrees of freedom in a more precise way. \\ \\ 
% abstrakt v rozsahu 80-200 slov v angličtině; nejedná se však o překlad
% zadání disertační práce
{\bf Keywords:}
% 3 až 5 klíčových slov v angličtině
stochastic thermodynamics, quasistatic process, steady state

\vss}

\newpage

%%% Strana s automaticky generovaným obsahem disertační práce. U matematických
%%% prací je přípustné, aby seznam tabulek a zkratek, existují-li, byl umístěn
%%% na začátku práce, místo na jejím konci.

\cleardoublepage

\addtocontents{toc}{\protect\thispagestyle{empty}}
\addtocontents{lof}{\protect\thispagestyle{empty}}
\addtocontents{lot}{\protect\thispagestyle{empty}}
\tableofcontents

\cleardoublepage
\pagestyle{fancy}
\setcounter{page}{1}
\pagenumbering{arabic}
\fancyhead[RO,LE]{\thepage}
\fancyhead[RE]{\slshape\small\MakeUppercase{\chaptername~\thechapter}}
\fancyhead[LO]{\slshape\small\leftmark}
\fancyfoot[L,C,R]{}
\renewcommand{\chaptermark}[1]{\markboth{#1}{}}

% Introduction
% a. general overview of the history of studies on thermodynamic processes
% b. motivations to go from equilibrium to steady nonequilibrium
% c. brief summary of the particular problems addressed in this thesis and an outline of the answers to come

\chapter{Introduction}
The statistical thermodynamics of non-equilibrium processes, as originated already by the founders of modern thermodynamic theory including L. Boltzmann, J. C. Maxwell or J. W. Gibbs, has experienced a large revival in last decades. 
This is mainly due to a great recent progress in micro- and nano-technologies which demands development of reliable theoretical methods to describe physical systems which are far from thermal equilibrium and in which fluctuations play a crucial role. 
This has led to an application of stochastic techniques in a fresh and deeper way, with a  better understanding of the role of time-reversal symmetry and its breaking as formulated in terms of the global and local detailed balance conditions, to the discovery of exact symmetry relations obeyed by the fluctuation even very far from equilibrium etc. 
The thermodynamics of open and typically rather small systems exhibiting non-negligible fluctuations has been coined the name ``stochastic thermodynamics'' \cite{Seifert2008,Seifert2012,Sekimoto1997,Speck2004,Sagawa2011,Komatsu2008-2,Komatsu2009,Esposito2010}.

Beyond the recently extensively studied properties of fluctuations in such open systems, there is a more conservative line of thoughts which tries to link the stochastic thermodynamics back to the standard thermodynamic framework as represented on the macroscopic level by the famous laws of thermodynamics \cite{Komatsu2008,Komatsu2010}. 
The main motivation is to formulate, if possible, general laws valid for large classes of thermodynamic process under inherently non-equilibrium conditions, similarly like this had been done with great success within the equilibrium framework. 
In contrast to purely thermodynamic concepts, the stochastic thermodynamics provides a far more reliable, systematic and in a good sense ``microscopic'' approach that older theories coming usually under the general name ``non-equilibrium thermodynamics''. 

Even restricting only to expected (or mean) values of relevant thermodynamic observables like work and heat, important and natural questions arise which have not yet been sufficiently answered: How to construct the fundamental elements of a standard thermodynamic theory, in particular the notion of quasistatic (= slow in a specific mathematical limit) processes above the inherently dissipative background (making the work or heat diverge in that limit)? 
Apparently, some ``renormalization'' scheme is needed here which many not be (and it is not!) unique, in general. 
Further, can the Clausius equality expressing the Second law for quasistatic processes be generalized to at least some large enough class of non-equilibrium systems? 
Intimately related questions are whether the operationally defined Clausius entropy has any natural non-equilibrium counterpart, whether there exist general relations between non-equilibrium response quantities in a way analogous to the Maxwell relations, what is the low-temperature asymptotics of (slow) far-from-equilibrium processes etc. 
It is exactly this sort of questions which are addressed in this thesis. 
Although we are still very far from providing complete answers, it is hoped that the results obtained in this work might provide another small step towards building a general and practicable scheme to describe strongly non-equilibrium systems and to reliably predict their properties.

\section{Outline of the thesis} 
Introduction and the necessary mathematical formalism used throughout the thesis is given in chapter \ref{chapter:models}. 
Then in chapter \ref{chapter:equilibrium_stat_phys} we explain the central notion of quasistatic limit within the framework of equilibrium stochastic thermodynamics. 
The main results of this work can be found in chapter \ref{chapter:non-equilibrium_thermodynamics} (quasistatic non-equilibrium process driven by non-potential forces), 
chapter \ref{chapter:periodically_driven_systems} (a generalization to systems driven by time-periodic forces) and chapter \ref{chapter:slow-fast_coupling} (the application of our methods to the problem of time scale separation). 
Finally, chapter \ref{chapter:conclusion} summarizes our results and shortly discusses open problems.

In the chapter \ref{chapter:models} we introduce the models widely used further within this thesis along with the necessary mathematical formalism. 
We define the general Markovian time evolution for an open system which we later specify as continuous-time Markov jump processes and diffusion. 
In order to describe the diffusions we review the Wiener process and consequently the It\^{o} and Stratonovich calculus and show some of its main results.

The chapter \ref{chapter:equilibrium_stat_phys} presents basic elements of (mostly equilibrium) stochastic thermodynamics.
We start with the introduction to Sekimoto stochastic energetics and its relation to the first law of thermodynamics. 
We define the local heat production and the local power which will play a crucial role later in the construction of the so called quasi-potential. 
Then we introduce the concepts of global and local detailed balance as the consequence of microscopic time reversibility 
and review the Crooks relation and the Jarzynski equality. 
We also show how the Second law inequality follows from the Crooks fluctuation relation.

The chapter \ref{chapter:non-equilibrium_thermodynamics} presents our first results. 
We start with the discussion of the consequences of the global detailed balance condition breaking  
and follow with the generalization of the stochastic energetics out of equilibrium. 
We decompose the quasistatic mean value of work and heat to what we will call a ``reversible'' \eqref{def:reversible_work} and ``housekeeping'' \eqref{def:housekeeping_work} work and heat and discuss their properties.
As our first result we show how the energetics on the level of mean values is solely governed by finite, geometric ``reversible'' components of the heat and work \eqref{equ:first_law_noneq}. 
We define the generalized heat capacity as the ``reversible'' heat associated with the quasistatic change of the temperature of the attached thermal bath \eqref{def:generalized_heat_capacity} and discuss its behaviour in various examples and show that it can be also negative, which is in contrast to its equilibrium counterpart.  
Our next result states that as the consequence of McLennan theorem the generalized version of the Clausius relation is verified in the close to equilibrium regime. 
We finish the chapter by brief illustration of low temperature behaviour of the heat capacity and conclude why no simple generalization of the Nernst theorem even for systems with finite number of non-degenerate states is to be expected.

In the chapter \ref{chapter:periodically_driven_systems} we study the behaviour of periodically driven systems in the quasistatic process, which was not previously studied in literature. 
By using the Floquet theory we are able again to identify the ``reversible'' and ``housekeeping'' components of the mean values of the total work and heat, see e.g. \eqref{equ:generalized_reversible_work}. 
We also define the generalized heat capacity and study its behaviour in various examples in order to better understand the behaviour of the ``reversible'' heat. 
In these examples we find that the generalized heat capacity becomes negative as it approaches intermediate regime where the relaxation time is comparable to the period of driving. 
We also discuss some fundamental interpretation problems as the ``housekeeping'' component contain a term which in general does not converge in the quasistatic limit.

In chapter \ref{chapter:slow-fast_coupling} we analyse the separation of time scales and the possibility to describe the evolution of such system by an effective Markovian dynamics. 
We briefly discuss the time evolution on the short time scale which in a special case is equivalent to system undergoing the quasistatic process. 
On the long time scale we provide more rigorous derivation of the Markovian dynamics and provide some estimates and bounds on precision of the Markovian approximation. 

% Stochastic models of open systems 
% a. Basic notation and formalism for "small" Markov systems

\chapter{Stochastic models of open systems}
\label{chapter:models}
For several centuries physicists have been trying to describe the real world using mathematics.  
During that time period they have established procedures how to describe any real physical system. 
Although the procedure can differ in technical aspects depending on the application, 
in general it always goes along similar lines. 
First we usually determine the set of all possible states of the system with respect to the level of description and observables of interest 
e.g. for mechanical particle system we determine all possible positions and momentums, for molecules we determine all possible conformations, 
then we relate these states to physical observables 
and finally we describe the time evolution, i.e. how the state changes in the course of time.
In this chapter we will introduce the mathematical formalism necessary to describe open thermodynamic systems in terms of continuous-time Markov stochastic processes, 
our presentation is rather informal as it is not the goal of this work to provide an exact mathematical treatment of the subject, cf. \cite{Feller2008}.
We introduce some of the models and examples used later in this thesis.

\section{Deterministic processes}
In classical mechanics we describe the \emph{state}\index{state} $x$ of the system by a complete collection of dynamical observables associated with all independent degrees of freedom.
For example in the case of single particle we need only to know its position and momentum $x=(\vec{q},\vec{p})$ to fully determine its state. 
Similarly mathematical pendulum is efficiently described by the angle and impulse momentum $x=(\varphi,l)$, 
or ratchet with $n$ tooths described by the position $x \in \mathbb Z_n$ of the pawl.
All possible states, i.e. configurations of the system, together make a \emph{configuration space}\index{configuration space} $\Omega$.

In this context physical \emph{observables}\index{observable!state}, which depends on the state in which the physical system is, are defined as functions 
\begin{equation}
A: \Omega \longrightarrow \reals,
\label{def:observable}
\end{equation}
where $A$ denotes some physical quantity. 
For example in the case of single particle its position is observable defined as $\vec{Q}(x)=\vec{q}$, similarly its velocity is $\vec{V}(x)= \frac{\vec{p}}{m}$ 
and also energy can be considered as an observable $E(x) = \frac{\vec{p}^2}{2m} + U(\vec{q})$, where $U$ is a potential in which presence the particle is.

Physical systems evolve with time and we denote $x_t$ the state of the system at time $t$.
One particular realization of the time evolution then creates a \emph{path}\index{path} $\omega$ in configuration space $\Omega$  
\[
\omega = \left\{ x_t \, \middle| \, t \in \reals \right\}.
\] 
The path can also be restricted over the time interval of our interest e.g. $[0,T]$, which we denote by  
\[
\omega^{[0,T]} = \left\{ x_t \, \middle| \, 0 \le t \le T \right\} .
\]

Introducing the path of the physical system enables us to describe another class of observables, so called \emph{path observables}\index{observable!path}, 
which do not depend only on the physical states of the system but also on its history. 
Typical representatives are heat and work, which we will define later, or first passage times $T_y(\omega^{[0,\infty)}) = \min_{s: x_s = y} s $.

In deterministic mechanics the history before the time $t$ represented by the path $\omega^{(-\infty,t)}$ uniquely determines the state $x_t$ at time $t$,
i.e. the \emph{time evolution}\index{time evolution!deterministic} in deterministic case is defined as 
\begin{equation}
x_t : \omega^{(-\infty,t)} \longrightarrow \Omega. 
\label{def:mechanics}
\end{equation}
In most cases we usually don't need a complete history but only a part of it, e.g. $\omega^{[0,t)}$.

\subsubsection{Example: Hamiltonian mechanics}
\index{Hamiltonian mechanics}
A typical example of such deterministic time evolution is the Hamiltonian mechanics of one particle described by Hamilton equations  
\begin{equation}
\begin{aligned}
\partial_t \vec{q}_t &= \left\{ \vec{q}_t , H_t\left(\vec{q}_t,\vec{p}_t\right) \right\}, \\
\partial_t \vec{p}_t &= \left\{ \vec{p}_t , H_t\left(\vec{q}_t,\vec{p}_t\right) \right\}, 
\end{aligned}
\label{equ:hamilton_equations}
\end{equation}
where $\{ \cdot , \cdot \}$ denotes Poisson bracket and $H_t(\vec{q},\vec{p})$ is time-dependent Hamiltonian.
The consequence of the Hamilton equations it that the full time evolution depends only on the initial condition, hence the dependence on the history is in a sense trivial. 
By integrating these differential equations from initial state $x_0=(\vec{q}_0,\vec{p}_0)$ at time $0$ over the time interval $[0,t)$ we obtain 
\begin{equation}
\begin{aligned}
\vec{q}_t &= \vec{q}_0 + \int\limits_0^t \rmd s \; \left\{ \vec{q}_s , H_s(\vec{q}_s,\vec{p}_s) \right\} , \\ 
\vec{p}_t &= \vec{p}_0 + \int\limits_0^t \rmd s \; \left\{ \vec{p}_s , H_s(\vec{q}_s,\vec{p}_s) \right\} ,
\end{aligned} 
\label{equ:hamiltonian_mechanics}
\end{equation}
where we can recognize the dependence on the path $\omega^{[0,t)}$ more clearly. 
Also the result \eqref{equ:hamiltonian_mechanics} more resembles the definition \eqref{def:mechanics} of the time evolution. 
The Hamiltonian mechanics is memory-less in the sense that the state at an arbitrary time $t$ contains enough information for the prediction of the system's behavior in the future with respect to the time $t$.

When modeling open systems coupled to their environment, we are forced to abandon the deterministic Hamiltonian structure and to replace with a more general stochastic law, yet a proper identification of relevant degrees of freedom often allows us to retain the memory-less property. 
Informally, this leads to the basic concept of \emph{Markovian dynamics}: 
The stochastic dynamical system is Markovian whenever the future time evolution started from state $x_t$ at any time $t$ (the "present") is statistically independent of the way the state $x_t$ was prepared (the "past").

\section{Stochastic processes}
%TODO: 
% + Stochasticka dynamika, -done 
% + pravdepodobnost na trajektoriich, - done
% + Markovovskost, - je treba jeste jednou predelat
% + charakterizace pomoci pravdepodobnostni distribuce, - done
% proudy, - done 
% rovnovaha, - done 
% detailni rovnovaha, - castecne
% lokalni detailni rovnovaha  - castecne
The detailed description of a large physical system in terms of all microscopic degrees of freedom becomes intractable since the number of available microscopic states grows exponentially with the number of degrees of freedom and due to the high complexity of the time evolution. 
Fortunately, we usually do not need the full information about the system yet we still want to make estimates about the values of relevant observables, the probabilistic approach proves to be useful. 
Within this approach we replace the notion of the state from microscopic level of description, now called \emph{microstate}\index{state!micro-}, by the \emph{configuration of the system}\index{state!configuration}, 
which corresponds to the maximum information about the real physical state in principle available on the coarse-grained level. 
While the configuration is determined only by the maximum information available to us, 
which does not necessarily corresponds to the full information contained on the microscopic level of description, 
we can conclude that one configuration can correspond to several microstates.  
Moreover in most cases we are not able to fully determine the exact configuration of the system either, 
but we are at least able to determine in which of them the system is more or less likely to be. 
Hence the physical state\index{state} is then described by probability distribution $\mu(x)$ over the configuration space $\Omega$, 
which members $x \in \Omega$ are all possible configurations of the system. 
Similarly as in the deterministic approach, 
we also define state-dependent observables\index{observable!state} as in this case \emph{measurable} functions \eqref{def:observable} from configuration space $\Omega$ to results represented by real numbers $\reals$,
as well as the path observables\index{observable!path} as \emph{measurable} functions on paths.

The remaining step is to adapt the time evolution\index{time evolution!stochastic} to fit into the current framework. 
We cannot determine the exact outcome of the state anymore, hence we describe the time-evolution as a probability measure $\rmd \mathbb P$ on paths $\omega$ with normalization to unity 
\[
\int \dprob{\omega} = 1 ,
\]
i.e. we assign a probability to each possible path, which also contain complete information about the initial state. 
The system's state at time $t$ is represented by the marginal probability 
\[
\mu_t (x) = \int \dprob{\omega} \; \delta_{X_t}(x), 
\]
where $\delta(\cdot)$ is the Dirac delta function in continuous case or the Kronecker's delta in discrete case and $X_t$ denotes the configuration in time $t$ from the path $\omega$, $X_t \in \omega$. 
Within this framework we can consider the deterministic time evolution as a special case in which each configuration is uniquely mapped on another.
Although we have described the time evolution in its entirety, we are usually interested only in its finite range. 
To address this requirement we define the probability measure up to the time $t$ as  
\[
\dprob[{(-\infty,t]}]{\omega^{(-\infty,t]}} = \int\limits_{\omega^{(-\infty,t]} \subset \varphi} \dprob{\varphi} ,
\]
where we integrate over all paths $\varphi$ containing the path $\omega^{(-\infty,t]}$, 
by which we can also fully characterize the time evolution up to arbitrary time. 
The state of the system at time $t$ is also determined by this probability measure
\[
\mu_t (x) = \int \dprob[{(-\infty,t]}]{\omega} \; \delta_{X_t}(x). 
\]

To correctly predict the time-evolution we in general need to know the full history of the system, which is usually not accessible to us, 
fortunately in most cases the dependency of the time evolution on the history weak enough so that in order to determine the plausibility of states at time $t$ we only need to know the path $\omega^{[t-T,t)}$ for not too large $T$.
In these cases we can restrict the description of the time evolution to the probability measure on more confined time interval $[t-T,t]$ 
\[
\dprob[{[t-T,t]}]{\omega^{[t-T,t]}} = \int\limits_{\omega^{[t-T,t]} \subset \varphi^{(-\infty,t]}} \dprob[{(-\infty,t]}]{\varphi^{(-\infty,t]}} 
= \int\limits_{\omega^{[t-T,t]} \subset \varphi} \dprob{\varphi} .
\]
This also allows us to exclude the information about the initial state from the probability measure 
and characterize the time-evolution by the conditional probability measure on the time interval $(t-T,t]$.
Hence the mean value over all paths of some path observable $A$ can equivalently be obtained as the mean value with respect to the conditional measure 
conditioned upon the initial configuration and then by averaging over the initial condition  
\begin{multline}
\int \dprob[{[t-T,t]}]{\omega} \; A(\omega) = \\
= \int \rmd y \; \mu_{t-T}(y) \int \dcprob[{(t-T,t]}]{\varphi}{X_{t-T}=y} \; A(\varphi \cup \{X_{t-T} = y \} ) . 
\label{def:conditional_measure}
\end{multline}
Using this conditional probability the state at time $t$ can be evaluated as 
\begin{equation}
\mu_t (x) = \int \rmd y \; \mu_{t-T}(y) \int \dcprob[{(t-T,t]}]{\omega}{X_{t-T}=y} \; \delta_{X_t}(x). 
\label{equ:state_by_conditional_measure}
\end{equation}

\section{Markov continuous-time stochastic processes}
\label{sec:Markov_processes} 
\index{Markov property}
\index{Markov stochastic process} 
In context of stochastic processes the Markov processes are those, which have extremely weak dependence on the history.  
Hence to obtain the full time evolution it is sufficient to know the conditional probability measure for arbitrarily short time.
As a consequence we can describe the full time evolution over the time interval $T$ by the time evolution over arbitrary number of shorter intervals. 

This feature can be exploited in order to show that the conditional mean value of some observable at time $t > t_k > t_{k-1} > \dots > t_0 $ conditioned on several values $x_0, \dots , x_k$ in the past does depend only on the last value $x_k$, 
\[
\left\langle X_t \middle| X_{t_k}=x_k , X_{t_{k-1}}=x_{k-1}, \dots , X_{t_0} = x_0 \right\rangle_{\mu_{t_0}} = 
\left\langle X_t \middle| X_{t_k}=x_k \right\rangle_{\mu_{t_k}} .
\]

\subsection{Forward Kolmogorov generator} 
\index{Kolmogorov generator!forward}
Another consequence of Markov property is the existence of autonomous time evolution for probability densities $\mu_t$ describing the state of the system, 
which is in particular useful when we are interested only in estimates of state dependent observables. 
However Markov property alone is not sufficient, we also need to \emph{assume} that the \emph{time evolution is smooth}, 
in particular that the difference between the state at time $t$ and time $t-\Delta t$ diminishes proportionally with $\Delta t$. 
Only then we can define the \emph{forward Kolmogorov generator} as the linear operator on probability densities 
\begin{multline}	
\gen^*_t [\mu_t] = 
\lim_{\Delta t \rightarrow 0^+} \frac{1}{\Delta t} \Bigl[ \int \rmd y \; \mu_{t-\Delta t}(y) \times \\
\times \int \dcprob[{(t-\Delta t,t]}]{\omega}{X_{t-\Delta t}=y} \; \delta_{X_t}(x) - \mu_{t-\Delta t}(x) \Bigr] ,
\label{def:forward_generator}
\end{multline}
and the time evolution of the state is described by the differential equation 
\begin{equation}
\partial_t \mu_t (x) = \gen_t^* [\mu_t] (x) , 
\label{equ:Markov_stochastic_time_evolution}
\end{equation}
compare with \eqref{equ:state_by_conditional_measure}. 
By integrating \eqref{equ:Markov_stochastic_time_evolution} over all possible states and by using the fact that the probability density $\mu_t$ has to be always normalized to unity, 
we obtain a condition
\begin{equation}
\int \rmd x \; \gen_t^* [\mu_t] (x) = 0 ,
\label{equ:normalization_condition}
\end{equation}
note that this condition can also be obtained directly from the definition of the forward Kolmogorov generator \eqref{def:forward_generator} and the normalization of probability densities. 
By explicit integration of \eqref{equ:Markov_stochastic_time_evolution} we obtain formal solution 
\[
\mu_t(x) = \overleftarrow{\exp} \left\{ \int\limits_{t_0}^t \rmd t \; \gen^*_t \right\} [\mu_{t_0}](x),
\] 
where $\overleftarrow{\exp}$ denotes time-ordered exponential.

\subsubsection{Example: Hamiltonian mechanics revisited}
\index{Hamiltonian mechanics}
To provide a simple example, we derive the forward Kolmogorov generator for the Hamiltonian mechanics.
Although the Hamiltonian mechanics is deterministic, it can also be described within the framework of stochastic time-evolution as was noted above.
The conditional probability measure is Dirac-like measure which gives zero for every trajectory which does not comply with solution of Hamilton equations \eqref{equ:hamiltonian_mechanics}.
Hence for the generator \eqref{def:forward_generator} we obtain
\[
\gen^*_t [\mu_t] (\vec{q},\vec{p}) = 
\lim_{\Delta t \rightarrow 0^+} \frac{1}{\Delta t} 
\left[ \mu_{t-\Delta t} \left( \vec{q}_{t-\Delta t}(\vec{q},\vec{p},t) , \vec{p}_{t-\Delta t}(\vec{q},\vec{p},t) \right) - \mu_{t-\Delta t}(\vec{q},\vec{p}) \right] ,
\]
where $( \vec{q}_{t-\Delta t}(\vec{q},\vec{p},t), \vec{p}_{t-\Delta t}(\vec{p},t) )$ denotes the configuration of the particle at time $t-\Delta t$ 
starting from which we reach the configuration $x=(\vec{q},\vec{p})$ at time $t$. 
Because the Hamiltonian mechanics is deterministic we can determine a past configuration of the system knowing the present configuration. 
Using \eqref{equ:hamiltonian_mechanics} we obtain 
\begin{align*}
\vec{q}_{t-\Delta t}(\vec{q},\vec{p},t) &= \vec{q} + \int\limits_t^{t-\Delta t} \rmd s \; \left\{ \vec{q}_s , H_s(\vec{q}_s,\vec{p}_s) \right\} , \\ 
\vec{p}_{t-\Delta t}(\vec{q},\vec{p},t) &= \vec{p} + \int\limits_t^{t-\Delta t} \rmd s \; \left\{ \vec{p}_s , H_s(\vec{q}_s,\vec{p}_s) \right\} .
\end{align*} 
By expanding the distribution density in $\Delta t$ and inserting the solution above we obtain the \emph{Liouville's equation}\index{Liouville's equation}  
\[
\gen^*_t [\mu_t] (\vec{q},\vec{p}) = \left\{ H_t(\vec{q},\vec{p}) , \mu_t(\vec{q}, \vec{p}) \right\} .
\]
Notice that in the context of stochastic Markov processes the condition \eqref{equ:normalization_condition} corresponds to the Liouville's theorem.

\subsection{Spectral properties of forward Kolmogorov generator}
The condition \eqref{equ:normalization_condition} has several mathematical implications related to the spectrum of the generator. 
The first one states that every \emph{eigenvector} $\nu_t$ of the generator $\gen^*_t$ at fixed time $t$ corresponding \emph{to the non-zero eigenvalue} $\lambda_t$ \emph{has zero integral} over all configurations
\[
0 = \int \rmd x \; \gen^*_t [\nu_t] (x) = \lambda_t \int \rmd x \; \nu_t (x). 
\] 
It means that all eigenvectors corresponding to non-zero eigenvalues are either zero up to the null set or have both negative and positive part. 
This necessarily leads to the condition that the \emph{real part of all eigenvalues has to be non-positive}, 
otherwise the negative part of the eigenvector will grow over all bounds causing the probability distribution to be negative $\mu_t (x) < 0$. 
In most of our applications we will also assume that if we fix the parameters of the generator at some fixed time $\widetilde{\gen}^* = \gen^*_s$ and let the system evolve with these fixed values, 
i.e. using the generator $\widetilde{\gen}^*$ instead of the original generator $\gen^*_t$, then there always exists a steady state to which the system will tend to converge.

\subsection{Steady state}
\index{steady state} 
The steady-sate of a Markov system described by the time-independent generator $\gen^*_t=\gen^*$ is described by the stationary distribution which is obtained as
\begin{equation}
\gen^* [ \rho ] (x) = 0 ,
\label{equ:stationary_condition}
\end{equation}
so the steady states $\rho$ lies in the kernel of the linear operator. 
In general the notion of stationarity can be extended to periodical time evolutions, 
where the steady state $\rho_t$, now explicitly depending on time $t$, is the \emph{periodic} solution of \eqref{equ:Markov_stochastic_time_evolution}.

\subsection{Backward Kolmogorov generator} 
\index{Kolmogorov generator!backward}
Until now we have described the time evolution in terms of time-dependent probability distribution $\mu_t$.
Equivalently, it can be described as the time-evolution on observables (cf. the quantum-mechanical Heisenberg picture),
where we can go from Schrödinger picture (representation) to Heisenberg picture. 
For this purpose we define the \emph{backward Kolmogorov generator} as the adjoint generator 
\begin{equation}
\int \rmd x \; A(x) \, \gen^*_t[\mu_t] (x) = \int \rmd x \; \gen_t [A] (x) \, \mu_t (x) .
\label{def:backward_generator}
\end{equation}
The time evolution using the backward Kolmogorov generator is then characterized by 
\[
\rmd_t A_t (x) = \partial_t A_t(x) + \gen_t [A_t] (x) ,
\]
where by $\rmd_t$ we denote the total time derivative, 
while by $\partial_t$ we denote the time derivative of explicit time-dependence of observable $A_t$ with respect of time. 
The time derivative of the mean value of some explicitly time-dependent observable $A_t$ can be written using the backward Kolmogorov generator as 
\begin{equation}
\partial_t \left\langle A_t \right\rangle_{\mu_t} = \left\langle \partial_t A_t + \gen_t [ A_t ] \right\rangle_{\mu_t} .
\label{equ:time_derivative_observable}
\end{equation}

\section{Continuous-time jump process}
\label{sec:jump_processes}
\begin{figure}[t]
\caption{Illustration of several paths $\omega^{[0,3]}$ all beginning at configuration $1$ for a system with three configurations (or ``levels''). }
\label{pic:illustration_of_path_jump_process}
\begin{center}
\includegraphics[width=.8\textwidth,height=!]{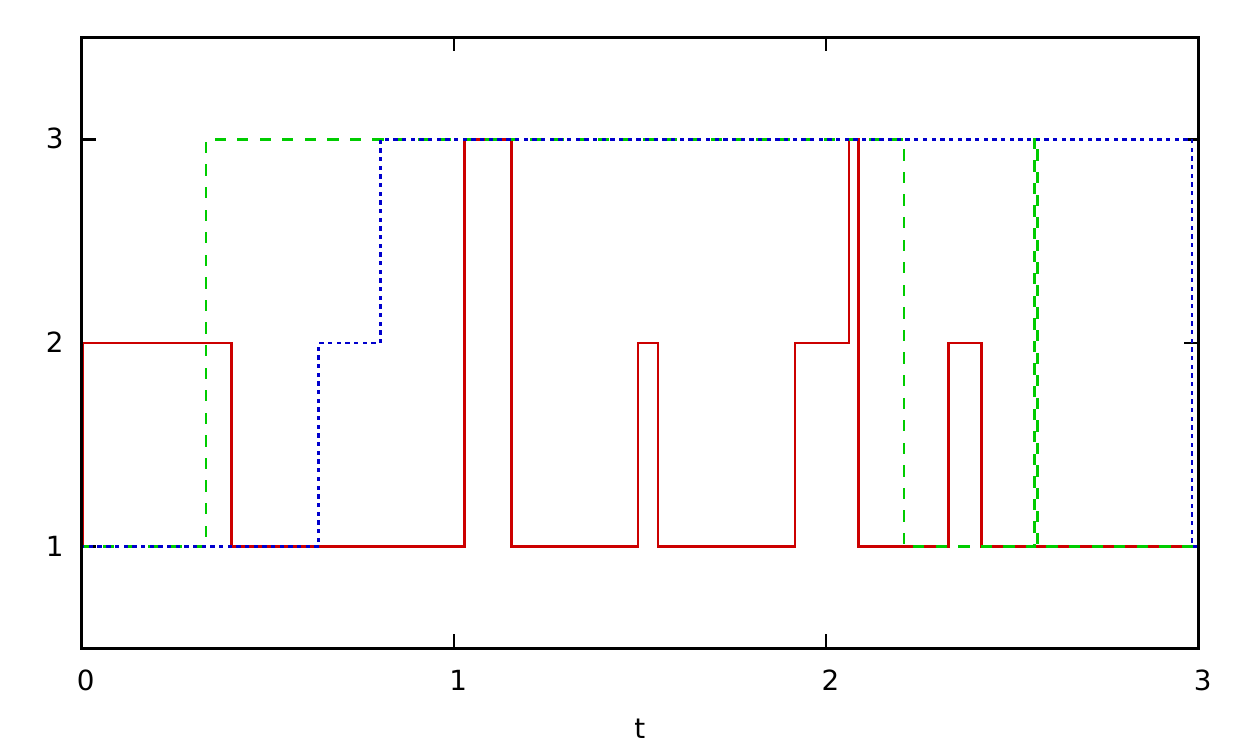}
\end{center}
\end{figure}

\index{jump processes}
Continuous-time jump processes represent a large class of the Markov stochastic processes with countable many configurations.  
They are used to model a huge variety of systems including lattice models, ratchets, chemical networks, biological systems and semi-classical description of quantum systems, etc. 
From a mathematical point of view the configuration space $\Omega$ is a countable set with a counting measure and the probabilities $\mu_t(x)$ are densities with respect to that measure. 
Hence integration over all possible configurations is represented as a summation 
\[
\int \rmd x \longrightarrow \sum\limits_x . 
\]
Typical paths in continuous-time jump processes consists of intervals where the configuration of the system does not change followed by the sudden change of the configuration (jump), 
for illustration see figure \ref{pic:illustration_of_path_jump_process}. 
Moreover we define all paths as right-continuous, i.e. the configuration at jump-time is the same as the configuration after the jump. 
By $x_{t^-}$ we denote the configuration just before the jump 
\[
x_{t^-} = \lim\limits_{s \rightarrow t^-} x_s . 
\]
The time evolution is fully characterized by \emph{transition rates}\index{transition rate} $\rate[t]{x}{y}$ defining the conditional probability measure on paths $\omega$
\begin{equation}
\dcprob[{(0,T]}]{\omega}{X_0=x_0}
= \exp \left[ - \int\limits_0^T \rmd s \; \lambda_s(x_s) \right] 
\prod_{i=1}^{m} \rate[t]{x_{t_i^-}}{x_{t_i}} \; \rmd t_i ,
\label{def:jump_process_path_probability}
\end{equation}
where $m$ is the number of jumps in the path $\omega$, $x_0$ is the initial configuration of the path $\omega$ ($x_0 \in \omega$), $t_i$ are the jump times, 
$\lambda_t(x)$ denotes the \emph{escape rate}\index{escape rate} from configuration $x$, 
\begin{equation}
\lambda_t(x) = \sum\limits_{y \neq x} \rate[t]{x}{y} .
\label{def:escape_rate}
\end{equation}
By definition impossible transitions are characterized by zero transition rate, $\rate[t]{x}{y}=0$.
Because the time-evolution is Markovian,
we can alternatively describe the time evolution using forward \eqref{def:forward_generator}\index{Kolmogorov generator!forward!Markov jump process} and backward \eqref{def:backward_generator}\index{Kolmogorov generator!backward!Markov jump process} Kolmogorov generator 
\begin{align}
\gen_t^* [\mu_t] (x) &= \sum\limits_{y \neq x} \left[ \mu_t(y) \rate[t]{y}{x} - \mu_t(x) \rate[t]{x}{y} \right] , 
\label{equ:forward_Kolmogorov_gen_jump} \\
\gen_t [A] (x) &= \sum\limits_{y \neq x} \rate[t]{x}{y} \left[ A(y) - A(x) \right] . 
\label{equ:backward_Kolmogorov_gen_jump} 
\end{align}
The derivation of Kolmogorov generators from the conditional path probability measure $\rmd \mathbb P$ is quite straightforward although technical 
and is provided in appendix \ref{ap:jump_processes_generators}.
The introduction of the forward Kolmogorov generator yields to the \emph{master equation}\index{master equation}
\begin{equation}
\partial_t \mu_{t}(x) = \sum\limits_{y \neq x} \left[ \mu_t(y) \rate[t]{y}{x} - \mu_t(x) \rate[t]{x}{y} \right] ,
\label{equ:Master_equation}
\end{equation}
representing here the time evolution equation \eqref{equ:Markov_stochastic_time_evolution}.

\subsection{Probability currents}
\index{current!probabilistic} 
In typical examples of continuous-time Markov jump processes like chemical networks, lattice models, etc., 
the jump is usually associated with transport or exchange of some quantity, e.g. electric charge, matter, with its surroundings. 
To describe the currents associated with such exchanges we define the \emph{probability current}\index{current!probabilistic} 
\begin{multline}
\curr[t]{x}{y} = \lim_{\Delta t \rightarrow 0^+} \frac{1}{\Delta t} \left[ \mu_{t-\Delta t} (x) \int \dcprob[{(t-\Delta t,t]}]{\omega}{X_{t-\Delta t} = x} \; \delta_{X_t}(y) - \right. \\
\left. - \mu_{t-\Delta t} (y) \int \dcprob[{(t-\Delta t,t]}]{\omega}{X_{t-\Delta t} = y} \; \delta_{X_t}(x) \right], 
\label{def:current}
\end{multline}
which characterizes the exchange of probability along one possible transition $x \rightarrow y$ at one particular time $t$. 
We can see, that by definition the current is antisymmetric 
\[
\curr[t]{x}{y} = - \curr[t]{y}{x}, 
\]
and is zero along the impossible transitions.

\paragraph{Example: Particle current}
To provide a specific example, let us consider a transport of single charged particle over the lattice. 
All possible configurations of the system are fully characterized by the position of the particle.
Hence the transition in the system corresponds to jump of the particle along the edge of the lattice. 
Consider a situation when we do not know a precise position of the particle, but only a probability $\mu_t(x)$ that the particle is at the particular vertex $x$ at time $t$. 
Then the average electrical current $\bar\jmath^{\,e}_{xy}(t)$ caused by the transition along the edge from $x$ to $y$ is the probability that the particle is at $x$ 
times the probability that the transition from $x$ to $y$ occurs times the charge $q$ of the particle. 
Moreover the transition from $y$ to $x$ can also occur producing the counter-current, hence reducing the total current along the edge.   
Altogether we obtain the average current along the edge from $x$ to $y$ being   
\[
\bar\jmath^{\,e}_{xy}(t) = q \curr[t]{x}{y}. 
\]
If we now consider having the huge amount of \emph{independent non-interacting} particles on the same lattice, then the effect of fluctuations will be largely reduced, 
and we will actually observe a macroscopic electrical current $j^e_{xy}(t)$ along the edge from $x$ to $y$ 
\[
j^e_{xy}(t) = N \bar\jmath^{\,e}_{xy}(t) = q N \curr[t]{x}{y},
\]
where $N$ is the total number of particles on the lattice.

In case of Markov jump processes the probability current are characterized by transition rates $\rate[t]{x}{y}$ and the actual probability distribution $\mu_t$ 
\begin{equation}
\curr[t]{x}{y} = \mu_t(x) \rate[t]{x}{y} - \mu_t(y) \rate[t]{y}{x} .
\label{equ:current_MJP}
\end{equation}
Comparing this expression for probability current with the master equation \eqref{equ:Master_equation} we obtain
\begin{equation}
\partial_t \mu_t(x) = \sum_y \curr[t]{y}{x} ,
\label{equ:continuity}
\end{equation}
which is the \emph{continuity equation} for probabilities.

From the continuity equation \eqref{equ:continuity} we obtain Kirchhoff's-like law for probability currents in the steady state 
\[
0 = \sum_y \curr{y}{x} 
\]
as a consequence of the stationary condition \eqref{equ:stationary_condition} in case of the system with time-independent transition rates.
Notice that this condition does not necessarily ensures the probability currents to be zero in the steady state.

\subsection{Global detailed balance condition}
\label{ssec:detailed_balance_jump}
\index{detailed balance condition!global}
In nature there exists a large class of systems in which all currents vanish in steady state, 
\begin{equation}
\curr{x}{y}=0, 
\label{equ:zero_current}
\end{equation}
namely equilibrium systems, see examples below. 
To be more specific, we restrict ourselves in this section to the class of systems with time-independent transition rates, $\rate[t]{x}{y} \equiv \rate{x}{y}$.
Under such assumptions the condition on all probability currents to be zero is equivalent to the \emph{global detailed balance condition}\index{detailed balance condition!global}
\begin{equation}
\frac{\rho(y)}{\rho(x)} = \frac{\rate{x}{y}}{\rate{y}{x}} ,
\label{equ:global_detailed_balance_jump}
\end{equation}
compare \eqref{equ:zero_current} and \eqref{equ:current_MJP}.
\footnote{The global detailed balance condition is also valid for systems with arbitrarily symmetric part $\psi(x,y)=\sqrt{\rate{x}{y} \, \rate{y}{x}}$ of the transition rates.}

The immediate consequence of global detailed balance condition is closely related to the time-reversal symmetry. 
Namely if there exists a sequence of configurations $x_0, x_1, \dots, x_n \equiv x_0 $ such that the transitions from $x_i$ to $x_{i+1}$ are possible,
then also transitions in the opposite direction are possible and the rates obeys 
\[
1 = \frac{\rate{x_0}{x_1} \, \rate{x_1}{x_2} \cdots \rate{x_{n-1}}{x_0} }{ \rate{x_0}{x_{n-1}} \cdots \rate{x_2}{x_1} \, \rate{x_1}{x_0}} .
\]
Put in words, the probability \eqref{def:jump_process_path_probability} of the closed path, $\omega^{(0,T]} : X_0 = X_T$, and its \emph{time reversal}\index{path!time reversed}
\begin{equation}
\Theta \omega^{(0,T]} = \left\{ X_{T-t} \middle| X_t \in \omega^{(0,T]} \right\}
\label{equ:time_reversal_jump}
\end{equation}
are equal.

\subsubsection{Equilibrium systems}
\label{ssec:equlibrium_systems_jump}
As was mentioned before, the equilibrium systems are the systems obeying the global detailed balance condition \eqref{equ:global_detailed_balance_jump},
which is ensured by the fact that in the equilibrium there are no macroscopic currents in the system. 
We consider the typical example is the system with discrete energy levels $E(x)$ in equilibrium with a single thermal bath at the inverse temperature $\beta$.
The stationary distribution in this case is given by canonical distribution of such system 
\[
\rho(x) = \frac{1}{Z} \rme^{-\beta E(x)},
\]
where $Z$ is the partition function.
As a consequence the antisymmetric part of the transition rates is closely related to the energy exchanged with the thermal bath and hence to the entropy production in the bath along the transition 
\[
\ln \frac{\rate{x}{y}}{\rate{y}{x}} = \beta \left[ E(x) - E(y) \right] = S^{\text{bath}}(x)-S^{\text{bath}}(y), 
\]
where $S^{\text{bath}}(x)$ denotes the entropy of the \emph{thermal bath},
here as a function of the system's configuration $x$.

Another example is the system attached to a single thermal and particle bath at the inverse temperature $\beta$ as well as to a single particle bath at the chemical potential $\mu$.
The stationary distribution in this case is given by the grand-canonical distribution 
\[
\rho(x) = \frac{1}{Z_G} \rme^{-\beta \left( E(x) - \mu N(x) \right)},
\]
where $N(x)$ is the number of particles present in the system and $Z_G$ denotes the grand-canonical partition function.
And similarly to the closed system the antisymmetric part corresponds to the entropy production within the thermal bath
\[
\ln \frac{\rate{x}{y}}{\rate{y}{x}} = \beta \left[ E(x) - E(y) \right] - \beta \mu \left[ N(x) - N(y) \right] = S^{\text{bath}}(x)-S^{\text{bath}}(y) . 
\] 
We can conclude that in equilibrium the antisymmetric part of the transition rates corresponds to the entropy production in thermal bath associated with such transition in general, 
which is known as the \emph{local detailed balance condition}\index{detailed balance condition!local}.

\subsection{Three-level model}
\label{ssec:three_level}
\begin{figure}[thb]
\caption{Three-level model with every transition possible and $E_0$ calibrated to zero.}
\label{fig:three_level}
\begin{center}
\includegraphics[width=.45\textwidth,height=!]{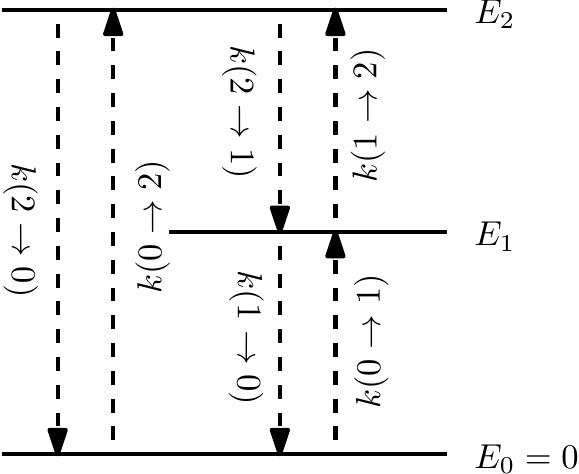}
\end{center}
\end{figure}

\index{discrete model!three-level}
\index{three-level model}
Our first model representing the class of Markov jump processes is the three-level model with fixed energy levels $E_0 \le E_1 \le E_2$, see figure \ref{fig:three_level}.
We consider the system to be connected to a single thermal bath at the inverse temperature $\beta$ 
and driven out of global detailed balance by the action of the non-potential external force performing work $W_{\text{ext}}$ along any jump in the direction $0 \to 2 \to 1 \to 0$.
Hence the transitions in the direction $0 \rightarrow 1 \rightarrow 2 \rightarrow 0$ are more likely to occur than those in the opposite direction, $ 0 \rightarrow 2 \rightarrow 1 \rightarrow 0$.  
As a typical example of such system can be considered a simplified classical model for three-level laser. 
The transition rates has to obey global detailed balance condition in case $W_{\text{ext}}=0$, hence 
\begin{align*}
\begin{aligned}
\rate{0}{1} &= \psi_{\beta}(0,1) \, \rme^{- \frac{\beta}{2} \left( E_1 - E_0 - W_{\text{ext}} \right) } , \\
\rate{1}{2} &= \psi_{\beta}(1,2) \, \rme^{- \frac{\beta}{2} \left( E_2 - E_1 - W_{\text{ext}} \right) } , \\
\rate{2}{0} &= \psi_{\beta}(0,2) \, \rme^{- \frac{\beta}{2} \left( E_0 - E_2 - W_{\text{ext}} \right) } , 
\end{aligned}
&& 
\begin{aligned}
\rate{1}{0} &= \psi_{\beta}(0,1) \, \rme^{ \frac{\beta}{2} \left( E_1 - E_0 - W_{\text{ext}} \right) } , \\
\rate{2}{1} &= \psi_{\beta}(1,2) \, \rme^{ \frac{\beta}{2} \left( E_2 - E_1 - W_{\text{ext}} \right) } , \\
\rate{0}{2} &= \psi_{\beta}(0,2) \, \rme^{ \frac{\beta}{2} \left( E_0 - E_2 - W_{\text{ext}} \right) } ,
\end{aligned}
\end{align*}
where $\psi_{\beta}(x,y)=\psi_{\beta}(y,x)$ denotes arbitrary symmetric parts in general dependent on $\beta$. 
Having the transition rates, we can directly obtain the escape rates 
\begin{align*}
\lambda(0) &= 
\rme^{ \frac{\beta}{2} E_0} \left[ \psi_{\beta}(0,1) \rme^{- \frac{\beta}{2} \left( E_1 - W_{\text{ext}} \right) } 
+ \psi_{\beta}(0,2) \rme^{- \frac{\beta}{2} \left( E_2 + W_{\text{ext}} \right)} \right] , \\
\lambda(1) &= 
\rme^{ \frac{\beta}{2} E_1} \left[ \psi_{\beta}(0,1) \rme^{- \frac{\beta}{2} \left( E_0 + W_{\text{ext}} \right) } 
+ \psi_{\beta}(1,2) \rme^{- \frac{\beta}{2} \left( E_2 - W_{\text{ext}} \right)} \right] , \\
\lambda(2) &= 
\rme^{ \frac{\beta}{2} E_2} \left[ \psi_{\beta}(0,2) \rme^{- \frac{\beta}{2} \left( E_2 - W_{\text{ext}} \right) } 
+ \psi_{\beta}(1,2) \rme^{- \frac{\beta}{2} \left( E_1 + W_{\text{ext}} \right)} \right] . 
\end{align*}

%\subsection{Four level model} 
%\label{ssec:four_level}
%\index{discrete model!four level}
%What we will consider as generic four level model is slightly extended three-level model, where exists another energy level $E_3 > E_2$,  
%directly accessible only from one configuration, namely 2, see figure \ref{fig:four_level}.  
%Furthermore we assume that the additional configuration $3$ has the highest lifetime, which goes to infinity with temperature going to zero, i.e. the lower the temperature the more the channel is closed. 
%Such model has a different low temperature behaviour compared to the three-level model introduced in the subsection before \ref{ssec:three_level}, as we will see in chapter \ref{chapter:low_temperature}.
%We define the transition rates in the channel $3 \leftrightarrow 2$ to be 
%\begin{align*}
%\rate{3}{2} &= \rme^{- \beta \Delta(2,3)} \rme^{- \frac{\beta}{2} \left[ E_2 - E_3 \right] } , & 
%\rate{2}{3} &= \rme^{- \beta \Delta(2,3)} \rme^{ \frac{\beta}{2} \left[ E_2 - E_3 \right] } , 
%\end{align*}
%where $\Delta(2,3)=\Delta(3,2)$ determines the symmetric part of transition rates. 
%\begin{figure}[h]
%\caption{ Graph of four level model with configuration $3$ connected only to the configuration $2$. }
%\label{fig:four_level}
%\begin{center}
%\includegraphics[width=.6\textwidth,height=!]{Stochastic_models_of_open_systems/4lvl-graph.pdf}
%\end{center}
%\end{figure}

\subsection{Multichannel two-level model}
\label{ssec:two_level}
\begin{figure}[tbh]
\caption{ Two-level model with two channels enabling transitions between configurations $0$ and~$1$, with the energy gap $\Delta E$. }
\label{fig:two_level}
\begin{center}
\includegraphics[width=.5\textwidth,height=!]{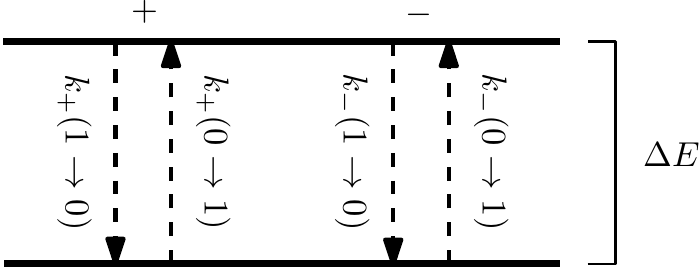}
\end{center}
\end{figure}

\index{discrete model!two-level}
Another way how to break the global detailed condition is by attaching the system to multiple thermal or particle baths. 
In this subsection we introduce the simplest of such models, which consists of a two-level system attached to two independent thermal baths 
in general at different inverse temperatures $\beta_+ > \beta_-$, see figure \ref{fig:two_level}.  
Having the system in contact with multiple thermal baths enforces us to modify the framework of continuous-time Markov jump processes (see beginning of the section \ref{sec:jump_processes}) 
to include also the information about which action of particular thermal bath caused which transition. 
Therefore a single realization of the time evolution of the system in contact with multiple thermal baths cannot be fully described only by the path on configurations $\omega$,
because it does not contain such information about the transition.
We extend the definition of the path and consequently conditional path measure \eqref{def:jump_process_path_probability}  
\[
\dcprob{\widetilde\omega}{X_0=x_0}  
= \exp \left[ - \int\limits_0^T \rmd s \; \sum\limits_i \lambda_i(x_s) \right] 
\prod_{i=1}^{m} \rate[(i)]{x_{t_i^-}}{x_{t_i}} \; \rmd t_i ,
\]
where $\lambda_i(x)$ denotes the escape rate of the $i$-th bath and $\rate[(i)]{x}{y}$ denotes the transition rate of the bath associated with $i$-th transition. 
As a consequence we obtain a modified master equation \eqref{equ:Master_equation} for the time evolution of this particular system  
\[
\partial_t \mu_t(x) 
= \sum\limits_{y \neq x} \mu_t(y) \left[ \rate[+]{y}{x} + \rate[-]{y}{x} \right] - \mu_t(x) \left[ \lambda_+(x) + \lambda_-(x) \right] ,
\]
where $k_+$ ($k_-$) are transition rates corresponding to the contact with thermal bath with the inverse temperature $\beta_+$ ($\beta_-$), 
and where it is also assumed that transitions governed by a single thermal bath obey detailed balance condition \eqref{equ:global_detailed_balance_jump}, hence
\begin{align*}
\rate[+]{0}{1} &= \psi_+ \, \rme^{- \frac{1}{2} \beta_+ \Delta E } , & 
\rate[-]{0}{1} &= \psi_- \, \rme^{- \frac{1}{2} \beta_- \Delta E } , \\
\rate[+]{1}{0} &= \psi_+ \, \rme^{ \frac{1}{2} \beta_+ \Delta E } , &
\rate[-]{1}{0} &= \psi_- \, \rme^{ \frac{1}{2} \beta_- \Delta E } .
\end{align*}
We can also define a probability current \eqref{equ:current_MJP} induced by a particular thermal bath as 
\begin{equation}
\curr[\pm]{x}{y} = \rate[\pm]{x}{y} \, \mu_t(x) - \rate[\pm]{y}{x} \, \mu_t(y) .
\label{equ:current_multiple_baths}
\end{equation}
A notable special case is when both baths are at the same temperature although having different symmetric parts of transition rates $\psi_+ \neq \psi_-$. 
This situation is formally equivalent to the case when the system is connected to two thermal baths at the same temperature with the same symmetric part of transition rates $\psi$, 
while the symmetry between the channels is broken by application of non-potential force $F$
\begin{align*}
\rate[+]{0}{1} &= \psi \, \rme^{- \frac{\beta}{2} \left( \Delta E - F \right) } , & 
\rate[-]{0}{1} &= \psi \, \rme^{- \frac{\beta}{2} \left( \Delta E + F \right) } , \\
\rate[+]{1}{0} &= \psi \, \rme^{ \frac{\beta}{2} \left( \Delta E - F \right) } , &
\rate[-]{1}{0} &= \psi \, \rme^{ \frac{\beta}{2} \left( \Delta E + F \right) } , 
\end{align*}
where we set 
\begin{align*}
\psi &= \sqrt{\psi_+ \, \psi_-}, &
F &= \frac{1}{\beta} \ln \frac{\psi_+}{\psi_-}. 
\end{align*}

\section{Diffusive systems}
\index{diffusion}
Diffusion are continuous-space Markovian models often used to describe the transport of matter or energy through a homogeneous environment, 
the typical examples being heat conduction or transport of diluted chemicals in liquids or gases.  
The behaviour of such systems is in general quite different from the systems described by continuous time Markov jump processes, 
although overdamped diffusion is the limiting case of discrete jump process on the lattice with properly rescaled time and space \cite{Itzykson_Drouffe:Statistical_field_theory}.

In this section we build the formalism necessary to describe diffusion.
At first we show the diffusion limit of the random walk and its connection to the Wiener process, 
then we will introduce the stochastic calculus, namely It\^o and Stratonowich calculus \cite{Oksendal2003,Evans2001}, as an universal tool to describe diffusion.  
We will conclude this section with an application of the introduced calculus to the under- and over-damped diffusion along with classical description of Josephson junction.

\subsection{Random walks and Wiener process}
\label{ssec:diffusion_limit}
Random walks are realizations of discrete time Markov process on some graph, 
as such they can be considered as a discrete versions of diffusion. 
They are also a simple tool to study a basic properties of diffusion or more precisely the properties of Wiener process, 
which is fundamental for stochastic calculus and hence for stochastic approach to diffusion.

\begin{figure}[htb]
\caption{Single random walker on square lattice with all possible transitions.}
\label{fig:random_walk}
\begin{center}
\includegraphics[width=.4\textwidth,height=!]{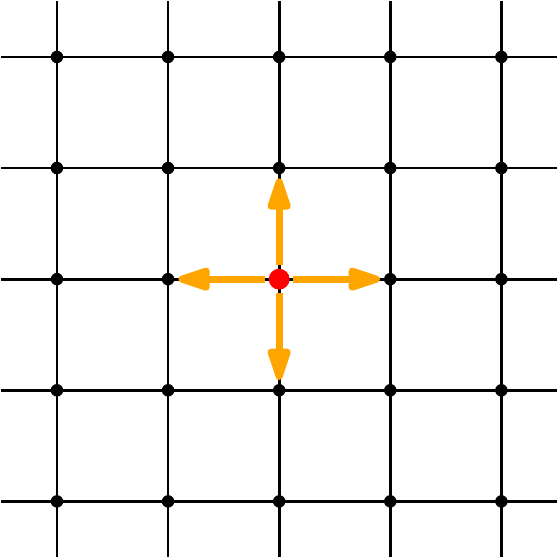}
\end{center}
\end{figure}

Let us consider a single \emph{random walker}\index{random walker} on the uniform regular $d$-dimensional hypercubic lattice with lattice constant $a$, see fig. \ref{fig:random_walk}  
The random walker at the beginning of each time step $\Delta t$ starts to move from the site of the lattice designed by its position $\vec{x}$ along an edge in the direction $\pm i$ 
and ends the move at the end of the time step in the site with position $\vec{x}^{(\pm i)} = ( x_0 , \dots , x_{i-1} , x_i \pm 1 , x_{i+1} , \dots , x_{d-1} )$.  
The process of random walker on uniform lattice can be then described by discrete time Markov jump process with time step $\Delta t$ and with the same constant transition probability $p$ for all transitions
\[
\cprob{\vec{X}_{t+\Delta t} = \vec{x}^{(\pm i)}}{\vec{X}_t = \vec{x}}
= \cprob{\vec{X}_{t+\Delta t} = \vec{x}}{\vec{X}_t = \vec{x}^{(\pm i)}}
= p,
\]
where the random variable $\vec{X}_t$ denotes the position at time $t$. 
Typically if we speak about the random walker, the random walker is not allowed to rest, hence the transition probability on the square lattice can be directly derived from the dimension of the lattice
\[
p = \frac{1}{2d} . 
\]

Before we introduce diffusion limit we show some of the properties of random walker. 
At first we can conclude that in this setup the probability distribution of positions of the random walker after $k$ time steps when starting from the position $\vec{x}_0$ is given by multinomial distribution
\begin{align*}
\cprob{\vec{X}_{k \Delta t} = \vec{x}_0 + \Delta \vec{x} }{ \vec{X}_0 = \vec{x}_0 } 
&= \sum\limits_{m^\pm_0, \dots , m^\pm_{d-1}} \binom{k}{m^+_0 \dots m^-_{d-1}} \left(\frac{1}{2d}\right)^{m^+_0} \dots \left(\frac{1}{2d}\right)^{m^-_{d-1}} \\
&= \frac{1}{(2d)^k} \sum\limits_{m^\pm_0, \dots , m^\pm_{d-1}} \binom{k}{m^+_0 \dots m^-_{d-1}} ,
\end{align*}
where we sum over all possible numbers of steps $m_i^\pm$ in the direction $\pm i$ such that we change the position by $\Delta \vec{x}$ and the total count of steps is $k$, i.e.  
\begin{gather*}
m^\pm_i \in \{0, 1, \dots, k \} , \\
\Delta x_i = a \left( m^+_i - m^-_i \right) , \\
k = \sum\limits_{i=0}^{d-1} \sum\limits_{s \in \{+,-\}} x^s_i .
\end{gather*} 
As a consequence of the symmetries of the lattice the probability distribution of the terminal position after $k$ steps is invariant with respect to mirroring the change of the position $\Delta \vec{x}$ along any axis.
Hence the mean position of the walker while starting from the position $\vec{0}$ at arbitrary time $t$ is zero 
\begin{equation}
\left\langle \vec{X}_t \right\rangle_{\delta_{\vec{0}}} 
= \frac{a}{(2d)^{\frac{t}{\Delta t}}} \sum\limits_{m^\pm_0, \dots, m^\pm_{d-1} } \binom{\frac{t}{\Delta t}}{m^\pm_0 \dots m^\pm_{d-1}} \left( m^+_0 - m^-_0 , \dots , m^+_{d-1} - m^-_{d-1} \right) 
= \vec{0} ,
\label{equ:average_position}
\end{equation}
where we also sum over all possible positions after $t/\Delta t$ steps, 
hence the imposed conditions now are 
\begin{gather*}
m^\pm_i \in \left\{ 0, 1, \dots , \frac{t}{\Delta t} \right\} , \\
\frac{t}{\Delta t} = \sum\limits_{i=0}^{d-1} \sum\limits_{s \in \{+, -\}} m^s_i .
\end{gather*}
In order to investigate the variance of the position at time $t$ it is convenient to introduce the \emph{moment-generating function}\index{function!moment-generating} 
\[
M_t(\vec{\alpha}) = \left\langle \rme^{\vec{\alpha} \cdot \vec{X}_t } \right\rangle_{\delta_{\vec{0}}} ,
\]
which in this particular case corresponds to 
\[
M_t(\vec{\alpha}) = \left[ \frac{1}{d} \sum\limits_{i=0}^{d-1} \cosh (\alpha_i a) \right]^{\frac{t}{\Delta t}} . 
\]
The various moments of position are obtained by taking derivatives with respect to $\vec{\alpha}$ at $\vec{0}$. 
By taking the gradient with respect to $\vec{\alpha}$ at point $\vec{0}$ we the result for mean position already obtained by taking notion of the symmetries of the lattice \eqref{equ:average_position}. 
We obtain the variance of position by applying the Laplace operator with respect to $\vec{\alpha}$ at point $\vec{0}$ 
\begin{equation}
\left\langle \vec{X}_t^2 \right\rangle_{\delta_\vec{0}} = \left. \Delta_\vec{\alpha} M_t(\vec{\alpha}) \right|_{\vec{\alpha}=\vec{0}} = \frac{a^2 t}{\Delta t} , 
\label{equ:random_walk_variance}
\end{equation}
which is proportional to the number of steps $t/{\Delta t}$.
The fact that the variance is proportional to the length of time interval $t$ is universal for all diffusive processes 
and we will see later, that this property ensures that the generator of diffusion depends only on the probability density and its first and second derivative, see subsections \ref{ssec:underdamped_diff} and \ref{ssec:overdamped_diff}.

\subsubsection{Wiener process}\index{Wiener process} 
Wiener process is a limiting case of random walker when we tend to go with the time step to zero, 
however in order to preserve the finite variance in position of the process \eqref{equ:random_walk_variance} 
we also need to scale the space in such a manner that the quantity $a^2/\Delta t$ is preserved. 
What we will call the Wiener process is the limiting process $\Delta t \to 0$ of the random walker with the choice 
\[
\frac{a^2}{\Delta t} = d .
\]
Such choice is also called \emph{diffusion limit}\index{diffusion limit}. 
Moreover by taking the limit of the moment-generating function
\[
M_t^{\text{Wiener}}(\vec{\alpha}) = \lim_{\Delta t \to 0} \left[ 1 + \frac{\Delta t}{2} \vec{\alpha}^2 + \err[2]{\Delta t} \right]^{\frac{t}{\Delta t}} = \exp\left[ \frac{1}{2} \vec{\alpha}^2 t \right] , 
\]
we see that the distribution of position displacement $\vec{x}-\vec{x}_0$ is Gaussian with the zero mean and variance equal to $\sqrt{t}$ 
\begin{equation}
\cprob{\vec{X}_t = \vec{x} }{\vec{X}_0 = \vec{x}_0 } = \frac{1}{\left(2 \pi t\right)^{\frac{d}{2}}} \exp \left[- \frac{\left( \vec{x} - \vec{x}_0 \right)^2}{2t} \right] .
\label{equ:Wiener_distribution}
\end{equation}
One of the most important properties of the Wiener process is that the combination of independent Wiener processes is also a Wiener process 
\begin{multline}
\cprob{\vec{X}_{t+s} = \vec{x} }{\vec{X}_0 = \vec{x}_0 } = \\
= \int\limits_{\reals^d} \rmd^d \vec{y} \; \cprob{\vec{X}_{t+s} = \vec{x} }{\vec{X}_s = \vec{y} } \, \cprob{\vec{X}_s = \vec{y} }{\vec{X}_0 = \vec{x}_0 } .
\label{equ:Wiener_combination}
\end{multline}
Notice also that it also means that the taking of the mean value over all realizations of the Wiener process can be replaced by taking the mean value over realizations of several independent subsequent Wiener processes. 
Consequently it is valid that for arbitrary times $0<t'<t$ the displacement $\vec{X}_t - \vec{X}_{t'}$ is Gaussian distributed random variable with zero mean value and variance equal to the length of the time interval  
\begin{align}
\left\langle \vec{X}_t - \vec{X}_{t'} \right\rangle_{\mu_0} &= \vec{0} , &
\left\langle \left( \vec{X}_t - \vec{X}_{t'} \right)^2 \right\rangle_{\mu_0} &= t-t' . 
\label{equ:Wiener_properties}
\end{align}
One can also find that another consequence of \eqref{equ:Wiener_combination} is that the covariance is proportional to lesser of the times $t'<t$, 
\[
\left\langle \vec{X}_t \vec{X}_{t'} \right\rangle_{\delta_\vec{0}} 
= \underbrace{ \left\langle \left( \vec{X}_t - \vec{X}_{t'} \right) \vec{X}_{t'} \right\rangle_{\delta_\vec{0}} }_{=0} + \left\langle \vec{X}_{t'} \vec{X}_{t'} \right\rangle_{\delta_\vec{0}} 
= t' { \mathbb I} .
\]

Another characteristic of the Wiener process is that it is almost surely continuous, 
this can be seen as a result of the property \eqref{equ:Wiener_combination} and the fact that the distribution \eqref{equ:Wiener_distribution} converges to delta function as $t \to 0^+$.

\subsection{It\^{o} versus Stratonovich calculus}
\label{ssec:stochastic_calculus}
In physics the diffusion can usually be handled on several level of description. 
On the macroscopic scale we usually use the fluid dynamics approach with diffusion or Fokker-Planck equation. 
However on the microscopic scale we describe the diffusion as a set of classical particles under the influence of the random force. 
These two approaches have to be compatible.
In this subsection we provide such a link in the form of stochastic calculus.
The stochastic calculus is a tool developed to solve such a type of equations. 
In the first part of this subsection we will provide some definitions and results of the It\^{o} calculus and later compare it with the approach of Stratonovich.  
Some of the details and proofs of some statements can be found in the appendix \ref{ap:stochastic_calculus}, 
for even more details see the standard textbooks \cite{Oksendal2003,Evans2001}.

\subsubsection{It\^{o} calculus} 
The basic idea behind the It\^{o} calculus is that we can handle the \emph{Gaussian white noise}\index{white noise} represented by the infinitesimal increment of the Wiener process $\rmd W_t$ as $\sqrt{\rmd t}$. 
Or informally 
\[
\rmd W_t^2 \approx \rmd t . 
\]
To make the statement more precise we start with the definition of the It\^{o} stochastic integral and show some of its properties.

The basic idea for the stochastic integral comes from the generalization of the Riemann integral to stochastic functions and variables. 
We define \emph{the It\^{o} stochastic integral}\index{stochastic integral!It\^{o}}\index{It\^{o} integral} of the vector field $\vec{f}(\vec{W},t)$ along the $d$-dimensional Wiener process $\vec{W}_t$ as 
\begin{equation}
\int\limits_0^T \vec{f}(\vec{W}_t,t) \cdot \rmd \vec{W}_t = \lim\limits_{N \to \infty} \sum\limits_{i=0}^{N-1} \vec{f}(\vec{W}_{t_i},t_i) \cdot \left[\vec{W}_{t_{i+1}} - \vec{W}_{t_i}\right] ,
\label{def:Ito_integral}
\end{equation}
where the times $0=t_0 < t_1 < \dots < t_N = T$ correspond to the partition of the time interval $[0,T]$  
and where the limit of the Riemann sum is taken over the decreasing length of the time interval 
\[
\lim_{N \to \infty} \max_i \left| t_{i+1} - t_i \right| = 0 .
\]
The result of the integral is also a random variable with the following properties.
The integral is linear in the integrand 
\[
\int\limits_0^T \left[ \alpha \vec{f}(\vec{W}_t,t) + \beta \vec{g}(\vec{W}_t,t) \right] \cdot \rmd \vec{W}_t 
= \alpha \int\limits_0^T \vec{f}(\vec{W}_t,t) \cdot \rmd \vec{W}_t 
+ \beta \int\limits_0^T \vec{g}(\vec{W}_t,t) \cdot \rmd \vec{W}_t ,
\]
which can be seen directly from definition \eqref{def:Ito_integral}. 
The mean value of the integral is zero when starting from arbitrary initial state $\mu_0$ at time $0$
\begin{equation}
\left\langle \int\limits_0^T \vec{f}(\vec{W}_t,t) \cdot \rmd \vec{W}_t \right\rangle_{\mu_0} 
= 0 .
\label{equ:Ito_zero_mean_value}
\end{equation}
It is also valid that the covariance of integrals corresponds to the time integral of the covariance 
\begin{equation}
\left\langle \int\limits_0^T \vec{f}(\vec{W}_t,t) \cdot \rmd \vec{W}_t \int\limits_0^T \vec{g}(\vec{W}_t,t) \cdot \rmd \vec{W}_t \right\rangle_{\mu_0} 
= \int\limits_0^T \left\langle \vec{f}(\vec{W}_t,t) \cdot \vec{g}(\vec{W}_t,t) \right\rangle_{\mu_0} \; \rmd t .
\label{equ:Ito_covariance}
\end{equation}
For further details see appendix \ref{ap:stochastic_calculus} sections \ref{sec:mean_value_Ito} and \ref{sec:covariance_Ito}.

Moreover the Riemann sum of the square of displacements alone converges to the length of the time interval $T$ \emph{almost surely}\index{almost surely}, i.e. the probability of the sum not being the length of the time interval converges to zero, 
\[
\lim_{N \to \infty} \sum\limits_{i=0}^{N-1} \left( \vec{W}_{t_{i+1}} - \vec{W}_{t_i} \right)^2 
= \lim_{N \to \infty} \sum\limits_{i=0}^{N-1} \left( t_{i+1} - t_i \right) 
= T \qquad \text{a.s.}  
\] 
For the sketch of the proof see appendix \ref{ap:stochastic_calculus} section \ref{sec:Riemann_sum}. 
Its generalized version can be used to connect the time integral of the random function over the time interval with the corresponding Riemann sum
\begin{equation}
\lim_{N \to \infty} \sum\limits_{i=0}^{N-1} f(\vec{W}_{t_i},t_i) \left( \vec{W}_{t_{i+1}} - \vec{W}_{t_i} \right)^2 
= \int\limits_0^T \rmd t \; f(\vec{W}_t,t) \qquad \text{a.s.}  
\label{equ:pre_Ito_lemma}
\end{equation}

We apply the theory of It\^{o} stochastic integral to solve the stochastic differential equations. 
Within the standard linear first-order differential equation theory we find the formal solution of the differential equation by associating it with an appropriate integral representation 
\[
\vec{\nabla} f(\vec{x}) = \vec{g}(\vec{x}) \qquad \Longrightarrow \qquad \rmd f(\vec{x}) = \vec{g}(\vec{x}) \cdot \rmd \vec{x} \qquad \Longrightarrow \qquad f(\vec{x}) = \int \rmd \vec{x} \cdot \vec{g}(\vec{x}) . 
\] 
In a similar fashion the formal solution $Y(\vec{W}_T,T)$ of \emph{the stochastic differential equation}\index{stochastic differential equation} in the form of It\^{o} total differential
\begin{equation}
\rmd Y(\vec{W}_t,t) = f(\vec{W}_t,t) \; \rmd t + \vec{g}(\vec{W}_t,t) \cdot \rmd \vec{W}_t  
\label{def:stochastic_differential_equation}
\end{equation}
is given by its integral representation
\[
Y(\vec{W}_T,T) = Y(\vec{W}_0,0) + \int\limits_0^T \rmd t \; f(\vec{W}_t,t) + \int\limits_0^T \rmd \vec{W}_t \cdot \vec{g}(\vec{W}_t,t) .
\]
Notice that consequently the solution $Y(\vec{W}_T,T)$ of stochastic differential equation is \emph{the random function}\index{random function} of the Wiener process.

Up to now we were developing the theory in order to solve a given stochastic differential equation. 
One can also be interested in how to determine the stochastic differential equation knowing the solution. 
The answer is given in the form of the \emph{It\^{o} lemma}\index{It\^{o} lemma} which states that the corresponding differential equation to the solution $Y(\vec{W}_t,t)$ with the initial condition $Y(\vec{W}_0,0)$ is given by
\begin{equation}
\rmd Y(\vec{W}_t,t) = \left[ \partial_t Y(\vec{W}_t,t) + \frac{1}{2} \left. \Delta_\vec{x} Y(\vec{x},t) \right|_{\vec{x} = \vec{W}_t} \right] \; \rmd t + \left. \vec{\nabla}_\vec{x} Y(\vec{x},t) \right|_{\vec{x} = \vec{W}_t} \cdot \rmd \vec{W}_t .
\label{equ:total_differential_Ito}
\end{equation}
The It\^{o} lemma is a direct consequence of the expansion of the solution in both the Wiener process $\vec{W}_t$ and the explicit time dependence while applying \eqref{equ:pre_Ito_lemma}. 
For further details see again appendix \ref{ap:stochastic_calculus} section \ref{sec:Ito_lemma}.

Notice that we usually do not describe physical systems with the differential equation in the form of the total differential \eqref{def:stochastic_differential_equation}, 
instead we are describing the system under the influence of the random force $\xi_t$, 
e.g. \emph{Langevin equation}\index{Langevin equation} is usually given in the form 
\[
m \ddot{\vec{x}}_t = \vec{f}(\vec{x}_t) + \vec{\xi}_t . 
\]
If the random force corresponds to \emph{the white Gaussian noise}\index{white noise} the connection is quite straightforward, 
we associate the random force with the Wiener process by $\vec{\xi}_t \; \rmd t = \rmd \vec{W}_t$.

\subsubsection{Stratonovich calculus}
The \emph{Stratonovich stochastic integral}\index{stochastic integral!Stratonovich}\index{Stratonovich integral} is defined as 
\begin{equation}
\int\limits_0^T \vec{f}(\vec{W}_t,t) \circ \rmd \vec{W}_t = \lim\limits_{N \to \infty} \sum\limits_{i=0}^{N-1} \vec{f}(\vec{W}_{\tau_i},\tau_i) \cdot \left[\vec{W}_{t_{i+1}} - \vec{W}_{t_i}\right] ,
\label{def:Stratonovich_integral}
\end{equation}
where $t_i$ again corresponds to the partition times of the interval $[0,T]$ and $\tau_i$ is a midpoint of the corresponding interval, $\tau_i = ( t_{i+1} + t_i )/2$. 
We can see that the difference between the Stratonovich calculus and the It\^{o} calculus in the choice of the point, where we evaluate the function. 
In case of It\^{o} integral we evaluate the function at the beginning of the interval while in case of Stratonovich integral we evaluate it in the middle.
From there also follows the different behaviour of It\^{o} and Stratonovich integral with respect to \emph{time inversion}\index{time inversion}, see below.

To show this difference, we define the time reversal of the particular realization of the Wiener process as 
\[
\vec{W}^\Theta_t = \vec{W}_{T-t} .
\]
Notice that such definition of the time reversal also corresponds to the substitution $s = T-t$. 
Also notice that the time reversal of the Wiener process is also a Wiener process with the same properties,
which is due to the fact that the probability of the random walker going along the given path fort and back is the same. 
Then if we take the It\^{o} integral over the particular realization of the Wiener process $\vec{W}_t$ and try to represent it with its time reversal $\vec{W}^\Theta_t$ we obtain
\begin{multline*}
\int\limits_0^T \vec{f}(\vec{W}_t) \cdot \vec{W}_t = \sum\limits_{i=0}^{N-1} \vec{f}(\vec{W}_{t_i}) \cdot \left[ \vec{W}_{t_{i+1}} - \vec{W}_{t_i} \right] = \\
= - \sum\limits_{i=0}^{N-1} \vec{f}(\vec{W}^\Theta_{T-t_i}) \left[ \vec{W}^\Theta_{T-t_i} - \vec{W}^\Theta_{T-t_{i+1}} \right] = \\
= - \sum\limits_{j=0}^{N-1} \vec{f}(\vec{W}^\Theta_{s_{j+1}}) \left[ \vec{W}^\Theta_{s_{j+1}} - \vec{W}^\Theta_{s_j} \right] ,
%= - \int\limits_0^T \vec{f}(\vec{W}_s) \cdot \rmd_\text{b} \vec{W}_s
\end{multline*} 
where we have already changed the notation of the partition $s_i = T - t_{N-i} $.
In the result we have obtained the increment of the Wiener process multiplied by the function evaluated at the terminal time $s_{j+1}$, 
hence the function and the increment of the Wiener process are no longer independent. 
By substitution $s=T-t$ we obtain a new object which fundamentally differs from the definition of the It\^{o} integral, 
which is usually called the \emph{backward It\^{o} integral}\index{stochastic integral!backward It\{o}}\index{backward It\^{o} integral} 
\begin{multline*}
\int\limits_0^T \vec{f}(\vec{W}_t) \cdot \rmd \vec{W}_t = \sum\limits_{i=0}^{N-1} \vec{f}(\vec{W}_{t_i}) \cdot \left[ \vec{W}_{t_{i+1}} - \vec{W}_{t_i} \right] \xrightarrow{s=T-t} \\
\xrightarrow{s=T-t} - \sum\limits_{j=0}^{N-1} \vec{f}(\vec{W}^\Theta_{t_{j+1}}) \cdot \left[ \vec{W}^\Theta_{t_{j+1}} - \vec{W}^\Theta_{t_j} \right] 
= - \int\limits_0^T \vec{f}(\vec{W}^\Theta_t) \cdot \rmd_{\text{b}}\vec{W}^\Theta_t . 
\end{multline*}
On the contrary if we proceed along the same lines with the Stratonovich integral we obtain again a Stratonovich integral 
\begin{multline}
\int\limits_0^T \vec{f}(\vec{W}_t) \circ \rmd \vec{W}_t = \sum\limits_{i=0}^{N-1} \vec{f}(\vec{W}_{\tau_i}) \cdot \left[ \vec{W}_{t_{i+1}} - \vec{W}_{t_i} \right] \xrightarrow{s=T-t} \\
\xrightarrow{s=T-t} - \sum\limits_{j=0}^{N-1} \vec{f}(\vec{W}^\Theta_{\tau_j}) \cdot \left[ \vec{W}^\Theta_{t_{j+1}} - \vec{W}^\Theta_{t_j} \right] 
= - \int\limits_0^T \vec{f}(\vec{W}^\Theta_t) \circ \rmd \vec{W}^\Theta_t , 
\label{equ:antisymmetry_Stratonovich}
\end{multline}
because the midpoint $\tau_i$ of the interval is preserved by the substitution $s=T-t$.

Although the It\^{o} and Stratonovich integrals behave differently under the time inversion, there is a way how to transform one to another  
\begin{equation}
\int\limits_0^T \vec{f}(\vec{W}_t,t) \circ \rmd \vec{W}_t 
= \int\limits_0^T \vec{f}(\vec{W}_t,t) \cdot \rmd \vec{W}_t 
+ \frac{1}{2} \int\limits_0^T \left. \vec{\nabla}_{\vec{x}} \cdot \vec{f}(\vec{x},t) \right|_{\vec{x}=\vec{W}_t} \; \rmd t ,
\label{equ:relation_Ito_Stratonovich}
\end{equation}
where $\vec{\nabla} \cdot \vec{f}$ denotes the divergence of $\vec{f}$.
This relation can also be used to express the total differential of the random process \eqref{equ:total_differential_Ito}  
\begin{equation}
\rmd Y(\vec{W}_t,t) = \partial_t Y(\vec{W}_t,t) \; \rmd t 
+ \left. \vec{\nabla}_\vec{x} Y(\vec{x},t) \right|_{\vec{x}=\vec{W}_t} \circ \rmd \vec{W}_t 
\label{equ:total_differential_Stratonovich}
\end{equation}
in terms of the Stratonovich calculus. 
From this point of view the Stratonovich approach can be considered as a more natural.

\subsection{Underdamped diffusion}
\label{ssec:underdamped_diff}
\index{diffusion!underdamped}
Although underdamped diffusion describes a large class of physical systems, 
the typical example of underdamped diffusion describes a heavy particle in an environment consisting of lighter particles, e.g. the droplet of oil in the water.
In order to effectively describe the movement of the heavy particle in such an environment it's impractical or almost impossible to keep the track of all light particles. 
Let us assume that there are no long range interaction, hence light particles interact with the heavy particle mostly by collisions. 
Collisions mainly cause two effects, the friction of the environment proportional to the velocity of the heavy particle characterized by a coefficient $\gamma(\vec{q})$, 
which in general depends on the position of the heavy particle $\vec{q}$, 
and the random force acting on the heavy particle described by Wiener process with an amplitude of $\sqrt{2\gamma(\vec{q_t})/\beta}$. 
The movement of the heavy particle with mass $m$ then can be effectively described through set of stochastic differential equations
\begin{align}
\rmd \vec{q}_t &= \frac{\vec{p}_t}{m} \; \rmd t , \label{def:position_underdamped} \\
\rmd \vec{p}_t &= \left[ \vec{F}(\vec{q}_t,\vec{p}_t) - \frac{\gamma(\vec{q}_t)}{m} \vec{p}_t \right] \; \rmd t + \sqrt{\frac{2\gamma(\vec{q}_t)}{\beta}} \circ \rmd \vec{W}_t , \label{def:momentum_underdamped}
\end{align}
where $\vec{F}(\vec{q},\vec{p})$ describes the deterministic part of the force, 
which in general depends on the position $\vec{q}$ and momentum $\vec{p}$ of the heavy particle 
and $\beta$ is the inverse temperature of the environment.
Directly from definitions \eqref{def:position_underdamped} and \eqref{def:momentum_underdamped} we can see that the position explicitly depends only on time $\vec{q}_t(t)$ while the momentum also explicitly depends on the realization of the Wiener process $\vec{p}_t(\vec{W}_t,t)$. 
Notice also that in most cases there is no difference between the It\^{o} or Stratonovich approaches in \eqref{def:momentum_underdamped}, 
because the friction coefficient $\gamma(\vec{q})$ does not depend on the momentum.
However it makes difference in case of the underdamped diffusion as we will see later. 
Since the It\^{o} approach is easier to handle we will use it in the rest of this subsection, hence 
\begin{equation}
\rmd \vec{p}_t = \left[ \vec{F}(\vec{q}_t,\vec{p}_t) - \frac{\gamma(\vec{q}_t)}{m} \vec{p}_t \right] \; \rmd t + \sqrt{\frac{2 \gamma(\vec{q}_t) }{\beta}} \cdot \rmd \vec{W}_t . 
\label{equ:momentum_underdamped}
\end{equation}

The solution of the stochastic differential equation is also a random function, hence one can be interested in the probability distribution of positions and momentums at a particular time
\begin{equation}
\mu_t(\vec{q},\vec{p}) = \left\langle \delta(\vec{q}-\vec{q}_t) \, \delta(\vec{p}-\vec{p}_t) \right\rangle_{\mu_0} ,
\label{def:probability_distribution_underdamped}
\end{equation}
where we take the mean value over all possible realizations of the random process $(\vec{q}_t,\vec{p}_t)$ with the initial condition given by $\mu_0(\vec{q},\vec{p})$. 
This approach proves to be useful in cases when we have set of experiments which we need to evaluate or if the initial position or momentum is indeterminate. 
It can be shown, see below, that the autonomous time evolution of the probability density is governed by 
\begin{multline}
\partial_t \mu_t(\vec{q},\vec{p}) 
= - \frac{\vec{p}}{m} \cdot \vec{\nabla}_{\vec{q}} \mu_t(\vec{q},\vec{p}) - \\
- \vec{\nabla}_{\vec{p}} \cdot \left[ \left( \vec{F}(\vec{q},\vec{p}) - \frac{\gamma(\vec{q})}{m} \vec{p} \right) \mu_t(\vec{q},\vec{p}) 
- \frac{\gamma(\vec{q})}{\beta} \vec{\nabla}_\vec{p} \mu_t(\vec{q},\vec{p}) \right] .
\label{equ:time_evolution_underdamped}
\end{multline}
We will use the proof of this statement to provide an example of the usage of previously introduced stochastic calculus. 
We start by expanding the definition of the distribution \eqref{def:probability_distribution_underdamped} using the It\^{o} lemma \eqref{equ:total_differential_Ito}  
\begin{multline*}
\rmd \mu_t(\vec{q},\vec{p}) 
= - \rmd t \; \left\langle \partial_t \vec{q}_t(t) \cdot \vec{\nabla}_\vec{q} \delta(\vec{q}-\vec{q}_t(t)) \, \delta(\vec{p}-\vec{p}_t(\vec{W}_t,t)) \right\rangle_{\mu_0} - \\
- \rmd t \; \left\langle \partial_t \vec{p}_t(\vec{W}_t,t) \cdot \vec{\nabla}_\vec{p} \delta(\vec{p}-\vec{p}_t(\vec{W}_t,t)) \, \delta(\vec{q}-\vec{q}_t(t)) \right\rangle_{\mu_0} + \\ 
- \left\langle \rmd \vec{W}_t \cdot \left. \nabla_{\vec{x}} \vec{p}_t(\vec{x},t) \right|_{\vec{x}=\vec{W}_t} \cdot \vec{\nabla}_\vec{p} \delta(\vec{p}-\vec{p}_t(\vec{W}_t,t)) \, \delta(\vec{q}-\vec{q}_t(t)) \right\rangle_{\mu_0} + \\ 
+ \frac{1}{2} \rmd t \; \tr \left\langle \left. \nabla_\vec{x} \vec{p}_t(\vec{x},t) \right|_{\vec{x}=\vec{W}_t} \cdot \vec{\nabla}^2_\vec{p} \delta(\vec{p}-\vec{p}_t(\vec{W}_t,t)) \cdot \left. \nabla_\vec{x} \vec{p}_t(\vec{x},t) \right|_{\vec{x}=\vec{W}_t} \, \delta(\vec{q}-\vec{q}_t(t)) \right\rangle_{\mu_0} ,
\end{multline*}
where we have explicitly denoted the dependencies on the time and the realization of the Wiener process. 
While the dependence on the $\vec{q}$ and $\vec{p}$ is only in the argument of delta functions, we can pull the derivatives with respect to them in front of the mean values. 
We also insert the terms from \eqref{def:position_underdamped} and \eqref{equ:momentum_underdamped} and obtain 
\begin{multline*}
\rmd \mu_t(\vec{q},\vec{p}) 
= - \rmd t \; \vec{\nabla}_\vec{q} \cdot \left\langle \frac{\vec{p}_t}{m} \, \delta(\vec{q}-\vec{q}_t) \, \delta(\vec{p}-\vec{p}_t) \right\rangle_{\mu_0} - \\
- \vec{\nabla}_\vec{p} \cdot \left\langle \left[ \left( \vec{F}(\vec{q}_t,\vec{p}_t) - \frac{\gamma(\vec{q}_t) \vec{p}_t }{m} \right) \; \rmd t + \sqrt{\frac{2\gamma(\vec{q}_t)}{\beta}} \; \rmd \vec{W}_t \right] \delta(\vec{q}-\vec{q}_t) \, \delta(\vec{p}-\vec{p}_t) \right\rangle_{\mu_0} + \\
+ \rmd t \; \Delta_\vec{p} \left\langle \frac{\gamma(\vec{q}_t)}{\beta} \, \delta(\vec{q}-\vec{q}_t) \, \delta(\vec{p}-\vec{p}_t) \right\rangle_{\mu_0} .
\end{multline*}
Now using the fact that the mean value of It\^{o} integral is zero \eqref{equ:Ito_zero_mean_value} along with 
\[
\left\langle f(\vec{q}_t,\vec{p}_t) \, \delta(\vec{q}-\vec{q}_t) \, \delta(\vec{p}-\vec{p}_t) \right\rangle_{\mu_0} 
= f(\vec{q},\vec{p}) \left\langle \delta(\vec{q}-\vec{q}_t) \, \delta(\vec{p}-\vec{p}_t) \right\rangle_{\mu_0} 
= f(\vec{q},\vec{p}) \mu_t(\vec{q},\vec{p}) 
\]
concludes the proof.

\subsubsection{Kolmogorov generators}
At the beginning of the section \ref{sec:Markov_processes} we have introduced Kolmogorov generators as an effective description of the time evolution of the system.
To provide the same level of description also for underdamped diffusion we use the time evolution equation \eqref{equ:time_evolution_underdamped} to define \emph{the forward Kolmogorov generator}\index{Kolmogorov generator!forward!underdamped diffusion} \eqref{equ:Markov_stochastic_time_evolution}
\begin{multline}
\gen^* \left[  \mu \right] (\vec{q},\vec{p}) 
= - \frac{\vec{p}}{m} \cdot \vec{\nabla}_{\vec{q}} \mu(\vec{q},\vec{p}) - \\
- \vec{\nabla}_{\vec{p}} \cdot \left[ \left( \vec{F}(\vec{q},\vec{p}) - \frac{\gamma(\vec{q})}{m} \vec{p} \right) \mu(\vec{q},\vec{p}) 
- \frac{\gamma(\vec{q})}{\beta} \vec{\nabla}_\vec{p} \mu(\vec{q},\vec{p}) \right] .
\label{equ:forward_generator_underdamped}
\end{multline}
It is easy to check that the normalization condition on forward Kolmogorov generator \eqref{equ:normalization_condition} is valid under the assumption that the probability distribution vanish at the boundary $\|\vec{q}\| \to \infty$ or $\|\vec{p}\| \to \infty$. 
From \eqref{def:backward_generator} we can also obtain \emph{the backward Kolmogorov generator}\index{Kolmogorov generator!backward!underdamped diffusion} for underdamped diffusion
\begin{multline}
\gen \left[ A \right] (\vec{q},\vec{p}) 
= \frac{\vec{p}}{m} \cdot \vec{\nabla}_{\vec{q}} A(\vec{q},\vec{p}) + \\
+ \left( \vec{F}(\vec{q},\vec{p}) - \frac{\gamma(\vec{q})}{m} \vec{p} \right) \cdot \vec{\nabla}_{\vec{p}} A(\vec{q},\vec{p})
+ \frac{\gamma(\vec{q})}{\beta} \Delta_\vec{p} A(\vec{q},\vec{p}) .
\label{equ:backward_generator_underdamped}
\end{multline}
If we examine the forward Kolmogorov generator, 
we can see that it has a structure of the generalized divergence,
hence the time evolution equation is similar continuity equation well known from the hydrodynamics. 
If we define a generalized probability current 
\[
\widetilde{\vec{\jmath}}(\vec{q},\vec{p}) = \begin{bmatrix}
\frac{\vec{p}}{m} \mu(\vec{q},\vec{p}) \\
\left( \vec{F}(\vec{q},\vec{p}) - \frac{\gamma(\vec{q})}{m} \vec{p} \right) \mu(\vec{q},\vec{p}) - \frac{\gamma(\vec{q})}{\beta} \vec{\nabla}_\vec{p} \mu(\vec{q},\vec{p})
\end{bmatrix}
\]
and a generalized gradient 
\[
\widetilde{\vec{\nabla}} = \begin{bmatrix}
\vec{\nabla}_\vec{q} \\ 
\vec{\nabla}_\vec{p}
\end{bmatrix} ,
\]
then the time evolution equation \eqref{equ:time_evolution_underdamped} can be written as 
\[
\partial_t \mu_t (\vec{q},\vec{p}) + \widetilde{\vec{\nabla}} \cdot \widetilde{\vec{\jmath}} (\vec{q},\vec{p}) = 0 . 
\]

\subsubsection{Equilibrium}
In equilibrium the force $\vec{F}(\vec{q},\vec{p})$ acting on particle is given by potential $U(\vec{q})$, which depends only on the position $\vec{q}$ of the particle.  
The stationary equilibrium distribution $\rho(\vec{q},\vec{p})$ is then given by \emph{the Maxwell-Boltzmann distribution}\index{Maxwell-Boltzmann distribution} 
\[
\rho(\vec{q},\vec{p}) = \frac{1}{Z} \exp \left[ - \beta \left( \frac{\vec{p}^2}{2m} + U(\vec{q}) \right) \right] 
\]
at the inverse temperature $\beta$.
As a consequence the generalized probabilistic current is given by 
\[
\widetilde{\vec{\jmath}} = \begin{bmatrix}
\frac{\vec{p}}{m} \\
- \vec{\nabla}_\vec{q} U(\vec{q}) 
\end{bmatrix} 
\rho(\vec{q},\vec{p}) .
\]

\subsection{Overdamped diffusion}
\label{ssec:overdamped_diff}
\index{diffusion!overdamped}
The overdamped diffusion is the limiting case of the underdamped diffusion, when the relaxation time for the velocities is much shorter then the relaxation time for positions, 
which we will show in chapter \ref{chapter:slow-fast_coupling}. 
Another possible approach is by taking the diffusion limit of the Markov jump process on the lattice with non-uniform jump probabilities along the edges \cite{Itzykson_Drouffe:Statistical_field_theory} as in the subsection \ref{ssec:diffusion_limit}.
For this purpose, we define \emph{the overdamped diffusion}\index{diffusion!overdamped} by the total differential in the Stratonovich form 
\begin{equation}
\rmd \vec{q}_t = \left[ \vec{\chi}(\vec{q}_t) \cdot \vec{F}(\vec{q}_t) + \frac{1}{2} \vec{\nabla}_\vec{q} \cdot \vec{D}(\vec{q_t}) \right] \; \rmd t + \sqrt{ 2 \vec{D}(\vec{q}_t) } \circ \rmd \vec{W}_t, 
\label{def:position_overdamped}
\end{equation}
which can be rewritten to the It\^{o} form by using the chain rule along with \eqref{equ:relation_Ito_Stratonovich} 
\begin{equation}
\rmd \vec{q}_t = \left[ \vec{\chi}(\vec{q}_t) \cdot \vec{F}(\vec{q}_t) + \vec{\nabla}_\vec{q} \cdot \vec{D}(\vec{q_t}) \right] \; \rmd t + \sqrt{ 2 \vec{D}(\vec{q}_t) } \cdot \rmd \vec{W}_t, 
\label{equ:position_overdamped}
\end{equation}
where $\rmd \vec{W}_t$ is again a multidimensional white noise 
and mobility matrix $\vec{\chi}(\vec{q})$ is related to the diffusion matrix $\vec{D}(\vec{q})$ by \emph{the Einstein relation}\index{Einstein relation} $\vec{\chi}(\vec{q}) =\beta \vec{D}(\vec{q})$.

We can again derive the autonomous time evolution equation for probability density in the same manner as in the subsection \ref{ssec:underdamped_diff} in case of the underdamped diffusion directly from the stochastic differential equation \eqref{equ:position_overdamped}, 
which yields to the Fokker-Planck equation\index{Fokker-Planck equation} 
\begin{equation}
\partial_t \mu_t(\vec{q}) = 
- \vec{\nabla}_{\vec{q}} \cdot \left[ \vec{\chi}(\vec{q}) \cdot \vec{F}(\vec{q}) \; \mu_t(\vec{q}) - \vec{D}(\vec{q}) \cdot \vec{\nabla}_{\vec{q}} \mu_t(\vec{q}) \right] .
\label{equ:time_evolution_overdamped}
\end{equation}

\subsubsection{Kolmogorov generators}
We can again associate the forward Kolmogorov generator with the operator on the right side of the Fokker-Planck equation \eqref{equ:time_evolution_overdamped}. 
Hence we define \emph{the forward Kolmogorov generator}\index{Kolmogorov generator!forward!overdamped diffusion} for overdamped diffusion as 
\[
\gen^* \left[ \mu \right] (\vec{q}) = 
- \vec{\nabla}_{\vec{q}} \cdot \left[ \vec{\chi}(\vec{q}) \cdot \vec{F}(\vec{q}) \; \mu(\vec{q}) - \vec{D}(\vec{q}) \cdot \vec{\nabla}_{\vec{q}} \mu(\vec{q}) \right] .
\]
In analogy with the underdamped diffusion we also define \emph{the backward Kolmogorov generator}\index{Kolmogorov generator!backward!overdamped diffusion} for overdamped diffusion 
\begin{equation}
\gen \left[ A \right] (\vec{q}) = 
\vec{F}(\vec{q}) \cdot \vec{\chi}(\vec{q}) \cdot \vec{\nabla}_{\vec{q}} A(\vec{q}) + \vec{\nabla}_{\vec{q}} \cdot \left[ \vec{D}(\vec{q}) \cdot \vec{\nabla}_{\vec{q}} A(\vec{q}) \right] .
\label{equ:bacward_generator_overdamped}
\end{equation}
In case of the overdamped diffusion the structure of the continuity equation in case of the time evolution equation is even more pronounced than in the underdamped case. 
Let us define the probability current 
\begin{equation}
\vec{j}(\vec{q}) = \left[ \vec{\chi}(\vec{q}) \cdot \vec{F}(\vec{q}) \; \mu(\vec{q}) - \vec{D}(\vec{q}) \cdot \vec{\nabla}_{\vec{q}} \mu(\vec{q}) \right] .
\label{def:current_overdamped}
\end{equation}
Then the Fokker-Planck equation \eqref{equ:time_evolution_overdamped} corresponds to
\[
\partial_t \mu_t(\vec{q}) + \vec{\nabla}_\vec{q} \cdot \vec{j}(\vec{q}) = 0 . 
\]

\subsubsection{Equilibrium}
In equilibrium the force $\vec{F}(\vec{q})$ is again determined by the potential $U(\vec{q})$,
hence the stationary distribution in equilibrium again corresponds to \emph{the Boltzmann distribution}\index{Boltzmann distribution}
\[
\rho(\vec{q}) = \frac{1}{Z} \exp \left[ - \beta U(\vec{q}) \right] 
\]
at the inverse temperature $\beta$. 
Moreover one can easily check that the probability current \eqref{def:current_overdamped} in equilibrium is zero. 

% Stochastic equilibrium thermodynamics 
% a. review of standard reversible (i.e. quasistatic) thermodynamics cleanly derived within the formalism of Markov models: heat, work, First law, the quasistatic limit, Secodn law, ...
% b. Jarzynski equality and its consequences - Second law rederived, corrections, ...
% c. /maybe/ structure of the leading correction from the quasistatic expansion

\chapter{Stochastic equilibrium thermodynamics}
\label{chapter:equilibrium_stat_phys}
There are several different approaches to classical equilibrium physics, for example standard textbook by Callen \cite{Callen1985} follows the argument of equal probability of microstates 
while for example Balian \cite{Balian1982} prefers information theory approach. 
In this chapter we will rehearse an alternative way how to build an equilibrium statistical physics more suitable for mesoscopic stochastic systems. 
Our starting point will be the concepts of conservation of the total energy on the microscopic level and the local and global detailed balance. 
We will use these principles to obtain the first law of thermodynamics for stochastic time evolution and to derive Crooks fluctuation relation as a weak formulation of the second law.

\section{First law of thermodynamics}
\index{first law of thermodynamics}
\label{sec:first_law}
The first law of thermodynamics is the mesoscopic version of the conservation law of total energy. 
In our systems we assume that the law of the total energy conservation is in general obeyed by the underlying microscopic dynamics for the total system, 
hence the energy of the system can only be altered by the interaction with a thermal bath or by the change of external conditions $\alpha$. 
Sekimoto associated \emph{the heat}\index{heat} $Q_{\alpha(t)}(\omega)$ with the energy dissipated to thermal baths and \emph{the work}\index{work} $W_{\alpha(t)}(\omega)$ with the energy change by varying the external conditions on the microscopic level for every possible microscopic path $\omega$ and thus introduced the concept of \emph{stochastic energetics}\index{stochastic energetics} \cite{Sekimoto1997,Sekimoto1998,Sekimoto2007,Seifert2008}
\begin{equation}
E_{\alpha(T)}(x_T) - E_{\alpha(0)}(x_0) = Q^{\text{sys}}_{\alpha(t)}(\omega) + W^{\text{sys}}_{\alpha(t)}(\omega),
\label{def:first_law_path}
\end{equation}
where $E_\alpha(x)$ is the energy of the system in configuration $x$ under external conditions $\alpha$ 
and where $Q^{\text{sys}}_{\alpha(t)}(\omega)$ is the heat dissipated to the system and $W^{\text{sys}}_{\alpha(t)}(\omega)$ is the work done on the system along the path $\omega$ from time $0$ to time $T$ with external conditions driven by protocol $\alpha(t)$. 
The heat and work strongly depend on the actual model, 
however usually the heat is the energy associated with the change of the configuration, 
which for example in case of Markov jump processes is given by 
\[
Q^{\text{sys}}_{\alpha(t)}(\omega) = \sum_{t_i} \left[ E_{\alpha(t_i)}(x_{t_i}) - E_{\alpha(t_i)}(x^-_{t_i}) \right], 
\]
while the work is associated with the action of the external agent via \emph{the work}
\begin{equation}
W^{\text{sys}}_{\alpha(t)}(\omega) = \int \rmd t \; \dot\alpha(t) \cdot \left. \nabla_\alpha E_\alpha(x_t) \right|_{\alpha=\alpha(t)} .
\label{def:work_external_parameters}
\end{equation}
For more examples see subsections \ref{ssec:work_heat_jump} and \ref{ssec:work_heat_diffusion}.

We can take advantage of the Markov property of time evolution and define \emph{the local power}\index{local power} $w_{\alpha(t)}(x)$ as an work done on the system per unit time if the system is starting from the initial configuration $x$ and freely evolve for timespan $s$ 
\begin{equation}
w_{\alpha(t)}(x) = \lim_{s \to 0^+} \frac{1}{s} \left\langle W^{\text{sys}}_{\alpha(t)}\left(\omega^{(0,s]}\right) \right\rangle^{\alpha(t)}_{\delta_x} 
\label{def:local_power}
\end{equation}
and similarly we define \emph{the local heat production}\index{local heat production} 
\begin{equation}
q_{\alpha(t)}(x) = \lim_{s \to 0^+} \frac{1}{s} \left\langle Q^{\text{sys}}_{\alpha(t)}\left(\omega^{(0,s]}\right) \right\rangle^{\alpha(t)}_{\delta_x} . 
\label{def:local_heat_production}
\end{equation}
Notice that in the most general case both of these quantities are functionals on the protocol of external parameters $\alpha(t)$, 
however because the mean value of the work and the heat are smooth in most cases with respect to protocol $\alpha(t)$ it depends only on the actual value of external parameters $\alpha$ and its first time-derivative $\dot\alpha$.  
A direct consequence is, that \emph{the mean total work} ${\mathcal W}_{\mu_0}(\alpha(t))$ and analogously \emph{the mean total heat} is equivalent to integrated mean value of the local power with respect to actual state $\mu_t$ at time $t$
when starting the time evolution from the initial state $\mu_0$ 
\begin{align}
{\mathcal W}_{\mu_0}(\alpha(t)) &= \left\langle W^{\text{sys}}_{\alpha(t)}(\omega) \right\rangle^{\alpha(t)}_{\mu_0} = \int\limits_0^T \rmd t \; \left\langle w_{\alpha(t)} \right\rangle_{\mu_t} , \label{equ:total_local_work_relation} \\
{\mathcal Q}_{\mu_0}(\alpha(t)) &= \left\langle Q^{\text{sys}}_{\alpha(t)}(\omega) \right\rangle^{\alpha(t)}_{\mu_0} = \int\limits_0^T \rmd t \; \left\langle q_{\alpha(t)} \right\rangle_{\mu_t}. \label{equ:total_local_heat_relation}
\end{align}

While we have relation between the change of the energy and work and heat for each possible realization of the path $\omega$ \eqref{def:first_law_path}, 
the similar relation has to be valid also on the level of mean values over all possible realizations of path $\omega$ 
\begin{equation}
{\mathcal U}_{\mu_T}(\alpha(T)) - {\mathcal U}_{\mu_0}(\alpha(0)) = {\mathcal W}_{\mu_0}(\alpha(t)) + {\mathcal Q}_{\mu_0}(\alpha(t)) ,
\label{equ:first_law_eq}
\end{equation}
where we have defined \emph{the internal energy}\index{internal energy} ${\mathcal U}_{\mu}(\alpha) = \left\langle E_\alpha \right\rangle_\mu $,
and which we associate with \emph{the first law of thermodynamics}\index{first law of thermodynamics}, as we will see later.    
From here we can also conclude that the mean local power and mean local heat production sums up to the change of the internal energy
\begin{equation}
\partial_t {\mathcal U}_{\mu_t}(\alpha(t)) = \left\langle w_{\alpha(t)} + q_{\alpha(t)} \right\rangle_{\mu_t} .
\label{equ:first_law_rates}
\end{equation}

An alternative approach to local power and local heat production can be by starting from the equation \eqref{equ:first_law_rates} and applying \eqref{equ:time_derivative_observable} 
\begin{equation}
\partial_t \left\langle E_{\alpha(t)} \right\rangle_\mu = \left\langle \partial_t E_{\alpha(t)} + \gen_{\alpha(t)} \left[ E_{\alpha(t)} \right] \right\rangle_\mu ,
\label{equ:time_evolution_energy}
\end{equation}
where we can identify the local power as the part connected with the change of external parameters 
\begin{equation}
w_{\alpha(t)}(x) = \partial_t E_{\alpha(t)}(x) = \dot\alpha(t) \cdot \left. \nabla_\alpha E_\alpha(x) \right|_{\alpha(t)} 
\label{equ:local_power}
\end{equation}
which can be also obtained from the definition \eqref{def:local_power} with the work defined by \eqref{def:work_external_parameters},
while the local heat production is the part corresponding to the change of the configuration of the system
\begin{equation}
q_{\alpha(t)}(x) = \gen_{\alpha(t)}\left[ E_{\alpha(t)} \right] (x) ,
\label{equ:local_heat_production}
\end{equation}
which is associated with the energy transfer to thermal bath.

\subsection{Work and heat in the Markov jump processes}
\label{ssec:work_heat_jump}
We now provide examples of definitions of the heat and work and their local production in two major classes of models starting with Markov jump processes,
where the system is in contact with single thermal bath and where the only time dependence lies in the protocol of the external parameters $\alpha(t)$. 
The simplest quantity to define in jump processes is the heat. 
The heat\index{heat!Markov jump processes} is the energy exchange associated with the change of configuration of the system, which occurs at jump points, hence we have 
\[
Q_{\alpha(t)}(\omega) = \sum_i \left[ E_{\alpha(t_i)}(x_{t_i}) - E_{\alpha(t_i)}(x_{t^-_i}) \right] ,
\]
where we sum over all jump times $t_i$ and where $x_{t^-_i}$ again denotes the configuration just before the jump. 
To obtain the local heat production\index{local heat production!Markov jump processes} we can use either definition \eqref{def:local_heat_production} or rather use the equation \eqref{equ:local_heat_production} along with the backward Kolmogorov generator for jump processes \eqref{equ:backward_Kolmogorov_gen_jump}
\[
q_\alpha(x) = \sum_y \rate[\alpha]{x}{y} \left[ E(y) - E(x) \right] .
\]
Notice that in case of Markov jump processes the local heat production depends only on the actual values of external parameters $\alpha$ but not on the velocity of its change.

Under such conditions the work\index{work!Markov jump processes} is solely associated with the change of the energy due to the change of external parameters $\alpha(t)$, hence we obtain
\[
W_{\alpha(t)}(\omega) = \sum_i \left[ E_{\alpha(t_{i+1})}(x_{t_i}) - E_{\alpha(t_i)}(x_{t_i}) \right] ,
\]
where we sum over all time intervals $[t_i,t_{i+1}]$ in the path $\omega$ where the configuration $x_{t_i}$ holds.
The local power\index{local power!Markov jump processes} in this case is directly obtained from the equation \eqref{equ:local_power}  
\[
w_{\alpha(t)}(x) = \dot\alpha(t) \cdot \left. \nabla_\alpha E_\alpha(x) \right|_{\alpha=\alpha(t)}.  
\]

\subsection{Work and heat in diffusion}
\label{ssec:work_heat_diffusion}
In diffusion we can determine the local power and the local heat production by decomposing the time evolution of the mean total energy \eqref{equ:time_evolution_energy} to the part which depends on the change of external parameters $\alpha$ \eqref{equ:local_power} and the part which depends on the change of state \eqref{equ:local_heat_production}.
However we can also obtain them directly from the pure mechanical considerations.
In this section we will show that these two approaches are consistent.

\subsubsection{Underdamped diffusion} 
In the underdamped diffusion the total force acting on the system \eqref{equ:momentum_underdamped} is 
\[
\vec{F}_\alpha(\vec{q},\vec{p}) = - \vec{\nabla}_\vec{q} U_\alpha (\vec{q}) - \frac{\gamma_\alpha(\vec{q})}{m} \vec{p} + \sqrt{\frac{2 \gamma_\alpha(\vec{q})}{\beta}} \frac{\rmd \vec{W}_t}{\rmd t} ,
\]
where the first term is the interaction with the potential $U_\alpha(\vec{q})$,
the second term corresponds to the friction
and the third term is the thermal random force acting on the particle, 
where by $\rmd \vec{W}_t / \rmd t$ we denote \emph{the white noise}\index{white noise}.
We we will handle the white noise formally in this subsection as the formal time derivative of the Wiener process, although it does not exists, 
because the Wiener process is in fact nowhere differentiable \cite{Evans2001}. 
The last two terms make together what we will later call heat.

In case there is no change in external parameters it is know from the classical mechanics, that the work done by action of the total forces corresponds to the change of the kinetic energy
\[
\Delta T = \int \vec{F}_\alpha(\vec{q}_t,\vec{p}_t) \circ \rmd \vec{q}_t 
= \int \left[ - \vec{\nabla}_\vec{q} U_\alpha (\vec{q}) - \frac{\gamma_\alpha(\vec{q})}{m} \vec{p} + \sqrt{\frac{2 \gamma_\alpha(\vec{q})}{\beta}} \frac{\rmd \vec{W}_t}{\rmd t} \right] \circ \rmd \vec{q}_t ,
\]
where $\circ$ denotes the Stratonovich integral \eqref{def:Stratonovich_integral}. 
The choice of the Stratonovich stochastic integral here is due to the antisymmetry with respect to the time inversion \eqref{equ:antisymmetry_Stratonovich}.
The first term in the work corresponds to the infinitesimal change of the potential $\rmd U_\alpha(\vec{q})$ 
in the Stratonovich sense \eqref{equ:total_differential_Stratonovich}
and hence can be integrated out and we obtain that the change of the total energy is given by thermal forces, i.e. friction and random force  
\begin{equation}
\Delta \left[ \frac{\vec{p}_t^2}{2m} + U_{\alpha}(\vec{q}) \right] = \int \left[ - \frac{\gamma_\alpha(\vec{q}_t)}{m} \vec{p}_t + \sqrt{\frac{2 \gamma_\alpha(\vec{q})}{\beta}} \frac{\rmd \vec{W}_t}{\rmd t} \right] \circ \rmd \vec{q}_t ,
\label{equ:energy_balance_underdamped}
\end{equation}
where we have written the kinetic energy explicitly. 
From thermodynamics we know that the change of total energy in case there is no action of external forces, i.e. no work done by the change of external parameters $\alpha$, corresponds to the heat, hence we define the heat in the underdamped diffusion as
\[
Q(\omega) = \int \left[ - \frac{\gamma_\alpha(\vec{q}_t)}{m} \vec{p}_t + \sqrt{\frac{2 \gamma_\alpha(\vec{q})}{\beta}} \frac{\rmd \vec{W}_t}{\rmd t} \right] \circ \rmd \vec{q}_t .
\]
To obtain the local heat production we need to express the mean total heat as the time integral.

In this case we will treat each of the two terms separately. 
The first term does not explicitly depend on the Wiener process, it depends on it only implicitly through the momentum $\vec{p}_t$, 
hence by applying the time evolution equation for the position \eqref{def:position_underdamped} we obtain an ordinary time integral 
\[
\left\langle Q^{(1)}_{\alpha(t)}(\omega) \right\rangle_{\mu_0} 
= - \int \left\langle \left[ \frac{\gamma_{\alpha(t)}(\vec{q}_t)}{m} \vec{p}_t \right] \cdot \frac{\vec{p}_t}{m} \right\rangle_{\mu_t} \; \rmd t ,
\] 
from where we can directly obtain the local  of that particular part of the total force  
\[
q^{(1)}_\alpha(\vec{q},\vec{p}) = - \frac{\gamma_\alpha(\vec{q})}{m^2} \vec{p}^2 .
\]
In the mean value for the random force  
\begin{multline*}
\left\langle Q^{(2)}_{\alpha(t)}(\omega) \right\rangle_{\mu_0} 
= \left\langle \int \sqrt{\frac{2 \gamma_{\alpha(t)}(\vec{q}_t)}{\beta}} \frac{\rmd \vec{W}_t}{\rmd t} \circ \rmd{\vec{q}_t} \right\rangle_{\mu_0} = \\
= \left\langle \int \sqrt{\frac{2 \gamma_{\alpha(t)}(\vec{q}_t)}{\beta}} \frac{\rmd \vec{q}_t}{\rmd t} \circ \rmd{\vec{W}_t} \right\rangle_{\mu_0}
= \left\langle \int \sqrt{\frac{2 \gamma_{\alpha(t)}(\vec{q}_t)}{\beta}} \frac{\vec{p}_t}{m} \circ \rmd{\vec{W}_t} \right\rangle_{\mu_0}
\end{multline*}
we have obtained the Stratonovich integral of a function of the momentum with respect to the Wiener process. 
At first we rewrite the Stratonovich stochastic integral by using the relation between Stratonovich and It\^{o} integral \eqref{equ:relation_Ito_Stratonovich} 
\[
\left\langle Q^{(2)}_{\alpha(t)}(\omega) \right\rangle_{\mu_0} = 
\left\langle \int \sqrt{\frac{2 \gamma_{\alpha(t)}(\vec{q}_t)}{\beta}} \frac{\vec{p}_t}{m} \cdot \rmd{\vec{W}_t} \right\rangle_{\mu_0}
+ \left\langle \int \frac{\gamma_{\alpha(t)}(\vec{q}_t)}{\beta} \frac{d}{m} \; \rmd t \right\rangle_{\mu_0} ,
\]
where $d$ is the dimension of the momentum space $\vec{p} \in {\mathbb R}^d$. 
Then we used that the mean value of the It\^{o} integral is zero \eqref{equ:Ito_zero_mean_value} 
and we have obtain the local heat production from the random force term as 
\[
q^{(2)}_\alpha(\vec{q},\vec{p}) 
= \frac{d \gamma_\alpha(\vec{q})}{m \beta} .
\]
Putting these two contribution together we obtain the total local power generated by the total local heat produced by the particle
\[
q_\alpha(\vec{q},\vec{p}) = - \frac{2 \gamma_\alpha(\vec{q})}{m} \left[ \frac{\vec{p}^2}{2m} - \frac{d}{2 \beta} \right],
\]
which is proportional to the difference between actual kinetic energy and the mean kinetic energy from the equipartition theorem.

In case there is also the change of the external parameters, there is an additional term in the total energy balance equation \eqref{equ:energy_balance_underdamped}, 
which corresponds to the action of external forces 
\[
\Delta \left[ \frac{\vec{p}_t^2}{2m} + U_{\alpha(t)}(\vec{q}_t) \right] 
= \int \rmd t \; \dot\alpha(t) \cdot \left. \nabla_\alpha U_\alpha(\vec{q}_t) \right|_{\alpha=\alpha(t)} + Q(\omega) ,
\]
and which defines the work \eqref{def:work_external_parameters}.

\subsubsection{Overdamped diffusion}
We again start with the case when the external parameters are kept constant by analysing the energy balance equation. 
In overdamped diffusion the total force acting on the system is given by \eqref{def:position_overdamped}
\[
\vec{F}_\alpha(\vec{q}) = - \vec{\nabla}_\vec{q} U_\alpha(\vec{q}) + \frac{1}{\beta} \chi^{-1}(\vec{q}) \cdot \left( \vec{\nabla}_\vec{q} \cdot \chi(\vec{q}) \right) + \sqrt{\frac{2}{\beta \chi(\vec{q})}} \cdot \frac{\rmd \vec{W}_t}{\rmd t} ,
\]
where we have used the Einstein relation to represent the expression in terms of mobility matrix $\chi$ (and its inverse $\chi^{-1}$) and the inverse temperature. 
In the overdamped diffusion there is no notion of the kinetic energy, hence the energy balance equation leads to 
\[
0 = \int \vec{F}_\alpha(\vec{q}_t) \circ \vec{q}_t 
= \int \left[ - \vec{\nabla}_\vec{q} U_\alpha(\vec{q}) + \frac{1}{\beta} \chi^{-1}(\vec{q}) \cdot \left( \vec{\nabla}_\vec{q} \cdot \chi(\vec{q}) \right) + \sqrt{\frac{2}{\beta \chi(\vec{q})}} \cdot \frac{\rmd \vec{W}_t}{\rmd t} \right] \circ \rmd \vec{q}_t ,
\] 
where the first term can be again as the total differential in the Stratonovich sense \eqref{equ:total_differential_Stratonovich} of the potential $U_\alpha(\vec{q}_t)$ 
\[ 
\Delta U_\alpha (\vec{q}_t) = \int \left[ \frac{1}{\beta} \chi^{-1}(\vec{q}) \cdot \left( \vec{\nabla}_\vec{q} \cdot \chi(\vec{q}) \right) + \sqrt{\frac{2}{\beta \chi(\vec{q})}} \cdot \frac{\rmd \vec{W}_t}{\rmd t} \right] \circ \rmd \vec{q}_t ,
\]
where the right hand side represent the heat 
\[
Q(\omega) = \int \left[ \frac{1}{\beta} \chi^{-1}(\vec{q}) \cdot \left( \vec{\nabla}_\vec{q} \cdot \chi(\vec{q}) \right) + \sqrt{\frac{2}{\beta \chi(\vec{q})}} \cdot \frac{\rmd \vec{W}_t}{\rmd t} \right] \circ \rmd \vec{q}_t .
\]

To obtain the local heat production we need again to obtain the mean value of the heat in the form of the integral over time. 
However in this case it is more convenient to use it equivalence to the energy change.
We start from 
\[
\left\langle Q(\omega) \right\rangle_{\mu_0} = \left\langle \int \left. \vec{\nabla}_\vec{q} U_\alpha (\vec{q}) \right|_{\vec{q}=\vec{q}_t} \circ \rmd \vec{q}_t \right\rangle_{\mu_0}, 
\]
where we apply the time evolution equation in the Stratonovich form \eqref{def:position_overdamped} 
\begin{multline*}
\left\langle Q(\omega) \right\rangle_{\mu_0} 
= \int \left\langle \vec{\nabla}_\vec{q} U_\alpha(\vec{q}) \cdot \left[ - \chi_\alpha(\vec{q}) \cdot \vec{\nabla}_\vec{q} U_\alpha(\vec{q}) + \frac{1}{2} \vec{\nabla}_\vec{q} \cdot D_\alpha(\vec{q}) \right] \right\rangle_{\mu_t} \; \rmd t + \\
+ \left\langle \int \vec{\nabla}_\vec{q} U_\alpha(\vec{q}_t) \cdot \sqrt{2 D_\alpha(\vec{q}_t)} \circ \rmd \vec{W}_t \right\rangle_{\mu_0} .
\end{multline*}
Now we can apply the relation between It\^{o} and Stratonovich integral \eqref{equ:relation_Ito_Stratonovich} 
\begin{multline*}
\left\langle Q(\omega) \right\rangle_{\mu_0} 
= \int \left\langle \vec{\nabla}_\vec{q} U_\alpha(\vec{q}) \cdot \left[ - \chi_\alpha(\vec{q}) \cdot \vec{\nabla}_\vec{q} U_\alpha(\vec{q}) + \frac{1}{2} \vec{\nabla}_\vec{q} \cdot D_\alpha(\vec{q}) \right] \right\rangle_{\mu_t} \; \rmd t + \\
+ \frac{1}{2} \int \left\langle \tr \left[ \sqrt{2 D_\alpha(\vec{q})} \, \vec{\nabla}_\vec{q} \cdot \left( \sqrt{2 D_\alpha(\vec{q})} \cdot \vec{\nabla}_\vec{q} U_\alpha(\vec{q}) \right) \right] \right\rangle_{\mu_t} \; \rmd t + \\ 
+ \left\langle \int \vec{\nabla}_\vec{q} U_\alpha(\vec{q}_t) \cdot \sqrt{2 D_\alpha(\vec{q}_t)} \cdot \rmd \vec{W}_t \right\rangle_{\mu_0} .
\end{multline*}
From where immediately follows
\[
q_\alpha(\vec{q}) = - \vec{\nabla}_\vec{q} U_\alpha(\vec{q}) \cdot \vec{\chi}_\alpha(\vec{q}) \cdot \vec{\nabla}_\vec{q} U_\alpha(\vec{q}) + \vec{\nabla}_{\vec{q}} \cdot \left[ \vec{D}_\alpha(\vec{q}) \cdot \vec{\nabla}_\vec{q} U_\alpha(\vec{q}) \right] .
\]

The reasoning in case when also external parameters $\alpha$ are time dependent is the same as for the underdamped diffusion, hence the work is also defined as 
\[
W(\omega) = \int \rmd t \; \dot\alpha(t) \cdot \left. \nabla_\alpha U_\alpha(\vec{q}_t) \right|_{\alpha=\alpha(t)} . 
\]

\subsubsection{Common properties}
We have verified that in both cases when we are in equilibrium, where there are no non-potential forces, the local heat production up to the sign reads
\begin{equation}
\gen_\alpha \left[ E_\alpha \right] = q_\alpha .
\label{equ:local_power_potential_underdamped}
\end{equation}

In non-equilibrium there can be in principle non-potential forces acting to the system. 
While the mechanical work dissipated by these forces can be considered as heat dissipated to the thermal bath at infinite temperature, 
because the stationary distribution is uniform with respect to the non-potential forces only, which corresponds to $\beta \to 0^+$ in Boltzmann distribution, 
we rather consider the system attached to the ``work'' reservoir and hence the local power for non-potential forces is given by the mean value of mechanical work.  
We can see that work and heat is then distinguished purely by convention, 
which in non-equilibrium situation leads to non-uniques of certain quantities, as we will see in the chapter \ref{chapter:non-equilibrium_thermodynamics}.

\section{Global and local detailed balance condition}
\label{sec:detailed_balance}
While investigating the equilibrium of the continuous time Markov jump processes we have introduced the concept of \emph{global detailed balance condition}\index{detailed balance condition!global},
see subsection \ref{ssec:detailed_balance_jump}.
The global detailed balance condition connects probabilities of transition forth and back with the equilibrium occupation probabilities under \emph{constant external conditions}. 
In general the statement can be written as 
\begin{equation}
\frac{ \dcprob[{(0,T]}][\alpha]{\omega}{X_0 = x_0 } }{ \dcprob[{(0,T]}][\pi \alpha]{\Theta \omega}{X_0 = \pi x_T }} = \frac{\rmd \rho_{\pi \alpha}(\pi x_T)}{\rmd \rho_\alpha(x_0)}, 
\label{def:global_detailed_balance}
\end{equation}
where $\alpha$ denotes the set of \emph{external parameters} and its \emph{fixed values}, 
$\omega$ denotes the path starting from configuration $x_0$ and ending in configuration $x_T$, 
$\rmd \rho_\alpha$ denotes the equilibrium probability measure given external conditions $\alpha$, 
$\pi$ is \emph{kinematic inversion}\index{kinematic inversion}, which changes the sign of the momentum, impulse momentum and similar quantities 
\begin{equation}
\pi: \quad (\vec{x},\vec{p}) \to (\vec{x},-\vec{p}),
\label{def:kinematic_inversion}
\end{equation}
or in case it is applied to external parameters it changes the sign of the magnetic fields and similar quantities
\[
\pi: \quad (\vec{E},\vec{B}) \to (\vec{E},-\vec{B}) ,
\]
and $\Theta \omega^{[0,T]}$ denotes \emph{time-reversed path}\index{path!time reversed} and is defined as 
\begin{equation}
\Theta \omega^{[0,T]} = \left\{ \pi X_{T-t} \middle| X_t \in \omega^{[0,T]} \right\}.
\label{def:time_reversal}
\end{equation}
Notice that the path probability measure $\rmd \mathbb P$ is a conditional probability with respect to initial configuration and values of external parameters.
If there are no momentum-like degrees of freedom in the system, like the typical case of Markov jump processes, the kinematic inversion reduces to identity, $\pi \equiv \id$, 
and hence the time reversal simplifies to what was introduced in \eqref{equ:time_reversal_jump}. 
Moreover by examining the paths with only one transition in case of continuous time Markov process one can obtain the global detailed balance condition \eqref{equ:global_detailed_balance_jump} 
as a result of the definition \eqref{def:global_detailed_balance}.

We also introduced the principle of \emph{local detailed balance}\index{detailed balance condition!local} at the end of subsection \ref{ssec:equlibrium_systems_jump}, 
which in general states that \emph{any} difference between the probability of transition forth and back can always be associated with \emph{entropy}\index{entropy} production in the environment 
\begin{equation}
k_B \ln \frac{ \dcprob[{(0,T]}][\alpha(t)]{\omega}{X_0 = x_0 } }{ \dcprob[{(0,T]}][\Theta \alpha (t)]{\Theta \omega}{X_0 = \pi x_T }} = S^{\text{bath}}_{\alpha(t)}( \omega ), 
\label{def:local_detailed_balance}
\end{equation}
where $\alpha(t)$ denotes the protocol describing the changes of external parameters, $\alpha(t) : \quad [0,T] \to \alpha$,
$S^{\text{bath}}_{\alpha(t)}(\omega)$ denotes the entropy production in the environment consisting of thermal and particle baths along the path $\omega$ given the protocol $\alpha(t)$ and $\Theta \omega$ again describes its time reversal \eqref{def:time_reversal}.
The $k_B$ is the \emph{Boltzmann constant}\index{Boltzmann constant} determining the physical dimension of entropy, which we will set to one in the rest of the text, $k_B \equiv 1$.

Although both these principles are consequence of \emph{microscopic time-reversibility}\index{time-reversibility}, 
we can see the fundamental difference between them in the fact that the local detailed balance condition connects path probabilities with entropy production, 
while the global detailed balance condition is a mathematical statement about invariance of the stationary density with respect to time reversal, 
which in case the configurations are invariant with respect to kinematic inversion leads to the condition of zero probabilistic currents along each possible transition in equilibrium. 
Moreover we consider the local detailed balance to be a more general principle, which is in general valid even out of equilibrium.

\subsubsection{Boltzmann distribution}
\index{Boltzmann distribution}
The Boltzmann distribution is the direct consequence of the simultaneous validity of the global \eqref{def:global_detailed_balance} and local \eqref{def:local_detailed_balance} detailed balance condition.
To simplify the situation we assume that system is attached to single thermal bath at inverse temperature $\beta$ and all force are given by potential $E_\alpha(x)$, which is invariant with respect to kinematic inversion. 
Moreover we restrict ourselves to the case when we hold the external parameters fixed $\alpha(t) \equiv \alpha$. 
As the first step we can see that the probability distribution of configurations in steady state is given by the entropy production in thermal bath 
\[
\frac{ \rho_{\pi \alpha}(\pi x_T) }{\rho_{\alpha}(x_0)} = \exp \left[ S^{\text{bath}}_{\alpha} (\omega ) \right] ,
\]
from where it immediately follows that the entropy production does not depend on the choice of the path $\omega$. 
Using also the fact that the entropy production in the thermal bath can be associated with heat transfered to the thermal bath from the system 
\[
S^{\text{bath}}_{\alpha}(\omega) = \beta Q^{\text{bath}}_{\alpha}(\omega) = - \beta Q^{\text{sys}}_{\alpha}(\omega) , 
\]
along with the fact that there is no work done while the external parameters are fixed \eqref{equ:local_power} and the fact that the change of the energy in case no work is done on the system can be associated with the heat \eqref{def:first_law_path}, we obtain the Boltzmann distribution 
\[
\frac{ \rho_{\pi \alpha}(\pi x_T) }{\rho_{\alpha}(x_0)} 
= \exp \left[ S^{\text{sys}}_\alpha(x_0) - S^{\text{sys}}_{\pi \alpha} (\pi x_T) \right] 
= \exp \left[ \beta \left( E_\alpha(x_0) - E_\alpha(x_T) \right) \right] .
\]

\subsection{Local detailed balance as a consequence of time-reversibility}
The reason why we can assume the validity of local detailed balance in most non-equilibrium situations is that the local detailed balance condition on mesoscopic scale is a consequence of time-reversibility of the underlying microscopic evolution. 
This claim was proved by Maes and Neto\v{c}n\'{y} in \cite{Maes2002} and later refined by Maes in \cite{Maes2003}, whom we will follow in this text.

In order to make arguments as clear as possible, we will restrict ourselves only to systems with classical Hamiltonian dynamics with fixed external conditions as an underlying dynamics, 
which simulates a mesoscopic open system in contact with single thermal path. 
We also restrict ourselves to the case where each configuration of the system corresponds to the single microstate of the mesoscopic system alone. 
Microstates of such system are points in the phase space $\Omega$ upon which the Hamiltonian dynamic acts.
The time-evolution from the time $0$ to time $T$ is represented by automorphism $\varphi_T$ on $\Omega$,
\[
x_T = \varphi_T (x_0), 
\]  
where $x_0$ is the initial microstate at time $0$ and $x_T$ is the final microstate. 
The time evolution has to be a semi-group
\[
\varphi_{T_2} \circ \varphi_{T_1} = \varphi_{T_1+T_2} 
\]
and in case of Hamiltonian dynamics it also preserves the \emph{Liouville measure}\index{Liouville measure} $\Gamma$, 
\begin{equation}
\Gamma \circ \varphi_T = \Gamma.
\label{equ:sym_of_Liouville_measure_time_evol}
\end{equation}
The microscopic reversibility of Hamiltonian dynamics states that 
if we reverse all the velocities of all particles, i.e. we apply kinematic inversion \eqref{def:kinematic_inversion} to the microstate, 
then let the system evolve over some period $T$ and then again inverse the velocities, we obtain a state before the time period $T$,
\[
\pi \circ \varphi_T \circ \pi = \varphi_{-T} .
\] 
With regard of kinematic inversion one can also observe that the Liouville measure is invariant with respect to kinematic inversion, 
\begin{equation}
\Gamma \circ \pi = \Gamma .
\label{equ:sym_of_Liouville_measure_kinematic}
\end{equation}

On the mesoscopic level of description the full information about the system is not accessible, hence we are not able to determine the exact microstate of the system, 
rather we represent the state of the system by its configuration on the mesoscopic level. 
We associate the configuration of the system with \emph{macrostate}\index{state!macro-} which corresponds to the specific values $a_i$ of macroscopic observables $A_i$, which are accessible. 
We denote by $\vec{A}$ the \emph{complete collection of the macroscopic observables} determining the macrostate $\vec{a}=\{\dots,a_i,\dots\}$. 
From the microscopic point of view macrostate is set of microstates with the same value of macroscopic observables,  
\[
\vec{A}^{-1}(\vec{a}) = \left\{ x \middle| x \in \Omega \land \vec{A} (x) = \vec{a} \right\} ,
\]
Because observables are functions, recall \eqref{def:observable}, each microstate also fully determines the macrostate, which belongs to $\vec{A}(x) = \vec{a}$.
Moreover we assume that macroscopic observables commute with kinematic inversion, $\vec{A} \circ \pi = \pi \circ \vec{A}$ and also $\pi \vec{A}^{-1}(\vec{a}) = \vec{A}^{-1}( \pi \vec{a})$.

Up to now we introduced the underlying microscopic dynamics and macrostates.
Furthermore we also need to introduce entropy to our description. 
We define \emph{the entropy}\index{entropy} for each macrostate $\vec{a}$ by variational principle as an extremal Shannon entropy \eqref{def:Shannon_entropy}
\begin{equation}
S(\vec{a}) = \sup_{\mu} \left[ - \int\limits_\Omega \rmd \Gamma(x) \; \mu(x) \ln \mu(x) \right] ,
\label{def:Shannon_entropy_extr}
\end{equation}
where $\mu$ denotes the probabilistic density of microstates representing the given macrostate $\vec{a}$. 
The supremum is taken over all probability densities $\mu$ under the conditions
\begin{align*}
\int\limits_\Omega \rmd \Gamma(x) \; \mu(x) &= 1 , & \supp \mu &= \vec{A}^{-1}(\vec{a}),
\end{align*}
where the first condition is the normalization and the second ensures that all microstates represents the given macrostate and hence 
\[
\int\limits_\Omega \rmd \Gamma(x) \; \mu(x) \, \vec{A}(x) = \vec{a} .
\]
We can see, that the supremal probability distribution $\mu^*_{\vec{a}}$ 
\[
S(\vec{a}) = - \int\limits_\Omega \rmd \Gamma(x) \; \mu^*_{\vec{a}}(x) \ln \mu^*_{\vec{a}}(x) 
\]
is uniform in $\vec{A}^{-1}(\vec{a})$ and hence the entropy of macrostate corresponds to the Boltzmann entropy 
\[
S(\vec{a}) = \ln \Gamma\left( \vec{A}^{-1}(\vec{a}) \right) . 
\]
Notice that the supremal probability distribution obeys the symmetry with respect to kinematic inversion
\begin{equation}
\mu^*_{\pi \vec{a}}(\pi x) = \mu^*_{\vec{a}}(x) .
\label{equ:sym_supremal_dist}
\end{equation}
%We associate the entropy of microstate with the entropy of the corresponding macrostate  
%\[
%S(x) = S(\vec{A}(x)) = \ln \Gamma\left( \vec{A}^{-1} (\vec{a}) \right) = - \ln \mu^*_\vec{a} (x) . 
%\]
For detailed discussion about the definition of entropy see \cite{Maes2003}.

Although we are not able to determine the exact microstate of the system we still can be asking 
what is the probability that we will observe a \emph{sequence of macrostates} while starting the evolution from a given \emph{macrostate} $\vec{a}_0$.
The probability of observing the sequence of macrostates $\omega=\{\vec{a}_t | t \in [0,T] \}$ when starting from macrostate $\vec{a}_0$ is given by 
\[
\cprob[{(0,T]}]{\omega}{X_0=\vec{a}_0}
= \frac{ \Gamma\left(\bigcap\limits_{t \in [0,T]} \varphi_{-t} \left( \vec{A}^{-1}(\vec{a}_t) \right) \right) }{ \Gamma\left( \vec{A}^{-1}(\vec{a}_0) \right) },
\]
where the denominator is the probability of observing macrostate $\vec{a}_0$ and the numerator corresponds to probability of observing the sequence of macrostates $\omega$. 
If we compare it with the probability of observing the reversed sequence of macroscopic observables $\Theta \omega$ starting from macrostate $\pi \vec{a}_T$ we obtain the local detailed balance 
\[
\ln \frac{\cprob[{(0,T]}]{\omega}{X_0=\vec{a}_0}}{\cprob[{(0,T]}]{\Theta \omega}{X_0 = \pi \vec{a}_T}} = \ln \Gamma\left( \vec{A}^{-1}(\vec{a}_T) \right) - \ln \Gamma\left( \vec{A}^{-1}(\vec{a}_0) \right) = S^{\text{bath}}(\vec{a}_T) - S^{\text{bath}}(\vec{a}_0),
\]
where we have used the invariance of the Liouville measure with respect to time evolution \eqref{equ:sym_of_Liouville_measure_time_evol} and kinematic inversion \eqref{equ:sym_of_Liouville_measure_kinematic}, 
and also the symmetry \eqref{equ:sym_supremal_dist}.
Notice that the change of Boltzmann entropy corresponds to the entropy change in the thermal bath, because each configuration corresponds to the single microstate of the observed system, 
but also to multiple microstates realizing the bath with inverse temperature $\beta$. 
Also notice that in this particular case we have not changed the external parameters of the system, so in this case we are in the transient regime.

\subsection{Local detailed balance for Markov jump processes} 
In the case of Markov jump process another line of thoughts can be followed. 
In subsection \ref{ssec:detailed_balance_jump} we have derived the local detailed balance condition for systems in equilibrium as a consequence of global detailed balance condition and Boltzmann distribution of microstates. 
In general case we assume that each jump, i.e. the change of microstate is associated with exchange of either particle or heat with a bath.
Furthermore we assume the baths in equilibrium and independent. 
In order to simplify the discussion we restrict ourselves only to the case of the system attached to multiple thermal baths. 
The independence of thermal baths ensures that the probability that two transitions coincide is zero and hence each transition in the system is associated with \emph{single} bath. 
The independence of thermal baths also ensures that the transition rates associated with the given thermal bath are independent while the other thermal baths are connected to the system or not.
Also along the single transition the system changes it's energy which is transfered to the corresponding reservoir and is directly associated with the production of the entropy there. 
Following these considerations we can conclude the local detailed balance condition is still even outside of equilibrium   
\[
\ln \frac{\rate[i]{x}{y}}{\rate[i]{y}{x}} = \beta_i \left( E(x) - E(y) \right) = S^{\text{bath}}(y) - S^{\text{bath}}(x) ,
\]
where $i$ denotes the thermal bath with inverse temperature $\beta_i$.

\section{Jarzynski and Crooks equalities}
In last twenty years the principle of local detailed balance condition or more generally the microscopical reversibility proved to be essential in developing fluctuation symmetries. 
First fluctuation symmetry was observed in numerical simulations of deterministic Hamiltonian evolution by Evans et al. \cite{Evans1993}, 
who also together with Searles provided first proof for such systems \cite{Evans1994}. 
In 1997 Jarzynski proved another fluctuation theorem the Jarzynski equality \cite{Jarzynski1997,Crooks1998}, 
which later Crooks showed as a special case of more general Crooks fluctuation relation \cite{Crooks1999,Crooks2000}. 
The first experimental verification of these relations were provided by Collin et al. in \cite{Ritort_ver_of_Crooks}. 
For review on the fluctuation theorems see \cite{Seifert2012}.

\emph{The Crooks fluctuation relation}\index{fluctuation relation!Crooks} states that the probability of the increase of the \emph{total entropy} $\Sigma$ along the given protocol of external parameters $\alpha(t)$ 
is exponentially larger then probability of the decrease of the total entropy $-\Sigma$ along the time reversed protocol $\Theta \alpha(t)$, 
\begin{equation}
\frac{\prob[{[0,T]}][\alpha(t)]{S^{\text{tot}}(\omega)=\Sigma}}{\prob[{[0,T]}][\Theta \alpha(t)]{S^{\text{tot}}(\omega)=-\Sigma}} = \rme^\Sigma , 
\label{equ:Crooks_relation}
\end{equation}
where \emph{the total entropy production}\index{entropy!total} consists of the change of the entropy of the system and of the entropy production in thermal bath 
\begin{equation}
S^{\text{tot}}_{\alpha(t)}(\omega) = \ln \mu_{\alpha(0)}(x_0) - \ln \mu_{\pi \alpha(T)}(\pi x_T) + S^{\text{bath}}_{\alpha(t)}(\omega) ,
\label{def:total_entropy}
\end{equation}
where $\mu_{\alpha(0)}(x_0)$ is the initial state under the external conditions $\alpha(0)$ 
and $\mu_{\pi \alpha(T)}(x_T)$ is the initial state for time-reversed process. 
Notice that the total entropy production is also antisymmetric with respect to kinematic inversion
\begin{equation}
S^{\text{tot}}_{\Theta \alpha(t)}(\Theta \omega) = - S^{\text{tot}}_{\alpha(t)}(\omega) . 
\label{equ:antisymmetry_total_entropy}
\end{equation}

As was stated before the proof of the Crooks fluctuation relation is centered around the local detailed balance condition \eqref{def:local_detailed_balance}.   
The probability to observe the total entropy production $\Sigma$ is given by 
\begin{align*}
\prob[{[0,T]}][\alpha(t)]{ S^{\text{tot}}(\omega) = \Sigma } 
&= \int \dprob[{[0,T]}][\alpha(t)]{\omega} \; \delta\left( S^{\text{tot}}_{\alpha(t)}(\omega) - \Sigma \right) \\
&= \int \rmd \mu_{\alpha(0)} (x_0) \, \dcprob[{(0,T]}][\alpha(t)]{\omega}{X_0 = x_0} \; \delta\left( S^{\text{tot}}_{\alpha(t)}(\omega) - \Sigma \right) , 
\end{align*}
where $\delta(\cdot)$ denotes the Dirac's delta function and where we have already extracted the initial condition from path probability. 
Inserting now the local detail balance condition and switching between the initial and final condition we obtain
\begin{multline*}
\prob[{[0,T]}][\alpha(t)]{ S^{\text{tot}}(\omega) = \Sigma } 
= \int \rmd \mu_{\pi \alpha(T)}(\pi x_T) \, \dcprob[{(0,T]}][\Theta \alpha(t)]{\Theta \omega}{X_0 = \pi x_T} \; \times \\ 
\times \rme^{S_{\alpha(t)}(\omega)} \frac{\mu_{\alpha(0)}(x_0)}{\mu_{\pi \alpha(T)}(\pi x_T)} \, \delta\left( S^{\text{tot}}_{\alpha(t)}(\omega) - \Sigma \right) .
\end{multline*}
Using the definition of total entropy \eqref{def:total_entropy} and its antisymmetry \eqref{equ:antisymmetry_total_entropy} along with the definition of Dirac's delta function concludes the proof.

\paragraph{Transient fluctuation relation}
The direct application of Crooks relation yields to the well known \emph{transient fluctuation symmetry}\index{fluctuation relation!transient-} for entropy 
\begin{multline}
\left\langle \rme^{-S^{\text{tot}}_{\alpha(t)}(\omega)} \right\rangle_{\mu_{\alpha(0)}}^{\alpha(t)} 
= \int \rmd \Sigma \; \left\langle \delta\left(S^{\text{tot}}_{\alpha(t)}(\omega)-\Sigma\right) \right\rangle_{\mu_{\alpha(0)}}^{\alpha(t)} \, \rme^{-\Sigma} = \\
= \int \rmd \Sigma \; \rme^{-\Sigma} \, \prob[{[0,T]}][\alpha(t)]{S^{\text{tot}}(\omega)=\Sigma} = \\
= \int \rmd \Sigma \; \prob[{[0,T]}][\Theta \alpha(t)]{S^{\text{tot}}(\omega)=-\Sigma} 
= 1 ,
\label{equ:transient_fluctuation_relation}
\end{multline}
where by $\langle \cdot \rangle_{\mu_{\alpha(0)}}^{\alpha(t)}$ we denote the mean value over all paths $\omega$ with the initial condition given by $\mu_{\alpha(0)}(x)$ and with the protocol of external parameters $\alpha(t)$.

\paragraph{Jarzynski equality}
When the system is connected to single thermal bath at constant inverse temperature $\beta$ and when the protocol $\alpha(t)$ connects two \emph{equilibrium} states then the total entropy reads
\[
S^{\text{tot}}_{\alpha(t)}(\omega) = \beta \left[ F_{\alpha(0)} - F_{\pi \alpha(T)} - E_{\alpha(0)}(x_0) + E_{\pi \alpha(T)}(\pi x_T) \right] + S^{\text{bath}}_{\alpha(t)}(\omega),
\]
where we have used the fact that the probability distribution of configurations in equilibrium in contact with single thermal bath is given by Boltzmann distribution
\[
\rho_\alpha(x) = \frac{1}{Z_\alpha} \rme^{- \beta E_\alpha(x)} = \rme^{\beta \left( {\mathcal F}_\alpha - E_\alpha(x) \right)},
\]
where ${\mathcal F}_\alpha$ denotes the equilibrium free energy and $E_\alpha(x)$ is the potential describing the energy landscape of the system, see e.g. \cite{Callen1985}. 
We also assume that the potential is symmetric with respect to kinematic inversion $E_{\pi \alpha(T)}(\pi x_T) = E_{\alpha(T)}(x_T)$ as well as the free energy is $ {\mathcal F}_{\pi \alpha(T)} = {\mathcal F}_{\alpha(T)} $.
Using also the fact that the entropy production in the thermal bath can be associated with heat transfered to the thermal bath from the system 
\[
S^{\text{bath}}_{\alpha(t)}(\omega) = \beta Q^{\text{bath}}_{\alpha(t)}(\omega) = - \beta Q^{\text{sys}}_{\alpha(t)}(\omega)  
\]
and the law of conservation of energy \eqref{def:first_law_path}, 
we obtain \emph{the Jarzynski equality}\index{fluctuation relation!Jarzynski equality} 
\begin{equation}
\left\langle \rme^{- \beta W^{\text{sys}}_{\alpha(t)}(\omega)} \right\rangle_{\mu_{\alpha(0)}}^{\alpha(t)} = \rme^{- \beta \left( {\mathcal F}_{\alpha(T)} - {\mathcal F}_{\alpha(0)} \right)} 
\label{equ:Jarzynski_equality}
\end{equation}
as a special case of the transient fluctuation relation \eqref{equ:transient_fluctuation_relation}.

\section{Second law inequality}
\index{second law of thermodynamics}
One of the most prominent results of the nineteenth century was the formulation of the second law of thermodynamics. 
One of those formulations of the second law states, that along any equilibrium process the total entropy is non-decreasing quantity, \cite{Callen1985}. 
Within our framework we cannot say such a definite statement without any other assumptions.

In most general case the Crooks relation \eqref{equ:Crooks_relation} states that the probability of the increase of the total entropy is exponentially larger than the probability of decrease by the same amount  
and hence can be considered as \emph{a weak formulation of the second law of thermodynamics}\index{second law of thermodynamics!weak formulation}. 
One of the consequences of the Crooks relation is the transient fluctuation relation \eqref{equ:transient_fluctuation_relation}, 
from which we can prove that also the mean total entropy production is non-decreasing quantity by using the \emph{Jensen's inequality}\index{Jensen inequality} 
\begin{equation}
\exp \left[ - \left\langle S^{\text{tot}}_{\alpha(t)}(\omega) \right\rangle_{\mu_{\alpha(0)}}^{\alpha(t)} \right] \le \left\langle \rme^{ - S^{\text{tot}}_{\alpha(t)}(\omega)} \right\rangle_{\mu_{\alpha(0)}}^{\alpha(t)} = 1 
\quad \Longrightarrow \quad 
\left\langle S^{\text{tot}}_{\alpha(t)}(\omega) \right\rangle_{\mu_{\alpha(0)}}^{\alpha(t)} \ge 0 .
\label{equ:second_law}
\end{equation}
The main feature of this particular formulation of the second law of thermodynamics is that it holds true for any initial condition $\mu$ and for an arbitrary protocol $\alpha(t)$ to the contrary of the classical thermodynamical formulation of the second law as seen for example in \cite{Callen1985}.

Using the definition of total entropy \eqref{def:total_entropy} we can see that the mean total entropy production consists of the difference of Shannon entropies \eqref{def:Shannon_entropy} for system and the entropy production in thermal bath
\[
S(\mu_{\pi \alpha(T)}) - S(\mu_{\alpha(0)}) + \left\langle S^{\text{bath}}_{\alpha(t)}(\omega) \right\rangle_{\mu_0} \ge 0 ,
\] 
which in case of the equilibrium process, which connects states described by Boltzmann distribution can be rewritten to the well known form of the second law of thermodynamics 
\[
{\mathcal W}^{\text{sys}}(\alpha(t)) \ge {\mathcal F}_{\alpha(T)} -  {\mathcal F}_{\alpha(0)} ,
\]
where $\mathcal F_\alpha$ denotes again the equilibrium free energy and ${\mathcal W}^\text{sys}(\alpha(t)$ is the mean value of the total work \eqref{equ:total_local_work_relation}.

\section{Quasistatic processes} 
\label{sec:quasistatic_processes_eq}
Up to now we have been discussing a general processes while we haven't made any specific assumptions about the trajectory in the external parameters space $\vec{\alpha}(t)$ nor about the system involved. 
At first we will restrict ourselves to equilibrium systems, hence we consider all the forces in the system to be of potential nature and also to be attached only to single thermal reservoir. 
The non-equilibrium quasistatic processes will be discussed in later chapters, while they are the main object of interest of this work.
In equilibrium thermodynamics the equilibrium or in other words quasistatic processes play a prominent role in the definition of thermodynamic entropy or thermodynamical potentials in general. 
The quasistatic process in general can be defined as the limiting process when we rescale the velocity of the changes of external parameters $\dot{\alpha}$ to zero, while preserving the path
\begin{align}
\alpha(t) & \longrightarrow \alpha(\epsilon t) , &
t \in [0,T] & \longrightarrow t \in \left[ 0, \frac{T}{\epsilon} \right] ,
\label{equ:time_rescaling}
\end{align}
where $\epsilon$ denotes the scaling parameter characterizing the magnitude of the velocity $\epsilon \, \dot{\alpha} (\epsilon t)$. 
The quasistatic process is then obtain by taking the limit $\epsilon \to 0^+$.

It is reasonable to further assume that the state of the system at any time $\mu_t$ during the quasistatic process is in the vicinity of the corresponding equilibrium state $\rho_{\alpha(\epsilon t)}$, 
also it is reasonable to assume that the difference is diminished with $\epsilon \to 0^+$, while the system has more time to relax closer to equilibrium. 
Moreover we can also assume that the difference between the equilibrium state and an actual state has to change on the same time scale for it not to exceed bounds as time goes on,
so in general we assume the solution of the time evolution \eqref{equ:Markov_stochastic_time_evolution} to be  
\[
\mu_t (x) = \rho_{\alpha(\epsilon t)} (x) + \epsilon \, \Delta \mu_{\epsilon t} (x) ,
\]
where $\Delta \mu_t$ denotes the difference. 
The time evolution equation \eqref{equ:Markov_stochastic_time_evolution} then gives us the prescription how to determine the $\Delta \mu_t$  
\[
\epsilon \, \dot\alpha(\epsilon t) \cdot \left. \nabla_\alpha \rho_\alpha (x) \right|_{\alpha=\alpha(\epsilon t)} + \epsilon^2 \left. \partial_s \Delta \mu_s (x) \right|_{s=\epsilon t} 
= \gen^*_{\alpha(\epsilon t)} \left[ \rho_{\alpha(\epsilon t)} + \epsilon \, \Delta \mu_{\epsilon t} \right] (x) ,
\]
which up to the first order in $\epsilon$ leads to 
\begin{equation}
\dot\alpha(\epsilon t) \cdot \left. \nabla_\alpha \rho_\alpha (x) \right|_{\alpha=\alpha(\epsilon t)} = \gen^*_{\alpha(\epsilon t)} \left[ \Delta \mu_{\epsilon t} \right] (x) ,
\label{equ:equation_first_order_density_correction}
\end{equation}
where we have used that the equilibrium state $\rho_\alpha$ is stationary under the equilibrium dynamic.

When we have determined the structure of the state of the system undergoing the quasistatic process we can proceed further. 
In the quasistatic limit we can directly see that the mean value of the total work in equilibrium \eqref{equ:local_power} is the \emph{geometric} integral over the path of external parameters 
\begin{align}
{\mathcal W}(\alpha) = \lim_{\epsilon \to 0^+} {\mathcal W}_{\rho_{\alpha(0)}}(\alpha(\epsilon t)) 
&= \lim_{\epsilon \to 0^+} \int\limits_0^{\frac{T}{\epsilon}} \rmd t \; \left\langle \epsilon \, \dot{\alpha}(\epsilon t) \cdot \left. \nabla_\alpha E_\alpha (x) \right|_{\alpha=\alpha(\epsilon t)} \right\rangle_{\mu_t} \nonumber \\
&= \int \rmd \alpha \cdot \left\langle \nabla_\alpha E_\alpha \right\rangle_{\rho_\alpha} , 
\label{equ:quasistatic_work_eq}
\end{align}
where $\langle \nabla_\alpha E_\alpha \rangle_{\rho_\alpha}$ is \emph{the thermodynamic force}\index{thermodynamic force}.   
As an example let us consider Heisenberg model in the presence of the constant external magnetic field $\vec{H}$ with the Hamiltonian 
\[
H(\{\vec{\sigma}_i\}) = H_{\text{int}}(\{\vec{\sigma}_i\}) + g \vec{H} \cdot \sum_i \vec{\sigma}_i ,
\]
where $H_{\text{int}}(\{\vec{\sigma}_i\})$ describes the interaction of the spins independent of the external field. 
One can see that the thermodynamic force corresponding to the quasistatic change of external magnetic field is the magnetization in agreement with thermodynamics  
\[
{\mathcal W}(\alpha) = - \int \rmd \vec{H} \cdot \vec{M} .
\]

For the heat the situation is bit more complicated. 
We start from the definition \eqref{equ:local_heat_production} by transferring the backward Kolmogorov generator from the potential to the total probability density by using \eqref{def:backward_generator}, 
then we can finally apply the first order correction \eqref{equ:equation_first_order_density_correction}
\begin{multline*}
{\mathcal Q}(\alpha) = \lim_{\epsilon \to 0^+} {\mathcal Q}_{\rho_{\alpha(0)}}(\alpha(\epsilon t))  
= \lim_{\epsilon \to 0^+} \int\limits_0^{\frac{T}{\epsilon}} \rmd t \; \left\langle \gen_{\alpha(\epsilon t)} \left[ E_{\alpha_{\epsilon t}} \right] \right\rangle_{\mu_t} = \\
= \lim_{\epsilon \to 0^+} \int\limits_0^{\frac{T}{\epsilon}} \rmd t \; \epsilon \, \dot\alpha(\epsilon t) \cdot \int \rmd \Gamma(x) \; \left. \nabla_\alpha \rho_\alpha(x) \right|_{\alpha=\alpha(\epsilon t)} E_{\alpha_{\epsilon t}}(x)  = \\
= \int \rmd \alpha \cdot \int \rmd \Gamma(x) \; \nabla_\alpha \rho_\alpha(x) \, E_\alpha(x) .
\end{multline*}
From where we can see that in the quasistatic limit \emph{the first law of thermodynamics}\index{first law of thermodynamics} can be written as 
\begin{multline*}
\int \rmd \alpha \cdot \nabla_\alpha \left\langle E_\alpha \right\rangle_{\rho_\alpha} = \\  
= \int \rmd \alpha \cdot \left\langle \nabla_\alpha E_\alpha \right\rangle_{\rho_\alpha}
+ \int \rmd \alpha \cdot \int \rmd \Gamma(x) \; E_\alpha(x) \, \nabla_\alpha \rho_\alpha(x) = \\ 
= {\mathcal W}(\alpha) + {\mathcal Q}(\alpha). 
\end{multline*}

In the case the equilibrium state is described by Boltzmann distribution at inverse temperature $\beta$ the mean value of the total heat can be expressed by covariance
\[
{\mathcal Q}(\alpha) = \int \rmd \alpha \cdot \left[ \left\langle \nabla_\alpha \left( \beta E_\alpha \right) \right\rangle_{\rho_\alpha} \left\langle E_\alpha \right\rangle_{\rho_\alpha} 
- \left\langle \nabla_\alpha \left( \beta E_\alpha \right) E_\alpha \right\rangle_{\rho_\alpha} \right] ,
\]
or we can obtain \emph{the second law of thermodynamics}\index{second law of thermodynamics} in the form of relation between the total heat $\mathcal Q$ and a total differential of entropy with the temperature $1/\beta$ being the Lagrange multiplier 
\begin{equation}
{\mathcal Q}(\alpha) 
= - \int \rmd \alpha \cdot \frac{1}{\beta} \nabla_\alpha \left\langle \ln \rho_\alpha \right\rangle_{\rho_\alpha} 
= \int \frac{1}{\beta} \; \rmd S(\alpha) 
\label{equ:Clausius_relation}
\end{equation}
where $S(\alpha)$ is \emph{the Shanon entropy}\index{entropy!Shanon} 
\begin{equation}
S(\alpha) = - \left\langle \ln \rho_\alpha \right\rangle_{\rho_\alpha} ,
\label{def:Shannon_entropy}
\end{equation}
and $\rho_\alpha$ is the Boltzmann distribution.

\subsection{Quasistatic work fluctuations}
The total work and the total heat are random quantities and as such one can be interested in the behaviour of the fluctuations in the quasistatic limit.
Although the problem is still open,
some partial results based upon the general validity of the Crooks relation \eqref{equ:Crooks_relation} and Jarzynski equality \eqref{equ:Jarzynski_equality} can be obtained. 
Namely the consequence of the validity of the Jarzynski equality in the quasistatic limit is that the difference between the free energy and the mean value of the total work is given by the variance of the total work in the leading order. 
To sketch the proof we start from the Jarzynski equality in the form
\[
\rme^{\beta \left( \left\langle W(\omega) \right\rangle - \Delta {\mathcal F} \right) } = \left\langle \rme^{-\beta \left[ W(\omega) - \left\langle W(\omega) \right\rangle \right] } \right\rangle ,
\]
where we assume that in the quasistatic limit the mean value of the total work is close to the free energy difference as well as the probability distribution of the total work is in the quasistatic limit narrow and hence $W(\omega) - \left\langle W(\omega) \right\rangle$ can be considered as small parameter. 
The proposition is then obtained by expanding the terms up to the leading order
\[
 \left\langle W(\omega) \right\rangle - \Delta {\mathcal F} \approx \frac{\beta}{2} \left\langle \left( W(\omega) - \left\langle W(\omega) \right\rangle \right)^2 \right\rangle , 
\]
from where it follows that the probability distribution of the total work has the zero variance in the quasistatic limit. 
Notice that it does not necessarily ensures the probability distribution to be asymptotically Gaussian,
although in case of overdamped diffusion Speck and Seifert \cite{Speck2004} proved that the distribution of the total work is asymptotically in the quasistatic limit Gaussian.

% Stochastic non-equilibrium thermodynamics 
% a. Intro - say what is new
% b. Detailed balance breaking, local detailed balance principle and its immediate consequences - non-uniques of energy function, gauge invariance, ... 
% c. review of previous methods i) Hatano-Sasa, ii) Komtatsu-Nakagawa-Sasa-Tasaki
% d. our basic results for systems driven by non-conservative forces: work and heat excess in terms of quasipotential, interpretation in terms of relaxation processes, "renormalized" quasistatic First law, discussion about generalized Clausius, general expression for heat capacity and other response quantities of interest
% e. model calculations  

\chapter{Stochastic non-equilibrium thermodynamics}
\label{chapter:non-equilibrium_thermodynamics}
At the beginning of this chapter we will briefly discuss consequences of the breaking of the global detailed balance such as the non-uniques of the energy function on the system or ``gauge'' freedom. 
Then we will review some of the approaches to the description of non-equilibrium stochastic systems, namely the approach by Hatano-Sasa \cite{Hatano2001} and Komatsu-Nakagawa-Sasa-Tasaki \cite{},
which will be followed by the introduction of our approach to the problem of non-equilibrium stochastic system in case the system is driven by non-potential forces or is attached to multiple thermal baths. 
In that section we will also show our first results \cite{result1,result2}.
At the end we will demonstrate the results on several models.

\section{Global detailed balance breaking}
\label{sec:broken_global_detailed_balance}
In the previous chapter \ref{chapter:equilibrium_stat_phys} in the section \ref{sec:detailed_balance} we have introduced the concepts of the global \eqref{def:global_detailed_balance} and local \eqref{def:local_detailed_balance} detailed balance. 
While the global detailed balance condition connects the probability of the path compared to the probability of its time reversal with probabilities of configurations in the steady state,
the local detailed balance condition relates the ratio of the probability of the path $\omega$ with the probability of its time reversal to the entropy production in thermal baths upon which the system is coupled to. 
As a consequence the global detailed balance along with local detailed balance condition ensures that there are no macroscopic currents in the system in the steady state, 
however this is no longer valid out of equilibrium.
As we assume that the underlying microscopic time evolution for the system together with the thermal baths is still time reversible, which is not necessarily valid for the system itself, thus ensuring the validity of local detailed balance condition and hence the global detailed balance condition has to be broken.

There are several natural ways how to drive system out of equilibrium. 
The first one is the introduction of time-dependent driving to the system, 
which causes that even in the steady state there is in general a current of energy of particles through the system as the system constantly changes its state to adjust itself to ever-changing external conditions. 
In these systems the notion of stationarity is quite different from other cases, the steady state if exists is such periodical state of the system to which the system converge from arbitrary initial state, 
if we assume the ergodicity of the dynamics.
While the steady state is periodical it is not necessary for the steady state to be also invariant with respect to the time reversal. 
We will discuss a class of these systems in separately in chapter \ref{chapter:periodically_driven_systems}.

The second way is by attaching the system to multiple thermal or particle baths. 
In general each thermal bath can be at different temperature and hence in the steady state we obtain a steady heat current through the system.
From mesoscopic point of view each possible transition between two configurations of the system can be associated with the action of any of the thermal baths.
In that case the full dynamics given by the collective action of all thermal baths together, 
so in general the steady state is given by the balance of the probability currents, 
hence global detailed balance is not necessarily valid for each particular transition associated with a single thermal bath.

The third way is introduction of non-potential force to the system, which may represent the effective mechanical action of surrounding media, e.g. rotational forces acting on colloidal particle in a suspension.  
In such case there is alway work being done on the system, which has to be dissipated out of the system to the thermal bath. 
From mesoscopic point of view it corresponds to the constant energy current through the system in the steady state.
As in the previous case the steady state is again determined by collective action of the thermal bath and with the non-potential force, which also means that there are non-zero probability currents present in the system.  
Hence by the same reasoning as in previous case the global detailed balance is no longer valid. 
Moreover by introducing the non-potential forces the energy function is also no longer unique. 
In the situation, when there were no non-potential forces acting on the system, it was natural to associate the energy of each particular configuration $x$ with the potential of the total force.  
However in the situation when there are non-potential forces acting on the system, 
we have the freedom in dividing of the total force acting on the system to the non-potential and the potential components, 
an arbitrary part of the potential force can be included in the non-potential force while the system's response, which is determined by the total force applied to the system, remains the same. 
This \emph{``gauge'' invariance}\index{gauge invariance}
\begin{equation}
\begin{aligned}
\vec{F}_\text{pot} = - \nabla U \quad & \longrightarrow \quad \widetilde{\vec{F}}_\text{pot} = - \nabla U - \nabla V \\
\vec{F}_\text{nonpot} \quad & \longrightarrow \quad \widetilde{\vec{F}}_\text{nonpot} = \vec{F}_\text{nonpot} + \nabla V 
\end{aligned}
\label{equ:gauge_symmetry}
\end{equation}
tells us that the potential, which we associated with the energy, is no longer unique and hence the energy function itself is also no longer unique.

%We have discussed the non-uniques of the energy function in the case, when the driving out of equilibrium is provided by the action of non-potential forces. 
%This also means that the entropy production associated with particular transition is no longer unique. 
%The entropy production in the thermal bath is given by the heat dissipated to the thermal bath divided by its temperature, 
%although the temperature is constant for each particular transition, the amount of the heat dissipated from the system can differ due to the non-uniques of the energy function. 
%In the case when the global detailed balance is broken by the system being attached to multiple thermal baths, there is also a non-uniques in the entropy production.  
%This time the amount of heat dissipated from the system is constant, what in general varies is the temperature of the thermal bath which triggered the transition. 
%We can conclude that in these systems the entropy production does no longer depend only on the initial and final configuration of the particular trajectory but on the whole trajectory itself containing the full information about the transition, which violates the global detailed balance condition \eqref{def:global_detailed_balance}. 

\section{Stochastic energetics}
In chapter \ref{chapter:equilibrium_stat_phys} in section \ref{sec:first_law} we have introduced the concept of stochastic energetics \eqref{def:first_law_path}. 
We have seen that the concept of stochastic energetics is microscopic version of the first law of thermodynamics. 
Taking the stochastic energetics given we have obtained the first law of thermodynamics \eqref{equ:first_law_eq} on the level of mean values of the total work and the heat follows. 
Moreover we have been able to fully characterize the mean values of the total work \eqref{equ:total_local_work_relation} and the heat \eqref{equ:total_local_heat_relation} by the local power \eqref{def:local_power} and the local heat production \eqref{def:local_heat_production}. 
All of these relations are direct consequences of the dynamic being Markovian and the validity of stochastic energetics,
which is reasonable to assume are to be valid also out of equilibrium, 
then we can consider them to be valid even out of equilibrium. 
The only essential difference there is in the particular definition of the work and heat along the given path and consequently in expressions for local power and local heat production.

In genuine non-equilibrium situation there is in general an additional contribution to the local power from non-potential forces or from the explicit dependence of the potential on the time 
\begin{equation}
w^{\text{tot}}_{\alpha(t)} = \dot\alpha(t) \cdot \left. \nabla_\alpha E_\alpha(x) \right|_{\alpha(t)} + w_{\alpha(t)}^{\text{nonp}}(x) .
\label{equ:total_local_power}
\end{equation}
In the situation where the system is connected to multiple thermal baths the situation become more complicated, 
because we need to track down the heat dissipated to each thermal bath respectively $Q_{\alpha(t)}^i(\omega)$, 
hence the total local heat production in all baths is the sum of all contributions over all baths 
\[
q^{\text{tot}}_{\alpha(t)}(x) = \sum_i q^i_{\alpha(t)}(x) . 
\]
Notice also that the local heat production for a single bath in case the system is attached to multiple thermal baths does not necessarily correspond to the local heat production when the bath is attached alone.

\section{Quasistatic processes}
\label{sec:quasistatic_processes_noneq}
In previous sections we have discussed several approaches to non-equilibrium phenomenona in small systems. 
In our attempt to approach the issue we have studied the quasistatic processes connecting various steady states,
which can be considered as a natural extension of equilibrium processes. 
In equilibrium the quasistatic or in other words equilibrium processes are essential in formulation of thermodynamics \cite{Callen1985}, 
e.g. the first law of thermodynamics states that the work done on the system along an arbitrary adiabatic equilibrium process corresponds to the change of the internal energy. 
Also one of the equivalent variant of the second law of thermodynamics connects the heat production with the entropy production, see chapter \ref{chapter:equilibrium_stat_phys} for further details. 
The fact that the work along any close adiabatic or isothermal trajectory is zero tells us about the existence of thermodynamical potentials, 
in those cases the internal energy $\mathcal U$ and the free energy $\mathcal F$. 
Similarly the fact that the heat along the closed isochoric trajectory is zero gives us enthalpy $\mathcal H$. 
The basic question which we ask is, to which extent some of these properties are still valid out of the equilibrium? 
Our approach is very similar to the approach of Komatsu-Nakagawa-Sasa-Tasaki, 
the main difference is that we rather focus on the energetic of quasistatic processes on the level of mean values and on the properties of generalized response functions, namely the generalized heat capacity,
while their main aim was to find the non-equilibrium version of Clausius theorem.

Komatsu-Nakagawa-Sasa-Tasaki addressed  the problem of the quasistatic limit by approximating the quasistatic process by a step process.  
In this section we will show in full a newly developed approach to the quasistatic limit based on an extension of the standard adiabatic theorem. 
which was already partially shown in the section \ref{sec:quasistatic_processes_eq}, where we have addressed the quasistatic processes in equilibrium.
In the quasistatic limit the mean values of path observables such as the total heat or work will prove to naturally decompose to the diverging ``housekeeping'' component and the finite ``reversible'' component. 
Where the ``reversible'' component will be responsible for the energetics of the system, and thus enables us to define generalized response functions, namely generalized heat capacity. 
We will briefly address the problem of general non-existence of the Clausius relation for the ``reversible'' components and its consequences.
We also discuss the consequences of the broken detailed balance condition such as non-uniques of the internal energy. 
These results will be illustrated on the series of examples. 
Most of these results can be found in \cite{result1,result2}.

\subsection{Probability distribution}
As the first step towards thermodynamics based on quasistatic processes we study the behaviour of the state of the system undergoing the quasistatic process.
To simplify the situation we will assume that further on the only explicit time dependence lies in external parameters $\alpha$,
hence the global detailed balance is broken either by attaching the system to multiple thermal bath or by the application of the non-potential force. 
In equilibrium the external parameters are usually quantities like volume, temperature or pressure, however we are not limited only to those. 
We only assume that the set of external parameters uniquely determines the steady state of the system $\rho_\alpha$ for all possible values of $\alpha$. 
The quasistatic process is the limiting process where the speed of the changes of external parameters is scaled to zero along with the time duration going to infinity \eqref{equ:time_rescaling}
\begin{align*}
\alpha(t) &\to \alpha(\epsilon t) & 
t \in [0,T] &\to t \in \left[ 0 , \frac{T}{\epsilon} \right] . 
\end{align*}  
The time evolution \eqref{equ:Markov_stochastic_time_evolution} of the state $\mu_t$ along the rescaled trajectory is given by  
\begin{equation}
\partial_t \mu_{t;\alpha(\epsilon t)} (x) = \gen^*_{\alpha(\epsilon t)} [\mu_{t;\alpha(\epsilon t)}] (x) ,
\label{equ:time_evolution_quasistatic}
\end{equation}
where we have explicitly denoted the dependence on external parameters $\alpha$. 
Being $\epsilon$ sufficiently small we can imagine, that the actual state can be considered as a perturbation of the steady state, 
as the system is in pursuit to reach the steady state.
We also assume that the small change of external parameters cannot cause a large change in the probability density neither in the stationary probability density,   
so we can assume the perturbation itself varies on the same time scale as external parameters
\[
\mu_{t;\alpha(\epsilon t)} (x) = \rho_{\alpha(\epsilon t)} (x) + \epsilon \Delta \mu_{\epsilon t} (x) .
\]
By expanding the time evolution \eqref{equ:time_evolution_quasistatic} up to the first order in $\epsilon$, which will prove to be sufficient, we obtain the equation for the perturbation
\[
\dot\alpha(\epsilon t) \cdot \left. \nabla_\alpha \rho_\alpha (x) \right|_{\alpha=\alpha(\epsilon t)} = \gen^*_{\alpha(\epsilon t)} \left[ \Delta \mu_t \right] (x) .
\]
Although the forward Kolmogorov generator is not in general invertible,
due to the fact that the steady state $\rho_\alpha$ corresponds to the zero eigenvalue of the forward Kolmogorov generator $\gen^*_\alpha$,
in this particular case the solution can be found
\begin{equation}
\mu_{t;\alpha(\epsilon t)} (x) = \rho_{\alpha(\epsilon t)} (x) 
+ \epsilon \dot\alpha (\epsilon t) \cdot \frac{1}{\gen^*_{\alpha(\epsilon t)}} \left[ \left. \nabla_\alpha \rho_\alpha \right|_{\alpha=\alpha(\epsilon t)} \right] (x) 
+ \err[2]{\epsilon} .
\label{equ:quasistatic_expansion_state}
\end{equation}
where we have introduced \emph{the forward pseudoinverse}\index{pseudoinverse!forward} 
\begin{equation}
\frac{1}{\gen^*} \left[ \mu \right] (x) = \int\limits_0^\infty \rmd t \; \left\{ \rho(x) \int \rmd \Gamma(y) \; \mu(y) - \rme^{t \gen^*} \left[ \mu \right] (x) \right\} .
\label{def:forward_pseudoinverse}
\end{equation}
From mathematician point of view the forward pseudoinverse is a linear operator which returns the argument of the forward Kolmogorov generator whenever can be found and zero otherwise,
i.e. the forward pseudoinverse is the extension of the inverse of the forward Kolmogorov generator from the subspace where the inverse exists up to the whole space.

\begin{figure}[ht]
\caption{Illustration of the transient contribution of the total work represented by pseudoinverse when relaxing toward steady state. 
The transient contribution is denoted by the filled area, while the total work also contains the hatched area.}
\begin{center}
\includegraphics[width=.7\textwidth,height=!]{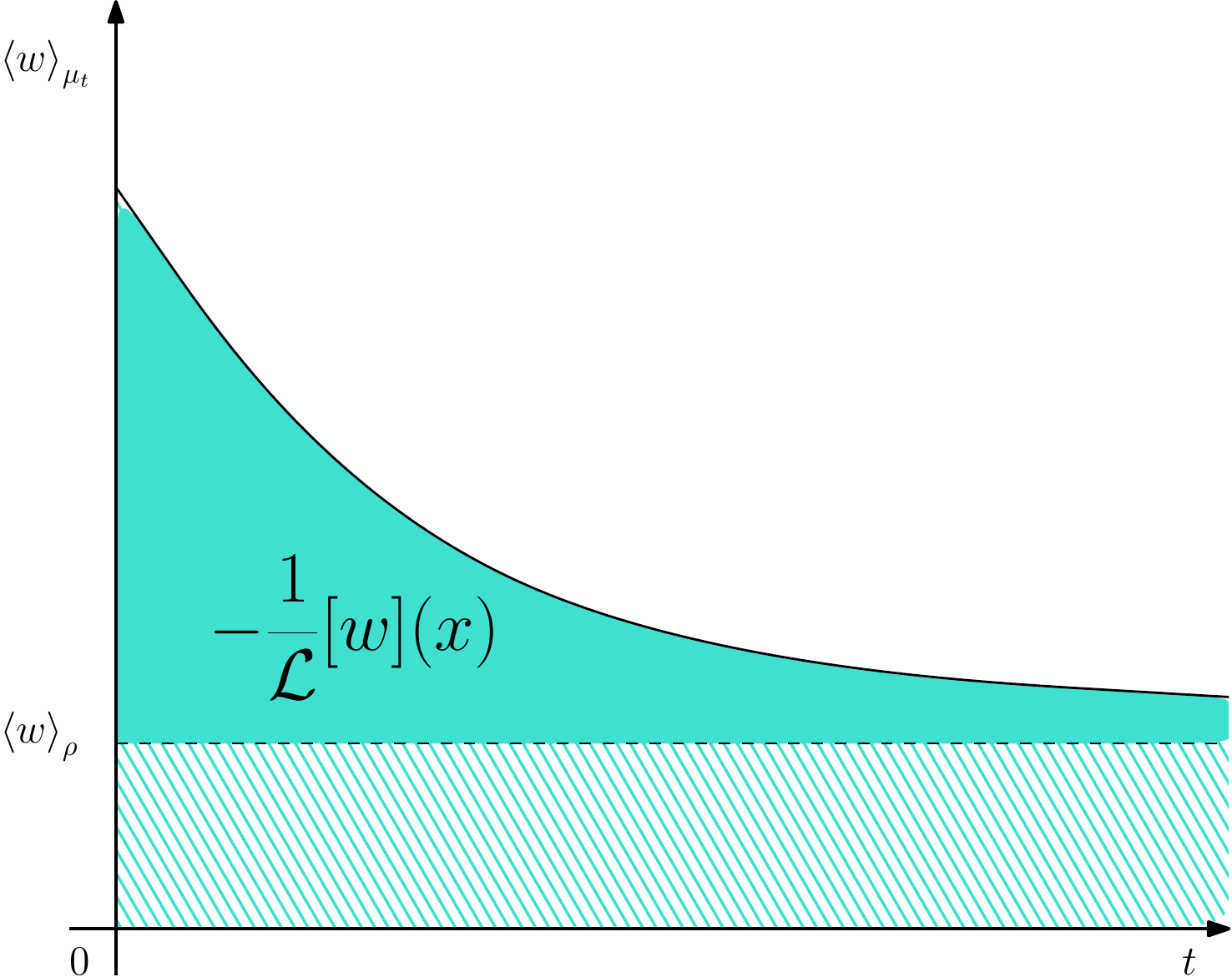}
\end{center}
\label{pic:backward_pseudoinverse}
\end{figure}

In order to understand the pseudoinverse from the physical viewpoint, 
we define at first \emph{the backward pseudoinverse}\index{pseudoinverse!backward} in a similar fashion how the backward Kolmogorov generator is related to the forward Kolmogorov generator \eqref{def:backward_generator} 
\[
\int \rmd \Gamma(x) \; w(x) \frac{1}{\gen^*}[\mu](x) = \left\langle \frac{1}{\gen}[w] \right\rangle_\mu ,
\]
which is equivalent to the definition  
\begin{equation}
\frac{1}{\gen}[w](x) = \int\limits_0^\infty \rmd t \; \left[ \left\langle w \right\rangle_\rho - \left\langle \rme^{t \gen}[w] \right\rangle_{\delta_x} \right] .
\label{def:backward_pseudoinverse}
\end{equation}
Now let us consider the $w$ being the local power, then the first term corresponds to the steady production of the work and the later term to the total work production along the relaxation process starting from the configuration $x$, 
hence the backward pseudoinverse can be interpreted as the transient or excess contribution to the total work, when the system relax from the initial configuration $x$ to the steady state $\rho$, see figure \ref{pic:backward_pseudoinverse}.  
The forward pseudoinverse can be then interpreted as the same transient contribution although in this case represented on states.

\subsection{Quasistatic work and heat}
\label{ssec:quasistatic_heat_and_work}
To obtain the quasistatic expansion of the mean value of the total work up to the zeroth order in $\epsilon$ 
we use the fact that the total mean value of the work can be expressed in terms of the local power \eqref{equ:total_local_work_relation}, 
to which we apply the quasistatic expansion of the state \eqref{equ:quasistatic_expansion_state} 
\begin{multline*}
{\mathcal W}(\alpha(t)) 
= \int\limits_0^{\frac{T}{\epsilon}} \rmd t \; \left\langle w^\epsilon_{\alpha(\epsilon t)} \right\rangle_{\mu_t} 
= \int\limits_0^{\frac{T}{\epsilon}} \rmd t \; \left\langle w^\epsilon_{\alpha(\epsilon t)} \right\rangle_{\rho_{\alpha(\epsilon t)}} + \\ 
+ \int\limits_0^{\frac{T}{\epsilon}} \rmd t \; \epsilon \dot\alpha(\epsilon t) \cdot \int \rmd \Gamma(x) \; \frac{1}{\gen^*_{\alpha(\epsilon t)}} \left[ \left. \nabla_\alpha \rho_\alpha \right|_{\alpha=\alpha(\epsilon t)} \right] (x) \, w^\epsilon_{\alpha(\epsilon t)} (x) + \err{\epsilon} = \\
= \frac{1}{\epsilon} \int\limits_0^T \rmd t \; \left\langle w^\epsilon_{\alpha(t)} \right\rangle_{\rho_{\alpha(t)}} 
- \int\limits \rmd \alpha \cdot \left\langle \nabla_\alpha \frac{1}{\gen_\alpha} [w^\epsilon_\alpha] \right\rangle_{\rho_\alpha}  
+ \err{\epsilon} .
\end{multline*} 
By inserting the total local power \eqref{equ:total_local_power}
\[
w^\epsilon_\alpha = \epsilon \dot \alpha \cdot \nabla_\alpha U_\alpha (x) + w^{\text{nonp}}_\alpha (x) ,
\]
where the non-potential part no longer depends on $\epsilon$, 
we obtain a quasistatic expansion of the mean value of the total work 
\[
{\mathcal W}(\alpha(t)) 
= \frac{1}{\epsilon} \int\limits_0^T \rmd t \; \left\langle w^\text{nonp}_{\alpha(t)} \right\rangle_{\rho_{\alpha(t)}} 
+ \int\limits \rmd \alpha \cdot \left\langle \nabla_\alpha \left( U_\alpha - \frac{1}{\gen_\alpha} [w^\text{nonp}_\alpha] \right) \right\rangle_{\rho_\alpha}  
+ \err{\epsilon} .
\]
The first term sometimes referred to as \emph{``housekeeping''}\index{housekeeping work}\index{work!housekeeping} or steady component of the work \cite{result1} corresponds to the work necessary to maintain the system in the non-equilibrium state 
and as such diverge in the quasistatic limit as the amount of the work necessary to maintain steady state over the infinite period of time goes to infinite. 
In equilibrium there is no work supply necessary to maintain the equilibrium state, 
hence there is no steady production of work and so forth this term is zero in equilibrium leading to the total work being determined only by the second term, which is usually finite. 
The second term usually referred to as \emph{``reversible''}\index{reversible work}\index{work!reversible} or ``excess''\index{excess work}\index{work!excess} work \cite{result1,result2} is the additional work associated with transitions between steady states along the process.

We can see that these two components differs in some key aspects,
the ``housekeeping'' work is symmetric with respect to the trajectory reversal $\Theta \alpha$ while the ``reversible'' is antisymmetric. 
This is also quite different from the equilibrium situation where in the quasistatic limit the total work and just the part of it is antisymmetric with respect to trajectory reversal. 
The second difference lies in the dependence on the trajectory $\alpha$, 
while the ``housekeeping'' component does not depend on the trajectory itself but rather on which steady states were visited and for how long,
the ``reversible'' component depends only on the ``shape'' of the trajectory thus being geometric, i.e. it does not depend on the actual parametrization of the trajectory. 
To summarize, we can see that the ``housekeeping'' component is 
\begin{enumerate}
\item extensive in time,
\item non-zero unless the global detailed balance condition is satisfied,
%\item present even if we does not alter any of the external parameters of the system $\alpha$,
\item invariant with respect to the reversal of the external protocol $\alpha(t) \to \alpha(T-t) $,
\end{enumerate}
while the ``reversible'' component is 
\begin{enumerate}
\item finite in the quasistatic limit,
\item geometric,
\item antisymmetric under the reversal of the external protocol $\alpha(t) \to \Theta \alpha(t)$.
\end{enumerate}

\begin{figure}[ht]
\caption{Demonstration of the experimental accessibility of the ``housekeeping'' and ``reversible'' work on the simulation of the system containing a single diffusing particle in the optical trap.
The parameters of the model in the simulation are set to be: mass of the particle $m=1$, friction coefficient $\gamma=1$, inverse temperature of the environment $\beta=0.1$, spring constant of the quadratic potential $k=10$, and the distance of the center of the potential well from the origin $R=5$, period of driving $\tau=20$.
Where the quasistatic process consists of the periodical changing the distance $R$ by 5\% with the period of $T=1280$.
The red line is the mean value of the total work over an statistical ensemble of $8192$ independent particles, 
the green line is the same although shifted by half the period $T/2$ 
and the blue line denotes the actual time dependence of the difference between, which in the times $kT+T/2$, $k \in {\mathbb N}$, coincide with the ``reversible'' work. 
}
\begin{center}
\includegraphics[width=.85\textwidth,height=!]{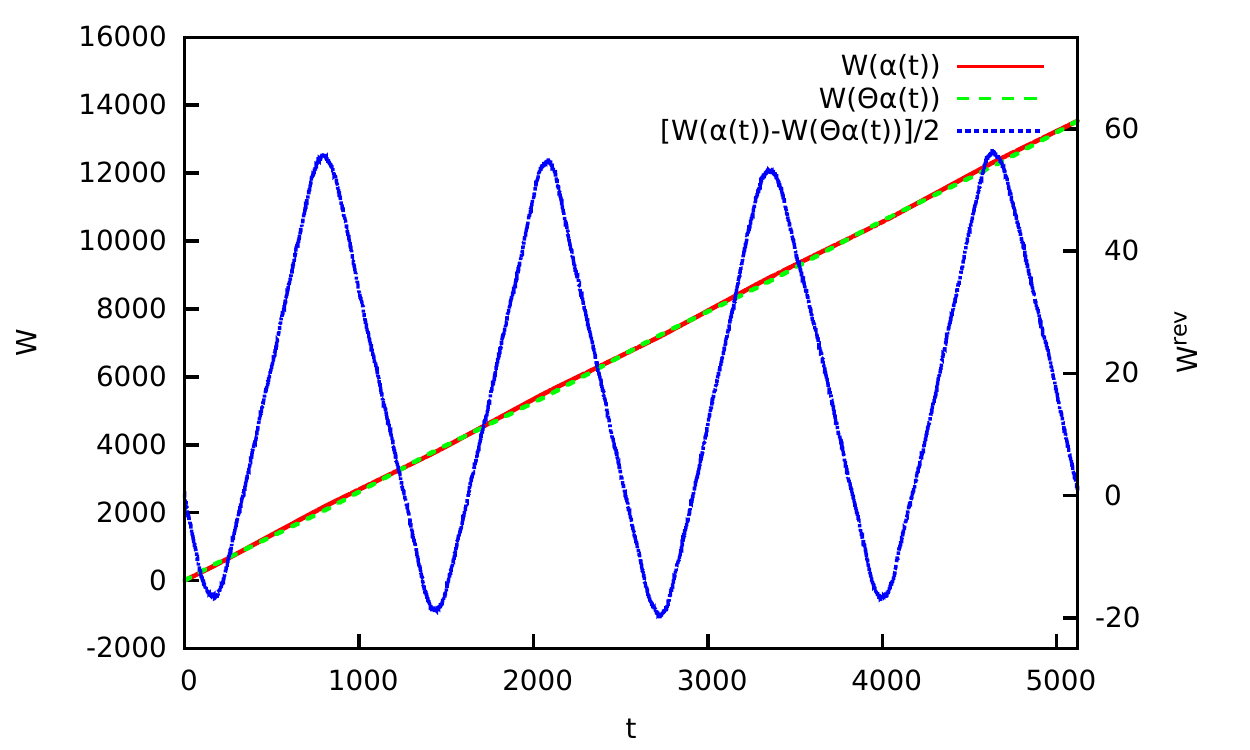}
\end{center}
\label{pic:work_decomposition}
\end{figure}

A typical experimentally easiest accessible quantity is the total work done on the system, 
however we can use the properties listed above namely the behaviour under trajectory reversal to separate these components apart. 
If the experimental setup can be made in such a way that we can measure the total work along some trajectory $\alpha(t)$ as well as along $\Theta \alpha(t)$ over the time interval $[0,T]$ 
then if the time interval $T$ is long enough to consider the process close to quasistatic, the ``housekeeping'' work can be estimated as the symmetric part of these works 
\begin{equation}
{\mathcal W}^{hk}(\alpha) 
\equiv \int\limits^T_0 \rmd t \; \left\langle w_{\alpha(t)} \right\rangle_{\rho_{\alpha(t)}} 
= \frac{1}{2} \left[ {\mathcal W}(\alpha(t)) + {\mathcal W}(\Theta \alpha(t)) \right] + \err{\epsilon}
\label{def:housekeeping_work}
\end{equation}
and the ``reversible'' work is estimated as the antisymmetric part
\begin{equation}
{\mathcal W}^{rev}(\alpha) 
\equiv \int \rmd \alpha \cdot \left\langle \nabla_\alpha \left( U_\alpha - \frac{1}{\gen} [ w_\alpha ] \right) \right\rangle_{\rho_\alpha} 
= \frac{1}{2} \left[ {\mathcal W}(\alpha(t)) - {\mathcal W}(\Theta \alpha(t)) \right] + \err{\epsilon}. 
\label{def:reversible_work}
\end{equation}
Although these estimates corresponds to the ``housekeeping'' and the ``reversible'' work respectively in the quasistatic limit, 
the fact that fluctuations of the total work usually diverge in the quasistatic limit along with the ``reversible'' work being \emph{non extensive} in time makes it hard to obtain,
for illustration see figure \ref{pic:work_decomposition}.

\paragraph{Example: Dragged particle}
To illustrate the experimental accessibility of the ``housekeeping'' and the ``reversible'' work we consider a numerical simulation of the underdamped diffusing particle in the optical trap in plane. 
The optical trap can be simulated by the quadratic potential. 
The system is driven out of equilibrium by dragging the particle by the optical trap in this particular case in circles with radius $R$ and period $\tau$. 
Such system can be equivalently described by non-potential angular force acting on the particle. 
The approximation of the quasistatic process is then achieved by periodically changing the radius $R$ on much longer time scale than the period of the driving $T \gg \tau$. 
The first half of the period $T$ is the radius $R$ linearly increasing, while the other half is linearly decreasing. 
Then if we compare the total mean work along such process with the same process shifted by half the period $T/2$, which can be considered as the reversed process,
we observe, see fig. \ref{pic:work_decomposition}, that in these two cases the lines showing the dependency of mean values of the total work on time almost coincide. 
We can see that the difference is of much smaller magnitude than the mean values of the total work, compare scales on left and right $y$ axis. 
The ``reversible'' work is then given as half of the difference of the mean values of the total heat over the half of the period $T/2$.  

The same computation can be repeated also for the mean value of the total heat dissipated to the system from the $i$-th thermal bath 
\[
{\mathcal Q}_i(\alpha(t)) 
= \frac{1}{\epsilon} \int\limits_0^T \rmd t \; \left\langle q^i_{\alpha(t)} \right\rangle_{\rho_{\alpha(t)}} 
- \int\limits \rmd \alpha \cdot \left\langle \nabla_\alpha \frac{1}{\gen_\alpha} [q^i_\alpha] \right\rangle_{\rho_\alpha}  
+ \err{\epsilon} , 
\]
where we identify \emph{the ``housekeeping'' heat}\index{housekeeping heat}\index{heat!housekeeping} 
\[
{\mathcal Q}_i^\text{hk} (\alpha) = \int\limits_0^T \rmd t \; \left\langle q^i_{\alpha(t)} \right\rangle_{\rho_{\alpha(t)}},
\]
and \emph{the reversible heat}\index{reversible heat}\index{heat!reversible} 
\begin{equation}
{\mathcal Q}_i^\text{rev} (\alpha) = - \int \rmd \alpha \cdot \left\langle \nabla_\alpha \frac{1}{\gen_\alpha} [ q^i_\alpha ] \right\rangle_{\rho_\alpha}. 
\label{def:reversible_heat}
\end{equation}
The ``housekeeping'' and the ``reversible'' heat have the same properties as the corresponding works. 
Again the ``housekeeping'' heat is to the heat necessary to maintain the steady states, which the system passed by. 
It is also symmetric with respect to trajectory reversal $\Theta \alpha$, is is extensive with time and also present even in case there is no change of external parameters. 
The ``reversible'' heat is antisymmetric with respect to trajectory reversal and is geometric.

\subsection{Generalized thermodynamics}
\label{ssec:generalized_thermodynamics}
\subsubsection{First law of thermodynamics}
One of the basic assumptions of our framework is that the underlying microscopic dynamic preserves the total energy on the microscopic level even under non-equilibrium conditions. 
This means that the first law of thermodynamic for each particular path \eqref{def:first_law_path} as well as on the level of mean values \eqref{equ:first_law_eq} is still valid 
\[
\left\langle E_{\alpha(T)} \right\rangle_{\rho_\alpha(T)} - \left\langle E_{\alpha(0)} \right\rangle_{\rho_\alpha(0)} = {\mathcal W}(\alpha) + \sum_i {\mathcal Q}_i(\alpha) ,
\]
where we sum over all baths attached to the system and the internal energy is represented by the steady mean value of the energy function ${\mathcal U}(\alpha) = \left\langle E_\alpha \right\rangle_{\rho_\alpha}$.

While the first law is valid for arbitrary trajectory $\alpha(t)$ it has to be also valid even in the quasistatic limit, 
even so that the mean value of the total heat and work diverge in the quasistatic limit. 
To be able to see that the diverging parts has to cancel each other out, we realize that in the steady state the total work done on the system is immediately dissipated to thermal baths
\[
0 = \left\langle w_\alpha \right\rangle_{\rho_\alpha} + \sum_i \left\langle q^i_\alpha \right\rangle_{\rho_\alpha} .
\] 
From there we can see that ``housekeeping'' parts in the quasistatic expansion cancel each other out, thus effectively renormalizing the mean values of the total work and heat 
\[
{\mathcal W}^\text{hk}(\alpha) + \sum_t {\mathcal Q}_i^\text{hk}(\alpha) = 
\int\limits_0^{\frac{T}{\epsilon}} \rmd t \; \left[ \left\langle w_{\alpha(\epsilon t)} \right\rangle_{\rho_{\alpha(\epsilon t)}} + \sum_i \left\langle q^i_{\alpha(\epsilon t)} \right\rangle_{\rho_{\alpha(\epsilon t)}} \right] 
= 0 .
\]
This means that in the quasistatic limit even out of equilibrium the energetics of the system is governed only by antisymmetric ``reversible'' components, 
\begin{equation}
{\mathcal U}(\alpha(T)) - {\mathcal U}(\alpha(0)) = {\mathcal W}^\text{rev}(\alpha) + \sum_i {\mathcal Q}_i^\text{rev}(\alpha) 
\label{equ:first_law_noneq} 
\end{equation}
which we will call the non-equilibrium version of \emph{the first law of thermodynamics}\index{first law of thermodynamics!non-equilibrium} in the quasistatic limit.
While the reversible components are usually finite, we have then found out the natural way how to renormalise the heat and work to obtain thermodynamically relevant quantities.

The ``reversible'' heat and work are given by geometric integral on configuration space, 
which we can use to introduce the differential version of the first law of thermodynamics \eqref{equ:first_law_noneq}
\[
\rmd_\alpha {\mathcal U}(\alpha) = \dbar {\mathcal W}^\text{rev}(\alpha) + \sum_i \dbar {\mathcal Q}^\text{rev}_i(\alpha) , 
\]
where $\dbar {\mathcal Q}^\text{rev}_i (\alpha) $ and $\dbar {\mathcal W}^\text{rev} (\alpha) $ denote \emph{inexact differentials}\index{inexact differential} 
\begin{align*} 
\dbar {\mathcal W}^\text{rev} (\alpha) &= \rmd \alpha \cdot \left\langle \nabla_\alpha \left( E_\alpha - \frac{1}{\gen_\alpha} [w^\text{nonp}_\alpha] \right) \right\rangle_{\rho_\alpha}, \\ 
\dbar {\mathcal Q}^\text{rev}_i (\alpha) &= - \rmd \alpha \cdot \left\langle \nabla_\alpha \frac{1}{\gen_\alpha} [q^i_\alpha] \right\rangle_{\rho_\alpha} . 
\end{align*}

\subsubsection{Generalized Clausius relation}
There exist in literature various proposals how to extend the (equilibrium) Clausius relation to non-equilibrium domains. 
A crucial point is that all these tentative generalizations heavily depend on the adopted scheme for the heat "renormalization", i.e., on the way how the finite component of the heat is actually defined. 
Whereas in some renormalization schemes the generalized Clausius equality is proved to be a mathematical identity, cf. \cite{Hatano2001,Esposito2010,trepagnier2004}, within the (supposedly more physical) scheme adopted here it posses a non-trivial problem.

In chapter \ref{chapter:equilibrium_stat_phys} we have shown the Clausius equality \eqref{equ:Clausius_relation} is the consequence of the local detailed balance condition \eqref{def:local_detailed_balance}. 
While we assume the local detailed balance condition to be still valid, one might naively expect the Clausius relation also to be valid. 
The main problem however lies in the fact that the total heat and hence the total entropy production along the quasistatic process diverge. 
This means that the second law \eqref{equ:second_law} \emph{does not} impose any condition on the ``reversible'' heat, which thus can in principle be arbitrary large. 
In general there is no Clausius relation relating the ``reversible'' heat \eqref{def:reversible_heat} to the entropy production,
which means that the existence of thermodynamical potentials and corresponding Maxwell relations known from equilibrium thermodynamics, or their generalized versions, is no longer guaranteed.

In the special case when the steady state can be written as Boltzmann-like distribution 
\begin{equation}
\rho_\alpha(x) = \frac{1}{Z_\alpha} \exp \left[ - \widetilde\beta(\alpha) \left( E_\alpha(x) - \frac{1}{\gen_\alpha} [ w^\text{nonp}_\alpha ] (x) \right) \right] 
\label{equ:generalized_Boltzmann}
\end{equation}
the Clausius relation can be again retrieved, where 
\begin{equation}
V_\alpha (x) = E_\alpha - \frac{1}{\gen_\alpha} [w^\text{nonp}_\alpha ]
\label{def:generalized_potential}
\end{equation}
is called \emph{the quasi-potential}\index{quasi-potential}
and $\widetilde\beta(\alpha)$ is an arbitrary function of external parameters representing \emph{the generalized inverse temperature}\index{generalized inverse temperature}. 
An example of such system is the diffusion on 2D plane with non-potential force in the angular direction and the potential force in the radial direction, for details see subsection \ref{ssec:diffusion_on_plane}. 
Another example is McLennan distribution in close to equilibrium situation, see section \ref{sec:first_order_expansion}.
The internal energy is thus given by the mean value of the quasi-potential 
\[
{\mathcal U}(\alpha) = \left\langle E_\alpha \right\rangle_{\rho_\alpha} 
= \left\langle E_\alpha - \frac{1}{\gen_\alpha} [ w^\text{nonp}_\alpha ] \right\rangle_{\rho_\alpha} 
= \left\langle V_\alpha \right\rangle_{\rho_\alpha} ,
\]
where we have used the identity \eqref{equ:backward_pseudoinverse_zero_mean} , 
and the ``reversible'' work is directly obtained from the definition \eqref{def:reversible_work}    
\[
\dbar {\mathcal W}^\text{rev} (\alpha) = \rmd \alpha \cdot \left\langle \nabla_\alpha V_\alpha \right\rangle_{\rho_\alpha} . 
\]
From there we can obtain the total ``reversible'' heat by using the first law of thermodynamics \eqref{equ:first_law_noneq} 
\begin{equation}
\dbar {\mathcal Q}^\text{rev}_\text{tot} (\alpha) = \rmd \alpha \cdot \left[ \nabla_\alpha \left\langle V_\alpha \right\rangle_{\rho_\alpha} - \left\langle \nabla_\alpha V_\alpha \right\rangle_{\rho_\alpha} \right] .
\label{equ:heat_capacity_quasi-potential}
\end{equation}
We can again introduce the Shannon entropy of the system \index{entropy!Shannon} 
\[
{\mathcal S} (\alpha) = - \left\langle \ln \rho_\alpha \right\rangle_{\rho_\alpha} ,
\]
which enables us to express the ``reversible'' heat as a generalized version of Clausius equilibrium relation, however in this case the Lagrange multiplier is generalized temperature $1/\widetilde\beta(\alpha)$, 
which in general does not depend only on the temperature of single or multiple thermal baths, but can depend also on other external parameters
\[
\dbar {\mathcal Q}^\text{rev}_\text{tot} (\alpha) = \frac{1}{\widetilde\beta(\alpha)} \; \rmd {\mathcal S}(\alpha) .  
\]

\subsubsection{Quasistatic response functions}
In equilibrium thermodynamics \emph{response functions}\index{response function} such as heat capacity or compressibility are related to the quasistatic change of a particular thermodynamical potential along the infinitesimal change of some external parameter \cite{Callen1985}, e.g. heat capacity at constant volume is the quasistatic change of the internal energy with respect to temperature, the compressibility can be related to the quasistatic isothermal change of the free energy with respect to pressure. 
Although in general there is no second law of thermodynamics out of equilibrium for ``reversible'' components and hence there are no thermodynamical potentials in general, 
it can be still meaningful to define response functions outside of equilibrium.

Response functions like the isothermal compressibility can be defined the same way as in equilibrium, if in this case the pressure $p$ and volume $V$ are also defined out of equilibrium, 
i.e. the isothermal compressibility is defined as  
\[
\kappa_t = - \frac{1}{V} \left. \partial_p V \right|_{T_i=\text{const.}} ,
\]
where by $\left. \partial_p V \right|_{T=\text{const.}}$ we denote the quasistatic change of the volume with respect to the pressure, 
while all temperatures $T_i$ of thermal baths attached to system are constant.

The key observation in case of the heat capacity and similar quantities is that the ``housekeeping'' heat (work) does not contribute to the change of internal energy \eqref{equ:first_law_noneq}.
Also the ``reversible'' heat is geometric and hence does not depend on the parametrization of the trajectory on external parameters $\alpha(t)$,
thus making the ``reversible'' heat the natural candidate to be used in the definition of the generalized heat capacity.  
We define \emph{the generalized heat capacity}\index{heat capacity!generalized} as 
\begin{equation}
C_i (\alpha) = \left. \frac{\dbar Q^\text{rev}_i (\alpha) }{\rmd T_i} \right|_\alpha = - \left\langle \partial_{T_i} \frac{1}{\gen_\alpha} [q^i_\alpha] \right\rangle_{\rho_\alpha} . 
\label{def:generalized_heat_capacity}
\end{equation}
Although the heat capacity defined in such way coincide with the standard heat capacity in equilibrium, as will be shown later, the non-existence of the second law of thermodynamics causes that the generalized heat capacity can be in general negative, as will be shown on examples.

We have already stated that the general non-validity of the Clausius equality on the level of ``reversible'' components cause the nonexistence of thermodynamical potentials, 
hence in general there are no relations between response functions as those of Mayer and Maxwell.

\subsubsection{Equilibrium case revisited} 
The equilibrium thermodynamics presented in the chapter \ref{chapter:equilibrium_stat_phys} especially in the section \ref{sec:quasistatic_processes_eq} can be considered as a special case of presented framework. 
To demonstrate it we derive again within our framework the results presented there by taking
\begin{align*} 
w^\text{nonp}_\alpha (x) &=0 , &
q_\alpha (x) &= \gen_\alpha [ E_\alpha ] (x) .
\end{align*} 
From where it immediately follows that ``housekeeping'' components are zero. 
Because the local power of non-equilibrium forces is zero, the ``reversible'' work is in this particular case given entirely by the change of the energy 
\[
{\mathcal W}^\text{rev}(\alpha) = \int \rmd \alpha \cdot \left\langle \nabla_\alpha E_\alpha \right\rangle_{\rho_\alpha} . 
\]
The ``reversible'' heat can be also simplified to 
\[
{\mathcal Q}^\text{rev}(\alpha) 
= - \int \rmd \alpha \cdot \left\langle \nabla_\alpha \frac{1}{\gen_\alpha} \gen_\alpha [ E_\alpha ] \right\rangle_{\rho_\alpha} 
= \int \rmd \alpha \cdot \left[ \nabla_\alpha \left\langle E_\alpha \right\rangle_{\rho_\alpha} - \left\langle \nabla_\alpha E_\alpha \right\rangle_{\rho_\alpha} \right] 
\]
by using the identity \eqref{equ:backward_pseudoinverse_identity}
\[
\frac{1}{\gen} \gen [A] (x) = A(x) - \langle A \rangle_\rho .  
\]

If we also assume that the equilibrium state is characterized by the Boltzmann or the Maxwell-Boltzmann distribution 
\[
\rho_\alpha (x) = \frac{1}{Z_\alpha} \rme^{- \beta E_\alpha (x) } 
\]
we obtain the ``reversible'' heat in the form of the Clausius relation 
\begin{equation}
{\mathcal Q}^\text{rev}(\alpha) 
= - \int \rmd \alpha \cdot \frac{1}{\beta} \nabla_\alpha \left\langle \ln \rho_\alpha \right\rangle_{\rho_\alpha} 
= \int \rmd \alpha \cdot \frac{1}{\beta} \nabla_\alpha {\mathcal S}(\alpha)
= \int \frac{1}{\beta} \; \rmd {\mathcal S}(\alpha) ,
\label{equ:reversible_heat_equilibrium}
\end{equation}
where ${\mathcal S}(\alpha) = - \langle \ln \rho_\alpha \rangle_{\rho_\alpha} $ denotes again the Shannon entropy.

Because the total energy does not depend on temperature the generalized heat capacity \eqref{def:generalized_heat_capacity} then simplifies to 
\[
C = \partial_T \left\langle E_\alpha \right\rangle_{\rho_\alpha} = \partial_T {\mathcal U}(\alpha) , 
\]
or by using expression \eqref{equ:reversible_heat_equilibrium} we obtain the standard definition 
\[
C = T \, \partial_T {\mathcal S}(\alpha) . 
\]

\subsection{``Gauge'' invariance}
\label{ssec:gauge_invariane}
\index{gauge transformation}
In the section \ref{sec:broken_global_detailed_balance} we have argued that the physics is invariant under the ``gauge'' transformation \eqref{equ:gauge_symmetry}, 
i.e. the physical results cannot depend on how we divide the total force to the potential and the non-potential component. 
As we have discussed in case of diffusion in chapter \ref{chapter:equilibrium_stat_phys} subsection \ref{ssec:work_heat_diffusion} the non-potential local power is given by the action of non-potential forces on the microscopic level.
If the non-potential power has also a component which can be characterized by some potential $U_\alpha(x)$, 
then the corresponding component of the local power is given by $w_\alpha(x) = - \gen[U_\alpha](x)$, see \eqref{equ:local_power_potential_underdamped}.
From there we can see that the ``gauge'' transformation \eqref{equ:gauge_symmetry} in terms of the local power is described by the transformation 
\begin{align*}
& E_\alpha(x) & 
&\longrightarrow&
& E_\alpha(x) + U_\alpha(x) , \\
w^\text{pot}_{\alpha(t)}(x) &= \dot \alpha (t) \cdot \left. \nabla_\alpha E_\alpha(x) \right|_{\alpha = \alpha(t)} &
&\longrightarrow&
\widetilde{w}^\text{pot}_{\alpha(t)}(x) &= \dot \alpha (t) \cdot \left. \nabla_\alpha \left[ E_\alpha(x) + U_\alpha(x) \right] \right|_{\alpha = \alpha(t)} , \\
& w^\text{nonp}_{\alpha(t)}(x) &
&\longrightarrow&
\widetilde{w}^\text{nonp}_{\alpha(t)}(x) &= w^\text{nonp}_{\alpha(t)}(x) + \gen_{\alpha(t)} [U_{\alpha(t)}] (x) . 
\end{align*}

The first thing to notice is that Kolmogorov generators and hence the steady state are invariant with respect to the ``gauge'' transformation. 
This is due to the fact, that the evolution of the system is determined by the total work and not by any particular decomposition to the potential and the non-potential force.

However how other physical quantities behave under the ``gauge'' transformation cannot be seen as easily. 
In this subsection we will focus on the behaviour of the work, heat and internal energy under the transformation further on. 
We can see that the internal energy is modified by the ``gauge'' transformation by the mean value of the additional potential $U_\alpha(x)$ 
\[
\widetilde{\mathcal U}(\alpha) = \left\langle E_\alpha + U_\alpha \right\rangle_{\rho_\alpha} = {\mathcal U}(\alpha) + \left\langle U_\alpha \right\rangle_{\rho_\alpha} 
\]
It is also easy to see that the ``housekeeping'' work is invariant with respect to the ``gauge'' transformation 
\[
\widetilde{\mathcal W}^\text{hk} (\alpha)
= \int\limits_0^T \rmd t \; \left\langle \widetilde{w}^\text{nonp}_{\alpha(t)} \right\rangle_{\rho_{\alpha(t)}}
= \int\limits_0^T \rmd t \; \left\langle {w}^\text{nonp}_{\alpha(t)} + \gen_{\alpha(t)} [ U_{\alpha(t)} ] \right\rangle_{\rho_{\alpha(t)}}
= {\mathcal W}^\text{hk} (\alpha) ,
\]
on the other hand the ``reversible'' part of the total work is not invariant with respect to the ``gauge'' transformation
\begin{multline*}
\dbar \widetilde{\mathcal W}^\text{rev}(\alpha) 
= \rmd \alpha \cdot \left\langle \nabla_\alpha \left( E_\alpha + U_\alpha - \frac{1}{\gen_\alpha} \left[ w^\text{nonp}_\alpha + \gen_\alpha [ U_\alpha ] \right] \right) \right\rangle_{\rho_\alpha} = \\
= \rmd \alpha \cdot \left\langle \nabla_\alpha \left( E_\alpha + \left\langle U_\alpha \right\rangle_{\rho_\alpha} - \frac{1}{\gen_\alpha} [ w^\text{nonp}_\alpha ] \right) \right\rangle_{\rho_\alpha} = \\
= \dbar {\mathcal W}^\text{rev}(\alpha)  + \rmd \alpha \cdot \nabla_\alpha \left\langle U_\alpha \right\rangle_{\rho_\alpha} 
= \dbar {\mathcal W}^\text{rev}(\alpha)  + \rmd \widetilde{\mathcal U}(\alpha) - \rmd {\mathcal U}(\alpha) ,
\end{multline*}
where we have again used the identity \eqref{equ:backward_pseudoinverse_identity}. 
We can see that the ``reversible'' work compensate the change of the internal energy under the ``gauge'' transformation. 
From where we can see that the ``reversible'' heat is invariant with respect to ``gauge'' transformation 
\[
\dbar \widetilde{\mathcal Q}^\text{rev} (\alpha) 
= \rmd \widetilde{\mathcal U} (\alpha) - \dbar \widetilde{\mathcal W}^\text{rev} (\alpha) 
= \rmd {\mathcal U} (\alpha) - \dbar {\mathcal W}^\text{rev} (\alpha) 
= \dbar {\mathcal Q}^\text{rev} (\alpha) ,
\]
where we have used the first law of thermodynamics \eqref{equ:first_law_noneq}. 
The invariance of the ``reversible'' heat is important because it tells us that the generalized heat capacity \eqref{def:generalized_heat_capacity} is also invariant with respect to the ``gauge'' transformation. 
Alternatively the invariance of the ``reversible'' heat can also be obtained from the fact, that the ``gauge'' transformation affects only the local power, while the local heat production is preserved. 
To be more precise the ``reversible'' heat depends only on the steady state, backward pseudoinverse and the local heat production, which are all invariant with respect to the ``gauge'' transform, which concludes the proof. 
The same reasoning can be made also for the ``housekeeping'' heat, from where we conclude that the ``housekeeping'' heat is also invariant with respect to the ``gauge'' transform 
\[
\widetilde{\mathcal Q}^\text{hk}(\alpha) = {\mathcal Q}^\text{hk}(\alpha) .
\]

To summarize the discussion, 
we can see that the mean value of the total heat as well as its components are invariant with respect to the ``gauge'' transformation, 
which means that the total entropy production associated with the heat does not depend on how we divide the total work to the non-potential and the potential component.  
Similarly the ``housekeeping'' work is also invariant with respect to the ``gauge'' symmetry, 
which reflects the fact, that the mean steady power, which is related to the physical state of the system, also does not depend on the choice of the non-potential work.

On the other hand the ``reversible'' work and the internal energy are not invariant with respect to the ``gauge'' symmetry as they are tightly bounded with the definition of the energy on the microscopic level. 
Thus we have there a certain freedom how to define the internal energy, in the extreme case the gauge can be fixed in such a way that the internal energy is uniformly zero. 
In equilibrium the notion of internal energy is also associated with the total energy which can be in principle extracted from the system by a quasistatic process. 
This is however no longer true in non-equilibrium steady, while there is constant energy current through the system. 
However we can asked slightly different question, how much \emph{additional} energy to the steady production are we able to extract from the system by any quasistatic process?
From this point of view the most natural gauge fixation is what is in electrodynamic called the Coulomb gauge\index{Coulomb gauge}, i.e. the divergence of the non-potential force is zero, 
which we will use in most cases in this thesis.

The last remark is that although the quasi-potential \eqref{def:generalized_potential} is also affected by the ``gauge'' transformation 
\begin{multline*}
\widetilde{V}_\alpha(x) 
= E_\alpha(x) + U_\alpha(x) - \frac{1}{\gen_\alpha} \left[ w^\text{nonp}_\alpha + \gen_\alpha [ U_\alpha ] \right] (x) = \\
= E_\alpha(x) - \frac{1}{\gen_\alpha} \left[ w^\text{nonp}_\alpha \right] (x) + \left\langle U_\alpha \right\rangle_{\rho_\alpha} 
= V_\alpha(x) + \left\langle U_\alpha \right\rangle_{\rho_\alpha} ,
\end{multline*}
because the effect of the transformation is given by the uniform shift of all energy levels only,  
the stationary distribution \eqref{equ:generalized_Boltzmann} as well as to the Shannon entropy are not affected by the ``gauge'' transformation.

\subsection{Example: Two-level model}
\label{ssec:two_level_nonp}
The first model we consider \cite{result2} is a system with two states `$0$' and `$1$' with energies $E(0)=0$ and $E(1) = \Delta E > 0$ that are connected by two distinct channels `$+$' and `$-$', see \ref{ssec:two_level}.
Each channel is associated with its respective thermal bath, in this particular case at the same inverse temperature $\beta$.
The asymmetry between the channels is provided by an additional driving force performing work 
\[
W^\text{nonp}\left(0 \xrightarrow{\pm} 1\right) = \pm F. 
\]
Hence, for the loop formed by the allowed transitions we have $W^\text{nonp}(0 \xrightarrow{+} 1 \xrightarrow{-} 0) = 2F$, 
manifesting a non-potential character of the driving force. 
From the expression for the heat \eqref{def:first_law_path}
\[
Q\left(0 \xrightarrow{\pm} 1\right) = \pm F - \Delta E
\]
we immediately see 
that the case $F > \Delta E$ (respectively $-F > \Delta E$) corresponds to a strong non-equilibrium regime in which the system dissipates a positive amount of energy along both transitions in the loop $0 \xrightarrow{+} 1 \xrightarrow{-} 0$ (respectively its reversal). 
Note that in this regime the (original) notion of energy gap separating both states and uniquely distinguishing between the ground and excited states becomes essentially meaningless.

In the most general case we can describe the system with transition rates 
\begin{align*}
\rate[\pm]{0}{1} &= A \exp \left[ \pm \frac{\Phi}{2} + \beta \frac{\pm F - \Delta E}{2} \right] , \\ 
\rate[\pm]{1}{0} &= A \exp \left[ \pm \frac{\Phi}{2} - \beta \frac{\pm F - \Delta E}{2} \right] ,
\end{align*}
where $A$ is the common symmetric part of the transition rates, which does not depends on the inverse temperature $\beta$, thus setting an overall time-scale, 
the $\Phi$ describes the direction-independent asymmetry between the symmetric parts of each channel. 
Although in general both parameters $A$ and $\Phi$ can possibly depend on other parameters like $\beta$, $F$ and $\Delta E$ in any non-trivial way,
we assume that they does not depend on the inverse temperature $\beta$. 
Such assumption isn't physically well motivated, 
however we use it simplify the situation and to better separate the non-potential and the additional channel-asymmetry effects.
While $A$ only sets an overall time-scale, it can be mostly ignored, hence for convenience, we set $A=1$ and always assume $\Phi \ge 0$, 
due to the symmetry of the dynamics to the dual exchange of $F \to -F$ and $\Phi \to - \Phi$.

\subsubsection{Steady state}
The steady state \eqref{equ:stationary_condition} coincides with steady state of the channel-unresolved two level system with the \emph{total} escape rates $\lambda(x) = \lambda_+(x) + \lambda_-(x)$ 
\begin{equation}
\frac{\rho(1)}{\rho(0)} = \frac{\lambda(0)}{\lambda(1)} = \rme^{- \beta \Delta E} \frac{1 + \zeta}{1 - \zeta} ,
\label{equ:2lvl_stat_occup}
\end{equation}
where 
\[
\zeta = \tanh \left( \frac{\Phi}{2} \right) \, \tanh \left( \frac{\beta F}{2} \right) .
\]
In case either of $F$ or $\Phi$ is zero we have a probability distribution corresponding to the Boltzmann equilibrium.

\paragraph{Stationary currents}
In the steady state we would expect to observe a non-zero constant probability current \eqref{equ:current_multiple_baths} in each particular channel 
\[
\curr[+]{0}{1} = \curr[-]{1}{0} = \frac{ \sinh\left( \frac{\beta F}{2} \right) }{ \cosh \left( \frac{ \Phi }{2} \right) \, \cosh \left( \frac{\beta \Delta E}{2} \right) \left[ 1 - \zeta \tanh\left( \frac{\beta \Delta E}{2} \right) \right] } ,
\]
from where it is evident that only the case $F = 0$ corresponds to the equilibrium. 
Notice that the current diverge when 
\[
\tanh \left( \frac{\Phi}{2} \right) \, \tanh \left( \frac{\beta F}{2} \right) \, \tanh\left( \frac{\beta \Delta E}{2} \right) = 1.
\]
It can be shown that the steady rate of dissipation from thermal bath associated with the $+$ channel to the system is the steady rate of dissipation from the system to the thermal bath associated with $-$ channel 
and it is also proportional to the steady probability current $\curr[+]{0}{j}$
\[
\left\langle q^+ \right\rangle_\rho = - \left\langle q^- \right\rangle_\rho = 2 F \curr[+]{0}{1} .
\]

In formula \eqref{equ:2lvl_stat_occup} we have noticed modifications with respect to the equilibrium Boltzmann statistics whenever $\Phi \neq 0$. 
It agrees with our intuition that relative throttling of the `$-$' with respect to the `$+$' channel under a positive $F > 0$ tends to increase the occupancy of the ``excited'' state `1'. 
Eventually in the limit $\Phi \to +\infty$ the `$-$' channel completely closes and the system is again found at thermal equilibrium ($\curr[+]{0}{1} = 0$) but now with the energy gap $\Delta E - F$. 
The resulting population inversion for $F > \Delta E$ is a most simple example of gauge transformation \eqref{equ:gauge_symmetry}, 
applied here to easily deal with the driving forces when they become derivable from a potential. 
From this point of view, the population inversion is a superficial concept here since $\Delta E - F = E(1) - E(0)$ is just the full energy gap after the transformation with $U(1) - U(0) = -F$ completely removing the driving force has been applied.

\paragraph{High temperature behaviour}
However, more important is how these simple observations carry over when both channels remain open to hold the system out of equilibrium: 
one checks that for an arbitrarily weak channel asymmetry the population inversion $\rho(1) > \rho(0)$ still occurs whenever the driving force is strong enough, and for large enough (but finite) temperature $T=1/\beta$. 
This follows from
\begin{equation}
\log \frac{\rho(1)}{\rho(0)} =
-\frac{{\Delta E} - F \tanh\left(\frac{\Phi}{2}\right)}{T} + \err{\frac{1}{T^2}}
\label{equ:2lvl_distribution_governing_potential} 
\end{equation}
In contrast to the limiting case $\Phi \to +\infty$, the driving force now does \emph{not} derive from a potential and hence cannot be transformed out.
Nevertheless, the leading term in the high-temperature expansion~\eqref{equ:2lvl_distribution_governing_potential} suggests that ${\Delta E} - F \tanh\bigl(\frac{\Phi}{2}\bigr)$ may take over the role of an effective energy gap, 
though we are now dealing with a genuine non-equilibrium system where the energy levels are ambiguously defined.

\paragraph{Low temperature behaviour}
In the low-temperature regime the relative occupation~\eqref{equ:2lvl_distribution_governing_potential} has the asymptotic 
\[
\log \frac{\rho(1)}{\rho(0)} = -\frac{\Delta E}{T} + \Phi \sgn(F) +\err{T}, 
\]
showing that independently of the driving force there is no population inversion at zero temperature. 
Nevertheless, the system undertakes a transition between ``insulator'' and ``conductive'' regimes at $F = \pm {\Delta E}$ as seen from the low-temperature current asymptotic,
\[
\curr[+]{0}{1} \simeq \sgn(F)\,e^{\frac{|F| - {\Delta E}}{2T} + \frac{\phi}{2} \sgn(F)}
\xrightarrow{T \to 0^+}
\begin{cases}
0 & \text{if } |F| < {\Delta E} \\
\pm\infty & \text{if } |F| > {\Delta E}
\end{cases}
\]
This can be understood by observing that in the low-driving (or insulator) regime, $|F| < {\Delta E}$, the state `$0$' remains a well defined ground state in the sense that in both channels 
\[
\log\frac{\rate[\pm]{1}{0}}{\rate[\pm]{0}{1}} \to +\infty \quad\text{for}\quad T \to 0, 
\]
whereas in the high-driving (or conductive) regime $F > {\Delta E}$ the system exhibits a limit cycle behavior, 
\begin{align*}
\log\frac{\rate[-]{1}{0}}{\rate[-]{0}{1}} &\to +\infty , &
\log\frac{\rate[+]{1}{0}}{\rate[+]{0}{1}} &\to -\infty ; 
\end{align*}
analogously for $-F > {\Delta E}$.

\subsubsection{Energetics}
In the pursuit of generalized heat capacity \eqref{def:generalized_heat_capacity},
we need to determine the reversible work and heat at first. 
We have seen that the reversible work can be expressed in terms of the quasi-potential \eqref{def:generalized_potential},
which depends on the total energy and local power of non-potential forces. 
Fixing the \emph{a priori} gauge the energy levels are given as $E(0) = 0$, $E(1) = {\Delta E}$ 
while the non-potential work being $W^\text{nonp}(0 \xrightarrow{\pm} 1) = \pm F$ and $W^\text{nonp}(1 \xrightarrow{\pm} 0) = \mp F$.
The local power can be obtained directly then from definition \eqref{def:local_power} 
\begin{align*}
w^\text{nonp}(0) 
&= W^\text{nonp}\left( 0 \xrightarrow{+} 1\right) \rate[+]{0}{1} + W^\text{nonp}\left( 0 \xrightarrow{-} 1\right) \rate[-]{0}{1} \\ 
&= F(\rate[+]{0}{1} - \rate[-]{0}{1}), \\ \\
w^\text{nonp}(1) 
&= W^\text{nonp}\left( 1 \xrightarrow{+} 0\right) \rate[+]{1}{0} + W^\text{nonp}\left( 1 \xrightarrow{-} 0\right) \rate[-]{1}{0} \\ 
&= F(\rate[-]{1}{0} - \rate[+]{1}{0})
\end{align*}
and hence the second term of the quasi-potential $V(i) = E(i) - 1/\gen[w^\text{nonp}](i)$ can be in general determined by looking for solution of $\gen[\breve V](i) = w^\text{nonp} (i) - \langle w^\text{nonp} \rangle_\rho$.

The generalized heat capacity \eqref{def:generalized_heat_capacity} in terms of quasi-potential 
\[
C_\text{noneq} = \partial_T \left\langle V \right\rangle_\rho - \left\langle \partial_T V \right\rangle_\rho 
= \left\langle V \right\rangle_{\partial_T \rho} 
\]
simplifies in case of two level model to 
\[
C_\text{noneq} 
= V(0) \, \partial_T \rho(0) + V(1) \, \partial_T \rho(1) 
= \beta^2 \rho(0) \, \rho(1) \, \Delta V \, G, 
\]
where $\beta$ is the inverse temperature, $\Delta V = V(1)-V(0)$ is the gap in the quasi-potential and the $G$ is shorthand for $G = \partial_\beta \log (\rho_0 / \rho_1)$.
We can see that to be able to determine the generalized heat capacity the knowledge of the gap in quasi-potential $\Delta V$ is sufficient.  
As a result we obtain
\[
\Delta V = {\Delta E} + F \frac{\tanh\left(\frac{{\Delta E}}{2T}\right) \tanh\left(\frac{F}{2T}\right)
- \tanh\left(\frac{\Phi}{2}\right)}{1 - \tanh\left(\frac{{\Delta E}}{2T}\right)
\tanh\left(\frac{F}{2T}\right) \tanh\left(\frac{\Phi}{2}\right)} . 
\]

\begin{figure}[ht]
\caption{The quasi-potential gap $\Delta V = V(1) - V(0)$ compared to the gap $G = \partial_\beta \ln [ \rho(0) / \rho(1) ]$ as a function of temperature $T=1/\beta$. 
The particular choice of parameters is $U=1$ and $\Phi=3$. }
\begin{center}
\includegraphics[width=.9\textwidth,height=!]{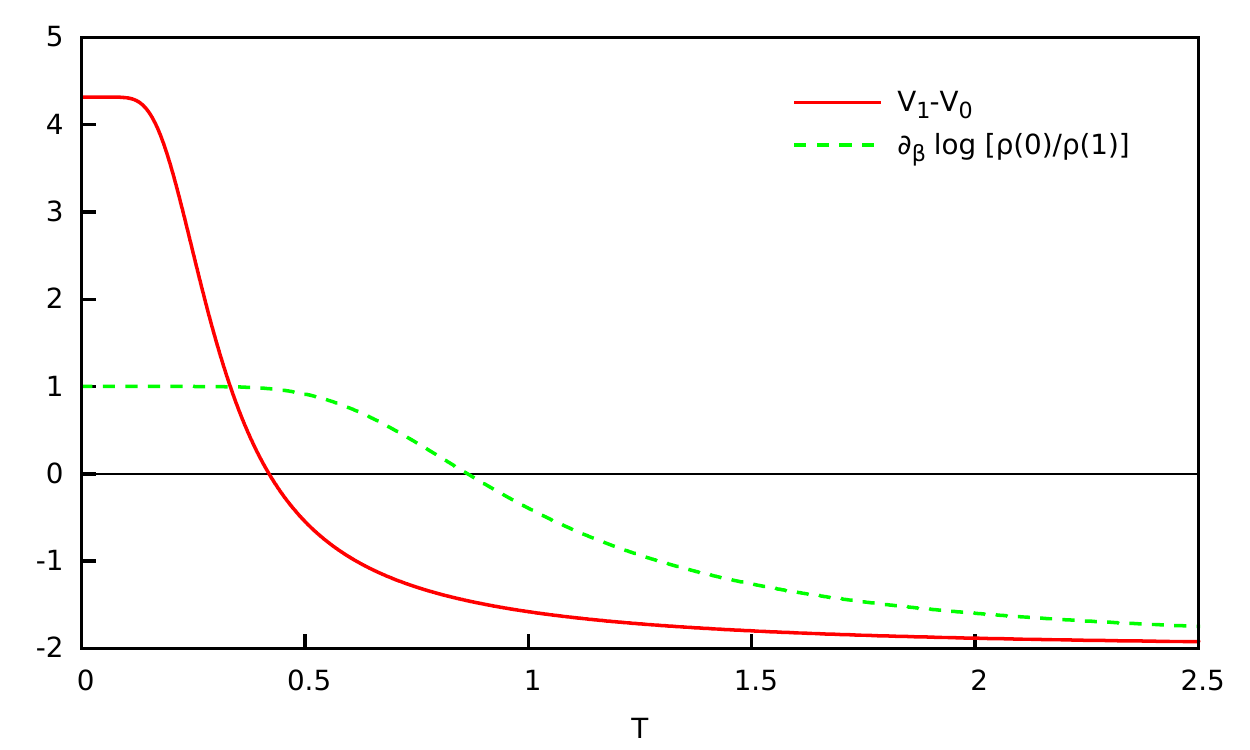}
\end{center}
\label{pic:2lvl-gap}
\end{figure}
\begin{figure}[ht]
\caption{The temperature dependence of the heat capacity for sub-critical, critical and supercritical driving. The model parameters are $U = 1$ and $\Phi = 3$.}
\begin{center}
\includegraphics[width=.9\textwidth,height=!]{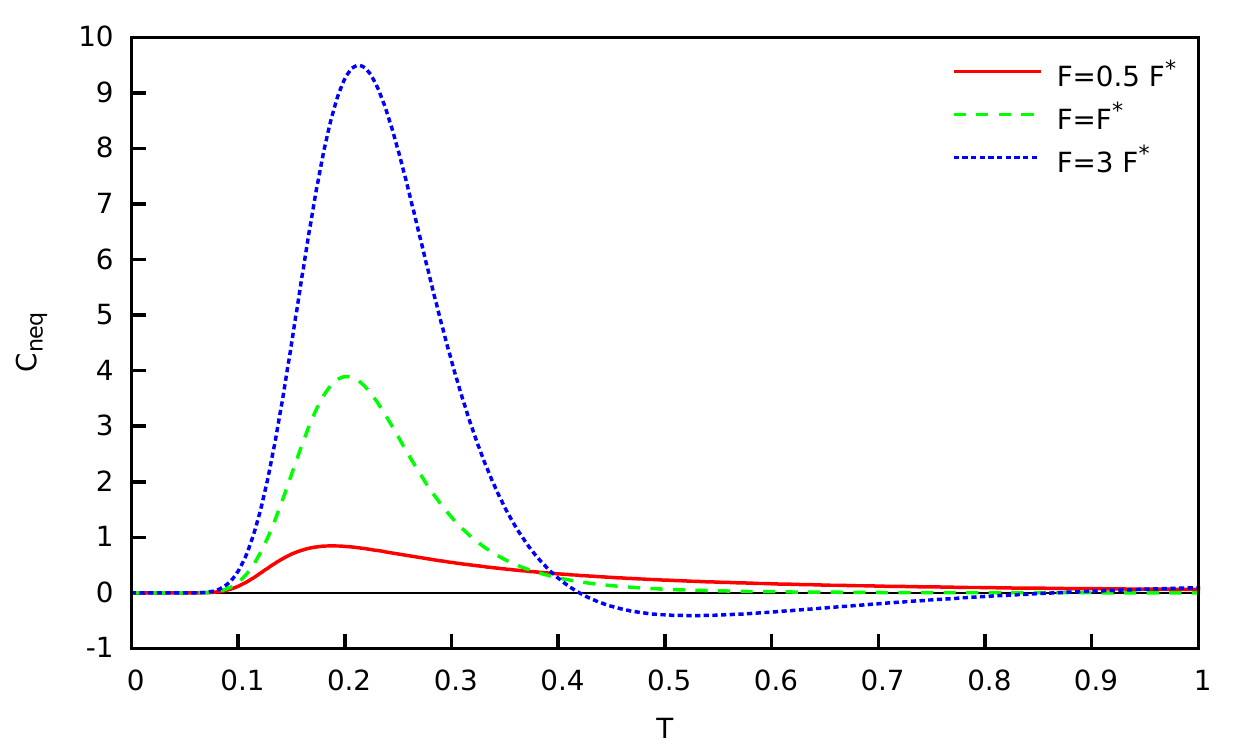}
\end{center}
\label{pic:2lvl-C_T}
\end{figure}

Note first that for $F = 0$ one gets $\Delta V = G = \Delta E$ and we obtain a well-known formula for the heat capacity of an equilibrium two-state model. 
Away from equilibrium the picture becomes far more complicated since both energy-dimensional quantities $G$ and $\Delta V$ are now different and generally not related in a simple way. 
Moreover, they can obtain opposite signs for large enough driving forces, which then results in negative values of the generalized heat capacity, see figures~\ref{pic:2lvl-gap}--\ref{pic:2lvl-C_T}. 
Next we separately analyze three asymptotic regimes.

\paragraph{High temperatures}
For large temperature values the quasi-potential gap is
\[
\Delta V = {\Delta E} - F \tanh(\Phi / 2) + \err{1/T^2},
\]
which coincides with the asymptotic of $G$ as obtained from equation~\eqref{equ:2lvl_distribution_governing_potential}. 
Hence the heat capacity equals
\[
  C_\text{neq} = \frac{\bigl[
  {\Delta E} - F \tanh \bigl( \frac{\Phi}{2} \bigr) \bigr]^2}{4 T^2} +
  \err{\frac{1}{T^3}}
\]
We see that $F^* \equiv {\Delta E} / \tanh(\Phi / 2)$ is a critical value of the driving, above which the system exhibits a population inversion and also the gap $\Delta V$ changes sign. 
As a result, for any $F \neq F^*$ the heat capacity is asymptotically strictly positive and decaying as $1 / T^2$, i.e. similarly as in equilibrium. 
Note that the asymptotic equality $\Delta V \simeq G$ remains true even for a driving force $F$ much larger than the model parameter ${\Delta E}$, 
the original meaning of which as an energy gap then becomes meaningless. 
Instead, there is another gauge that becomes natural here: by making the transformation~\eqref{equ:gauge_symmetry} with $U(0) = 0$ and $U(1) = -F \tanh(\Phi / 2)$, 
we obtain ``renormalized'' energy levels with the gap $\widetilde{E}(1) - \widetilde{E}(0) = {\Delta E} - F \tanh(\Phi / 2)$, 
which is directly seen in the leading asymptotic $\Delta V \simeq G \simeq \widetilde{E}(1) - \widetilde{E}(0)$. 
After this transformation, the residual non-potential forces contribute to the heat capacity only by correction ${\mathrm o}(1 / T^2)$. 
In this sense the full high-temperature regime away from the critical value $F^*$ is to be understood as essentially close to equilibrium, 
but with the renormalized energy levels $\widetilde{E}(i)$ and the corresponding Boltzmann stationary distribution. 
From this point of view the observed population inversion at high temperatures and strong driving is only an artifact of describing the model in terms of ``unphysical'' energy levels $E(i)$.

\begin{figure}[ht]
\caption{Steady heat capacity as a function of the driving, with the parameters
  ${\Delta E} = 1$, $\Phi = 3$, and $T = 1$. The tiny region of negative heat capacity in the vicinity of the critical driving is zoomed in.}
\begin{center}
\includegraphics[width=.9\textwidth,height=!]{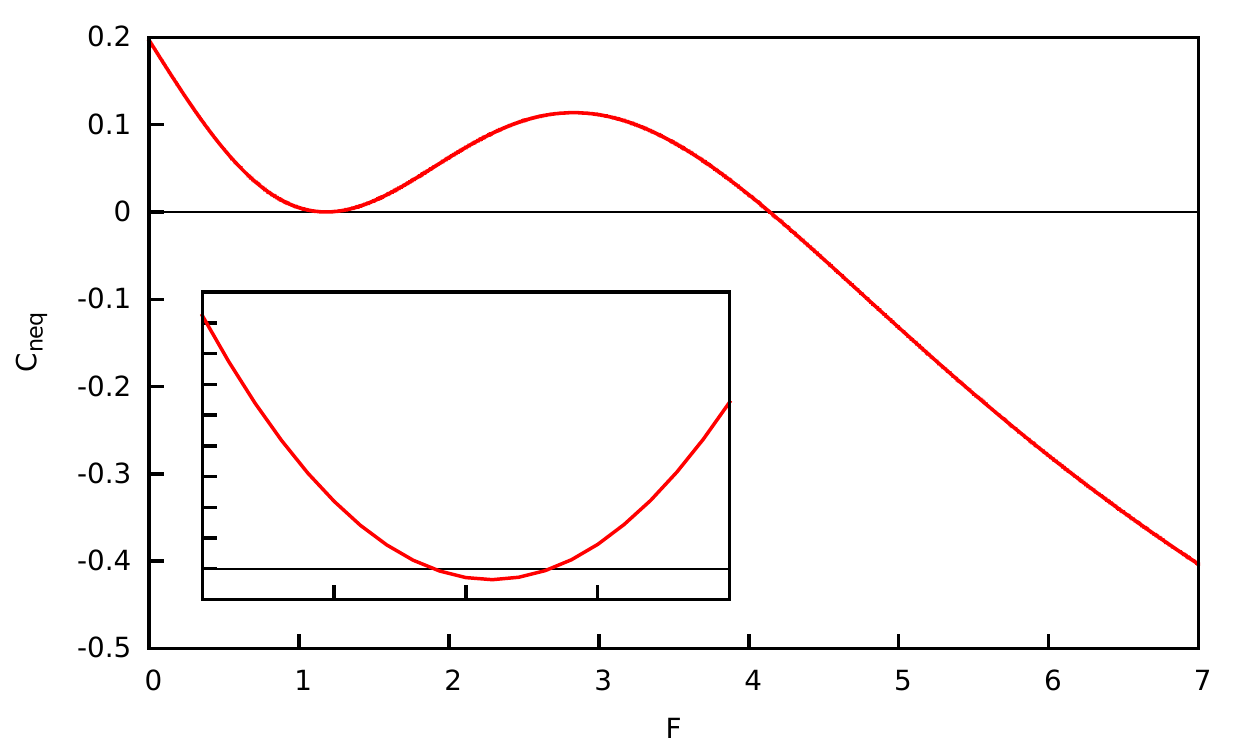}
\end{center}
\label{pic:2lvl-C_F}
\end{figure}

\paragraph{High temperatures --- critical}
We have seen that the value $F = F^*$ plays a special role since in this case the above gauge transformation leads to degenerate energy levels, 
and therefore the heat capacity becomes zero up to order $1 / T^2$. 
More detailed calculations reveal that both gaps $\Delta V$ and $G$ are of order $1/T^2$ which yields an anomalously fast-decaying heat capacity,
\begin{equation}
  C_\text{neq} = \frac{{\Delta E}^6}{64 T^6 \sinh^4 \frac{\Phi}{2} }
  + {\mathrm o}\left(\frac{1}{T^6}\right)
\end{equation}
The existence of the high-temperature critical driving leads to the following subtle phenomenon: 
there is a temperature curve $T = T^1(F)$ along which $\Delta V = 0$ and another one, 
$T = T^2(F) > T^1(F)$, on which $G = 0$. 
Both curves have the identical leading asymptotic, for $F > F^*$,
\[
  \frac{1}{T^{1,2}(F)} = 2 \sinh \left( \frac{\Phi}{2} \right) \sqrt{\frac{2(F - F^*)}{{\Delta E}^3 \sinh \Phi}} + {\mathrm o}\left((F - F^*)^{\frac{1}{2}}\right) .
\]
On both curves the heat capacity vanishes and they form the boundary of a tiny region in the $(T,F)-$space inside which $C_\text{neq}$ exhibits negative values.
Its full dependence on the driving for a fixed intermediate temperature is depicted on figure~\ref{pic:2lvl-C_F} where the above mentioned region has been zoomed in. 
Notice the negative values of the heat capacity for large $F$; 
this is a strong-non-equilibrium effect and we may expect that no gauge transformation would significantly simplify the thermodynamic description in this region due to a strong temperature-dependence of the quasi-potential gap $\Delta V$.

\paragraph{Low temperatures}
The quasi-potential gap $\Delta V$ has the low-temperature asymptotic
\begin{equation}
\Delta V = {\Delta E} + |F| + \err{|F|\,\rme^{- \beta \min\{{\Delta E},|F|\} }}
\label{equ:2lvl_gap_lowT}
\end{equation}
in which the temperature dependence emerges only in the exponentially small correction (along the limit $T \to 0^+$). 
This suggests that it is appropriate to make the gauge transformation with $U(0) = 0$ and $U(1) = F$, 
to define the ``renormalized'' energy levels $\widetilde{E}(0) = 0$ and $\widetilde{E}(1) = {\Delta E} + |F|$ through which the heat capacity gets the simplified approximate form 
\[
C_\text{neq} \simeq \partial_T \langle \widetilde{E} \rangle_\rho , 
\]
i.e. with only a negligible contribution from the second term in~\eqref{equ:heat_capacity_quasi-potential}. 
One checks by comparing with the exact result that this intuition is indeed correct.
The apparent disagreement between the low-temperature asymptotic of the gaps $\Delta V$ and $G$, cf.~\eqref{equ:2lvl_stat_occup} and \eqref{equ:2lvl_gap_lowT}, 
indicates that the low-temperature regime corresponds to strong non-equilibrium with non-Boltzmannian statistics. 
Formally, it can be described by an effective temperature defined by
\[
\log (\rho_1 / \rho_0) = -(\widetilde{E}(1) - \widetilde{E}(0)) / T^\text{eff}, 
\]
explicitly $T^\text{eff} = T (1 + |F| / {\Delta E}) > T$.
Using the fact that $C_\text{neq} \simeq (1 + |F|/{\Delta E})\,\partial \langle \widetilde{E} \rangle_\rho / \partial T^\text{eff}$, 
we can trace back the exponential decay of the heat capacity for $T \to 0^+$ to the exponential suppression of thermal excitation, which is analogous to the equilibrium third law of thermodynamics.
Since $\Delta V \simeq {\Delta E} + |F| > 0$, and recalling that our model exhibits no population inversion in the zero-temperature limit, 
we conclude that the generalized heat capacity remains strictly positive at low temperatures. 
In particular, it does not exhibit any transition at $F = {\Delta E}$ where the system undertakes a change between the ``insulator'' and the ``conductive'' transport regimes.

We finish this example by indicating how to extend the above approximate description of the low-temperature behavior to arbitrary temperatures and driving forces. 
\emph{Formally} defining the effective temperature $T^\text{eff} = \Delta V / \log(\rho_0 / \rho_1)$, 
we can write the reversible heat~\eqref{def:reversible_heat} in the form of a Clausius equality
\begin{align*}
  \dbar Q^\text{rev} &= T^\text{eff} \rmd S , &
  S &= -\sum_{i=0,1} \rho(i) \log \rho(i)
\end{align*}
with $S$ the Shannon entropy of the stationary distribution $\rho$.
In this framework the heat capacity obtains the form $C_\text{neq} = T^\text{eff} \rmd S / \rmd T$.
In contrast with the above low-temperature regime, the effective temperature now becomes a nontrivial function of $T$; for example, it becomes zero on the critical line $T = T^1(F)$.
Obviously, such a representation in terms of a (single) effective temperature has no straightforward extension to models with a larger number of states, and significant modifications are needed. 
Some extensions of the Clausius relation to non-equilibrium and its limitations have been studied in~\cite{Komatsu2008,Komatsu2009,Sagawa2011}.

\subsection{Example: Three-level model}
\label{ssec:three_level_nonp}
Now we consider a three-level version of the above model, introduced in subsection \ref{ssec:three_level} also presented in \cite{result2}, 
with states `$0$', `$1$' and `$2$' mutually connected by single channels. 
The system is driven out of equilibrium by a force acting along the loop $0 \rightarrow 1 \rightarrow 2 \rightarrow 0$ and performing equal non-potential work,
\[
W^\text{nonp}_{0 \rightarrow 1} = W^\text{nonp}_{1 \rightarrow 2} = W^\text{nonp}_{2 \rightarrow 0} = W_\text{ext}, 
\]
along all those transitions, see 
The dynamics are defined by the transition rates
\[
 \rate{i}{i\pm} =
 A_{i,i\pm}\,\exp \left[ \frac{\beta}{2} \left( E(i) - E(i\pm) \pm W_\text{ext} \right) \right]
\]
where $i+$ (respectively $i-$) is the succeeding (respectively the preceding) state along the oriented loop $0 \rightarrow 1 \rightarrow 2 \rightarrow 0$; e.g., $0+ = 1$, $0- = 2$ etc.
The prefactors $A_{i j} = A_{j i} > 0$ are symmetric in order to satisfy the local detailed balance condition~\eqref{equ:global_detailed_balance_jump} but arbitrary otherwise.
We only assume that they can be kept constant and independent of other parameters like the temperature or forces.
To be specific, we assume that $E(2) > E(1) > E(0) = 0$ and $W_\text{ext} > 0$.

The present model exhibits a rich collection of different zero-temperature phases which are summarized in table~\ref{table:3lvl}.
In particular, it demonstrates a population inversion between levels `$0$' and `$1$' in the case $E(2) > 2E(1)$ and $W_\text{ext} > E(1)$. 
Later we will see that \emph{at} the ``critical'' driving $W_\text{ext} = E(1)$, where the zero-temperature population inversion occurs, the model exhibits an anomalous low-temperature behavior.
\begin{table}
\caption{Zero-temperature phases of the driven three-level model.
In all cases $\log(\rho(2)/\rho(0 \text{ or } 1)) \to -\infty$ for $T \to 0^+$.}
\label{table:3lvl}
\begin{center}
\shorthandoff{-} 
\begin{tabular}{|cc|c|c|}
\hline
\multicolumn{2}{|c|}{Zero-temperature phases}& $\log (\rho_0 / \rho_1)$ & $J$ \\ \hline \hline
\multicolumn{1}{|c|}{\multirow{2}{*}{$E(2) < 2 E(1)$}} & $W_\text{ext} < E(1)$ & $+\infty$ & $0$ \\ \cline{2-4} 
\multicolumn{1}{|c|}{}& $W_\text{ext} > E(1)$ & $+\infty$ & $+\infty$ \\ \hline
\multicolumn{1}{|c|}{\multirow{3}{*}{$E(2) > 2 E(1)$}} & $W_\text{ext} < E(1)$ & $+\infty$ & $0$ \\ \cline{2-4}
\multicolumn{1}{|c|}{}& $E(1) < W_\text{ext} < E(2) - E(1)$ & $-\infty$ & $0$ \\ \cline{2-4}
\multicolumn{1}{|c|}{}& $E(2) - E(1) < W_\text{ext} $ & $-\infty$ & $+\infty$ \\ \hline
\end{tabular}
\shorthandon{-} 
\end{center}
\end{table}

\begin{figure}[ht]
\caption{The heat capacity landscape in $(T,W_\text{ext})$-plane with curves of constant heat capacity. 
The particular choice of parameters is $E(1) = 1$, $E(2) = 3$ and $A_{0,1} = 1$, $A_{1,2} =2$ and $A_{2,0}=4$.}
\begin{center}
\includegraphics[width=.9\textwidth,height=!]{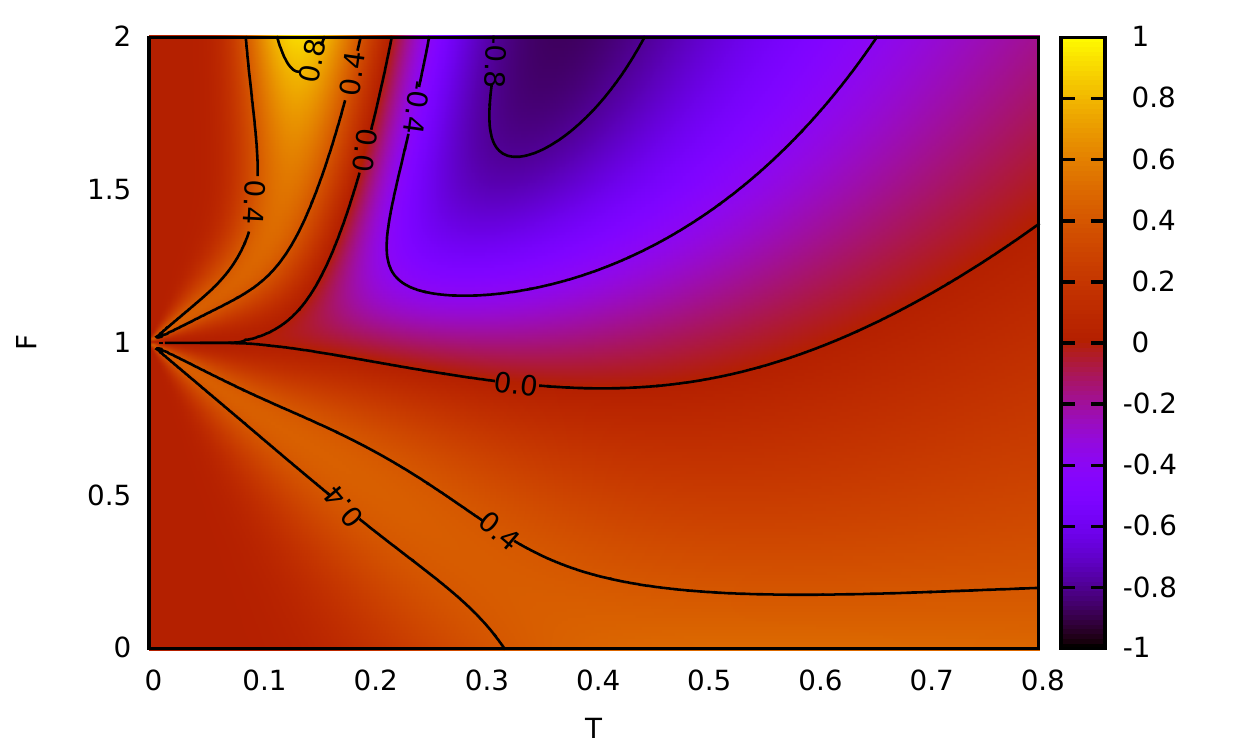}
\end{center}
\label{pic:3lvl-iso}
\end{figure}

For the three-level model we skip the detailed analysis and only concentrate on some new features that were not seen in the previous two-level example. 
Therefore we consider the case $E(2) > 2E(1)$ in which the system exhibits a zero-temperature transition in the stationary occupations, see table \ref{table:3lvl}.
The generalized heat capacity in the $(T,W_\text{ext})-$plane is depicted in figure~\ref{pic:3lvl-iso} where we see that $(T=0,F=E(1))$ is an accumulation point of the curves of zero generalized heat capacity.

\begin{sidewaysfigure}
  \caption{The properties three-level model in various regimes.} 
  \begin{center}
    \begin{subfigure}[hb]{0.31\textheight}
      \caption{Sub-critical regime.}
      \begin{center}
      \includegraphics[width=\textwidth,height=!]{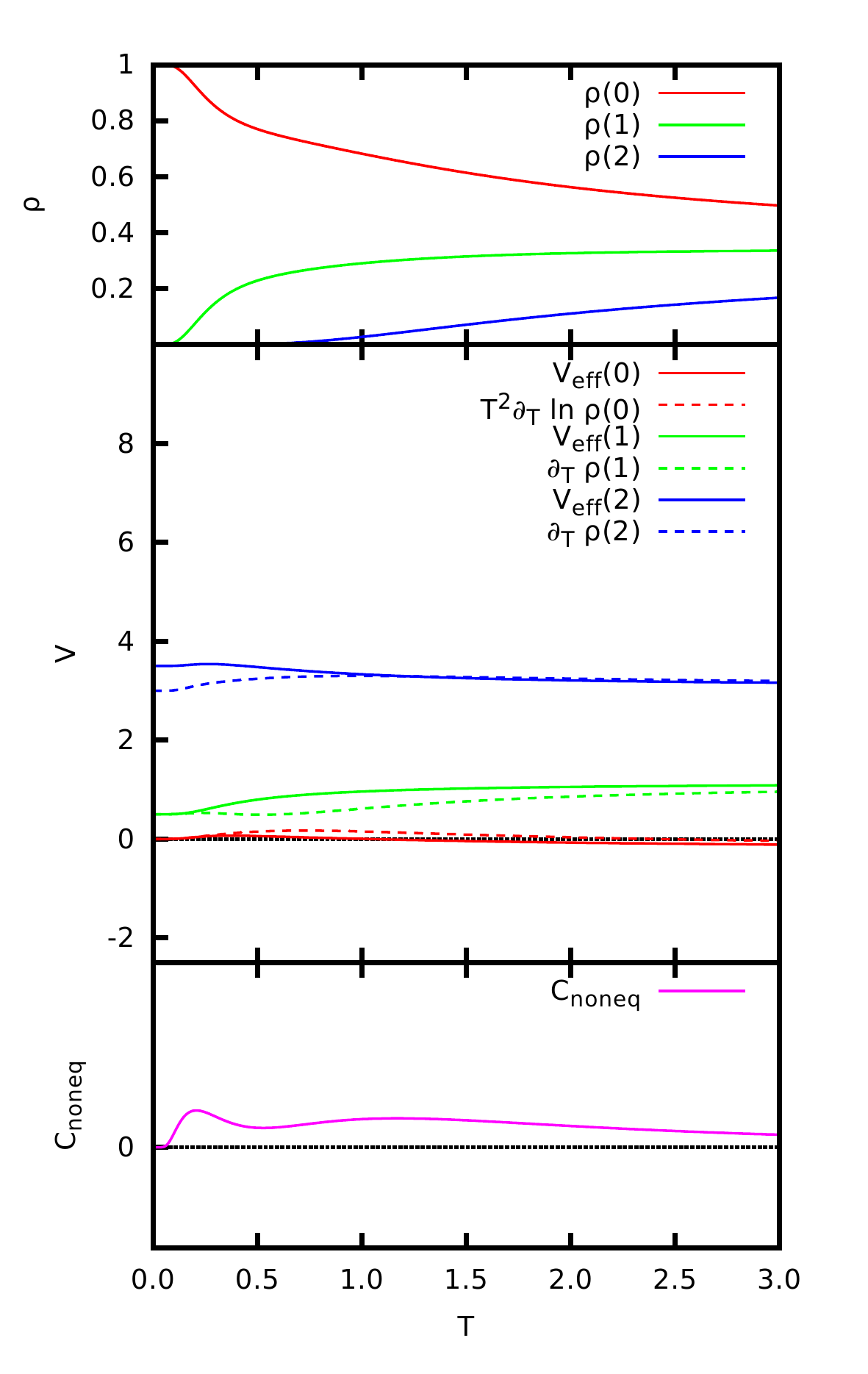}
      \end{center}
      \label{pic:subcritical-3lvl}
    \end{subfigure} ~
    \begin{subfigure}[hb]{0.31\textheight}
      \caption{Critical regime.}
      \begin{center}
      \includegraphics[width=\textwidth,height=!]{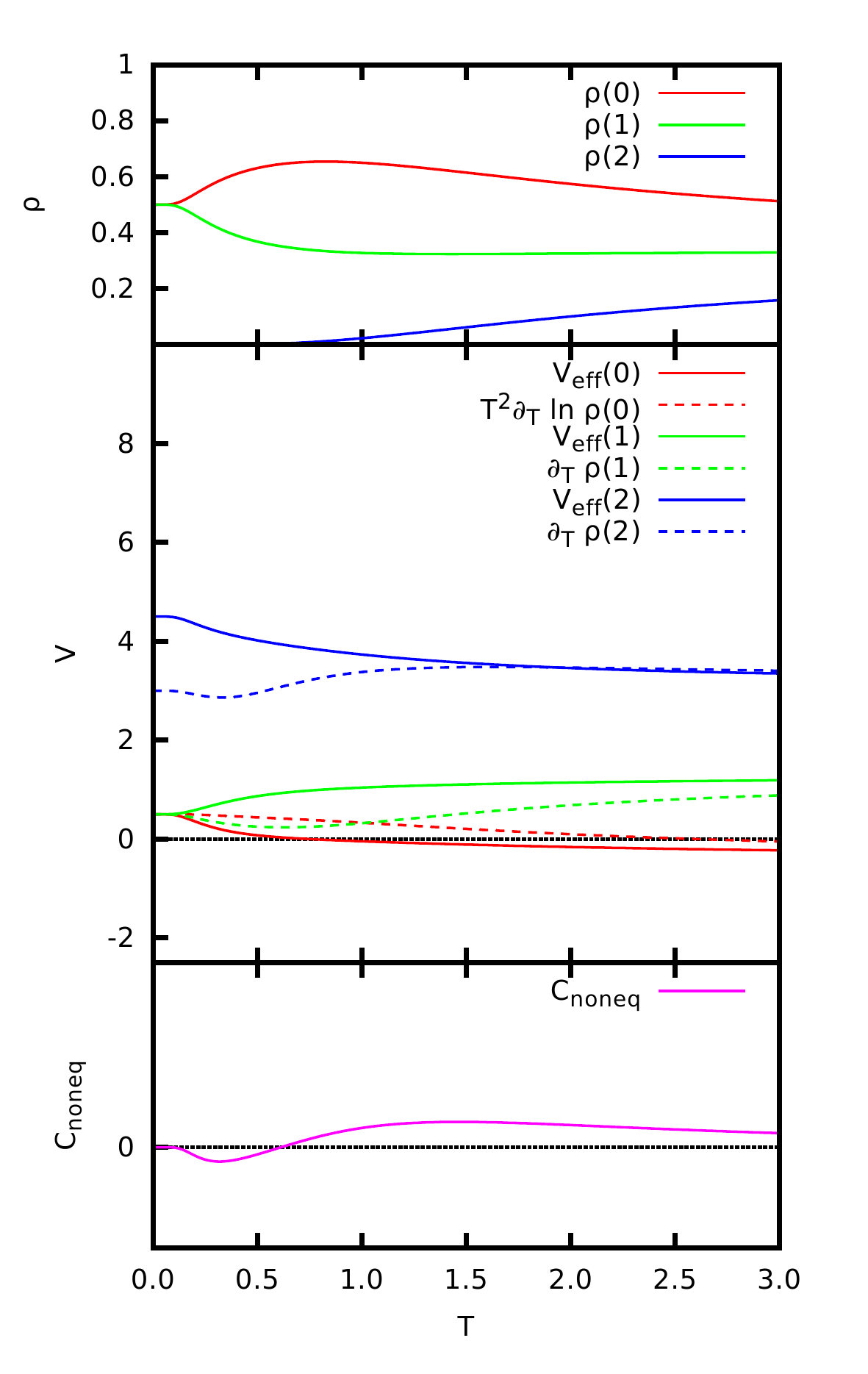}
      \end{center}
      \label{pic:critical-3lvl}
    \end{subfigure} ~
    \begin{subfigure}[hb]{0.31\textheight}
      \caption{Super-critical regime.}
      \begin{center}
      \includegraphics[width=\textwidth,height=!]{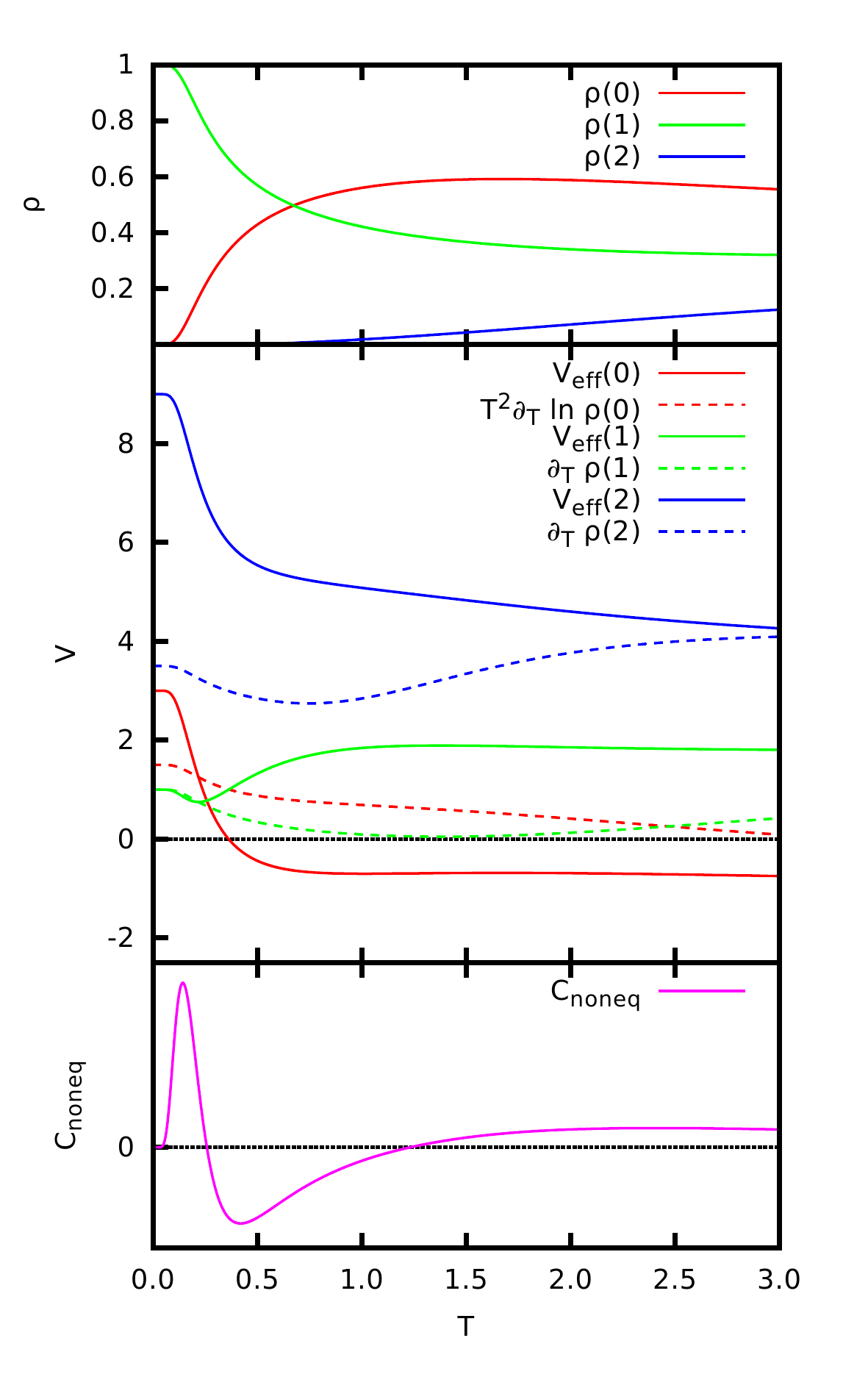}
      \end{center}
      \label{pic:supcritical-3lvl}
    \end{subfigure}
  \end{center}
  \label{pic:properties-3lvl}
\end{sidewaysfigure}
In order to better understand the behavior we look into the critical case $W_\text{ext} = E(1)$ in more detail and compare it with sub- and super-critical case, the results are summarized in figure~\ref{pic:properties-3lvl}.
In the critical scenario at zero temperature the states `$0$' and `$1$' degenerate into a single energy level, both in the sense of stationary occupations, $\rho(0) = \rho(1) = 1/2$, and in the sense of the quasi-potential, $V(0) = V(1)$. 
Increasing the temperature, the occupation of state `$0$' also increases (in this way behaving like an excited state), whereas the quasi-potential satisfies $V(1) > V(0)$, 
meaning that the degeneracy gets removed and a positive energy gap opens between states `$0$' (lower) and `$1$' (higher). 
As a direct consequence of these opposite tendencies the generalized heat capacity becomes negative at $W_\text{ext} = E(1)$ and low temperatures, 
with an anomalously fast decay to zero for $T \to 0^+$ due to the zero-temperature degeneracy of both states. 
Note that although the presence of state `$3$' is essential for breaking the detailed balance and for the non-equilibrium features of our model, it does not directly enter low-temperature energetics. 
It also does not substantially contribute to the heat capacity until high enough temperatures where its occupation becomes relevant.

\subsection{Example: Diffusion in 2D-plane}
\label{ssec:diffusion_on_plane}
As an example for diffusions let us consider the driven two-dimensional rotationally symmetric overdamped diffusion in an homogeneous environment at the inverse temperature $\beta=1/T$, 
see subsection \ref{ssec:overdamped_diff} in chapter \ref{chapter:models} for additional details about overdamped diffusions, 
which is described by time evolution \eqref{def:position_overdamped}
\[
\rmd {\vec{q}}_t = \left[ \vec{F}(\vec{q}_t) - \vec{\nabla} U(\vec{q}_t) \right] \; \rmd t + \left( \frac{2}{\beta} \right)^{\frac{1}{2}}\, \rmd \vec{W}_t ,
\]
where $U(\vec{q}) = \frac{\lambda}{2}\,\|\vec{q}\|^2$ is the quadratic potential
and $\vec{F}(\vec{q}) = v(\|\vec{q}\|)\,\vec{e}_\theta(\vec{q})$ is the angular driving force, where $\vec{e}_\theta(\vec{q})$ is the unit vector in the angular direction at the position $\vec{q}$.
Note that both forces are mutually orthogonal $\vec{F}(\vec{q}) \cdot \vec{\nabla}U(\vec{q}) = 0$. 
Furthermore we will assume that the magnitude of the driving is proportional to the power of radius $v(r) = \kappa\, r^\alpha$ with some fixed exponent $\alpha > -1$.

Because of the angular symmetry of the system we will further proceed in the polar coordinates $(r,\theta)$.
The stationary distribution and the steady probability current \eqref{def:current_overdamped} are then obtained from the stationarity equation
\[
\partial_r \left(r j_r(r,\theta) \right)+ \partial_\theta j_\theta(r,\theta) = 0
\]
with the components or the probability current being 
\begin{align*}
  j_r(r,\theta) &= -\rho(r,\theta) \, \partial_r U(r) - T\,\partial_r \rho(r,\theta), \\
  j_\theta(r,\theta) &= \rho(r,\theta)\, v(r) - \frac{T}{r}\,\partial_\theta \rho(r,\theta) . 
\end{align*}
The explicit solution can be found
\begin{align*}
\rho(r,\theta) &= \frac{1}{Z}\,\rme^{-\beta U(r)}, &
Z &= \frac{2\pi T}{\lambda} , &
\vec{j}(r,\theta) &\equiv (j_r,j_\theta) = (0, \rho(r,\theta)\, v(r)), 
\end{align*}
where we can see that steady state has the same probability distribution as in equilibrium, which is due to the orthogonality of the potential $-\vec{\nabla} U$ and non-potential driving $\vec{F}$ forces. 
Moreover the steady probability current is proportional only to the non-potential forces and as such it has only an angular component, thus creating a vortex around the origin.

As there is a non-zero steady probabilistic current, there exists a steady heat current from the system to the environment which is given by the mean value of the local power, 
where the local power of the driving force $F$ is given by the function 
\[
w(\vec{q}) = \vec{F}(\vec{q}) \cdot ( \vec{F}(\vec{q}) - \vec{\nabla} U(\vec{q}) ) + T \, \vec{\nabla} \cdot \vec{F}(\vec{q})  = v^2(\|\vec{q}\|) ,
\]
from where it follows that the mean steady dissipated power is 
\[
\langle w \rangle_\rho \equiv \int\limits_0^\infty \rmd r \; \int\limits_0^{2\pi} \rmd \theta \; w(r,\theta) \, \rho(r,\theta) =
\Gamma(\alpha + 1)\,\kappa^2\,\left( \frac{2T}{\lambda} \right)^\alpha,
\]
where $\Gamma$ is the gamma function. 
Notice that $\langle w \rangle_\rho$ diverges when $\alpha \to -1$.
Moreover the mean local power is increasing function of temperature $T$ for $\alpha > 0$ while it is decreasing for $-1 < \alpha < 0$, 
this associated with the fact that the increase of the temperature always increase the probability of observation a particle further from the origin, 
which is in case $\alpha > 0$ is associated with higher dissipation as the driving force increase with the radius, while in the other case $-1 < \alpha < 0$ the dissipation diminishes as the driving force is weaker.

\subsubsection{Non-equilibrium heat capacity}
Such dependence on the parameter $\alpha$ is also reflected on quasistatic response functions, namely the heat capacity. 
The generalized heat capacity \eqref{def:generalized_heat_capacity} as defined via the quasistatic ``reversible'' heat \eqref{equ:heat_capacity_quasi-potential} is given by the general formula
\begin{align*}
C_F &= \partial_T \langle U \rangle_\rho + \Delta C_F, \\
\Delta C_F &= \left\langle \partial_T\,\frac{1}{\gen}[w] \right\rangle_\rho ,
\end{align*}
where with the subscript $F$ we denote the that the heat capacity is given with respect to constant driving force, namely $\kappa = {\mathrm {const.}}$ 
For this particular model, the first ``classical'' term is in accordance with the equilibrium equipartition theorem $\partial_T \langle U \rangle_\rho = 1$ (using the convention $k_B \equiv 1$). 
For the second genuinely non-equilibrium term we need to compute the function
$G(r,\theta) = \frac{1}{\gen}[w](r,\theta)$ satisfying the equations \eqref{equ:backward_pseudoinverse_identity} and \eqref{equ:backward_pseudoinverse_zero_mean}
\begin{gather*}
\langle G \rangle_\rho = 0 , \\
\gen[G](r,\theta) = \gen \frac{1}{\gen} [w] (r,\theta) \equiv w(r,\theta) - \langle w \rangle_\rho . 
\end{gather*}
where the backward generator generator \eqref{equ:bacward_generator_overdamped} is in polar coordinate system given by 
\begin{align*}
\gen [A] (r,\theta) 
=& - \partial_r U(r) \, \partial_r A(r,\theta) + \frac{v(r)}{r}\,\partial_\theta A(r,\theta) + \\ 
&+ \frac{T}{r}\, \partial_r \left( r \partial_r A(r,\theta) \right)
+ \frac{T}{r^2}\, \partial^2_\theta A(r,\theta) .
\end{align*} 
We will assume that the angular symmetry of the system also ensures the pseudoinverse will be again symmetric $G(r,\theta) \equiv G(r)$, which yields to 
\[
-U'(r)\,G'(r) + \frac{T}{r}\,(r\,G'(r))' = w(r) - \langle w \rangle_\rho , 
\]
where we used the notation $f' \equiv \partial_r f(r)$.
By changing the variable $G(r) = g(z)$, $z = \frac{\lambda}{2T}\,r^2$, it simplifies to
\begin{gather*}
2 z\, \dot{g}(z) = h(z) , \\
\int\limits_0^\infty \rmd z \; g(z) \, \rme^{-z} = 0 ,
\end{gather*}
where we denote $\dot g(z) \equiv \partial_z g(z)$ 
and $h(z)$ satisfy
\begin{align*}
\lambda \left[ \dot{h}(z) - h(z) \right] &= w(r) - \langle w \rangle_\rho , \\ 
h(0) &= 0 .
\end{align*}  
The latter equation has the solution
\[
h(z) = \frac{\langle w \rangle_\rho}{\lambda}\,\left[1 - \bar\Gamma(\alpha + 1,z)\,\rme^z \right]  
\]
where 
\[
\bar\Gamma(s,z) = \Gamma^{-1}(s) \int\limits_z^\infty \rmd t \; t^{s-1}\,\rme^{-t} 
\]
is the regularized incomplete gamma function. 
Finally, the non-equilibrium correction to the equilibrium heat capacity is
\begin{equation}
\Delta C_F 
= \int\limits_0^\infty \rmd z \; \left[ \partial_T g(z) - \dot g(z)\, \frac{z}{T} \right] \rme^{-z}
= -\frac{1}{2T}\,\int\limits_0^\infty \rmd z \; h(z)\,\rme^{-z}
= \frac{\alpha\langle w \rangle_\rho}{2\lambda\,T}
\label{equ:2D_diffusion_noneq_heat_capacity}
\end{equation}
We observe that the non-equilibrium correction is linearly growing with the steady entropy production rate, $\langle w \rangle_\rho / T$, and therefore that it depends on temperature as $\err[{\alpha-1}]{T}$. 
It is positive for $\alpha > 0$ and negative for $-1 < \alpha < 0$. 
The latter means that the full generalized heat capacity $C_F$ becomes negative whenever $\alpha < 0$ and $v$ large enough or $T$ small enough. 
\begin{figure}[ht]
\caption{Non-equilibrium correction for the heat capacity for different values of $\alpha$.}
\label{pic:heat_capacity_2d_diffusion}
\begin{center}
\includegraphics[width=.9\textwidth,height=!]{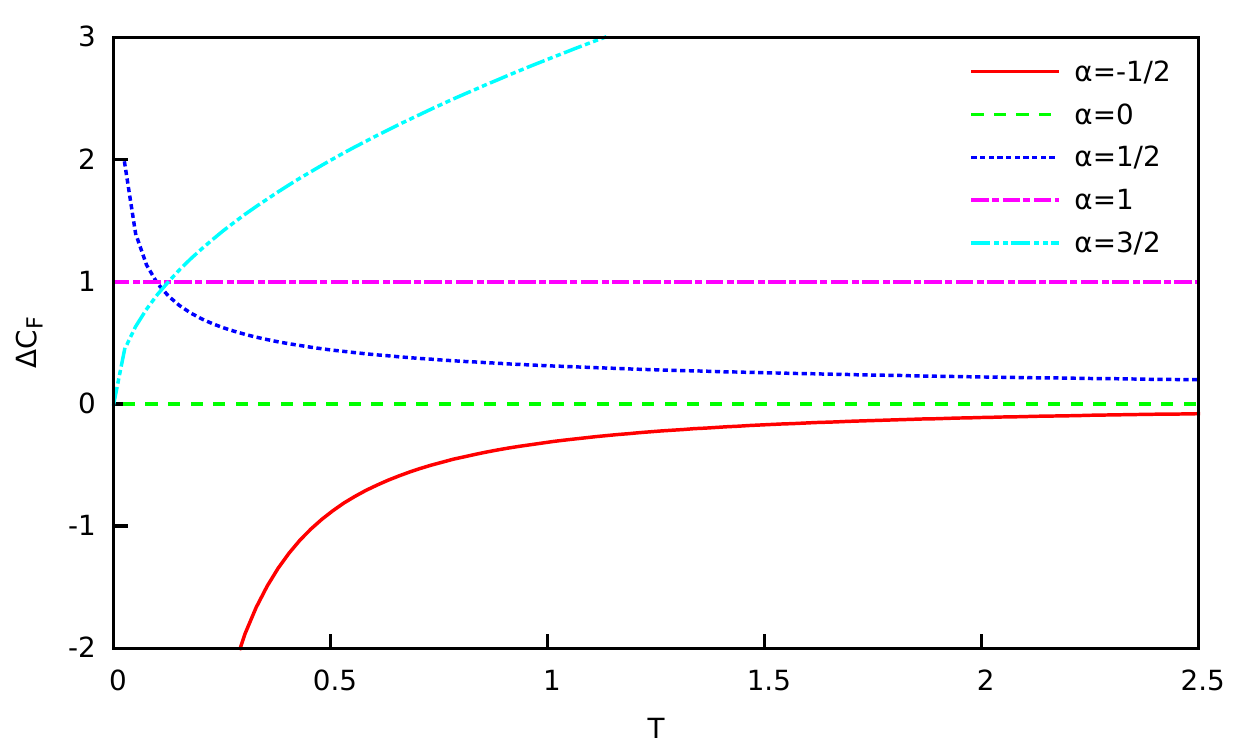}
\end{center}
\end{figure}

Instead of the heat capacity $C_F$ at constant driving, here meaning $\kappa = {\mathrm {const.}}$, we can consider the heat capacity at constant dissipation, $C_W$, as defined under the constraint $\langle w \rangle_\rho = {\mathrm {const.}}$. 
For the model under consideration, $\Delta C_W = \Delta C_F$ and also $C_W = C_F$ as an immediate consequence of the driving-independence of the stationary distribution $\rho$. 
Nevertheless, the constant-dissipation heat capacity has a different temperature dependence $\Delta C_W \propto 1/T$, independently of the exponent $\alpha$.

One also checks that the heat capacity essentially does not depend on the mobility as long as the latter is homogeneous and isotropic. 
Indeed, assuming a more general (scalar but possibly temperature-dependent) mobility $\chi(T)$, the formula~\eqref{equ:2D_diffusion_noneq_heat_capacity} gets only slightly modified
\begin{equation}
\Delta C_F =  \frac{\alpha \langle w \rangle_\rho}{2\lambda\,T\,\chi(T)} \propto T^{\alpha - 1}
\label{equ:2D_diffusion_noneq_heat_capacity_mobility}
\end{equation}
Since $D = T \chi$ coincides with the diffusion parameter, we can read the result in the way that $\Delta C_F$ comes from the mutual ratio between the injected/dissipated energy $\langle w \rangle_\rho t$ and $\lambda D t$ which is a typical scale of system's energy changes by the diffusion process, both within same small time interval $t$.

One possible physical interpretation of the non-equilibrium correction in this particular model comes out by rewriting the formula~\eqref{equ:2D_diffusion_noneq_heat_capacity_mobility} in the form
\[
\Delta C_F = \frac{1}{2\lambda}\,\partial_T \left[ \chi^{-1} \langle w \rangle_\rho \right] .
\]
Since the relaxation mechanism is (at least in the $r-$sector) determined by the equilibrium $\vec{F} \equiv 0$ process, its characteristic time of relaxation is $\tau = \chi \lambda$. 
Hence, the heat capacity correction goes like $\Delta C_F \propto \partial_T [\tau \langle w \rangle_\rho ]$, with only a numerical proportionality factor $1/2$. 
Here the quantity $\tau \langle w \rangle_\rho$ reads the total amount of dissipated energy within the time needed to run through a relaxation process --- in some sense, one can consider that it measures the energy `available in the non-equilibrium surroundings' the changes of which contribute to the renormalized heat exchange; a bit analogously as the $pV-$term in the equilibrium enthalpy. 
In this interpretation of the negative sign of the non-equilibrium correction term appears whenever the increase of temperature moves the system into a lower dissipation regime or, more precisely, into the regime with a lower ``available energy'' from the non-equilibrium driving --- typically when higher temperature effectively corresponds to weaker non-equilibrium as measured on the relaxation time-scale.

\subsection{Example: Diffusion in quadratic potential}
\label{ssec:diffusion_quadratic_potential}
\begin{figure}[ht]
\caption{Scheme of the 2D model with quadratic potential.}
\label{pic:2D_perturb}
\begin{center}
\includegraphics[width=.6\textwidth,height=!]{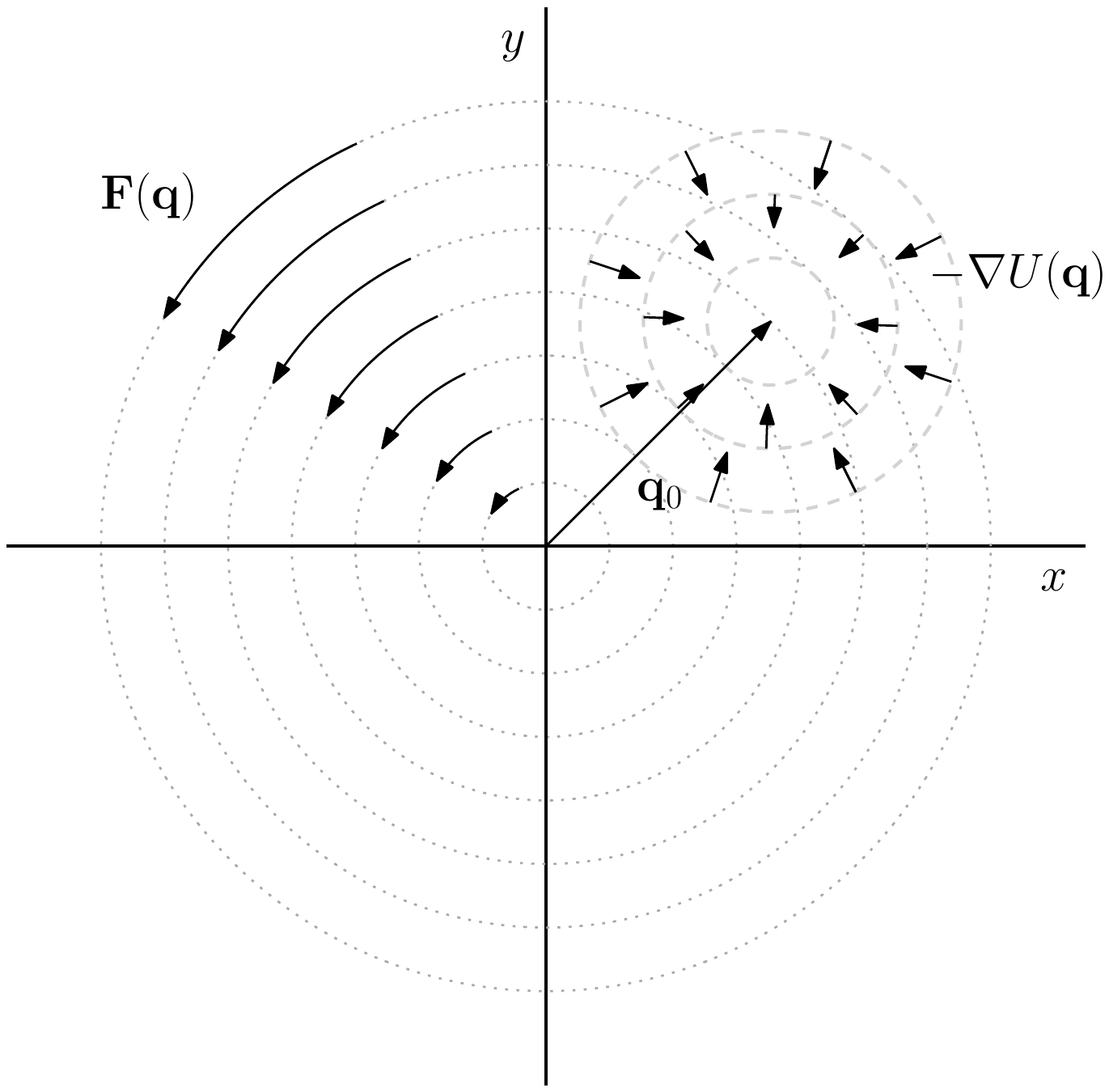}
\end{center}
\end{figure}

As an example of the system where the Clausius relation holds true even far of equilibrium is the particle in quadratic potential in homogeneous media undergoing the driven overdamped diffusion in 2D, see also \ref{ssec:overdamped_diff},
here if compared with the example \ref{ssec:diffusion_on_plane} the quadratic potential is not in the center of the vortex created by non-potential force, see picture \ref{pic:2D_perturb}. 
In order to simplify the analysis we assume that the mobility matrix $\chi$ does not depend on the temperature of the environment, in particular we can set the mobility to be one, $\chi = \mathbb I$. 
The system's dynamics is then fully determined by the quadratic potential 
\[
U(\vec{q}) = \frac{1}{2} \omega^2 \left( \vec{q} - \vec{q}_0 \right)^2
\]
and by the driving force 
\[
\vec{F} (\vec{q}) = F {\mathbb J} \cdot \vec{q}
\]
where the matrix $\mathbb J$ creates an orthogonal vector to $\vec{q}$ and hence is given 
\[
{\mathbb J} = \begin{bmatrix}
0 & -1 \\
1 & 0 
\end{bmatrix} .
\]
The total force is then given by
\[
\vec{G}(\vec{q}) = - \omega^2 \left( \vec{q} - \vec{q}_0 \right) + F {\mathbb J} \cdot \vec{q} 
= - {\mathbb A} \cdot \left( \vec{q} - \widetilde{\vec{q}}_0 \right) 
\]
where the matrix $\mathbb A$ and the vector $\widetilde{\vec{q}}_0$ are given by  
\begin{align*}
{\mathbb A} &= \begin{pmatrix} 
\omega^2 & F \\
-F & \omega^2 
\end{pmatrix} , & 
\widetilde{\vec{q}}_0 &= \omega^2 {\mathbb A}^{-1} \cdot \vec{q}_0 . 
\end{align*}
Notice that the matrix $\mathbb A$ is symmetric ${\mathbb A}^T = \mathbb A$ if and only if there is no non-potential driving.

One can check that whenever there exists a positively definite matrix $\mathbb K$ solving 
\[
\mathbb K \cdot \mathbb A + \mathbb A^T \cdot \mathbb K = 2 \mathbb K^2
\]
then the steady state \eqref{equ:stationary_condition} is given by 
\[
\rho(\vec{q}) = \frac{1}{Z} \rme^{-\frac{\beta}{2} \left( \vec{q} - \widetilde{\vec{q}}_0 \right) \cdot \mathbb K \cdot \left( \vec{q} - \widetilde{\vec{q}}_0 \right) } . 
\]
In this particular case the matrix $\mathbb K$ is given by 
\[
\mathbb K = \omega^2 \mathbb I.
\]
Similarly the quasi-potential \eqref{def:generalized_potential} has the quadratic form
\[
V(\vec{q}) = \frac{1}{2} \left( \vec{q} - \widetilde{\vec{q}}_0 \right) \cdot \mathbb U \cdot \left( \vec{q} - \widetilde{\vec{q}}_0 \right) ,
\]
where $\mathbb U$ solves the equation 
\[
\mathbb U \cdot \mathbb A + \mathbb A^T \cdot \mathbb U = 2 \mathbb A^T \cdot \mathbb A,
\]
which in this particular case is 
\[
\mathbb U = \frac{F^2 + \omega^4}{\omega^2} \mathbb I. 
\]
We can see that in this particular case the energy function governing the steady state is the multiple of the quasi-potential,
hence the stationary distribution is Boltzmann-like \eqref{equ:generalized_Boltzmann} with generalized inverse temperature 
\[
\widetilde{\beta}(\alpha) = \beta \frac{\omega^4}{\omega^4 + F^2}. 
\]
Hence we can see that in this particular model the Clausius relation holds true for arbitrary choice of parameters.

\section{Near equilibrium regime} 
\label{sec:first_order_expansion}
It is well known fact the analysis of most non-equilibrium systems simplifies when we approach the equilibrium. 
In this section we will analyse systems in the vicinity of equilibrium by providing a systematic expansion of the ``reversible'' heat and work up to the first order in presumably small non-equilibrium driving. 
The underlying bases for this expansion was provided by McLennan, when he showed that the steady state of the system in the vicinity of equilibrium is approximately Boltzmann-like \cite{McLennan1959}.  
Later Komatsu and Nakagawa proved the close relation of the proposition of McLennan to the transient fluctuation theorem \cite{Komatsu2008-2}.
In this section we however mostly follow the work of Maes and Neto\v{c}n\'{y} \cite{Maes2010}.
As we have already seen the Boltzmann-like steady state with an appropriate quasi-potential is a sufficient condition to obtain an extended Clausius relation, see subsection \ref{ssec:quasistatic_heat_and_work}.
We will show that in the linear order around equilibrium the Clausius equation is indeed valid.

\subsection{Steady state}
In order to simplify the discussion we restrict ourselves only to systems without any explicit time dependence in contact with the single thermal bath without velocity-like degrees of freedom, e.g. overdamped diffusion and Markov jump processes, 
driven out of equilibrium only by a non-potential force $\vec{F}$.
Under these condition the magnitude of such force $\|\vec{F}\|$ can be used as the small parameter measuring the strength of the non-equilibrium driving.

By setting the $\vec{F}=\vec{0}$ we obtain an equilibrium dynamics represented by the forward Kolmogorov generator $\gen^*_0$ with the steady state $\rho_0$ given by the equilibrium Boltzmann distribution, i.e. 
\begin{align*}
0 &= \gen^*_0 [ \rho_0 ] (x) , &
\rho_0(x) &= \frac{1}{Z_0} \rme^{-\beta E(x)} ,
\end{align*}
which provides us the leading order of the expansion. 
We formally expand the forward Kolmogorov generator and the steady state around the equilibrium ones $\gen^*_0$ and $\rho_0$ up to the first order in the magnitude of driving and obtain 
\begin{align*} 
\gen^* [\mu](x) &= \gen^*_0 [\mu](x) + \gen^*_1 [\mu](x) + \err[2]{\|\vec{F}\|} , \\
\rho(x) &= \rho_0(x) + \rho_1(x) + \err[2]{\|\vec{F}\|} ,
\end{align*}
where by index $1$ are denoted first order corrections. 
Notice that in case of the overdamped diffusion the expansion of the forward Kolmogorov generator up to the first order is exact,
because the forward Kolmogorov generator \eqref{equ:time_evolution_overdamped} is a linear function of the applied force. 
The steady state up to the first order in the magnitude of the non-equilibrium driving is then given by the solution of the stationary condition \eqref{equ:stationary_condition}, i.e. by
\[
0 = \gen^*[\rho] (x) = \gen^*_0 [\rho_1] (x) + \gen^*_1[ \rho_0] (x) + \err[2]{\|\vec{F}\|} ,
\]
from where the first order correction can be found to be 
\begin{equation}
\rho_1 (x) = - \frac{1}{\gen^*_0} \gen^*_1 \left[ \rho_0 \right] (x) ,
\label{equ:first_order_density}
\end{equation}
where $1/\gen^*_0$ is the forward pseudoinverse \eqref{def:forward_pseudoinverse}.

The first step in order to associate the correction \eqref{equ:first_order_density} with an observable quantities, 
is to realize that it is valid
\[
\gen^*_1 [ \rho_0 ] (x) = - \beta \, w_1(x) \, \rho_0(x) + \err[2]{\| \vec{F} \|} ,
\]
where the $w_1(x)$ is the first order contribution the local power of non-potential forces. 
One can show this either directly for each particular model, e.g. for jump processes see \cite{Maes2010}, or one can follow the general heuristic argument. 
The $\gen^*_1 [\rho_0]$ can be considered to be the time evolution of the equilibrium probability distribution with respect to the full dynamics, 
i.e. we consider it to represent an initial stage of the relaxation process towards steady state starting from the equilibrium. 
As the probability density has to be normalized at each time, the change of the probability distribution for particular configuration $x$ has to be compensated by induced probability currents from $x$. 
Each probability current can also be related to the heat flux associated with the transition and also to the probability that given configuration is going to be occupied. 
On the other hand the heat flux is solely induced by the action of non-potential forces during the relaxation process, while the potential forces are balanced in equilibrium, and hence can be directly related to the local power of these forces. 
Thus concluding the heuristic argument. 
The first order correction \eqref{equ:first_order_density} is hence given by 
\[
\rho_1 (x) = \beta \frac{1}{\gen^*_0} [ w \rho_0 ] (x) .
\]

One of the consequences of the global detailed balance \eqref{def:global_detailed_balance} for the equilibrium dynamics is the symmetry in correlation function with respect to equilibrium state 
\begin{multline*}
\left\langle \rme^{t \gen_0}[A] \, B \right\rangle_{\rho_0} 
= \iint \rmd \rho_0(x_0) \, \dcprob[{(0,T]}]{\omega}{X_0=x_0} \; A(x_T) \, B(x_0) = \\
= \iint \rmd \rho_0(x_T) \, \dcprob[{(0,T]}]{\Theta \omega}{X_0=x_T} \; A(x_T) \, B(x_0) = \\
= \iint \rmd \rho_0(x'_0) \, \dcprob[{(0,T]}]{\omega}{X_0=x'_0} \; A(x'_0) \, B(x'_T) 
= \left\langle A \, \rme^{t \gen_0} [ B ] \right\rangle_{\rho_0} ,
\end{multline*}
which yields to the identity   
\[
\rme^{t \gen_0} [ A ] (x) \, \rho_0 (x) = \rme^{t \gen_0^*} \left[ A \rho_0 \right] (x) 
\]
or equivalently to 
\[
\gen^*_0 [ A \, \rho_0 ] (x) = \gen_0 [A] (x) \, \rho_0(x) .
\]
As a consequence of this identity we express the forward pseudoinverse in the first order correction \eqref{equ:first_order_density} in terms of backward pseudoinverse and obtain 
\[
\rho_1 (x) = \beta \frac{1}{\gen_0} [ w_1 ] (x) \, \rho_0(x) .
\]
By putting all the terms of the stationary distribution altogether we obtain 
\begin{equation}
\rho(x) = \frac{1}{Z_0} \exp \left[ - \beta \left( E(x) - \frac{1}{\gen_0} [ w_1 ] (x) \right) \right] + \err[2]{\|\vec{F}\|} ,
\label{equ:first_order_rho}
\end{equation}
where we can see that the stationary probability distribution up to the first order in non-equilibrium driving effectively corresponds to the equilibrium system with an energy function given by  
\[
\widetilde{E}(x) = E(x) - \frac{1}{\gen_0} [w_1] (x) . 
\]

\subsection{Work and heat}
We have seen that the steady state is given by the stationary distribution \eqref{equ:first_order_rho}, 
where the effective energy function resembles the quasi-potential present in the definition of the reversible work \eqref{def:reversible_work}. 
In order to verify whether these terms are the same or if there is some fundamental difference we need to expand the quasi-potential up to the first order in the magnitude of non-equilibrium driving. 
For that purpose we introduce the Dyson like series for backward pseudoinverse, which reads
\begin{equation}
\frac{1}{\gen} [A] (x) = \frac{1}{\gen_0} [A] (x) - \frac{1}{\gen_0} \gen_1 \frac{1}{\gen} [A] (x) + \left\langle \frac{1}{\gen} [A] \right\rangle_{\rho_0} . 
\label{equ:Dyson_expansion}
\end{equation}
Notice that the main difference between the Dyson series and our extension is the presence of the last term which is associated with the difference between the kernels of $\gen$ and $\gen_0$.
It is quite straightforward that up to the first order only the first order term of the local power contributes, hence the quasi-potential simplifies to 
\[
V(x) = E(x) - \frac{1}{\gen} [w] (x) = E(x) - \frac{1}{\gen} [w_1] (x) + \err[2]{\|\vec{F}\|} .
\]
We apply the Dyson-like expansion \eqref{equ:Dyson_expansion} and obtain 
\[
V(x) = E(x) - \frac{1}{\gen_0} [w_1] (x) - \left\langle \frac{1}{\gen} [w_1] \right\rangle_{\rho_0} + \err[2]{\|\vec{F}\|} .
\]
Obviously, 
\begin{multline*}
\left\langle \frac{1}{\gen} [w_1] \right\rangle_{\rho_0} = \left\langle \frac{1}{\gen} [w_1] - \frac{1}{\gen_0} [w_1 ] \right\rangle_{\rho_0} = \\ 
= \int\limits_0^\infty \rmd t \; \left\{ \left( \left\langle w_1 \right\rangle_\rho - \left\langle w_1 \right\rangle_{\rho_0} \right) - \left\langle \left( \rme^{t \gen} - \rme^{t \gen_0} \right) [ w_1] \right\rangle_{\rho_0} \right\} = \\
= - \int\limits_0^\infty \rmd t \; \left\langle \left( \rme^{t \gen} - \rme^{t \gen_0} \right) [ w_1] \right\rangle_{\rho_0} + \err[2]{\|\vec{F}\|} ,
\end{multline*}
moreover it is also valid that
\[
\left( \rme^{t \gen} - \rme^{t \gen_0} \right) [ w_1] (x) = \int\limits_0^t \rmd s \; \rme^{(t-s) \gen_0} \gen_1 \rme^{s \gen} [w_1 ] (x) = \err[2]{\|\vec{F}\|} 
\]
hence the term $\left\langle \frac{1}{\gen} [A] \right\rangle_{\rho_0}$ is at least of second order in non-equilibrium driving.

We can conclude that up to the first order in the non-equilibrium driving the ``reversible'' work is given by the quasi-potential which corresponds to the effective energy $\widetilde{E}(x)$. 
As we have already seen if the ``reversible'' work is given by the same quasi-potential as the steady state, we have shown that the generalized Clausius relation \eqref{equ:Clausius_relation} is valid up to the first order in non-potential driving around equilibrium.  
In this case the temperature is the actual temperature of the attached thermal bath and Shannon entropy is given as 
\[
{\mathcal S} = \beta \left\langle E - \frac{1}{\gen_0} [ w_1 ] \right\rangle_{\rho} + \ln Z_0 . 
\]
Consequently the generalized heat capacity \eqref{def:generalized_heat_capacity} is always positive and given by 
\[
C = \frac{1}{T^2} \left[ \left\langle V^2 \right\rangle_\rho -  \left( \left\langle V \right\rangle_\rho \right)^2 \right] + \err[2]{\|\vec{F}\|} . 
\]

\section{Low-temperature behaviour} 
At beginning of the 20th century Nernst and Planck \cite{Callen1985} provided the last building block of the classical thermodynamics in the form of so called Nernst theorem or more commonly as the third law of thermodynamics. 
The third law states that the for temperature going to absolute zero also the entropy needs to converge to zero, i.e. close to zero temperature the isothermal and adiabatic processes became indistinguishable. 
It's precisely the third law of thermodynamics that ensures that many quasistatic response functions go to zero at zero temperature, e.g. the heat capacity under any constraint, thus ensuring practical inaccessibility of the absolute zero. 
Although the third law seems to be valid in real systems there are models which does not obey the third law, e.g. the classical ideal gas. 
In equilibrium the breaking of the third law of thermodynamics is closely related to the degeneracy of the ground state usually associated with the simplified description of the real physical system, which is not adequate at very low temperatures. 
There is no definite answer to the question 
whether the Third law of thermodynamics has a meaningful non-equilibrium generalization, e.g. in terms of vanishing generalized heat capacity in the zero temperature limit.
We have seen that the answer in case of the two- and three-level model was positive.
However in this sections we will show examples which violates the third law even among the systems with finite number of configurations and discuss the reasons behind such behaviour. 
Let us stress once more that the classification of systems according to their low temperature behaviour is still an open question.

\subsection{Example: ``Merry-go-round'' model}
\label{ssec:merry-go-around}
\begin{figure}[ht]
\caption{Illustration of the ``Merry-go-round'' model for $N=6$ with denoted possible transitions.}
\label{pic:merry-go-round}
\begin{center}
\includegraphics[width=.5\textwidth,height=!]{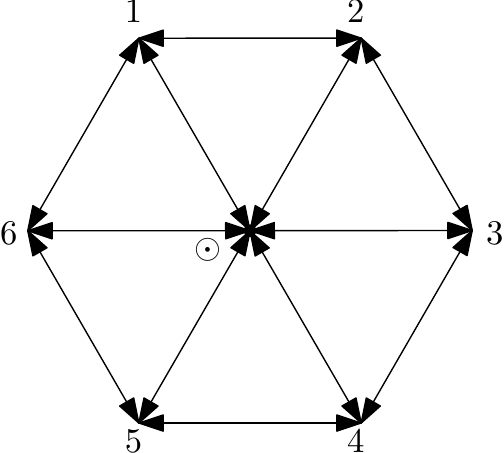} 
\end{center}
\end{figure}

The first example which we will study is a discrete model with $N$ states on the ring denoted as $x=1,2,\ldots,N$ ($N \geq 3$ and $N+1 \equiv 1$) and an extra state connected to all others denoted by $\odot$, see figure \ref{pic:merry-go-round}. 
Such system can represent a driven ratchet with the possibility to hurdle over several tooths at once. 
The dynamics is fully determined by transition rates 
\begin{align*}
\rate{x}{x\pm 1} &= \rme^{\pm \beta F / 2} , \\
\rate{\odot}{x} &= \rme^{-\frac{\beta(\Delta + U)}{2}} , \\
\rate{x}{\odot} &= \rme^{-\frac{\beta(\Delta - U)}{2}} ,
\end{align*}
where $F$ denotes the work associated with driving forces, $U$ is the energy associated with the middle state $\odot$ and $\Delta>0$ is the height of the barrier which separates the middle state from the ring. 
Given these transition rates, the stationary distribution is given by 
\begin{align*}
\rho(\odot) &= \left(1 + N \rme^{-\beta U} \right)^{-1} , \\ 
\rho(x) &= \left(N + \rme^{\beta U} \right)^{-1} ,
\end{align*}
from where we can see that it is independent of the driving force $F$ as well as the height of the barrier $\Delta$. 
On the level of occupations, we can interpret the situation as to have the `ground state' $\odot$ and $N$ identical `excitations' with energy gap $U$.

Despite its simplicity, our model exhibits a low-temperature crossover between prevailing conductor- and insulator-like behavior. 
We compare the steady probabilistic current on the ring 
\[
\jmath \equiv \curr{x}{x+1} 
= \rho(x)\,[\rate{x}{x+1} - \rate{x+1}{x}] 
= \frac{\rme^{\frac{\beta F}{2}} - \rme^{-\frac{\beta F}{2}}}{N + \rme^{\beta U}}
\]
with the overall activity of the system represented by the overall ``number of transitions per second''
\begin{align*}
  {\mathcal A} &= \sum_{x=1}^N \rho(x) \left[\rate{x}{x+1} + \rate{x}{x-1} + \rate{x}{\odot}\right]
  + \rho(\odot) \rate{\odot}{x}
\\
  &= \frac{\rme^{-\frac{\beta\Delta}{2}} + N(\rme^{\frac{\beta F}{2}} + \rme^{-\frac{\beta F}{2}} + \rme^{\frac{\beta(U - \Delta)}{2}})}{N + \rme^{\beta U}} ,
\end{align*}
i.e. we calculate the relative number of those transitions contributing to transport and dissipation
\[
\frac{N \jmath}{{\mathcal A}}
= \frac{N \left( \rme^{\frac{\beta F}{2}} - \rme^{-\frac{\beta F}{2}} \right)}{\rme^{-\frac{\beta\Delta}{2}} + N \left( \rme^{\frac{\beta F}{2}} + \rme^{-\frac{\beta F}{2}} + \rme^{\frac{\beta(U - \Delta)}{2}} \right)} .
\]
With $N$ fixed, this ratio tends along $\beta \to +\infty$ 
\begin{enumerate}
\item to zero provided $|F| < U - \Delta$ (\emph{insulator}), 
\item to unity on the condition $0 \neq |F| > U - \Delta$ (\emph{conductor}); in particular, it is always conductor in the case $\Delta > U$.
\end{enumerate}

This result can be also concluded from the heuristic picture: 
For $|F| < U - \Delta$ the state $\odot$ is a preferred successor, i.e. transition rate $\rate{x}{\odot}$ is by far larger then transition rates $\rate{x}{x\pm1}$, 
for all states on the ring and hence whenever an excitation from ground state occurs it is most likely to be suppressed by de-excitation within short time interval with marginal contribution to the current on the ring. 
On the other hand, for $|F| > U - \Delta$ the preferred successor lies on the ring for every state and hence an excitation from the ground state $\odot$ is typically followed by a large number of rounds over the ring before it eventually vanishes, like when on a fast-enough rotating merry-go-round, thus inducing a large current.

In order to study the behaviour of the generalized heat capacity \eqref{def:generalized_heat_capacity} in low temperature asymptotic we need to find the quasi-potential $V_0 = V(\odot)$, $V_1 = V(1) = \ldots = V(N)$, 
i.e. we need to solve the equation $\gen[ V ] = q - \langle q \rangle_\rho$ with the backward Kolmogorov generator \eqref{equ:backward_Kolmogorov_gen_jump} being
\begin{align*}
\gen[V]_0 &\equiv \gen[V](\odot) = N\,\rme^{\frac{-\beta(\Delta + U)}{2}} (V_1 - V_0) , \\
\gen[V]_1 &\equiv \gen[V](1) = \ldots = \gen[V](N) = -\rme^{\frac{-\beta(\Delta - U)}{2}} (V_1 - V_0) 
\end{align*}
and the local heat production 
\begin{align*}
q_0 & \equiv q(\odot) = -N U \rme^{\frac{-\beta(\Delta + U)}{2}} , \\
q_1 & \equiv q(1) = \ldots = q(N) = U \rme^{-\frac{\beta(\Delta - U)}{2}} + F \left(\rme^{\frac{\beta F}{2}} - \rme^{-\frac{\beta F}{2}} \right)
\end{align*}
which yields 
\[
V_0 - V_1 = U + F \rme^{\frac{\beta(\Delta - U)}{2}} \frac{\rme^{\frac{\beta F}{2}} - \rme^{-\frac{\beta F}{2}}}{1 + N \rme^{-\beta U}} .
\]
Hence, $V_0 - V_1 = U + \err{ |F| \rme^{\frac{\beta(|F| - U + \Delta)}{2}} }$ where the non-equilibrium correction is exponentially damped in $\beta$ for the insulator $(|F| < U - \Delta)$, 
whereas it exponentially diverges in the conductor regime. 
Heuristically, the asymptotics follows by observing that in the conduction regime the particle started anywhere in the ring typically performs $\err{ \rate{x}{x+1} / \rate{x}{\odot} } \approx \rme^{\beta(|F| - U + \Delta)}$ jumps along the ring (each contributing to the heat by $|F|$) before it finally jumps to $\odot$.

Finally, the generalized heat capacity \eqref{def:generalized_heat_capacity} is
\[
C =  - \left\langle \frac{\partial V}{\partial (\beta^{-1})} \right\rangle_\rho
  = N\beta^2 (V_0 - V_1) U \rme^{-\beta U} \rho^2(\odot)
\]
In the conduction regime it goes asymptotically like
$C \asymp \rme^{\frac{\beta(|F| - 3U + \Delta)}{2}}$ and hence it exhibits exponential divergence whenever $0 \neq |F| > 3U - \Delta$.
In particular, for $\Delta > 3U$ we have exponential damping for $F = 0$ (detailed balance), whereas already for an arbitrarily small $F \neq 0$ the heat capacity becomes divergent, see summary in table \ref{tab:merry-go-round}.
\begin{table}
\caption{Two regimes of the ``merry-go-round'' model.}
\label{tab:merry-go-round}
\begin{center}
\begin{tabular}{l|l|c}
insulator & 
$|F| < U - \Delta$ & 
$C \asymp \rme^{-\beta U}$ \\ \hline 
conductor & 
$|F| > U - \Delta$ & 
$C \asymp \rme^{\frac{\beta(|F| - 3U + \Delta)}{2}}$ if $F \neq 0$ \\
& 
& 
$ \phantom{C}  \asymp \rme^{-\beta U}$ \phantom{******} if $F = 0$
\end{tabular}
\end{center}
\end{table}

We have seen that this simple model exhibits a rich behaviour in the zero temperature limit, where there is not only the transition between insulator and conductor, 
but also a transition within the conductor regime, where the heat capacity becomes divergent in the zero temperature limit. 
While the transition between conducting and insulator regime is easy to explain on the level of physical reality, 
we have no simple physical explanation of the transition within the conducting regime yet.

\subsection{Example: Driven diffusion on the ring}
\label{ssec:diffusion_on_the_ring}
As a second example exhibiting the non-trivial low temperature behaviour we will investigate the diffusion on one-dimensional ring described by the stochastic differential equation \eqref{def:position_overdamped}
\begin{align*}
\rmd q_t &= f(q_t) \; \rmd t + \sqrt{2T}\,\rmd W_t , &
q_t &\in {\mathbb S}_1  
\end{align*}
where the total force $f(q) = F - U'(q)$ contains the constant driving force $F$ and the $\mathbb S_1 $ is the unit length interval with periodical boundaries representing the ring. 
The forward \eqref{equ:time_evolution_overdamped} and backward \eqref{equ:bacward_generator_overdamped} Kolmogorov generators are then given by 
\begin{align*}
\gen [A](q) &= f(q) A'(q) + T A''(q) , \\
\gen^* [\mu](q) &= - (f(q) \mu(q))' + T \mu''(q) .
\end{align*}
We want to compute the generalized heat capacity \eqref{def:generalized_heat_capacity}
\begin{align*}
C_F &= \int \rmd q \; \partial_T \rho(q) \, \left( U(q) - \frac{1}{\gen}[w](q) \right) , &
w(q) &= F f(q) ,
\end{align*}
with the driving $F$ constant. 
From the stationarity equation \eqref{equ:stationary_condition} and linearity of the froward Kolmogorov generator we have
\[
\left(\partial_T \gen^*\right) [\rho] (q) + \gen^* [\partial_T \rho] (q) = 0
\]
and by using the normalization condition 
\[
\int \rmd q \; \partial_T \rho(q) = 0, 
\]
we obtain
\[
\partial_T \rho (q) = -\frac{1}{\gen^*} [\rho''] (q) 
\]
and hence
\begin{equation}
C_F = \int \rmd q \; \left\{ F f(q) \left(\frac{1}{\gen^*}\right)^2 [\rho''] (q) - U(q) \frac{1}{\gen^*} [\rho''] (q) \right\} .
\label{equ:1D_heat_capacity}
\end{equation}

Before we proceed to the analysis of the behaviour of the system under low temperatures we define the notion of the stable point as whether they are present will prove to be crucial in low temperature limit. 
We say that $q^*$ is a stable point of the dynamics if the total force in such a point is zero and if forces in the vicinity of the stable point points towards it, 
i.e. $U'(q*) = F$ and $U''(q^*) > 0$. 
Notice that in case there is no random thermal force acting on the particle, 
the position of the particle placed in the stable point does not evolve in time. 
However the presence of random thermal forces ensures in the stochastic dynamics that the particle will eventually leave the point.

\subsubsection{``Insulator'' regime}
%The first case which we will study is the case when there is a single stable point in the system. 
%The heuristic argument follows the idea that the dynamics of the system in the proximity of the stable point can be approximated by the dynamics locally governed by an effective quadratic potential.  
%As we decreases the temperature the probability of leaving the effective potential well, 
%typically exponentially with the depth of the effective potential well, 
%also decreases.
%Thus the biggest contribution to the heat capacity comes from an effective quadratic potential and hence the heat capacity in the zero temperature limit goes to  
%\[
%C_F^{T=0} = \frac{1}{2} . 
%\]
%
%
The first case which we will study is the case when there is a single stable point in the system. 
We skip a rigorous treatment of the problem and only give a simple heuristic argument. 
On the assumption that there exist a unique stable point $q*$, the low-temperature dynamics of the system is localized and hence it can be well approximated by the effective diffusion dynamics in the quadratic potential $U''(q^*) (q-q^*)^2/2$ on the full real line. 
Hence, it becomes asymptotically indistinguishable from an equilibrium dynamics under the quadratic potential force, for which the equipartition theorem yields $C = 1/2$. 
More detailed analysis reveal corrections exponentially damped when $T -> 0$.

\subsubsection{``Conducting'' regime}
In case there is no stable point where the forces can be balanced there is always a steady probability current even in case of low temperatures, 
hence we speak about the conducting regime of the system, 
i.e. the conducting regime then occurs whenever $| f(q) | > 0$ along the entire circle. 
In this case we can evaluate the heat capacity \eqref{equ:1D_heat_capacity} in the zero temperature limit explicitly 
and we reveal a remarkably different behaviour with respect to the insulator regime.  
Using that the zero-temperature stationary density equals $\rho_0(q) = j_0 / f(q)$, with $j_0$ the stationary current, and by employing the identity
\begin{multline*}
\frac{1}{\gen^*_0}[\mu'](q) = -\frac{1}{f(q)} \left[ \mu(q) - \left( \int \rmd \bar{q} \; \frac{1}{f(\bar{q})} \right)^{-1} \int \rmd \widetilde{q} \; \frac{\mu(\widetilde{q})}{f(\widetilde{q})} \right] = \\
= -\frac{1}{f(q)} \left[ \mu(q) - j_0 \int \rmd \widetilde{q} \; \frac{\mu(\widetilde{q})}{f(\widetilde{q})} \right] 
= - \frac{\mu(q) - \left\langle \mu \right\rangle_{\rho_0}}{f(q)},
\end{multline*}
where by $\gen_0^*$ we denote the zero temperature limit of the forward Kolmogorov generator $\lim_{T \to 0^+} \gen^*$,
along with the periodic boundary conditions we get
\[
\left(\frac{1}{\gen_0^*}\right)^2[\rho_0''] (q) 
= -\frac{1}{\gen_0^*} \left[ \frac{(\rho_0^2)'}{2j_0} \right] (q) 
= \frac{\rho_0^3(q)}{2 j_0^2} - \frac{\rho_0(q)}{2 j_0^2} \int \rmd \bar{q} \; \rho_0^3(\bar{q})
\]
and finally
\begin{align*}
C_F^{T=0} &= \frac{F}{2j_0} \, \left[ \int \rmd q \; \rho_0^2(q) - \int \rmd q \; \rho_0^3(q) \right]
-\frac{1}{2j_0} \int \rmd q \; U'(q) \, \rho_0^2(q) \\
&= \frac{1}{2} - \frac{F}{2j_0} \int \rmd q \; \rho_0^3(q) .
\end{align*}

\begin{figure}[ht]
\caption{$C_F^{T=0}$ for $U(q) = \sin(2\pi q)$.}
\label{pic:1D_ring_lowT}
\begin{center}
\includegraphics[width=.9\textwidth,height=!]{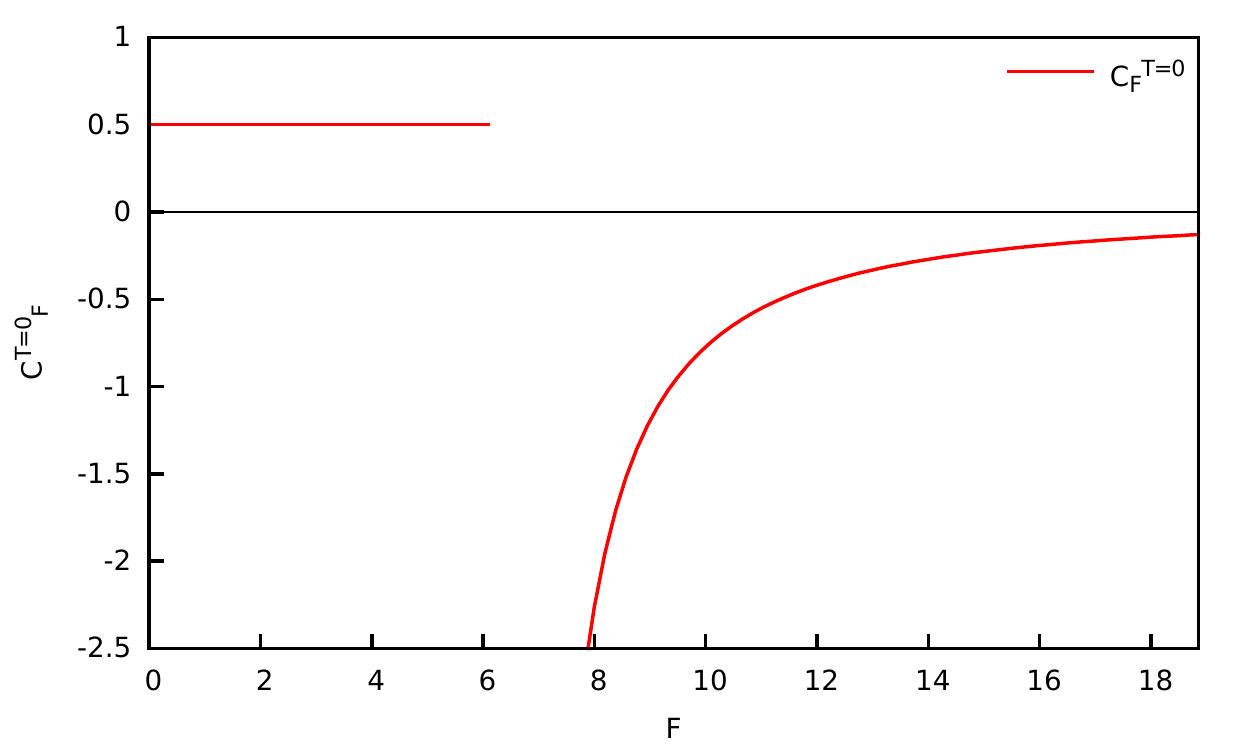}
\end{center}
\end{figure}

We can see that there is a clear distinction between the conduction and insulator regime represented as a zero temperature phase transition at $F= \max_{q} | U'(q) |$, where the zero temperature heat capacity suddenly changes from the (finite) equilibrium value to a divergent pattern, see for example the figure \ref{pic:1D_ring_lowT}. 
Moreover we can see that in the large driving limit, $|F| \to \infty$, the low temperature heat capacity converges to zero as the effect of increased temperature is overwhelmed by the driving force and hence the stationary distribution converge to the uniform distribution over the ring. 
Notice also that this behaviour corresponds to the physical intuition that in the singular driving limit the thermal force is negligible to the driving force and thus the system does not react to the changes of temperature.  

Let us also remark that the formula~\eqref{equ:1D_heat_capacity} can be written in the ``invariant'' form for arbitrary $T$,
\[
C_F = \int \rmd q \; (f^2(q) + T f'(q))\,\left(\frac{1}{\gen^*}\right)^2[\rho''] (q)  
= \left\langle \left[ \left(\frac{1}{\gen}\right)^2 [ f^2 + T f' ] \right]'' \right\rangle_\rho 
\]
which manifests that the generalized heat capacity is independent of how the forces are decomposed into the potential and non-potential components; 
one may also note that $f^2 + T f' = \gen[ V ]$ whenever $f = V'$.

\section{Conclusion}
In this chapter we have studied quasistatic processes in non-equilibrium system driven mostly by non-potential forces. 
Our main result states that the heat and work can be naturally decomposed in the quasistatic limit to what we call ``housekeeping'' and ``reversible'' components, see \eqref{def:reversible_heat}, \eqref{def:reversible_work}, \eqref{def:housekeeping_work}. 
We have identified the diverging ``housekeeping'' components with the steady heat and work production which is a consequence of the system being out of equilibrium. 
On the other hand we have shown that the ``reversible'' components are geometric in the sense that they don't depend on the actual parametrization of the external protocol $\alpha$. 
We have also shown that the ``reversible'' component of total heat/work is a natural generalization of the equilibrium concept of reversible heat/work.
By construction the ``housekeeping'' and ``reversible'' parts behave differently under the protocol inversion $\alpha \mapsto \Theta \alpha$: while the ``housekeeping'' part is symmetric, the ``reversible'' part is antisymmetric. 
This enables us in principle to experimentally distinguish these components of the total heat by considering the cycle process composed of the forward protocol $\alpha$ and the backward protocol $\Theta \alpha$.

Another result is that the sum of ``reversible'' heat and work corresponds to the change of the internal energy defined as the mean value of the energy function, 
which is to be interpreted as a (renormalized) quasistatic form of the law of energy conservation, see \ref{ssec:generalized_thermodynamics}. 
We have also shown that the ``reversible'' heat is invariant with respect to the ``gauge'' transformation, see \ref{ssec:gauge_invariane}, 
which also enabled us to define the generalized heat capacity \eqref{def:generalized_heat_capacity} as generalized quasistatic response function. 
We have studied the behaviour of the generalized heat capacity in several models \ref{ssec:two_level_nonp}, \ref{ssec:three_level_nonp} and \ref{ssec:diffusion_on_plane},
and seen that in strong non-equilibrium situations the heat capacity can be also negative.

A drawback of the presented renormalization scheme for the (quasistatically divergent) heat and work is that the Clausius relation \eqref{equ:Clausius_relation} isn't valid in general,
which is in good agreement with the fact that the generalized heat capacity can have negative values. 
Nevertheless up to the first order in non-equilibrium driving \ref{sec:first_order_expansion} one proves the Clausius relation to be still valid as a consequence of the McLennan theorem.  
We have found the general sufficient condition \eqref{equ:generalized_Boltzmann} for the system to obey the Clausius relation beyond the close-to-equilibrium regime, the question whether we can extend the class of such systems any further is still open. 
A related open question is whether there exist general relations between (quasistatic) response functions analogous to equilibrium Maxwell relations.

Another open question concerns the possible extensions of the Nernst theorem to non-equilibrium systems, which we studied only briefly.  
We have provided examples exhibiting a zero temperature phase transition and even a divergent low-temperature asymptotics of the steady heat capacity, see \ref{ssec:diffusion_on_plane}, \ref{ssec:merry-go-around} and \ref{ssec:diffusion_on_the_ring}, 
however the nature of such low temperature behaviour is still not well understood.

Notice also that we have been focused on mean values of the heat and work only, 
whereas the detailed structure of fluctuations in the non-equilibrium quasistatic regime still remains a largely unexplored field.
This is also related to the question on the role of non-equilibrium response functions in the stability analysis of non-equilibrium steady states.

Main results of this chapter were published in \cite{result1,result2}.

% Quasistatic transformations of periodically driven systems 
% a. Adapting the previous framework
% b. Model calculations on periodic two-level system

\chapter{Quasistatic transformations of periodically driven systems}
\label{chapter:periodically_driven_systems}
In previous chapter we have described the behaviour of various physical systems with broken detailed balance undergoing quasistatic transformations of external parameters. 
The global detailed balance in the previous chapter was broken by either the effect of non-potential forces or by attaching the system to multiple thermal or particle baths. 
In this chapter we further extend the formalism of quasistatic processes to also describe systems driven by periodic forces.
We will study the quasistatic transformations connecting periodic steady states and corresponding to slow changes of arbitrary system parameters, except for the period $\tau$ of the driving which in the sequel is always assumed constant.
Typical physical example of such system is diffusion of ions in the presence of periodical electric field.

The first part of this chapter will be more mathematical as we will introduce the description of these systems, 
first in terms of explicitly time dependent quantities, 
then by using the results of Floquet theory in terms of Fourier components. 
The second part of this chapter will then focus on how to generalize thermodynamics to these systems, 
followed by an analysis of specific limiting cases and finished by numerical results for some particular examples.

\section{Description of periodically driven system}
The time evolution of a system driven by periodical external forces is again characterized by the forward Kolmogorov generator\index{Kolmogorov generator!forward} $\gen^*_t$
\begin{equation}
\partial_t \mu_t = \gen^*_t \left[ \mu_t \right], 
\label{equ:time_evolution_2}
\end{equation} \index{time evolution!periodical driving}
where the generator does not depend on time only via external parameters $\alpha(t)$ but it also explicitly depends on time, $\gen^*_t \equiv \gen^*_{t;\alpha(t)}$. 
We assume that for fixed external parameters $\alpha$ the generator is periodical with period $\tau$   
\[
\gen^*_{t+\tau;\alpha} = \gen^*_{t;\alpha} ,
\]
where the period does not depend on $\alpha$, $\tau(\alpha) \equiv \tau$. 
As we will see this assumption is essential, because it will allow us to reformulate the problem by using the Floquet theory
and thus effectively remove the explicit time dependence.

\subsection{The Floquet theory} 
\label{ssec:Floquet_theory}
The main result of the Floquet theory is the \emph{Floquet theorem}\index{Floquet theorem} \cite{Ward:Floquet_theory}, 
which states that every solution of \eqref{equ:time_evolution_2} can be written in the form
\begin{equation}
\mu_t(x) = \sum_k \rme^{-\lambda_k t} p^k_t(x), 
\label{equ:Floquet_theorem}
\end{equation}
where $p^k_t(x)$ are real $\tau$-periodic functions, $p^k_{t+\tau}(x) = p^k_t(x)$, and $\lambda_k$ are non-negative numbers, 
which can be interpreted as the reciprocal of relaxation times associated with each mode.
Furthermore the number of linearly independent periodic functions $p^k_t(x)$ is given by the dimension of the space, for proof see \cite{Ward:Floquet_theory}.

Although the Floquet theory gives us the explicit form of the solution, it gives us no simple algorithm how to find the $\tau$-periodic functions nor the relaxation times $1/\lambda_k$. 
To provide such an algorithm we use the periodicity of the functions $p^k_t(x)$ and the forward Kolmogorov generator $\gen^*_{t;\alpha}$ to expand them in the Fourier series 
and we will construct the generalized Kolmogorov generator describing the time evolution of Fourier components. 
We will further show that $p^k_t(x)$ correspond to eigenvectors of generalized forward Kolmogorov generator associated with real eigenvalues.

\subsubsection{Generalized forward Kolmogorov generator} 
We start with the construction of the generalized forward Kolmogorov generator by expansion of the periodic functions $p^k_t(x)$ to Fourier series
\begin{equation}
p^k_t(x) = \sum\limits_{n \in \integers} \rme^{\frac{2 \pi \rmi n t}{\tau}} \hat{p}^k_n(x), 
\label{equ:Fourier_decomposition_p}
\end{equation}
where by $\hat{p}^k_n(x)$ we denote the Fourier components. 
The fact that functions $p^k_t(x)$ are real implies a symmetry 
\[
\hat{p}^k_{-n}(x) = \overline{\hat{p}^k_n(x)},
\]
where $\bar{f}$ denotes the complex conjugation of $f$.
Inserting the decomposition \eqref{equ:Fourier_decomposition_p} into \eqref{equ:Floquet_theorem} we obtain Fourier expansion of probability density 
\begin{equation}
\mu_t(x) = \sum\limits_{n \in \integers} \rme^{\frac{2 \pi \rmi n t}{\tau}} \sum_k \rme^{- \lambda_k t} \hat{p}^k_n(x) 
= \sum\limits_{n \in \integers} \rme^{\frac{2 \pi \rmi t}{\tau}} \hat{\mu}_{n;t}(x) ,
\label{equ:mu_decomposition}
\end{equation}
where
\[ 
\hat{\mu}_{n;t}(x) = \sum_k \rme^{- \lambda_k t} \hat{p}^k_n (x) 
\]
denote time-dependent Fourier components of the probability density $\mu_t(x)$. 
Negative-order Fourier components of probability density are also by definition connected to the positive ones by the symmetry 
\begin{equation}
\hat{\mu}_{-n;t}(x) = \overline{\hat{\mu}_{n;t}(x)}.
\label{equ:complex_conjugation_of_density}
\end{equation}
Similarly we expand the forward Kolmogorov generator 
\[
\gen^*_{t;\alpha} \left[ \mu \right] (x) = \sum\limits_{n \in \integers} \rme^{\frac{2 \pi \rmi n t}{\tau}} \hat\gen^*_n \left[ \mu \right] (x), 
\]
where $\hat\gen^*_n$ denotes the Fourier component which also obeys
\begin{equation}
\overline{\hat\gen^*_{-n}} = \hat\gen^*_n .
\label{equ:complex_conjugation_of_generator}
\end{equation}
By putting all these decompositions together into the time evolution equation \eqref{equ:time_evolution_2} we obtain  
\[
0 = \sum\limits_{n \in \integers} \rme^{\frac{2 \pi \rmi n t}{\tau}} 
\left[ \partial_t \hat\mu_{n;t} (x) - \sum\limits_{m \in \integers} \hat\gen^*_{n-m} \left[ \hat\mu_{m;t} \right] (x) + \frac{2 \pi \rmi n t}{\tau} \hat\mu_{n;t} (x) \right] ,
\]
which has to be valid for each possible time $t$, hence each Fourier component can be considered as independent and henceforth the equation itself is equivalent to the set of equations   
\begin{equation}
\partial_t \hat\mu_{n;t} (x) = \sum\limits_{m \in \integers} \hat\gen^*_{n-m} \left[ \hat\mu_{m;t} \right] (x) - \frac{2 \pi \rmi n t}{\tau} \hat\mu_{n;t} (x) .
\label{equ:time_evolution_of_Fourier_components}
\end{equation}
Notice that equation for $(-n)$-th component and $n$-th component are connected by complex conjugation, 
which preserve the symmetry \eqref{equ:complex_conjugation_of_density} during the time evolution. 
By introduction of \emph{vectors of Fourier components}, 
\[
\hat{\vec\mu}_t (x) = \left( \dots , \hat\mu_{-1;t} (x), \hat\mu_{0;t} (x), \hat\mu_{1;t} (x), \dots \right), 
\]
we can rewrite the set of equations \eqref{equ:time_evolution_of_Fourier_components} into the form of \emph{generalized time evolution}\index{time evolution!periodical driving} 
\begin{equation}
\partial_t \hat{\vec\mu}_t (x) = \gensc^*_\alpha \left[ \hat{\vec\mu}_t \right] (x) , 
\label{equ:generalized_time_evolution}
\end{equation}
where $\gensc^*_\alpha$ denotes the \emph{generalized forward Kolmogorov generator}\index{Kolmogorov generator!forward!generalized}
\begin{equation}
\left[ \gensc^*_\alpha \right]_{mn} = \hat\gen^*_{m-n} - \frac{2 \pi \rmi m}{\tau} \delta_{mn} . 
\label{equ:generalized_forward_generator}
\end{equation}
Notice that the generalized forward Kolmogorov generator is \emph{not explicitly time-dependent} and the only time-dependence left lies in $\alpha$. 
Also notice that the generalized forward Kolmogorov generator is not self adjoint 
\[
\overline{\left[ \gensc^*_\alpha \right]}_{nm} = \left[ \gensc^*_\alpha \right]_{mn} - \frac{4 \pi \rmi m}{\tau} \delta_{mn} , 
\]
neither is symmetric
\[
\left[ \gensc^*_\alpha \right]_{nm} \neq \left[ \gensc^*_\alpha \right]_{mn} .
\]

\subsubsection{Spectral properties of generalized Kolmogorov generator}
As was noted before we want to show that vectors $\hat{\vec{p}}^k(x)$ are eigenvectors with real eigenvalue of the generalized Kolmogorov generator. 
We start by inserting the decomposition \eqref{equ:mu_decomposition} into the time evolution equation \eqref{equ:generalized_time_evolution}
\begin{equation}
0 = \sum\limits_k \left( \gensc^*_\alpha \left[ \hat{\vec{p}}^k \right] (x) + \lambda_k \hat{\vec{p}}^k (x) \right) \rme^{-\lambda_k t} , 
\label{equ:generalized_time_evolution_for_p}
\end{equation}
where by $\hat{\vec{p}}^k (x)$ we denote the vector of Fourier components. 
One of the statements of the Floquet theorem is that the periodic functions are linearly independent, 
which ensures that also Fourier components and so the vectors of Fourier components are linearly independent. 
Also all the operators in the equation \eqref{equ:generalized_time_evolution_for_p} are linear, which yields to 
\[
\gensc^*_\alpha \left[ \hat{\vec{p}}^k \right] (x) = - \lambda_k \hat{\vec{p}}^k (x) ,
\]
where we recognize the eigenvector problem. 
Thus every $p^k_t(x)$ gives us an eigenvector $\vec{p}^k(x)$ of the generalized forward Kolmogorov generator $\gensc^*$ with the real eigenvalue $-\lambda_k$.

If we shift the eigenvector's $\hat{\vec{p}}^k$ components by $m$ 
\begin{equation}
\left[ \hat{\vec{q}}^{k;m} (x) \right]_n = \hat{p}^k_{m+n} (x), 
\label{equ:shifted_state}
\end{equation}
we again obtain an eigenvector although with shifted eigenvalue $- \lambda_k + 2 \pi \rmi m/\tau$
\begin{multline*}
\left[ \gensc^*_\alpha \left[ \hat{\vec{q}}^{k;m} \right] (x) \right]_l
= \sum_{j \in \integers} \hat\gen^*_{l+m-j} [ \hat{p}^k_j ] (x) - \frac{2 \pi \rmi l}{\tau} \hat{p}^k_{l+m} (x) = \\
= - \lambda_k \hat{p}^k_{l+m}(x) + \frac{2 \pi \rmi m}{\tau} \hat{p}^k_{l+m} (x)   
= \left( - \lambda_k + \frac{2 \pi \rmi m}{\tau} \right) \left[ \hat{\vec{q}}^{k;m} \right]_l .
\end{multline*}
Thus we can generate an infinite number of eigenvectors from a single one. 
Hence it is reasonable to assume that the maximum number of eigenvectors with real eigenvalues corresponds dimension of the space $x$, i.e. it corresponds to the count of states at fixed time.
From where it follows that all linear independent solutions \eqref{equ:Floquet_theorem} can be found as the eigenvectors corresponding to real eigenvalues. 
e.g. for two-level system we have two eigenvectors with real eigenvalues, one presumably being steady state thus corresponding to the zero eigenvalue.

We can also conclude that the steady state corresponds to the steady state of the generalized time evolution equation \eqref{equ:generalized_time_evolution},
\begin{equation}
0 = \gensc^* \left[ \hat{\vec{\rho}} \right] (x) . 
\label{equ:generalized_stationary_distribution}
\end{equation}
Let us also notice here that by shifting the steady state \eqref{equ:shifted_state} we obtain an infinite number of eigenvectors with zero real part.

\subsubsection{Time-dependence in external parameters}
%Up to now we have considered the external parameters $\alpha$ to be fixed. 
%In case they explicitly depend on time we would like to be able to parametrize the generalized forward Kolmogorov generator by them
%and also to parametrize linearly independent solutions $\vec{p}^k_\alpha(x)$ by them.
%The necessary assumptions to be able to do it is that by a manipulation with external parameters $\alpha$ we are not able converge out of the function space given by $p^k_t(x)$,
%or in other words we are not able to obtain unphysical solutions as the limiting cases. 
%This leads to the requirement that the forward Kolmogorov generator is smooth function of external parameters $\alpha$,
%and that for almost every possible value of $\alpha$ the functions $p^k_t(x)$ are non-zero and linearly independent. 
%
%
Up to now we have considered the external parameters $\alpha$ to be fixed. 
In systems driven by non-potential forces the change from fixed external parameters to external parameters explicitly dependent on the time 
occurred only in the change 
\begin{align*}
\alpha &\to \alpha(t) & 
&\Longrightarrow & 
\partial_t\mu_t (x) = \gen^*_\alpha [ \mu_t ] (x) & 
\to \partial_t \mu_t (x) = \gen^*_{\alpha(t)} [ \mu_t ] (x) ,
\end{align*}
which proved to be crucial for further considerations in the quasistatic limit. 
Although the same can be done on the level of the time dependent forward Kolmogorov generator $\gen^*_{t;\alpha(t)}$ even in systems driven by periodic forces,
it is not clear whether we can assume the same to be true also for generalized forward Kolmogorov generator. 
In order to ensure the same behaviour of the generalized forward Kolmogorov, 
\begin{align*}
\alpha &\to \alpha(t) & 
&\Longrightarrow & 
\partial_t\hat{\vec{\mu}}_t (x) = \gensc^*_\alpha [ \hat{\vec{\mu}}_t ] (x) & 
\to \partial_t \hat{\vec{\mu}}_t (x) = \gensc^*_{\alpha(t)} [ \hat{\vec{\mu}}_t ] (x), 
\end{align*}
we need to assume that we are able to trace each eigenvector as we change the external parameters, i.e. 
\begin{align*}
\lim_{t' \to t } \hat{\vec{p}}^k_{\alpha(t')}(x) &= \hat{\vec{p}}^k_{\alpha(t)}(x) &
&\Longleftrightarrow &
\lim_{t' \to t } p^k_{t',\alpha(t')}(x) &= p^k_{t,\alpha(t)}(x)
\end{align*}
and also 
\[
\lim_{t' \to t} \lambda^k_{\alpha(t')} = \lambda^k_{\alpha(t)} .
\]

\subsubsection{Initial condition}
We start by the construction of the initial state in the generalized time evolution \eqref{equ:generalized_time_evolution}. 
Our starting point is the assumption that the functions $p^k_t(x)$ are linearly independent in an arbitrary time $t$, 
hence the initial state described by the probability distribution $\mu_{t_0}(x)$ can be decomposed into these functions
\[
\mu_{t_0}(x) = \sum_k \beta_k \rme^{- t_0 \lambda_k} p^k_{t_0}(x),
\]
where $\beta_k$ represent real coefficients in the decomposition. 
This decomposition enables us to rewrite the initial state in the form of the vector of Fourier components by 
\[
\hat{\vec{\mu}}_{t_0}(x) = \sum_k \beta_k \hat{\vec{p}}^k (x) .
\] 
We can see that the determination of the initial condition suitable for the generalized time evolution is at least as hard as to solve the original time evolution \eqref{equ:time_evolution_2} with the initial condition $\mu_{t_0}(x)$,
because in order to obtain the initial state we need to determine the basis of $\hat{\vec{p}}^k(x)$ as well as its eigenvalues $\lambda_k$ 
and hence at that point we are able to directly describe the solution of the time evolution at arbitrary time starting from the arbitrary initial condition. 
Let us note here, that our aim is not to provide any simple method how to solve the time evolution with periodical driving,
but rather to reformulate it as an explicitly time-independent evolution at cost of enlarging the state space. 
Moreover, we will see that the ``reversible'' component of the heat and work in the quasistatic limit will again be independent of the initial condition.

\subsubsection{The generalized backward Kolmogorov generator}
In the chapter \ref{chapter:non-equilibrium_thermodynamics} we have introduced the backward Kolmogorov generator as the adjoint generator to the forward Kolmogorov generator.
We would like to define the generalized backward Kolmogorov generator in the similar fashion. 
We start by representing the mean value of some observable $A_t$ which can also periodically depend on time in terms of Fourier components
\[
\left\langle A_t \right\rangle_{\mu_t} = \int \rmd \Gamma(x) \; A_t(x) \, \mu_t(x) 
= \sum_{n \in \integers} \rme^{\frac{2 \pi \rmi n t}{\tau}} \int \rmd \Gamma(x) \; \left[ \hat{\vec{A}}(x) * \hat{\vec{\mu}}_t(x) \right]_n ,
\]
where $\hat{\vec{a}} * \hat{\vec{b}}$ is a shorthand for discrete convolution 
\begin{equation}
[\hat{\vec{a}} * \hat{\vec{b}}]_n = \sum_{m \in \integers} \hat{a}_{n-m} \hat{b}_m . 
\label{def:konvolution}
\end{equation}
The steady state is now time-periodic and so is the mean value of an arbitrary (possibly periodic as well) observable $A_t$.
Then the time derivative of the mean value of $A_t$ is 
\begin{align*}
\partial_t \left\langle A_t \right\rangle_{\mu_t} 
=& \sum_{n \in \integers} \frac{2 \pi \rmi n}{\tau} \rme^{\frac{2 \pi \rmi n t}{\tau}} \int \rmd \Gamma(x) \; \left[ \hat{\vec{A}}(x) * \hat{\vec{\mu}}_t(x) \right]_n + \\
&+ \sum_{n \in \integers} \rme^{\frac{2 \pi \rmi n t}{\tau}} \int \rmd \Gamma(x) \; \left[ \hat{\vec{A}}(x) * \gensc^*[\hat{\vec{\mu}}_t](x) \right]_n , \\ 
=& \sum_{n \in \integers} \rme^{\frac{2 \pi \rmi n t}{\tau}} \int \rmd \Gamma(x) \; \left[ \hat{\Omega} \cdot \left( \hat{\vec{A}}(x) * \hat{\vec{\mu}}_t(x) \right) + \hat{\vec{A}}(x) * \gensc^* [ \hat{\vec{\mu}}_t ] (x) \right]_n ,
\end{align*}
where we have introduced \emph{the frequency matrix}\index{frequency matrix}
\begin{equation}
\left[ \hat{\Omega} \right]_{mn} = \frac{2 \pi \rmi n}{\tau} \delta_{mn} .
\label{def:frequency_matrix}
\end{equation}
We see that the time derivative of the mean value consists of two terms, 
the first one reflecting the periodic nature of the stationary distribution (together with the possible periodicity of the observable itself), 
and the second containing the relaxation of the probability distribution towards the steady state $\rho_t$. 
Notice that in the steady state only the first term contributes
\[
\partial_t \left\langle A_t \right\rangle_{\rho_t}
= \sum_{n \in \integers} \rme^{\frac{2 \pi \rmi n t}{\tau}} \int \rmd \Gamma(x) \; \left[ \hat{\Omega} \cdot \left( \hat{\vec{A}}(x) * \hat{\vec{\rho}}(x) \right) \right]_n .
\]

By using the definition of backward Kolmogorov generator \eqref{def:backward_generator} the time derivative can also be expressed as 
\[
\partial_t \left\langle A_t \right\rangle_{\mu_t} 
= \sum_{n \in \integers} \rme^{\frac{2 \pi \rmi n t}{\tau}} \int \rmd \Gamma(x) \; \sum_{m \in \integers} \left[ \hat{\vec{\mu}}_t(x) \right]_{n-m} 
\left\{ \sum_{k \in \integers} \hat\gen_{m-k} [ \hat{A}_k ] (x) + \frac{2 \pi \rmi m}{\tau} \hat{A}_m(x) \right\} ,
\]
where we can identify \emph{the generalized backward Kolmogorov generator}\index{backward Kolmogorov generator!generalized} 
\begin{equation}
\left[ \gensc [ \hat{\vec{A}} ] (x) \right]_n = \sum_{m \in \integers} \hat\gen_{n-m} [ \hat{A}_m ] (x) + \frac{2 \pi \rmi n}{\tau} \hat{A}_n(x) ,
\label{def:generalized_backward_generator}
\end{equation}
which yields to more compact expression  
\[
\partial_t \left\langle A_t \right\rangle_{\mu_t} 
= \sum_{n \in \integers} \rme^{\frac{2 \pi \rmi n t}{\tau}} \int \rmd \Gamma(x) \; \left[ \hat{\vec{\mu}}_t (x) * \gensc[ \hat{\vec{A}} ] (x) \right]_n .
\]
We can see that when the observable is periodic function of time, i.e. there is no explicit aperiodic time dependence for example in external parameters, 
the time derivative of the mean value is solely given by the generalized backward Kolmogorov generator,
this corresponds in case when there is no periodical driving to the situation when the observable does not depend on time at all, 
then there also the time derivative of the mean value is solely given by backward Kolmogorov generator. 
Notice that from the linear independence of Fourier components we also obtained the relation between generalized forward and backward generators 
\begin{equation}
\int \rmd \Gamma(x) \; \hat{\vec{\mu}}_t (x) * \gensc[ \hat{\vec{A}} ] (x) =
\int \rmd \Gamma(x) \; \left\{ \hat{\Omega} \cdot \left( \hat{\vec{A}}(x) * \hat{\vec{\mu}}_t(x) \right) + \hat{\vec{A}}(x) * \gensc^* [ \hat{\vec{\mu}}_t ] (x) \right\} .
\label{equ:generalized_generators_identity}
\end{equation}

\section{Quasistatic processes} 
In the previous section we have developed a mathematical description of the time evolution, which reformulates the problem with explicit time dependence of the forward Kolmogorov generator in terms of the generalized forward Kolmogorov generator acting on vector of components in the Fourier basis. 
Since that we have formally obtained the time evolution equation, which does not explicitly depends on time \eqref{equ:generalized_time_evolution},
hence we can follow the same pattern as in section \ref{sec:quasistatic_processes_noneq} in order to study the heat and the work in the quasistatic limit.

\subsection{Probability distribution}
We start with the Fourier components of the probability distribution on configurations,
where we assume that the system is at any time close to the stationary state, 
\[
\hat{\vec{\mu}}_t (x) = \hat{\vec{\rho}}_{\alpha(\epsilon t)} (x) + \Delta \hat{\vec{\mu}}_{\epsilon t}(x) .
\]   
By expanding the time evolution equation \eqref{equ:generalized_time_evolution} up to the first order in $\epsilon$ we immediately obtain the equation for the correction $\Delta \hat{\vec{\mu}}$ 
\[
\epsilon \dot\alpha(\epsilon t) \cdot \left. \nabla_\alpha \hat{\vec{\rho}}_\alpha (x) \right|_{\alpha=\alpha(\epsilon t)} 
= \gensc^*_{\alpha(\epsilon t)} [ \Delta \hat{\vec{\mu}}_{\epsilon t} ] (x) + \err[2]{\epsilon} ,
\]
which can be again formally solved 
\begin{equation}
\Delta \hat{\vec{\mu}}_t(x) = \epsilon \dot\alpha(\epsilon t) \cdot \frac{1}{\gensc^*_{\alpha(\epsilon t)}} \left[ \left. \nabla_\alpha \hat{\vec{\rho}}_\alpha \right|_{\alpha=\alpha(\epsilon t)} \right] (x) 
+ \err[2]{\epsilon} 
\label{equ:quasistatic_expansion_generalized_density}
\end{equation}
with \emph{the generalized forward pseudoinverse}\index{forward pseudoinverse!generalized} 
\begin{equation}
\frac{1}{\gensc^*_\alpha} [ \hat{\vec{\mu}} ] (x) = \lim_{k \to \infty} \int\limits_0^{k \tau} \rmd s \; \left\{ \projsc^*_{0;\alpha} [ \hat{\vec{\mu}} ] (x) - \rme^{s \gensc^*_\alpha} [ \hat{\vec{\mu}} ] (x) \right\} ,
\label{def:generalized_forward_pseudoinverse}
\end{equation}
where by $\projsc^*_{0;\alpha}$ we denote the projection to the steady state $\hat{\vec{\rho}}_\alpha$ and we integrate over multiples of periods, i.e. the $k$ is an integer $k \in \mathbb N$.
Notice that in the case the argument of pseudoinverse is physically acceptable solution, i.e. it is the decomposition to Fourier series of real function and hence does not contain shifted steady states, 
the pseudoinverse is simply given by the integral from $0$ to $\infty$. 
The integration over the multiple of periods is there to provide a correct definition for unphysical arguments, where we need to suppress indeterminate oscillating integrals 
from the imaginary eigenvalues of shifted steady states.

\subsubsection{Stationary projection}
In the section \ref{sec:quasistatic_processes_noneq} the projector to the steady state was given simply by the stationary distribution multiplied by the normalization of distribution \eqref{def:projector}, 
in this case in general we expect the stationary projection to have a similar form of the product of the steady state and some normalization 
\[
\left[ \hat\projsc^*_{0;\alpha} [ \hat{\vec{\mu}} ] (x) \right]_n = \left[ \hat{\vec{\rho}}_\alpha (x) \right]_n \sum_{m \in \mathcal I} \int \rmd \Gamma(y) \; \hat\mu_m(y) , 
\] 
where $\mathcal I$ is an index set, which yet need to be determined. 
The projection has to obey 
\begin{align*}
\hat\projsc^*_{0;\alpha} [\hat{\vec{\rho}}_\alpha ] (x) &= \hat{\vec{\rho}}_\alpha (x) , && \\
\hat\projsc^*_{0;\alpha} [\hat{\vec{\mu}}_\alpha ] (x) &= 0 , & \forall \hat{\vec{\mu}} (x) & \text{ linearly independent of } \hat{\vec{\rho}}(x) . 
\end{align*}
For the first condition to be valid the index $0$ has to member of the index set $\mathcal I$, 
because only the zeroth component of the probability distribution is normalized to unity, while the other components are normalized to zero. 
For the second condition, 
we recall that shifted eigenvector \eqref{equ:shifted_state} of the generalized forward Kolmogorov is again an eigenvector of the generalized forward Kolmogorov generator which is linearly independent of the original one. 
If we now take the steady state $\hat{\vec{\rho}}_\alpha(x)$ as the basis for generating the shifted solution,
we can see that the component normalized to unity also shifts, and hence for the second condition on the projector to be valid, any other component than zero cannot contribute. 
Hence the projector has to be defined as 
\[
\left[ \hat\projsc^*_{0;\alpha} [ \hat{\vec{\mu}} ] (x) \right]_n = \left[ \hat{\vec{\rho}}_\alpha (x) \right]_n \int \rmd \Gamma(y) \; \hat\mu_0(y) . 
\]

\subsection{Heat and work}
Now we can proceed to the quasistatic expansion of path observables as heat and work. 
Although we will proceed along the same lines as in chapter \ref{chapter:non-equilibrium_thermodynamics}, hence we would expect to obtain similar results, 
we will see that the system driven by periodic action of external forces differs in several aspects from what we have introduced before.

\subsubsection{Local power and local heat production}
The first step is to represent the local power and the local heat production in terms of Fourier components. 
In case there are no non-potential forces acting on the system the local power is given by the total time derivative of the energy $E_{t;\alpha}(x)$
\begin{equation}
w_{t;\alpha(\epsilon t)}(x) = \frac{\rmd E_{t;\alpha(\epsilon t)}(x)}{\rmd t} 
= \epsilon \dot\alpha(\epsilon t) \cdot \left. \nabla_\alpha E_{t;\alpha}(x) \right|_{\alpha=\alpha(\epsilon t)} + \partial_t E_{t;\alpha(\epsilon t)}(x) ,
\label{def:local_power_t}
\end{equation}
where the first term correspond to the action of external forces, the second term corresponds to the action of driving forces. 
Fourier components of the local power are then obtained by expanding the explicit time dependence in to the Fourier series 
\[
\left[ \hat{\vec{w}}_{\alpha(\epsilon t)}(x) \right]_n 
= \epsilon \dot\alpha(\epsilon t) \cdot \left. \nabla_\alpha \left[ \hat{\vec{E}}_\alpha(x) \right]_n \right|_{\alpha=\alpha(\epsilon t)} + \frac{2 \pi \rmi n}{\tau} \left[ \hat{\vec{E}}_{\alpha(\epsilon t)}(x) \right]_n ,
\]
which by using the definition of frequency matrix \eqref{def:frequency_matrix} can be written more compactly as 
\begin{equation}
\hat{\vec{w}}_{\alpha(\epsilon t)}(x) 
= \epsilon \dot\alpha(\epsilon t) \cdot \left. \nabla_\alpha \hat{\vec{E}}_\alpha(x) \right|_{\alpha=\alpha(\epsilon t)} + \hat{\Omega} \cdot \hat{\vec{E}}_{\alpha(\epsilon t)}(x) ,
\label{def:work_nonp_gen}
\end{equation}
where the first term represents again the work of external forces, or in terminology of the chapter \ref{chapter:non-equilibrium_thermodynamics} potential forces $\hat{\vec{w}}^\text{pot}_\alpha(x)$, and the second term corresponds to the work of driving forces $\hat{\vec{w}}^\text{drv}_\alpha(x)$, which cannot be described in terms of the derivative of the potential along an external parameters hence corresponding to the work of non-potential forces. 
Similarly the local heat production given by 
\begin{equation}
q_{t;\alpha(\epsilon t)}(x) = \gen^*_{t;\alpha(\epsilon t)} \left[ E_{t;\alpha(\epsilon t)} \right] (x) 
\label{def:local_heat_production_t}
\end{equation}
is in terms of Fourier components represented by
\[
\left[ \hat{\vec{q}}_{\alpha(\epsilon t)}(x) \right]_n 
= \sum_{m \in \integers} \hat{\gen}_{n-m;\alpha(\epsilon t)} \left[ \left[\hat{\vec{E}}_{\alpha(\epsilon t)}\right]_m \right] (x) 
= \left[ \gensc_{\alpha(\epsilon t)} \left[ \hat{\vec{E}}_{\alpha(\epsilon t)} \right] (x) \right]_n 
- \frac{2 \pi \rmi n}{\tau} \left[ \hat{\vec{E}}_{\alpha(\epsilon t)} (x) \right]_n ,
\]
where we have used the definition of the generalized backward Kolmogorov generator \eqref{def:generalized_backward_generator}, 
or in more compact form as 
\begin{equation}
\hat{\vec{q}}_{\alpha(\epsilon t)}(x) 
= \gensc_{\alpha(\epsilon t)} \left[ \hat{\vec{E}}_{\alpha(\epsilon t)} \right] (x) - \hat{\Omega} \cdot \hat{\vec{E}}_{\alpha(\epsilon t)} (x) .
\label{def:heat_gen}
\end{equation}

\subsubsection{Steady currents}
Having the local power and local heat production we can verify that there is a steady energy current through the system. 
The mean steady local power is given by 
\begin{align*}
\left\langle w_{t;\alpha} \right\rangle_{\rho_{t;\alpha}} 
&= \sum_{n \in \integers} \rme^{\frac{2 \pi \rmi n t}{\tau}} \int \rmd \Gamma(x) \; \left[ \hat{\vec{w}}_\alpha(x) * \hat{\vec{\rho}}_\alpha(x) \right]_n \\
&= \sum_{n \in \integers} \rme^{\frac{2 \pi \rmi n t}{\tau}} \int \rmd \Gamma(x) \; \left[ \left( \hat{\Omega} \cdot \hat{\vec{E}}_\alpha (x) \right) * \hat{\vec{\rho}}_\alpha (x) \right]_n ,
\end{align*}
and similarly we obtain the mean steady local heat production as 
\begin{multline*}
\left\langle q_{t;\alpha} \right\rangle_{\rho_{t;\alpha}} 
= \sum_{n \in \integers} \rme^{\frac{2 \pi \rmi n t}{\tau}} \int \rmd \Gamma(x) \; \left[ \hat{\vec{q}}_\alpha (x) * \hat{\vec{\rho}}_\alpha (x) \right]_n = \\
= \sum_{n \in \integers} \rme^{\frac{2 \pi \rmi n t}{\tau}} \int \rmd \Gamma(x) \; \left[ \hat{\Omega} \cdot \left( \hat{\vec{E}}_\alpha (x) * \hat{\vec{\rho}}_\alpha (x) \right) - \left( \hat{\Omega} \cdot \hat{\vec{E}}_\alpha (x) \right) * \hat{\vec{\rho}}_\alpha (x) \right]_n = \\ 
= \sum_{n \in \integers} \rme^{\frac{2 \pi \rmi n t}{\tau}} \int \rmd \Gamma(x) \; \left[ \hat{\vec{E}}_\alpha (x) * \left( \hat{\Omega} \cdot \hat{\vec{\rho}}_\alpha (x) \right) \right]_n ,
\end{multline*}
where we have used the identity \eqref{equ:generalized_generators_identity}, the fact the $\hat{\vec{\rho}}$ is steady state and the distributivity of the frequency matrix \eqref{def:frequency_matrix} 
\begin{equation}
\hat{\Omega} \cdot \left( \hat{\vec{A}} * \hat{\vec{B}} \right)  
= \left( \hat{\Omega} \cdot \hat{\vec{A}} \right)  * \hat{\vec{B}} 
+ \hat{\vec{A}} * \left( \hat{\Omega} \cdot \hat{\vec{B}} \right) .  
\label{equ:spectral_matrix_property}
\end{equation}
We can see that together the mean local heat production and the mean local power sum up to the time derivative of the mean steady value of the energy, as one would have suspected,  
\begin{multline*}
\partial_t \left\langle E_t \right\rangle_{\rho_{t;\alpha}} 
= \sum_{n \in \integers} \rme^{\frac{2 \pi \rmi n t}{\tau}} \int \rmd \Gamma(x) \; \left[ \hat{\Omega} \cdot \left( \hat{\vec{E}}_\alpha (x) * \hat{\vec{\rho}}_\alpha (x) \right) \right]_n = \\ 
= \sum_{n \in \integers} \rme^{\frac{2 \pi \rmi n t}{\tau}} \int \rmd \Gamma(x) \; \left[ \left( \hat{\Omega} \cdot \hat{\vec{E}}_\alpha (x) \right) * \hat{\vec{\rho}}_\alpha (x) 
+ \hat{\vec{E}}_\alpha (x) * \left( \hat{\Omega} \cdot \hat{\vec{\rho}}_\alpha (x) \right) \right]_n .
\end{multline*}
From there we can also see that the time-average change of the energy over the period $\tau$ is equal to zero, while the average local power and heat is in general not,
hence there is the steady energy current through the system even on the scales longer than the period. 
This is due to the fact that external periodic force is responsible for the changes of the total energy, hence it is constantly powering the system, 
while the same amount of energy need to be dissipated from the system to the thermal bath in order to maintain the steady state.

\subsubsection{Quasistatic expansion}
The first step in obtaining the quasistatic expansion of the mean work or heat is then rewriting it in terms of Fourier components, 
where we will take the advantage of the characterization of the total work by the local power
\begin{multline*}
{\mathcal W}(\alpha(\epsilon t)) 
= \int\limits_0^{\frac{T}{\epsilon}} \rmd t \; \left\langle w_{t;\alpha(\epsilon t)}(x) \right\rangle_{\mu_t} = \\
= \int\limits_0^{\frac{T}{\epsilon}} \rmd t \; \int \rmd \Gamma(x) \; \sum_{n \in \integers} \rme^{\frac{2 \pi \rmi n t}{\tau}} \left[ \hat{\vec{w}}_{\alpha(\epsilon t)} (x) \right]_n 
\sum_{m \in \integers} \rme^{\frac{2 \pi \rmi m t}{\tau}} \left[ \hat{\vec{\mu}}_{t;\alpha(\epsilon t)} (x) \right]_m = \\
= \sum_{n \in \integers} \int\limits_0^{\frac{T}{\epsilon}} \rmd t \; \rme^{\frac{2 \pi \rmi n t}{\tau}} \int \rmd \Gamma(x) \; 
\left[ \hat{\vec{w}}_{\alpha(\epsilon t)} (x) * \hat{\vec{\mu}}_{t;\alpha(\epsilon t)} (x) \right]_n . 
\end{multline*} 
Now we can insert the quasistatic expansion of the probability distribution \eqref{equ:quasistatic_expansion_generalized_density} and obtain 
\begin{multline*}
{\mathcal W}(\alpha(\epsilon t)) 
= \sum_{n \in \integers} \int\limits_0^{\frac{T}{\epsilon}} \rmd t \; \rme^{\frac{2 \pi \rmi n t}{\tau}} \int \rmd \Gamma(x) \; 
\left[ \hat{\vec{w}}_{\alpha(\epsilon t)} (x) * \hat{\vec{\rho}}_{\alpha(\epsilon t)} (x) \right]_n + \\
+ \sum_{n \in \integers} \int\limits_0^{\frac{T}{\epsilon}} \rmd t \; \epsilon \, \dot\alpha(\epsilon t) \cdot \rme^{\frac{2 \pi \rmi n t}{\tau}} \int \rmd \Gamma(x) \; 
\left[ \hat{\vec{w}}_{\alpha(\epsilon t)} (x) * \frac{1}{\gensc^*_{\alpha(\epsilon t)}} \left[ \left. \nabla_\alpha \hat{\vec{\rho}}_\alpha \right|_{\alpha=\alpha(\epsilon t)} \right] (x) \right]_n + \\ 
+ \err{\epsilon} , 
\end{multline*}
where the first term is the integrated mean local steady power which contains an extensive in time component of the work and hence in the leading order is proportional to $1/\epsilon$, 
while the second term is finite, which can be better seen after the substitution $\epsilon t \to t$. 
By inserting the definition of the work \eqref{def:work_nonp_gen} we can separate the terms according the power of $\epsilon$ and we obtain 
\begin{multline*}
{\mathcal W}(\alpha(\epsilon t)) 
= \int\limits_0^{\frac{T}{\epsilon}} \rmd t \; \left\langle w^\text{drv}_{t;\alpha(\epsilon t)} \right\rangle_{\rho_{\alpha(\epsilon t)}}  
+ \sum_{n \in \integers} \int\limits_0^T \rmd t \; \dot\alpha(t) \cdot \rme^{\frac{2 \pi \rmi n t}{\epsilon \tau}} \int \rmd \Gamma(x)  \times \\
\times \left[ \left. \nabla_\alpha \hat{\vec{E}}_\alpha (x) \right|_{\alpha=\alpha(\epsilon t)} * \hat{\vec{\rho}}_{\alpha(t)} (x) + \hat{\vec{w}}^\text{drv}_{\alpha(t)} (x) * \frac{1}{\gensc^*_{\alpha(t)}} \left[ \left. \nabla_\alpha \hat{\vec{\rho}}_\alpha \right|_{\alpha=\alpha(t)} \right] (x) \right]_n + \\ 
+ \err{\epsilon} .
\end{multline*}
Moreover if we assume that the all quantities in the second term are smooth with respect to time, 
we can apply the Riemann-Lebesgue lemma and see that only the zeroth Fourier component is preserved up to the linear order in $\epsilon$, 
and while the zeroth component depends on time only via external parameters $\alpha$ it can be represented as the geometric integral 
\begin{multline*}
{\mathcal W}(\alpha(\epsilon t)) 
= \int\limits_0^{\frac{T}{\epsilon}} \rmd t \; \left\langle w^\text{drv}_{t;\alpha(\epsilon t)} \right\rangle_{\rho_{\alpha(\epsilon t)}} + \\ 
+ \int \rmd \alpha \cdot \int \rmd \Gamma(x) \; 
\left[ \nabla_\alpha \hat{\vec{E}}_\alpha (x) * \hat{\vec{\rho}}_\alpha (x) +\hat{\vec{w}}^\text{drv}_\alpha (x) * \frac{1}{\gensc^*_\alpha} \left[ \nabla_\alpha \hat{\vec{\rho}}_\alpha \right] (x) \right]_0 
+ \err{\epsilon} ,
\end{multline*}
where we will again identify the ``housekeeping'' component of the work\index{housekeeping work} as the first term and the ``reversible''\index{reversible work} component of work as the second geometric term. 
Directly from the definition we can see that the ``housekeeping'' component is
\begin{enumerate} 
\item extensive in time in the leading order,
\item present even if the external parameters $\alpha$ are kept constant, 
\item invariant with respect to external protocol reversal $\alpha(t) \to \alpha(T-t)$ if $T/\epsilon$ is multiple of period~$\tau$,
\item equivalent to the ``housekeeping'' component \eqref{def:housekeeping_work} in case there are no periodical forces acting on the system,
\end{enumerate}
while the ``reversible'' component is
\begin{enumerate}
\item geometric in the sense that it is independent of parametrization of the trajectory of external parameters, 
\item finite, 
\item antisymmetric with respect to the external protocol reversal $\alpha(t) \to \alpha(T-t)$,
\item equivalent to the ``reversible'' component \eqref{def:reversible_work} in case there are no periodical forces acting on the system,
\end{enumerate}
which justifies our choices.

\subsubsection{The ``reversible'' work and the generalized backward pseudoinverse}
Up to now we didn't try to provide any physical description of the generalized forward pseudoinverse \eqref{def:generalized_forward_pseudoinverse} present in the ``reversible'' component of the total work 
as we have done for other types of driving in the section \ref{sec:quasistatic_processes_noneq}. 
This is due to the fact that in general we cannot associate the generalized backward pseudoinverse with the generalized forward pseudoinverse due to the extra term in the relation \eqref{equ:generalized_generators_identity} if compared with \eqref{def:backward_generator}. 
However in the ``reversible'' component of the total work there is only the zeroth Fourier component present, hence the relation \eqref{equ:generalized_generators_identity} simplifies to
\[
\int \rmd \Gamma(x) \; \left[ \hat{\vec{\mu}}_t (x) * \gensc[ \hat{\vec{A}} ] (x) \right]_0 =
\int \rmd \Gamma(x) \; \left[ \hat{\vec{A}}(x) * \gensc^* [ \hat{\vec{\mu}}_t ] (x) \right]_0 .
\]
In this particular case we can effectively define \emph{the generalized backward pseudoinverse}\index{backward pseudoinverse!generalized} as 
\begin{equation}
\frac{1}{\gensc} [ \hat{\vec{w}} ] (x) = \int\limits_0^\infty \rmd t \; \left\{ \hat{\vec{1}}_0 \int \rmd \Gamma(y) \; \left[ \hat{\vec{w}} (y) * \hat{\vec{\rho}} (y) \right]_0 - \rme^{t \gensc} [ \hat{\vec{w}} ] (x) \right\} ,
\label{def:generalized_backward_pseudoinverse}
\end{equation}
where the vector $[ \hat{\vec{1}}_m ]_n = \delta_{mn}$ select the $m$-th Fourier component. 
Because the zeroth component is obtained by time-averaging over the period, the generalized backward pseudoinverse can be equivalently defined as 
\[
\frac{1}{\gensc} [ \hat{\vec{w}} ] (x) = \lim_{k \to \infty} \int\limits_0^{k \tau} \rmd t \; \left\{ \hat{\vec{1}}_0 \left\langle w_t \right\rangle_{\rho_t} - \rme^{t \gensc} [ \hat{\vec{w}} ] (x) \right\} ,
\]
which corresponds to the additional work done on the system to the steady work production along the relaxation process starting from the configuration $x$, 
when taking the limit in such a way, that we suppress the oscillations.
If we also introduce the notation for time-averaged mean values with respect to the steady state 
\[
\frac{1}{\tau} \int\limits_0^\tau \rmd t \; \left\langle A_t \right\rangle_{\rho_t} 
= \int \rmd \Gamma(x) \; \left[ \hat{\vec{A}}(x) * \hat{\vec{\rho}}(x) \right]_0 
\equiv \left\llangle \hat{\vec{A}} \right\rrangle_{\hat{\vec{\rho}}} ,
\]
the ``reversible'' work can be written more compactly as 
\begin{equation}
{\mathcal W}^\text{rev}(\alpha) = \int \rmd\alpha \cdot \left\llangle \nabla_\alpha \left( \hat{\vec{E}}_\alpha - \frac{1}{\gensc_\alpha} [ \hat{\vec{w}}^\text{drv}_\alpha ] \right) \right\rrangle_{\hat{\vec{\rho}}_\alpha} .
\label{equ:generalized_reversible_work}
\end{equation}

\subsubsection{The properties of the ``housekeeping'' work}
We have seen that the ``reversible'' component of work is well defined in the limit for all rescaled smooth time evolutions. 
As we will see this is not always true for the ``housekeeping'' component of the work. 
At first we will rewrite the ``housekeeping'' work in terms of Fourier components with rescaled time scale 
\begin{equation}
{\mathcal W}^\text{kh}(\alpha(t)) = \frac{1}{\epsilon} \int\limits_0^T \rmd t \; \sum_{n \in \integers} \rme^{\frac{2 \pi \rmi n t}{\epsilon \tau}} \int \rmd \Gamma(x) \; \left[ \hat{\vec{w}}^\text{drv}_{\alpha(t)} (x) * \hat{\vec{\rho}}_{\alpha(t)}(x) \right]_n .
\label{equ:generalized_housekeeping_work}
\end{equation}
Then we treat the zero component separately from the rest of the terms and rewrite it to the form more suitable for further analysis 
\begin{multline}
{\mathcal W}^\text{kh}(\alpha(t)) 
= \frac{1}{\epsilon} \int\limits_0^T \rmd t \; \left\llangle \hat{\vec{w}}^\text{drv}_{\alpha(t)} \right\rrangle_{\hat{\vec{\rho}}_{\alpha(t)}} - \\
- \sum_{n \in \integers \setminus \{0\} } \frac{\rmi \tau}{2 \pi n} \Biggl\{ \rme^{\frac{2 \pi \rmi n T}{\epsilon \tau}} \int \rmd \Gamma(x) \; \left[ \hat{\vec{w}}^\text{drv}_{\alpha(T)} (x) * \hat{\vec{\rho}}_{\alpha(T)}(x) \right]_n - \\
- \int \rmd \Gamma(x) \; \left[ \hat{\vec{w}}^\text{drv}_{\alpha(0)} (x) * \hat{\vec{\rho}}_{\alpha(0)}(x) \right]_n \Biggr\} + \\
+ \int\limits_0^T \rmd t \; \sum_{n \in \integers \setminus \{0\} } \frac{\rmi \tau}{2 \pi n} \rme^{\frac{2 \pi \rmi n t}{\epsilon \tau}} \int \rmd \Gamma(x) \; \partial_t \left[ \hat{\vec{w}}^\text{drv}_{\alpha(t)} (x) * \hat{\vec{\rho}}_{\alpha(t)}(x) \right]_n .
\label{equ:generalized_housekeeping_work_rew}
\end{multline}
We can see that in the quasistatic limit $\epsilon \to 0^+$ the last term tends to go to zero because of Riemann-Lebesgue lemma, 
the first term is responsible for the divergence of the ``housekeeping'' component of the work in the quasistatic limit.
The second term has the most interesting properties, 
in general it is of the same order of the ``reversible'' component and hence in principle it can be comparable,
it is also responsible for the ``housekeeping'' component not being invariant with respect of the external parameters trajectory reversal $\alpha (t) \to \alpha(T-t)$ unless taken the limit over the multiples of period $\tau$. 
Moreover the second term heavily depends on the initial and final condition in the sense that it is oscillating and hence does not converge in general in the quasistatic limit.
From physical point of view this term corresponds to the choice of the initial and final ``phase'' or in other words to the choice of the initial and final time within the scope of one period. 
It also means that the second term then can play an important role in the evaluation of experimental results.
To illustrate it consider two different experimental setups, in whose the quasistatic limit is approached by extending the time interval, 
in the first setup we assume we are able to determine the times inside the period quite precisely and the fluctuations are small across ale runs of the experiment, then the second term can contribute significantly. 
In the second setup lets assume that we are not able to control at all the initial and final time within the scope of one period, 
then we can assume the ``phases'' are uniformly distributed and then if we average over the ``phases'' before taking the quasistatic limit, we can see that the second term does not contribute.

\subsection{Quasistatic energetics} 
In previous chapter \ref{chapter:non-equilibrium_thermodynamics} we have obtained the firs law of thermodynamics by analyzing the difference of the mean value of the total energy in the quasistatic limit. 
Here we choose a bit different approach, we start with combination of ``reversible'' components only and then discuss how it is related to the change of energy and compare it with sum of the full quasistatic expansion of heat and work together. 
In the previous subsection we have mainly discussed the quasistatic expansion of the work, 
however in a similar fashion we can also obtain the ``housekeeping'' and ``reversible'' component of the heat 
\begin{align}
{\mathcal Q}^\text{hk}(\alpha) &= \int\limits_0^{\frac{T}{\epsilon}} \rmd t \; \left\langle q_{t;\alpha(\epsilon t)} \right\rangle_{\rho_{t;\alpha(\epsilon t)}} , \label{equ:generalized_housekeeping_heat} \\
{\mathcal Q}^\text{rev}(\alpha) &= - \int \rmd \alpha \cdot \left\llangle \nabla_\alpha \frac{1}{\gensc_\alpha}[\hat{\vec{q}}_\alpha] \right\rrangle_{\hat{\vec{\rho}}_\alpha} \label{equ:generalized_reversible_heat} .
\end{align}

If we combine the ``reversible'' components and apply the definitions of the local heat production \eqref{def:heat_gen} and the local power \eqref{def:work_nonp_gen} we obtain 
\begin{multline*}
{\mathcal W}^\text{rev}(\alpha) + {\mathcal Q}^\text{rev}(\alpha) 
= \int \rmd\alpha \cdot \left\llangle \nabla_\alpha \left( \hat{\vec{E}}_\alpha - \frac{1}{\gensc_\alpha} [ \hat{\vec{w}}^\text{drv}_\alpha + \hat{\vec{q}}_\alpha ] \right) \right\rrangle_{\hat{\vec{\rho}}_\alpha} = \\
= \int \rmd\alpha \cdot \left\llangle \nabla_\alpha \left( \hat{\vec{E}}_\alpha - \frac{1}{\gensc_\alpha} \gensc_\alpha [ \hat{\vec{E}}_\alpha ] \right) \right\rrangle_{\hat{\vec{\rho}}_\alpha} ,
\end{multline*}
which corresponds to the change of time-averaged mean value of the energy 
\begin{equation}
\rmd \llangle \hat{\vec{E}}_\alpha \rrangle_{\hat{\vec{\rho}}_\alpha} 
= \dbar {\mathcal W}^\text{rev}(\alpha) + \dbar {\mathcal Q}^\text{rev}(\alpha) .
\label{equ:first_law_t}
\end{equation}
This equation we will consider to be a candidate for the generalized first law of thermodynamics\index{first law of thermodynamics!generalized}, with ${\mathcal U}(\alpha) = \llangle \hat{\vec{E}}_\alpha \rrangle_{\hat{\vec{\rho}}_\alpha}$ being a generalizes internal energy\index{internal energy!generalized}.
Notice that in case there is no periodical time dependence the generalized internal energy is already equal to the internal energy introduced in chapter \ref{chapter:non-equilibrium_thermodynamics} and hence the \eqref{equ:first_law_t} corresponds to the equilibrium first law of thermodynamics \eqref{equ:first_law_eq}.

On the other hand if we took the total work and the total heat in the quasistatic expansion we obtain the difference of the total mean energy as a consequence of \eqref{def:local_power_t} and \eqref{def:local_heat_production_t} 
\[
{\mathcal Q}(\alpha) + {\mathcal W}(\alpha) 
= \int\limits_0^{\frac{T}{\epsilon}} \frac{\rmd \left\langle E_{t;\alpha(\epsilon t)} \right\rangle_{\mu_t} }{\rmd t} 
= \left\langle E_{\frac{T}{\epsilon};\alpha(T)} \right\rangle_{\mu_{\frac{T}{\epsilon}}} - \left\langle E_{0;\alpha(0)} \right\rangle_{\mu_0} ,
\]
which in general oscillates in the quasistatic limit $\epsilon \to 0^+$.
As a consequence the quasistatic limit of the difference of mean values of energies is not defined. 
This reflects the fact that we are not able to uniquely associate a fixed energy with the steady state as it is a periodic function of time. 
However we have seen that from thermodynamical point of view it is reasonable to associate the internal energy with the mean steady energy averaged over the period,
as the time-averaged quantity represents the overall energy present in the system independent of the fluctuations on the level of single period 
hence the equation \eqref{equ:first_law_t} represents the first law of thermodynamics in systems driven by periodical force.

\subsubsection{Heat capacity} 
The fact that the ``reversible'' component of the heat does not depend on the exact parametrization of the path of external parameters as well as its antisymmetry with respect to the trajectory inverse $\alpha(t) \to \alpha(T-t)$, 
enables us to extend the definition of the generalized heat capacity\index{generalized heat capacity!periodically driven systems} from the section \ref{sec:quasistatic_processes_noneq} to systems with periodical driving. 
We define the generalized heat capacity as  
\[
C = \frac{\dbar {\mathcal Q}^\text{rev} (\alpha)}{\rmd T} 
= - \left\llangle \partial_T \frac{1}{\gensc_\alpha}[\hat{\vec{q}}_\alpha] \right\rrangle_{\hat{\vec{\rho}}_\alpha} ,
\]
which is equivalent in case there are present no non-potential forces to
\[
C = \partial_T \left\llangle \hat{\vec{E}}_\alpha \right\rrangle_{\hat{\vec{\rho}}_\alpha} - \left\llangle \partial_T \hat{\vec{E}}_\alpha \right\rrangle_{\hat{\vec{\rho}}_\alpha} 
+ \left\llangle \partial_T \frac{1}{\gensc_\alpha} \left[ \hat{\Omega} \cdot \hat{\vec{E}}_\alpha \right] \right\rrangle_{\hat{\vec{\rho}}_\alpha} ,
\]
where the first two terms are presented also in the equilibrium thus the third term being genuine non-equilibrium contribution.

If we introduce a quasi-potential
\[
\hat{\vec{V}}_\alpha (x) = \hat{\vec{E}}_\alpha(x) - \frac{1}{\gensc_\alpha} \left[ \hat{\Omega} \cdot \hat{\vec{E}}_\alpha \right] (x) ,
\]
the heat capacity is then given in the similar fashion as in equilibrium although there is quasi-potential instead of the total energy
\[
C = \partial_T \left\llangle \hat{\vec{V}}_\alpha \right\rrangle_{\hat{\vec{\rho}}_\alpha} - \left\llangle \partial_T \hat{\vec{V}}_\alpha \right\rrangle_{\hat{\vec{\rho}}_\alpha} ,
\]
notice that the steady state does not necessarily need to function of the quasi-potential and hence the generalized heat capacity is not necessarily positive.

\subsection{Clausius relation}
Again as in other cases of non-equilibrium driving the second law of thermodynamics \eqref{equ:second_law} poses no limitations to the ``reversible'' component of the heat as the ``housekeeping'' component diverge in the quasistatic limit.

Although there is no direct consequence of the second law of thermodynamics, one can be still interested in which special cases there is a generalized Clausius relation relating the heat with the entropy production. 
In section \eqref{sec:quasistatic_processes_noneq} we have seen that in the case of the system is driven out of equilibrium by attaching the system to multiple thermal baths or by action on non-potential forces the sufficient condition has been that the steady state is function of pseudo-potential. 
The same argument is here a bit complicated by the explicit time dependence of all quantities. 
To avoid these complications we represent the ``reversible'' component of the heat by time dependent quantities
\[
\dbar {\mathcal Q}^\text{rev} = \rmd \alpha \cdot \frac{1}{\tau} \int\limits_0^\tau \rmd t \; \left[ \nabla_\alpha \left\langle V_{t;\alpha} \right\rangle_{\rho_{t;\alpha}} - \left\langle \nabla_\alpha V_{t;\alpha} \right\rangle_{\rho_{t;\alpha}} \right] ,
\]
where the time dependent potential is constructed as $V_{t;\alpha} (x) = \sum_{n \in \integers} \left[ \hat{\vec{V}}_\alpha(x) \right]_n \exp \frac{2 \pi \rmi n t}{\tau}$.

If we now assume the steady state probability distribution is given as 
\[
\rho_{t;\alpha} (x) = \frac{1}{Z_{t;\alpha}} \rme^{-\beta(\alpha) V_{t;\alpha} (x)},
\] 
where $\beta(\alpha)$ is an arbitrary function of external parameters representing the generalized temperature,
the ``reversible'' component of the heat is given by 
\[
\dbar {\mathcal Q}^\text{rev} = \rmd \alpha \cdot \frac{1}{\tau \beta(\alpha)} \int\limits_0^\tau \rmd t \; \nabla_\alpha \left\langle - \ln \rho_{t;\alpha} \right\rangle_{\rho_{t;\alpha}} ,
\]
where we can recognize the generalized Clausius relation and where the generalized thermodynamical entropy is given by the time-averaged Shannon entropy
\[
{\mathcal S}(\alpha) = - \frac{1}{\tau} \int\limits_0^\tau \rmd t \; \left\langle \ln \rho_{t;\alpha} \right\rangle_{\rho_{t;\alpha}} .
\]

\section{Slow driving limit}
\label{ssec:slow}
\index{slow driving limit}
We have seen that in fact we have three time scales present in the system. 
The first time scale is given by the evolution of external parameters, the second by the period of the driving and the third by the typical relaxation time of the system $\tau_\text{relax}$. 
While in the quasistatic limit the evolution of the external parameters has to be always on the longest time scale 
\begin{align*}
\frac{T}{\epsilon} &\gg \tau , &
\frac{T}{\epsilon} &\gg \tau_\text{relax} \quad \forall k ,
\end{align*}
the other two time scales can be arbitrary, 
i.e. we can have $\tau > \tau_\text{relax}$ as well as $\tau < \tau_\text{relax}$. 
The two limiting cases then being the slow driving limit $\tau \gg \tau_\text{relax}$ and the singular driving limit $\tau \ll \tau_\text{relax}$. 
In this section we will focus on the slow driving limit while in the other section we will analyse the singular driving limit.

The slow driving limit is the limiting case when the period of the driving forces is much larger then the relaxation time of the system, however is is still much smaller than the characteristic time of the change of external parameters $\alpha$. 
Technically we characterize the slow driving regime by the small scaling parameter $\eta \ll 1$, 
which scales the period of the driving $\tau \to \tau/\eta $,
and which in the slow driving limit goes to zero $\eta \to 0^+$ while scaling the period of the driving up to the infinity.  
To avoid dealing with infinities we also rescale the time $t'=t/\eta$, which correspond to observing the system on the time scale of the characteristic relaxation time.
Then on this particular time scale the time evolution of the system is described by 
\[
\eta \, \partial_{t'} \mu_{t'}(x) = \gen^*_{t'} [\mu_{t'}] (x), 
\]
which yields to the generalized forward Kolmogorov generator being
\begin{equation}
\left[ \gensc^*_\eta \left[\hat{\vec{\mu}}\right] (x) \right]_n 
= \sum_{m \in \integers} \hat{\gen}^*_{n-m} \left[ \hat{\mu}_m \right] (x) - \eta \frac{2 \pi \rmi n}{\tau} \hat{\mu}_n (x) . 
\label{equ:generator_slow}
\end{equation}
We can see that the effect of the slow driving regime is reduced to formal substitution of the spectral matrix $\hat{\Omega} \to \hat{\Omega}' = \eta \hat{\Omega}$ in the definition of generalized forward Kolmogorov generator \eqref{equ:generalized_forward_generator}, 
where now the spectral matrix $\hat{\Omega}'$ is the small parameter. 
The same can be done for the local power \eqref{def:work_nonp_gen} and the local heat production \eqref{def:heat_gen}, from where we obtain
\begin{align*}
\hat{\vec{w}}_{\alpha(t)}^\eta (x) &= \dot\alpha(t) \cdot \left. \nabla_\alpha \hat{\vec{E}}_\alpha(x) \right|_{\alpha=\alpha(t)} + \eta \, \hat{\Omega} \cdot \hat{\vec{E}}_{\alpha(t)}(x), \\
\hat{\vec{q}}_{\alpha(t)}^\eta (x) &= \gensc_\eta \left[ \hat{\vec{E}}_{\alpha(t)} \right] (x) - \eta \, \hat{\Omega} \cdot \hat{\vec{E}}_{\alpha(t)}(x).
\end{align*}

\subsection{Steady state} 
In the slow driving limit one expects the steady state of the system to be such that at each time it is in the steady state of the dynamics corresponding to that particular time as the driving is so slow that it provides the system enough time to relax there. 
To prove this proposition we start need to assume that for sufficiently small $\eta$ the steady state can be expanded into the power series 
\[
\hat{\vec{\rho}}^\eta (x) = \sum_{i=0}^\infty \eta^i \, \hat{\vec{\rho}}^{(i)} (x) .
\]
Then by inserting this assumption to the steady state condition \eqref{equ:generalized_stationary_distribution} 
we obtain a set of equations determining the steady state
\begin{align*}
\sum_{m \in \integers} \hat{\gen}^*_{n-m} \left[ \hat{\rho}^{(0)}_m \right] (x) &= 0, \\
\sum_{m \in \integers} \hat{\gen}^*_{n-m} \left[ \hat{\rho}^{(i+1)}_m \right] (x) &= \frac{2 \pi \rmi n}{\tau}  \hat{\rho}^{(i)}_n (x) , \qquad \forall i>0 .
\end{align*}
From where we can see that the leading order component $\hat{\vec{\rho}}^{(0)}$ given by the first equation is equivalent to the Fourier series of the solution of the equation 
\[
\gen^*_t \left[ \rho^{(0)}_t \right] (x) = 0 ,
\]
i.e. the leading order of the steady state is really given by steady states \eqref{equ:stationary_condition} at each time with respect to fixed dynamics at that particular time.

In case the system is attached to the single thermal bath at constant inverse temperature and when there are no non-potential forces acting on the system, the driving is present only in the explicit time dependence of the potential 
and hence the leading term is given by the Boltzmann or Maxwell-Boltzmann distribution with time dependent potential 
\[
\rho^{(0)}_t (x) = \frac{1}{Z_t} \rme^{- \beta E_t(x) } .
\]

The corrections beyond the slow driving limit are then obtained by application of the recurrent relation 
\begin{equation}
\hat{\vec{\rho}}^{(n+1)} (x) = \frac{1}{\gensc^*_0}\left[ \hat\Omega \cdot \hat{\vec{\rho}}^{(n)} \right] (x) , 
\label{equ:recurent_relation_steady_state}
\end{equation}
where we have denoted by
\[
\gensc^*_0 [\hat{\vec{\mu}}] (x) = \sum_{m \in \integers} \gen^*_{n-m} [ \hat{\mu}_m ] (x) 
\]
the generalized forward Kolmogorov generator in the slow driving limit $\eta \to 0^+$, 
and where $1/\gensc^*_0$ is the pseudoinverse \eqref{def:generalized_forward_pseudoinverse} although this time with the projector to the zero eigenvalue subspace is given by 
\[
\projsc^*_0 [\hat{\vec{\mu}}] (x) = \sum_{n \in \integers} \hat{\vec{\rho}}^{(0)}_{+n} (x) \int \rmd \Gamma(y) \; \hat\mu_n (y) ,
\]
where by $\hat{\vec{\rho}}^{(0)}_{+n}$ we denote shifted zeroth component \eqref{equ:shifted_state}. 
This is due to the fact that no only the zeroth component of the steady state is the eigenvector corresponding to the zero eigenvalue of this limiting generator $\gensc^*_0 [ \hat{\vec{\rho}}^{(0)} ] = 0$,
but also the vectors obtained by shifting the zero component are also the eigenvectors to the zero eigenvalue in the slow driving limit. 
I.e. the zero eigenvalue is degenerate and hence we need to project out the whole subspace in order for pseudoinverse to exist.

We can conclude that the steady state in the slow driving regime is then explicitly given by a series
\[
\hat{\vec{\rho}}i^\eta (x) = \hat{\vec{\rho}}^{(0)} (x) + \sum_{n=1}^\infty \eta^n \; \underbrace{\frac{1}{\gensc^*_0} \biggl[ \hat\Omega \cdot \frac{1}{\gensc^*_0} \biggl[ \hat{\Omega} \cdot \dots \cdot \frac{1}{\gensc^*_0} \biggl[ \hat{\Omega}}_{n \times} \cdot \hat{\vec{\rho}}^{(0)} \biggr] \dots \biggr] \biggr] (x) .
\]

\subsection{The heat and work}
Having the expansion of the steady state around the slow driving limit we can proceed further to investigate the behaviour of the ``reversible'' work and heat in the slow driving limit. 
We start with the ``reversible'' work which we already expand up to the first order in $\eta$  
\begin{multline*}
{\mathcal W}^\text{rev} (\alpha) 
= \int \rmd \alpha \cdot \left\llangle \nabla_\alpha \hat{\vec{E}}_\alpha \right\rrangle_{\hat{\vec{\rho}}^{(0)}_\alpha} + \\
+ \eta \left\{ \int \rmd \alpha \cdot \left\llangle \nabla_\alpha \hat{\vec{E}}_\alpha \right\rrangle_{\hat{\vec{\rho}}^{(1)}_\alpha} 
- \int \rmd \alpha \cdot \left\llangle \nabla_\alpha \frac{1}{\gensc_{\alpha;\eta}} \left[ \hat{\Omega} \cdot \hat{\vec{E}}_\alpha \right] \right\rrangle_{\hat{\vec{\rho}}^{(0)}_\alpha} \right\}
+ \err[2]{\eta} ,
\end{multline*}
from where we can see that the leading order in the expansion is an equilibrium-like term, 
which in case when there are no non-potential forces correspond to the work done in equilibrium averaged over the period $\tau$. 
Moreover if the trajectory over the external parameters is isothermal it corresponds to the change of the equilibrium free energy $\mathcal F$ averaged over the period  
\[
\left. \dbar {\mathcal W}^\text{rev}_0 (\alpha) \right|_{T=\text{const.}} 
= \rmd \alpha \cdot \frac{1}{\tau} \int\limits_0^\tau \rmd t \; \nabla_\alpha {\mathcal F}_{\alpha;t}
= \rmd \alpha \cdot \frac{1}{\tau} \int\limits_0^\tau \rmd t \; \left[ - \frac{1}{\beta} \nabla_\alpha \ln Z_{\alpha;t} \right] .
\]

In order to refine the first order correction to more compact form we expand the pseudoinverse in $\eta$ in a similar fashion as in the section \ref{sec:first_order_expansion} and up to the first order we obtain 
\begin{multline*}
{\mathcal W}^\text{rev} (\alpha) 
= \int \rmd \alpha \cdot \left\llangle \nabla_\alpha \hat{\vec{E}}_\alpha \right\rrangle_{\hat{\vec{\rho}}^{(0)}_\alpha} + \\
- \eta \int \rmd \alpha \cdot \left\llangle \hat{\Omega} \cdot \frac{1}{\gensc_{\alpha;0}} \left[ \nabla_\alpha \hat{\vec{E}}_\alpha \right] 
+ \nabla_\alpha \frac{1}{\gensc_{\alpha;0}} \left[ \hat{\Omega} \cdot \hat{\vec{E}}_\alpha \right] \right\rrangle_{\hat{\vec{\rho}}^{(0)}_\alpha} 
+ \err[2]{\eta} ,
\end{multline*}
where we have explicitly used the expression for the first order correction in the steady state \eqref{equ:recurent_relation_steady_state} and the definition of generalized backward Kolmogorov generator \eqref{def:generalized_backward_pseudoinverse} along with the behaviour of the spectral matrix $\hat{\Omega}$ under the mean value \eqref{equ:spectral_matrix_property}, 
from where we can see that the first order correction is proportional to the frequency of the driving or to be more precise to the ration of the characteristic relaxation time and the period of driving.

Similarly for the ``reversible'' heat we obtain 
\begin{multline*}
{\mathcal Q}^\text{rev} (\alpha) 
= \int \rmd \alpha \cdot \left[ \nabla_\alpha \left\llangle E_\alpha \right\rrangle_{\hat{\vec{\rho}}^{(0)}} - \left\llangle \nabla_\alpha E_\alpha \right\rrangle_{\hat{\vec{\rho}}^{(0)}} \right] + \\
- \int \rmd \alpha \cdot \left\{ \eta \left\llangle \hat{\Omega} \cdot \frac{1}{\gensc_{\alpha;0}} \left[ \nabla_\alpha \hat{\vec{E}}_\alpha \right] \right\rrangle_{\hat{\vec{\rho}}_\alpha^{(0)}} 
+ \nabla_\alpha \left\llangle \frac{1}{\gensc_{\alpha;\eta}} \gensc_{\alpha;0} \left[ \hat{\vec{E}}_\alpha \right] \right\rrangle_{\hat{\vec{\rho}}^{(0)}} \right\} 
+ \err[2]{\eta} ,
\end{multline*} 
where we can see the first term being again equilibrium-like as it is determined only by the energy function $\hat{\vec{E}}_\alpha$ 
and from where it follows that the heat generalized heat capacity in the slow driving limit in case there are no non-potential forces corresponds to the equilibrium heat capacity averaged over the period of the driving $\tau$ 
\[
C_\alpha = \frac{\dbar {\mathcal Q}^\text{rev}(\alpha)}{\rmd T} = \frac{1}{\tau} \int\limits_0^\tau \rmd t \; C_{\alpha;t} + \err{\eta} .
\]
Notice that the generalized heat capacity is positive as in the leading order it is given by the time average of equilibrium heat capacities, which are known to be positive \cite{Callen1985}.

At last let us remark, that in case there are no non-potential forces acting on the system the extensive part of the ``housekeeping'' heat and work over the time interval $T$ are given by 
\[
{\mathcal W}^\text{hk} (\alpha) = - {\mathcal Q}^\text{hk} (\alpha) = \frac{\eta}{\epsilon} \int\limits_0^T \rmd t \; \left\llangle \hat{\Omega} \cdot \hat{\vec{E}}_{\alpha(t)} \right\rrangle_{\hat{\vec{\rho}}_{\alpha(t);\eta}} ,
\]
which magnitude is of order of $\eta$. 
What need to noticed is that although this term is of order $\eta$ it cannot be in slow driving limit neglected, because the external parameters are evolving on much longer time-scale, $\epsilon \ll \eta$.

\section{Singular driving limit}
\label{ssec:fast}
\index{singular driving limit}
Another limit can be obtained, when the system is driven so fast, that the relaxation time $\tau_R$ of the system is much longer than the period of the driving $\tau$.
The limit when the ratio of relaxation time by period of driving is going to infinity is called a \emph{singular driving limit}, $\tau_R/\tau \rightarrow \infty$. 
In this section we want to inspect the system in the singular driving limit as well as obtain first non-zero corrections.  
In usual physical situation we have the period of the external force under control not the relaxation time, hence we induce the singular driving limit in the system by scaling the period to zero.
However this approach has serious disadvantage from mathematical point of view. 
Especially to avoid dealing with terms of $1/\tau$ in \eqref{equ:generalized_forward_generator} and corresponding quantities, 
we rather rescale both the physical time $t$ and the period $\tau$ of the driving by the same scaling factor $\eta \ll 1$, 
\begin{align}
\tau \longrightarrow \eta \tau, && t \longrightarrow \eta t,
\label{equ:rescaling}
\end{align}
thus fixating the period and effectively sending the relaxation time to infinity. 
The singular driving limit is then obtained by taking the limit $\eta \rightarrow 0^+$. 
Notice that the hierarchy of time scales in the singular driving limit follows 
\[
\tau \ll \frac{\tau_{\text{relax}}}{\eta} \ll \frac{T}{\varepsilon},
\]
where by $T/\varepsilon$ we denote the characteristic time scale of the quasistatic transformations.

This setup yields to the same time evolution as in \eqref{equ:generalized_time_evolution} with $\eta$-dependent generator 
\[
\left[ \gensc^*_\eta \right]_{mn} = \eta \hat\gen^*_{m-n} - \frac{2 \pi \rmi m}{\tau} \delta_{mn} , 
\]
compare with \eqref{equ:generalized_forward_generator} or with \eqref{equ:generator_slow}.
Although the scaling also affects other aspects of the system like local heat power, 
for now we will solely focus on the consequences of this particular structure of generalized forward generator.

\subsection{Steady state} 
At first we will inspect the stationary distribution and its corrections. 
The stationary distribution is given by \eqref{equ:generalized_stationary_distribution}, which in this particular case yields to set of equations
\begin{equation}
0 = \eta \sum\limits_{n \in \integers } \hat\gen^*_{m-n} \left[ \hat\rho_n \right] (x) - \frac{2 \pi \rmi m}{\tau} \hat\rho_m (x).
\label{equ:fast_stationary_distribution}
\end{equation}
To proceed further we expect that the typical physical system in the singular driving limit does not ``feel'' the oscillations of the driving force 
and that the oscillations became more important as the system is further from singular driving limit. 
This expectation leads us to the assumption that the stationary distribution is analytical in $\eta$, 
and hence we can expand the stationary distribution in powers of $\eta$ with zeroth leading order 
\[
\hat\rho (x) = \hat\rho^{(0)} (x) + \eta \hat\rho^{(1)} (x) + \eta^2 \hat\rho^{(2)} (x) + \dots .
\]
Inserting the expansion of the stationary distribution to the \eqref{equ:fast_stationary_distribution} 
and inspecting the zeroth order for $m \neq 0$ we immediately find that the non-zero components of the zeroth order contribution vanish  
\[
\forall n \neq 0 : \quad \hat\rho^{(0)}_n (x) = 0,
\]
i.e. only the zeroth component of the stationary distribution is essentially non-zero.  
Applying the same technique but inspecting higher orders for $m \neq 0$ we also obtain a recurrent relation for non-zero components of the stationary distribution
\begin{equation}
\forall n \neq 0, k \in \mathbb N : \quad \hat\rho^{(k+1)}_n (x) = - \frac{\rmi \tau}{2 \pi n} \sum\limits_{m \in \integers} \hat\gen^*_{n-m} \left[ \hat\rho^{(k)}_m \right] (x).
\label{equ:fast_stationary_density_nonzero}  
\end{equation}

The only remaining independent set of equations left in \eqref{equ:fast_stationary_distribution} corresponds to $m=0$,  
\[
\forall k \in \mathbb N : \quad 0 = \sum\limits_{n \in \integers} \hat\gen^*_{-n} \left[ \hat\rho^{(k)}_{n} \right] (x), 
\]
which determine the zeroth component. 
In the zeroth order we obtain the zeroth component as a solution of equation 
\begin{equation}
0 = \hat\gen^*_0 \left[ \hat\rho^{(0)}_0 \right] (x) ,
\label{equ:steady_state_fast}
\end{equation}
while for higher order contribution of the zeroth component we obtain a recurrent relation  
\begin{equation}
\hat\rho^{(k)}_0 (x) = - \frac{1}{\hat\gen^*_0} \sum\limits_{n \neq 0} \hat\gen^*_{-n} \left[ \hat\rho^{(k)}_n \right] (x) ,
\label{equ:fast_stationary_density_zero}  
\end{equation}
where $1/\hat\gen^*_0$ is the pseudo-inverse\index{pseudo-inverse} \eqref{def:forward_pseudoinverse} as in chapter \ref{chapter:non-equilibrium_thermodynamics} with $\hat\gen^*_0$ as the explicitly time independent generator, i.e. 
\[
\frac{1}{\hat\gen^*_0} \left[ \mu \right] = \int\limits_0^\infty \rmd s \; \left\{ \hat\rho^{(0)}_0 (x) \int \rmd y \; \mu(y) - \rme^{s \hat\gen^*_0} \left[ \mu \right] (x) \right\}.
\]

We can see, that in the singular driving limit the system's steady state is explicitly time independent as expected, $\rho_t (x) \equiv \hat\rho^{(0)}_0 (x)$, 
and is given by time-averaged generator, 
\[
0 = \frac{1}{\tau} \int\limits_0^\tau \rmd t \; \gen^*_t [ \rho ] = \hat\gen^*_0 [ \rho ].
\]
Notice that even in the case the periodic time dependence of the system is given only by the time periodic potential, 
the stationary distribution with respect to time-averaged generator is not necessarily the stationary distribution obtained by implying the time-averaged potential, 
as we will briefly discuss further. 

We can also obtain corrections beyond the singular driving limit by using the recurrent relations \eqref{equ:fast_stationary_density_nonzero} and  \eqref{equ:fast_stationary_density_zero}.  
Combining these equations together we can explicitly obtain the first order correction
\begin{align*}
\hat\rho^{(1)}_{n \neq 0} (x) &= - \frac{\rmi \tau}{2 \pi n} \hat\gen^*_n \left[ \hat\rho^{(0)}_0 \right] (x), & 
\hat\rho^{(1)}_0 (x) &= \frac{\rmi \tau}{2 \pi} \frac{1}{\hat\gen^*_0} \left[ \sum\limits_{n \neq 0} \frac{1}{n} \hat\gen^*_{-n} \hat\gen^*_n \left[ \hat\rho^{(0)}_0 \right] \right] (x),  
\end{align*}
where the zeroth component of the first order correction can be also rewritten using \eqref{equ:complex_conjugation_of_generator} as 
\[
\hat\rho^{(1)}_0 (x)  = - \frac{\tau}{\pi} \frac{1}{\hat\gen^*_0} \left[ \sum\limits_{n = 1}^\infty \frac{1}{n} \, \Im \left( \overline{\hat\gen^*_n} \hat\gen^*_n \right) \left[ \hat\rho^{(0)}_0 \right] \right] (x),  
\]
where $\Im$ denotes the imaginary part and $\bar{x}$ denotes the complex conjugation.

\subsection{Generalized pseudo-inverse} 
Before we proceed further to the analysis of the work and heat in the singular driving limit, 
we need to investigate the behaviour of the pseudoinverse in the quasistatic limit. 
The basic problem here is that the limit of the forward Kolmogorov generator $\lim_{\eta \to 0^+} \gensc^*_\eta $ does not evolve towards any steady state 
and hence the generalized forward pseudoinverse \eqref{def:generalized_forward_pseudoinverse} as well as generalized backward pseudoinverse \eqref{def:generalized_backward_pseudoinverse} 
cannot be defined in that way as the integrals in definitions does not converge. 
To at least partially avoid the problem, let us then define the regularized \emph{backward} generator as 
\[
\left[ \gensc_\epsilon [\hat{\vec{A}} ] (x) \right]_n = \left[ \hat\Omega \cdot \hat{\vec{A}} (x) \right]_n + 
\begin{cases}
\epsilon \hat\gen_0 \left[ \hat{A}_0 \right] (x) & n = 0 \\
- \epsilon \hat{A}_n (x) & n \neq 0 
\end{cases} ,
\]
which in limit $\epsilon \to 0^+$ corresponds to the limit of the generalized forward Kolmogorov generator,
otherwise ensures the observable to converge toward its steady mean value averaged over the period $\tau$ in the singular driving limit. 
Then the generalized pseudoinverse in the limit is given as the limit of the pseudoinverse for regularized generator 
\[
\left[ \frac{1}{\gensc_\epsilon} [ \hat{\vec{A}} ] (x) \right]_n 
= \begin{cases}
\displaystyle \frac{1}{\epsilon \, \hat{\gen}_0 } \left[ \hat{A}_0 \right] (x) & n = 0 \\ \\
\displaystyle \frac{\tau}{2 \pi \rmi n - \tau \epsilon} \hat{A}_n (x) & n \neq 0  
\end{cases} ,
\]
which diverges in case of the zeroth component while it converge for any other component.  
Although the zeroth component of the pseudoinverse of regularized generator, 
in the ``reversible'' component of the work and the heat the pseudoinverse is applied to the quantity which zeroth components is zero $A_0(x) = 0$.

\subsection{Heat and work}
In previous subsections we have analysed the structure of the steady state in the singular driving limit and prepared the regularized backward pseudoinverse as the leading order of the generalized backward generator in the singular driving expansion.
Now we will proceed to the expansion of the work and heat in the singular driving limit. 
We start with the analysis of the local power and the local heat production. 
The local power in the singular driving limit is given by the equation \eqref{def:work_nonp_gen}
\[ 
\hat{\vec{w}}_{\alpha(t)} (x) = \dot\alpha(t) \cdot \left. \nabla_\alpha \hat{\vec{E}}_\alpha (x) \right|_{\alpha = \alpha(t)} + \hat{\Omega} \cdot \hat{\vec{E}}_{\alpha(t)} ,
\]
as it is independent of the backward Kolmogorov generator, 
on the other hand we can deduce that the local heat production \eqref{def:heat_gen} is entirely of the first order in $\eta$ 
\[
\left[ \hat{\vec{q}}_\alpha (x) \right]_n = \left[ \gensc_\alpha [ \hat{\vec{E}}_\alpha ] (x) \right]_n - \left[ \hat{\Omega} \cdot \hat{\vec{E}}_\alpha (x) \right]_n 
= \eta \sum\limits_{m \in \integers} \hat\gen_{n-m;\alpha} [ \hat{E}_{m;\alpha} ] (x) .
\]

The leading order of the ``reversible'' work is then obtained by application of the regularized pseudoinverse on the second term in the local power which yields to 
\begin{multline*}
{\mathcal W}^\text{rev} (\alpha) = \int \rmd \alpha \cdot \left\llangle \nabla_\alpha \left( \hat{\vec{E}}_\alpha - \frac{1}{\gensc_\alpha} \left[ \hat{\Omega} \cdot \hat{\vec{E}}_\alpha \right] \right) \right\rrangle_{\hat{\vec{\rho}}_{\alpha}} = \\
= \int \rmd \alpha \cdot \left\langle \nabla_\alpha \hat{E}_{0;\alpha} \right\rangle_{\hat{\rho}^{(0)}_0} + \err{\eta} .
\end{multline*} 
Similarly for the heat we have 
\begin{multline*}
{\mathcal Q}^\text{rev} (\alpha) = \int \rmd \alpha \cdot \left\{ \nabla_\alpha \left\llangle \hat{\vec{E}}_\alpha \right\rrangle_{\hat{\vec{\rho}}_{\alpha}} - \left\llangle \nabla_\alpha \left( \hat{\vec{E}}_\alpha - \frac{1}{\gensc_\alpha} \left[ \hat{\Omega} \cdot \hat{\vec{E}}_\alpha \right] \right) \right\rrangle_{\hat{\vec{\rho}}_{\alpha}} \right\} = \\
= \int \rmd \alpha \cdot \left\{ \nabla_\alpha \left\langle \hat{E}_{0;\alpha} \right\rangle_{\hat{\rho}^{(0)}_0} - \left\langle \nabla_\alpha \hat{E}_{0;\alpha} \right\rangle_{\hat{\rho}^{(0)}_0} \right\} + \err{\eta} ,
\end{multline*}
from where under the assumption that the energy function $\hat{E}_0$ is independent of temperature $T$ we obtain the generalized heat capacity as 
\[
C = \frac{\rmd}{\rmd T} \left\langle \hat{E}_{0;\alpha} \right\rangle_{\hat{\rho}^{(0)}_0} + \err{\eta} .
\]
Notice that we have obtained in the leading order the generalized heat capacity to be given by the temperature derivative of the internal energy. 
Moreover in case the driving of the system is given by periodic time dependence of the energy function and if the generator depends on the potential linearly we obtain the equilibrium heat capacity with respect to the energy function averaged over the period $\tau$ and thus the positivity of the generalized heat capacity is guaranteed. 
Notice that also in this case the strongly non-equilibrium behaviour of the system can be expected only in the intermediate regime where the period of driving $\tau$ and the typical relaxation time are comparable. 
Although the Kolmogorov generators for diffusions \eqref{equ:forward_generator_underdamped} and \eqref{equ:time_evolution_overdamped} are linear in the potential, 
the Kolmogorov generator for jump processes is in general not \eqref{equ:forward_Kolmogorov_gen_jump}, 
thus the generalized heat capacity in case of jump processes in the singular driving limit does not necessarily correspond to any equilibrium heat capacity.

The extensive part of the ``housekeeping'' component of the total heat and the total work simplifies in the singular driving limit to 
\[
{\mathcal W}^\text{hk} (\alpha(t)) = - {\mathcal Q}^\text{hk} (\alpha(t)) 
= \frac{1}{\epsilon} \int\limits_0^T \rmd t \; \left\llangle \hat{\Omega} \cdot \hat{\vec{E}}_{\alpha(t)} \right\rrangle_{\hat{\vec{\rho}}_{\alpha(t)}}
= \frac{\eta}{\epsilon} \int\limits_0^T \rmd t \; \left\llangle \hat{\Omega} \cdot \hat{\vec{E}}_{\alpha(t)} \right\rrangle_{\hat{\vec{\rho}}^{(1)}_{\alpha(t)}} + \err[2]{\eta},
\]
where we have used the fact that in the leading order only the zeroth component of the steady state is non-zero while the zeroth component of the local power of the driving forces is zero, thus effectively canceling each other out. 
Again as in the slow driving limit, although the extensive parts of ``housekeeping'' heat and work are of linear order in $\eta$ they are in general not negligible as the external parameters are again changed on much longer time scale.

\section{Example: Periodically driven two-level models}
\label{sec:two_level_models_t}
\index{discrete model!two-level}
\index{discrete model!two-level!time dependent}
In this section we provide a simple example of periodically driven system with only two configurations and for two physically different scenarios.
The first scenario corresponds, e.g., to an "incoherent" hoping between two levels in a quantum dot. 
In the second scenario the two configurations can represent a pair of metastable states of a complex system, mutually separated by an energy barrier.
In both cases the local detailed balance condition related the time-dependent transition rates to the time-dependent energy levels. 
We mostly provide numerical results only.

\subsection{Scenario A: no barrier}
\begin{figure}[ht]
\caption{Illustration of the two-level model with periodical driving.} 
\label{pic:two_level_illustration}
\begin{center}
\includegraphics[width=.8\textwidth,height=!]{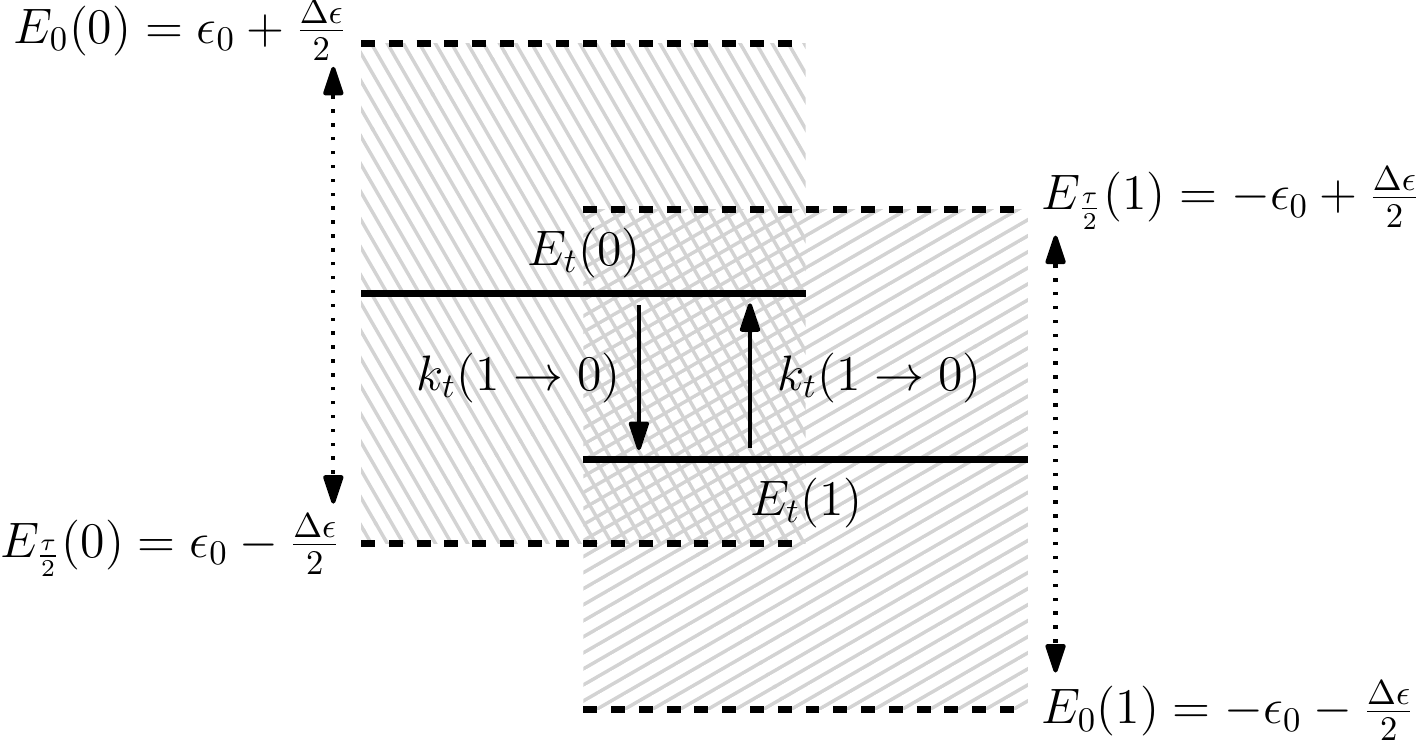}
\end{center}
\end{figure}
The standard two-level model, see subsection \ref{ssec:two_level}, is determined by energy levels,
which in this case periodically depends on the time  
\[
E_t = \begin{bmatrix} 
\epsilon_0 + \frac{\Delta \epsilon}{2} \cos \frac{2 \pi t}{\tau} \\
- \epsilon_0 - \frac{\Delta \epsilon}{2} \cos \frac{2 \pi t}{\tau} ,
\end{bmatrix}
\]
where $2 \epsilon_0$ is the basic gap between the states and $\Delta \epsilon$ is the amplitude of the energy levels movement.
The periodic transition rates are then given by 
\[
\rate[t]{x}{y} = \psi \exp\left[ -\frac{\beta}{2} \left( E_t(y) -E_t(x) \right) \right]. 
\]
Notice that they obey detailed balance condition \eqref{equ:global_detailed_balance_jump} at each time $t$. 
Given the transition rates we can construct the forward Kolmogorov generator, which also explicitly depends on time 
\[
\gen^*_t = \begin{bmatrix}
 - \psi \exp \left[ \beta \left( \epsilon_0 + \frac{\Delta \epsilon}{2} \cos \frac{2 \pi t}{\tau} \right) \right] & 
 \psi \exp \left[ - \beta \left( \epsilon_0 + \frac{\Delta \epsilon}{2} \cos \frac{2 \pi t}{\tau} \right) \right] \\ 
 \psi \exp \left[ \beta \left( \epsilon_0 + \frac{\Delta \epsilon}{2} \cos \frac{2 \pi t}{\tau} \right) \right] & 
 - \psi \exp \left[ - \beta \left( \epsilon_0 + \frac{\Delta \epsilon}{2} \cos \frac{2 \pi t}{\tau} \right) \right] \\ 
\end{bmatrix} 
\]
and equivalently similarly we can define the backward Kolmogorov generator, which is in this case equivalent to the transpose of matrix representation of forward Kolmogorov generator.
\[
\gen_t = \left( \gen^*_t \right)^T = \begin{bmatrix}
 - \psi \exp \left[ \beta \left( \epsilon_0 + \frac{\Delta \epsilon}{2} \cos \frac{2 \pi t}{\tau} \right) \right] & 
 \psi \exp \left[ \beta \left( \epsilon_0 + \frac{\Delta \epsilon}{2} \cos \frac{2 \pi t}{\tau} \right) \right] \\ 
 \psi \exp \left[ - \beta \left( \epsilon_0 + \frac{\Delta \epsilon}{2} \cos \frac{2 \pi t}{\tau} \right) \right] & 
 - \psi \exp \left[ - \beta \left( \epsilon_0 + \frac{\Delta \epsilon}{2} \cos \frac{2 \pi t}{\tau} \right) \right] \\ 
\end{bmatrix} .
\]

Before we show on this particular example the construction of the generalized forward Kolmogorov generator and local power in terms of Fourier components we have to obtain the local power and the local heat production in the time domain first. 
The non-potential part of the local power is given by the time derivative of the total energy with external parameters fixed, in this particular case $\Delta \epsilon$ and $\epsilon_0$ is sufficient 
\[
w^\text{drv}_t = 
\dot{E_t} = \frac{\Delta \epsilon \pi}{\tau} \sin \left( \frac{2 \pi t}{\tau} \right) \cdot 
\begin{bmatrix} 
- 1 \\
1
\end{bmatrix} ,
\]
whether the local heat production is then given by \eqref{def:local_heat_production_t} 
\[
q_t = \gen_t \left[ E_t \right] = 2 \psi \left( \epsilon_0 + \frac{\Delta \epsilon}{2} \cos \frac{2 \pi t}{\tau} \right) \cdot
\begin{bmatrix}
- \exp \left[ \beta \left( \epsilon_0 + \frac{\Delta \epsilon}{2} \cos \frac{2 \pi t}{\tau} \right) \right] \\ 
\exp \left[ - \beta \left( \epsilon_0 + \frac{\Delta \epsilon}{2} \cos \frac{2 \pi t}{\tau} \right) \right]  
\end{bmatrix} . 
\]
Now we can proceed further by finding the Fourier components of the energy 
\begin{align*}
\left[ \hat{E} \right]_0 &= \epsilon_0 \cdot 
\begin{bmatrix}
1 \\
- 1 
\end{bmatrix} & 
\left[ \hat{E} \right]_{\pm 1} &= \frac{\Delta \epsilon}{4} \cdot 
\begin{bmatrix}
1 \\
- 1  
\end{bmatrix}  
\end{align*}
The local power in terms of Fourier components is 
\begin{align*}
\left[ \hat{\vec{w}} \right]_{-1} &= -\frac{\rmi \pi \Delta \epsilon}{2 \tau} \begin{pmatrix} 1 \\ -1 \end{pmatrix} &&&
\left[ \hat{\vec{w}} \right]_1 &= \frac{\rmi \pi \Delta \epsilon}{2 \tau} \begin{pmatrix} 1 \\ -1 \end{pmatrix} &&&
\left[ \hat{\vec{w}} \right]_n &= \begin{pmatrix} 0 \\ 0 \end{pmatrix} && n \neq \pm 1
\end{align*}
while the local heat power is represented by 
\[
\left[ \hat{\vec{q}} \right]_n = \frac{\rmi^n \psi \Delta \epsilon}{2} 
\begin{pmatrix}
J_{n+1} \left( \frac{\rmi \beta \Delta \epsilon}{2} \right) + J_{n-1} \left( \frac{\rmi \beta \Delta \epsilon}{2} \right) \\
J_{n+1} \left( - \frac{\rmi \beta \Delta \epsilon}{2} \right) + J_{n-1} \left( - \frac{\rmi \beta \Delta \epsilon}{2} \right) 
\end{pmatrix},
\]
where $J_n$ is Bessel function of the first order. 
Similarly one can obtain diagonal 
\[
\left[\gensc^* \right]_{nn} = \rmi^n \psi 
\begin{pmatrix}
\frac{2 \pi \rmi n}{\tau} - J_0 \left( \frac{\rmi \beta \Delta \epsilon}{2} \right) & J_0 \left( - \frac{\rmi \beta \Delta \epsilon}{2} \right) \\
J_0 \left( \frac{\rmi \beta \Delta \epsilon}{2} \right) & \frac{2 \pi \rmi n}{\tau} - J_0 \left( - \frac{\rmi \beta \Delta \epsilon}{2} \right) 
\end{pmatrix}
\]
as well as of-diagonal elements of effective generator $\gensc$ 
\begin{align*}
\left[\gensc^* \right]_{nm} &= \rmi^{n-m} \psi 
\begin{pmatrix}
- J_{n-m} \left( \frac{\rmi \beta \Delta \epsilon}{2} \right) & J_{n-m} \left( - \frac{\rmi \beta \Delta \epsilon}{2} \right) \\
J_{n-m} \left( \frac{\rmi \beta \Delta \epsilon}{2} \right) & - J_{n-m} \left( - \frac{\rmi \beta \Delta \epsilon}{2} \right) 
\end{pmatrix}
&& n \neq m.
\end{align*}
One can then easily check that relations \eqref{def:heat_gen} and \eqref{def:work_nonp_gen} are valid. 

%\begin{figure}[ht]
%\caption{Heat capacity of the driven two-level system as a function of temperature $T$ and driving frequency $\frac{1}{\tau}$. }
%\label{pic:two_level_C_T}
%\begin{center}
%\includegraphics[width=.95\textwidth,height=!]{Quasistatic_transformations_of_periodically_driven_systems/2lvl-C(T).pdf}
%\end{center}
%\end{figure}

\begin{figure}[ht]
\caption{Heat capacity of the driven two-level system as a function of inverse temperature $\beta$ and driving frequency $\frac{1}{\tau}$ on the logarithmic scale, with the average relaxation time denoted by the green dashed line. 
$\psi=25$, $\Delta \epsilon = 1$, $\epsilon_0 = 1/8$
}
\label{pic:two_level_C_beta}
\begin{center}
\includegraphics[width=.95\textwidth,height=!]{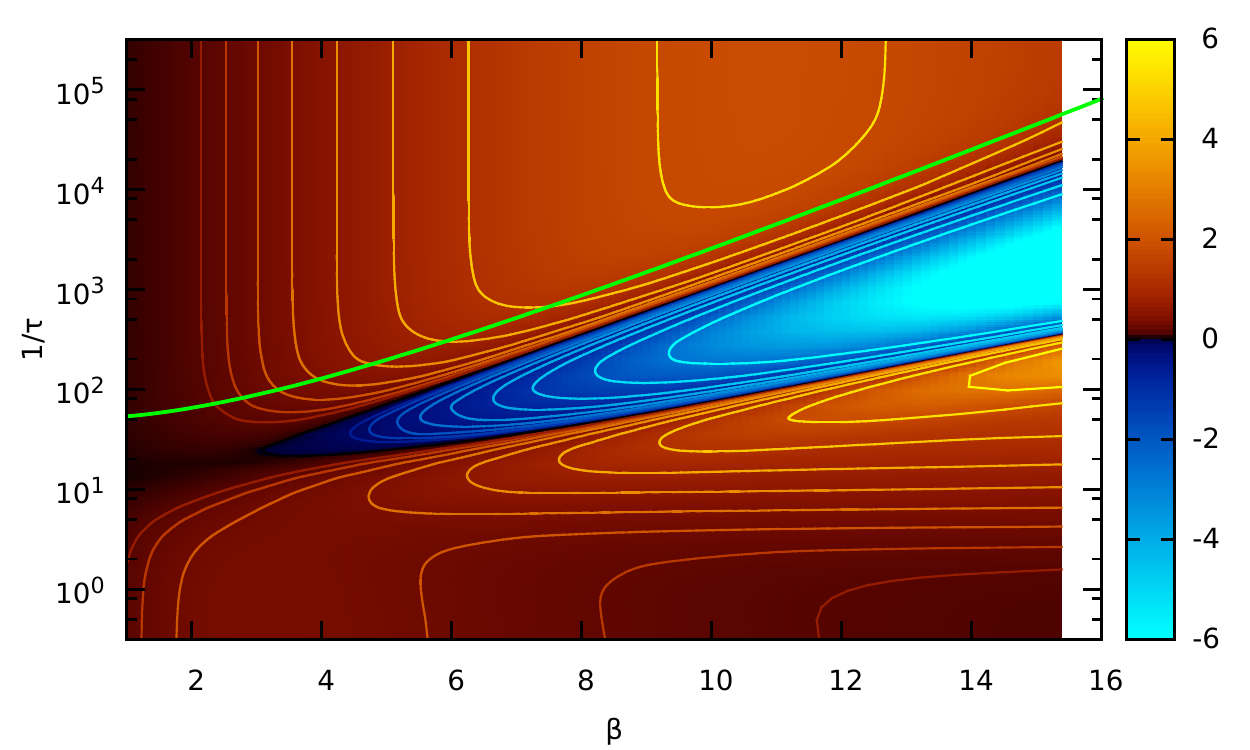}
\end{center}
\end{figure}

The generalized heat capacity $C$ is numerically evaluated with respect to inverse temperature $\beta$ and frequency of the driving $1/\tau$ on exponential scale, see figure~\ref{pic:two_level_C_beta}.  
As one can see, the system exhibits slow driving limit behaviour as well as singular driving limit behaviour, depending on the driving frequency with respect to characteristic relaxation time $\tau_{relax}$. 
Although we cannot determine characteristic relaxation time analytically, we can heuristically estimate the temperature dependence by analysing the relaxation time for each fixed time, i.e. fixed value of the potential. 
The heuristic estimate of the characteristic time may be obtained by looking at the typical magnitude of the relaxation times corresponding to the dynamics "frozen" at each time $t$, which has the form
\begin{align*}
\ln \left[ \frac{1}{\tau_{relax} (t)} \right] 
&= \ln \left[  k_t ( 1 \rightarrow 2 ) + k_t ( 2 \rightarrow 1 ) \right] \\
&= \ln \psi + \ln \left[ 2 \cosh \left( \beta \left( \epsilon_0 + \frac{\Delta \epsilon}{2} \cos \frac{2 \pi t}{\tau} \right) \right) \right]. 
\end{align*}
In low temperature region $\beta \gg 1$ the logarithm of relaxation time can be easily approximated by linear dependence on temperature 
\[
\ln \left[ \frac{1}{\tau_{relax} (t)} \right] 
\sim \beta \left| \epsilon_0 + \frac{\Delta \epsilon}{2} \cos \frac{2 \pi t}{\tau} \right| \ge 0 , 
\]
while in the high temperature region $\beta \ll 1$ the temperature dependence of logarithm of characteristic relaxation time vanish  
\[
\ln \left[ \frac{1}{\tau_{relax} (t)} \right] 
= \ln \psi + \left( \epsilon_0 + \frac{\Delta \epsilon}{2} \cos \left( \frac{2 \pi t}{\tau} \right) \right)^2 \beta^2 + \err[4]{\beta} ,
\]
which is in good qualitative agreement with results shown in the figure \ref{pic:two_level_C_beta}. 
Notice also that the region with negative generalized heat capacity is in the vicinity of the characteristic relaxation time, hence it is in the vicinity of the transition regime between fast and slow driving.

\subsection{Scenario B: with barrier}
\begin{figure}[ht]
\caption{Illustration of the two-level model with barrier with periodical driving.} 
\label{pic:two_level_barrier_illustration}
\begin{center}
\includegraphics[width=.8\textwidth,height=!]{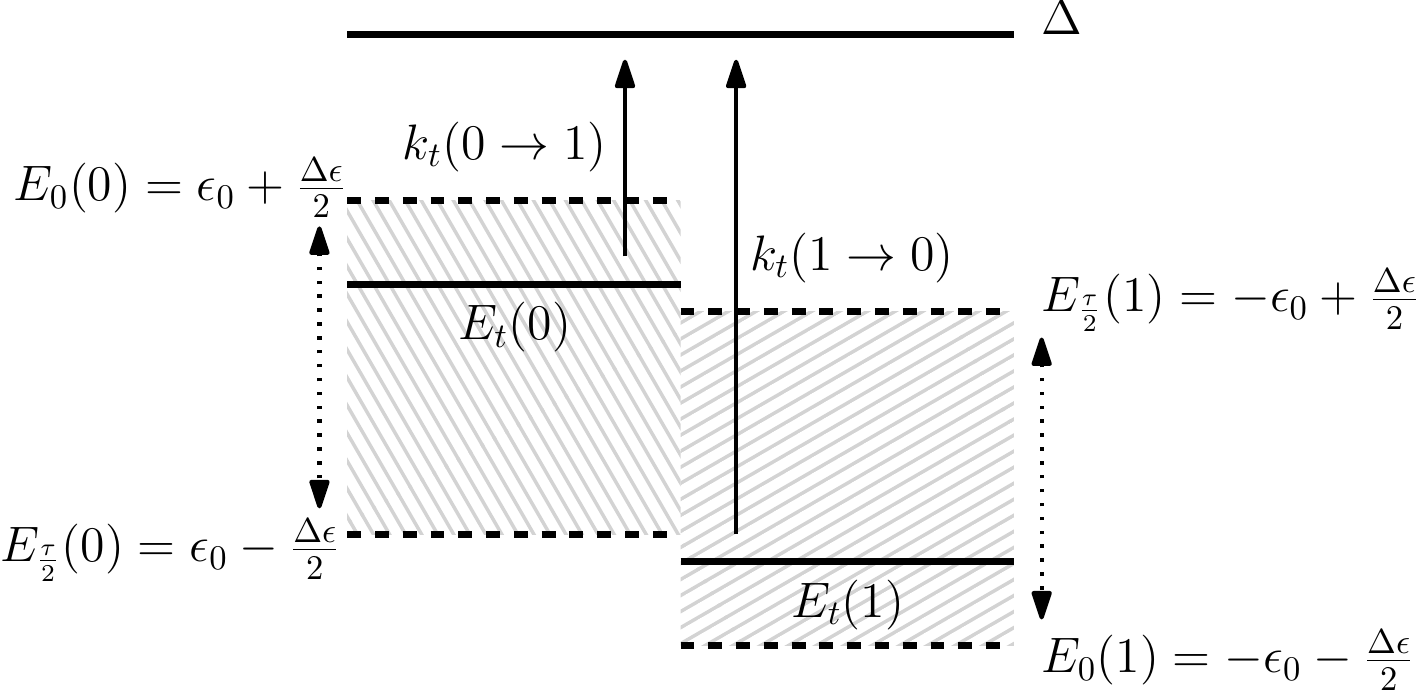}
\end{center}
\end{figure}
The main difference between these two models is the presence of the barrier in this model, hence energy levels are the same as in previous case 
\[
E_t = \begin{bmatrix} 
\epsilon_0 + \frac{\Delta \epsilon}{2} \cos \frac{2 \pi t}{\tau} \\
- \epsilon_0 - \frac{\Delta \epsilon}{2} \cos \frac{2 \pi t}{\tau} 
\end{bmatrix}
\]
while the transition rates are given by 
\[
\rate[t]{x}{y} = \psi \exp\left[ - \beta \left( \Delta - E_t(x) \right) \right],
\]
where $\Delta$ is the height of barrier, also one can verify that even this choice of transition rates obey the detailed balance condition \eqref{equ:global_detailed_balance_jump}.
Given the transition rates we can again construct the forward Kolmogorov generator
\[
\gen^*_t = \begin{bmatrix}
 - \psi \exp \left[ - \beta \left( \Delta - \epsilon_0 - \frac{\Delta \epsilon}{2} \cos \frac{2 \pi t}{\tau} \right) \right] & 
 \psi \exp \left[ - \beta \left( \Delta + \epsilon_0 + \frac{\Delta \epsilon}{2} \cos \frac{2 \pi t}{\tau} \right) \right] \\ 
 \psi \exp \left[ - \beta \left( \Delta - \epsilon_0 - \frac{\Delta \epsilon}{2} \cos \frac{2 \pi t}{\tau} \right) \right] & 
 - \psi \exp \left[ - \beta \left( \Delta + \epsilon_0 + \frac{\Delta \epsilon}{2} \cos \frac{2 \pi t}{\tau} \right) \right] \\ 
\end{bmatrix} 
\]
and equivalently similarly we can define the backward Kolmogorov generator, which is in this case equivalent to the transposed matrix of the forward Kolmogorov generator
\[
\gen_t = \left( \gen^*_t \right)^T = \psi \rme^{- \beta \Delta} \begin{bmatrix}
 - \exp \left[ \beta \left( \epsilon_0 + \frac{\Delta \epsilon}{2} \cos \frac{2 \pi t}{\tau} \right) \right] & 
 \exp \left[ \beta \left( \epsilon_0 + \frac{\Delta \epsilon}{2} \cos \frac{2 \pi t}{\tau} \right) \right] \\ 
 \exp \left[ - \beta \left( \epsilon_0 + \frac{\Delta \epsilon}{2} \cos \frac{2 \pi t}{\tau} \right) \right] & 
 - \exp \left[ - \beta \left( \epsilon_0 + \frac{\Delta \epsilon}{2} \cos \frac{2 \pi t}{\tau} \right) \right] \\ 
\end{bmatrix} .
\]
We can see that the local power is the same
\[
w_t = 
\dot{E_t} = \frac{\Delta \epsilon \pi}{\tau} \sin \left( \frac{2 \pi t}{\tau} \right) \cdot 
\begin{bmatrix} 
- 1 \\
1
\end{bmatrix}
\]
although what differs is the expression for the local heat production
\[
q_t = \gen_t \left[ E_t \right] = 2 \psi \rme^{-\beta \Delta} \left( \epsilon_0 + \frac{\Delta \epsilon}{2} \cos \frac{2 \pi t}{\tau} \right) \cdot
\begin{bmatrix}
- \exp \left[ \beta \left( \epsilon_0 + \frac{\Delta \epsilon}{2} \cos \frac{2 \pi t}{\tau} \right) \right] \\ 
\exp \left[ - \beta \left( \epsilon_0 + \frac{\Delta \epsilon}{2} \cos \frac{2 \pi t}{\tau} \right) \right]  
\end{bmatrix} .
\]

%\begin{figure}[ht]
%\caption{Heat capacity of the driven two-level system with barrier as a function of temperature $T$ and driving frequency $\frac{1}{\tau}$. }
%\label{pic:two_level_barrier_C_T}
%\begin{center}
%\includegraphics[width=.95\textwidth,height=!]{Quasistatic_transformations_of_periodically_driven_systems/2lvl-barrier-C(T).pdf}
%\end{center}
%\end{figure}

\begin{figure}[ht]
\caption{Heat capacity of the driven two-level system with barrier as a function of inverse temperature $\beta$ and driving frequency $\frac{1}{\tau}$, with the mean relaxation time denoted by the green dashed line. }
\label{pic:two_level_barrier_C_beta}
\begin{center}
\includegraphics[width=.95\textwidth,height=!]{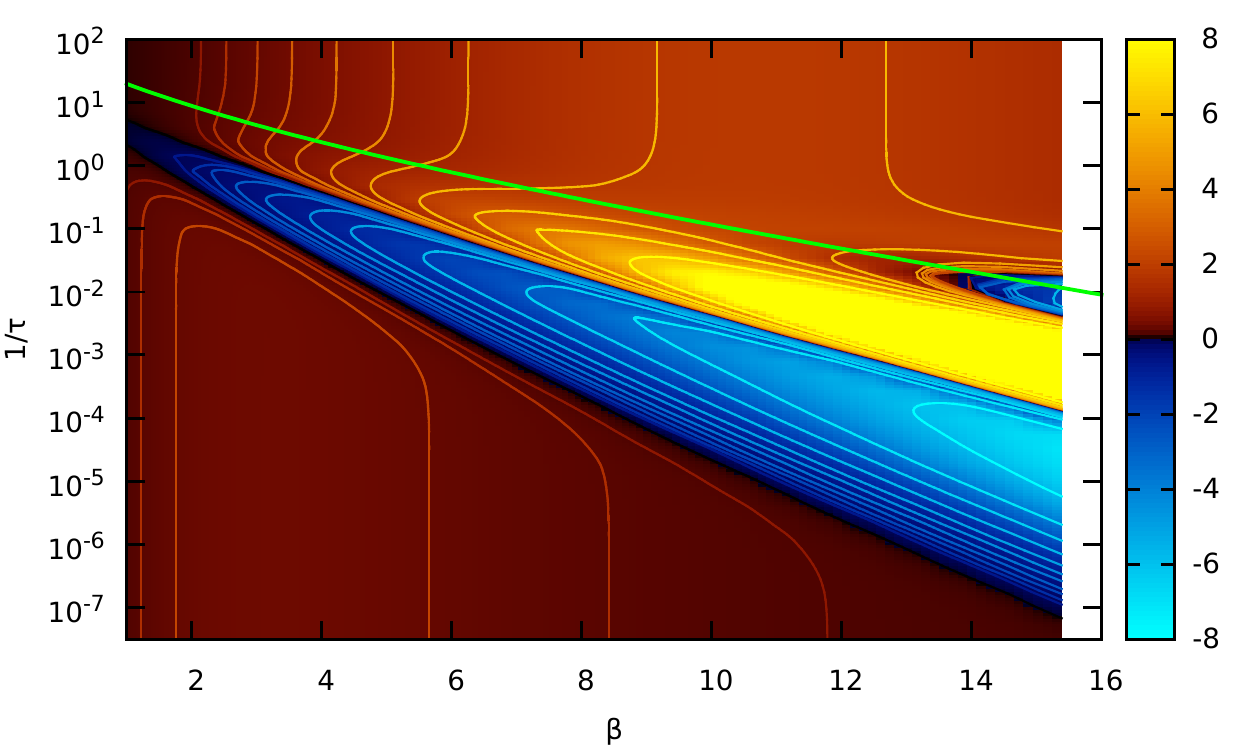}
\end{center}
\end{figure}

We again evaluate numerically the generalized heat capacity $C$ with respect to inverse temperature $\beta$ and frequency of the driving $1/\tau$ on exponential scale, see figure~\ref{pic:two_level_barrier_C_beta}.  
As one can see, the system exhibits slow driving limit behaviour as well as singular driving limit behaviour, depending on the driving frequency with respect to characteristic relaxation time $\tau_{relax}$. 
Moreover we can see that the system again has a rich behaviour on the time scale comparable with the typical relaxation time. 
Again, we are not able to determine characteristic relaxation time analytically, 
however we can heuristically estimate the temperature dependence by analysing the relaxation time for each fixed time, i.e. fixed value of potential. 

Again, the characteristic time of the dynamics can be estimated via the relaxation times of the "frozen" system,
\begin{align*}
\ln \left[ \frac{1}{\tau_{relax} (t)} \right] 
&= \ln \left[  k_t ( 1 \rightarrow 2 ) + k_t ( 2 \rightarrow 1 ) \right] \\
&= \ln \psi - \beta \Delta + \ln \left[ 2 \cosh \left( \beta \left( \epsilon_0 + \frac{\Delta \epsilon}{2} \cos \frac{2 \pi t}{\tau} \right) \right) \right]. 
\end{align*}
In low temperature region $\beta \gg 1$ the logarithm of relaxation time can be easily approximated by linear dependence on temperature, approximatly  
\[
\ln \left[ \frac{1}{\tau_{relax} (t)} \right] 
\sim \beta \left[ \left| \epsilon_0 + \frac{\Delta \epsilon}{2} \cos \frac{2 \pi t}{\tau} \right| - \Delta \right] \le 0 , 
\]
while in the high temperature region $\beta \ll 1$ the temperature dependence of logarithm of characteristic relaxation time vanish  
\[
\ln \left[ \frac{1}{\tau_{relax} (t)} \right] 
= \ln \psi - \beta \Delta + \left( \epsilon_0 + \frac{\Delta \epsilon}{2} \cos \left( \frac{2 \pi t}{\tau} \right) \right)^2 \beta^2 + \err[4]{\beta}. 
\]
We can see that there is an additional term when compared with classical model which largely influence the transition region.

\section{Conclusions} 
We have developed the theory of quasistatic processes in systems driven by periodical action of force. 
Our starting point were results of the Floquet theory \eqref{equ:Floquet_theorem}, which we used to reformulate the problem such that the explicit time dependence is removed \eqref{equ:generalized_time_evolution}. 
Then we have been able to show that also in this type of driving the quasistatic heat and work can be decomposed to so-called ``housekeeping'' part \eqref{equ:generalized_housekeeping_work} or \eqref{equ:generalized_housekeeping_heat} responsible for maintaining the steady state out of equilibrium
and to the ``reversible'' part \eqref{equ:generalized_reversible_work} or \eqref{equ:generalized_reversible_heat} which is of geometrical nature and corresponds to what we in classical thermodynamics call heat and work, which is our main result of this chapter. 
While ``housekeeping'' components of the work \eqref{equ:generalized_housekeeping_work_rew} and heat are in general extensive in time, it also contains fluctuating term tightly bounded with the initial and the terminal condition, which in general does not need to converge in the quasistatic limit and hence can affect the result of potential measurements.

Similarly to the systems driven out by non-potential forces or by attaching the system to multiple thermal bath presented in the chapter \ref{chapter:non-equilibrium_thermodynamics}, 
the systems driven by periodical forces has also a rich non-equilibrium behaviour such as the general non-validity of the Clausius relation, negative generalized heat capacities, 
as was shown on two level models, see section \ref{sec:two_level_models_t}.
However as our results for slow, see subsection \ref{ssec:slow}, and singular, see subsection \ref{ssec:fast}, driving limit suggests, 
this behaviour is typically limited to the intermediate regime where the period of the driving is comparable to the relaxation time, as was also illustrated in the aforementioned models.

The article summarizing these results is still in preparation \cite{result3}. 

% Slow-fast coupling
% a. Introduction to the problem why it naturally fits this thesis]
% b. Slow-fast separation - both for the slow and the fast subsystems: the markovianess up to the first order corretion, the connection to the quasistatic processes before
% c. Thermodynamics of slow-fast system

\chapter{Slow-fast coupling} 
\label{chapter:slow-fast_coupling}
Until now we have been trying to describe thermodynamical processes of the system in contact with several thermal and chemical baths and in general in the presence of external non-potential force fields, 
e.g. electromagnetic field, which time-dependence is periodical. 
To be more precise, we have been studying Markovian systems in environment described by few fully manageable external parameters,
which entirely determine the Markovian dynamics of the system as well as the steady state of the system and which were subjects of the quasistatic transformations.
Although these conditions seem to be very general, not every possible physical situation fits into this description even in the simplified case of quasistatic processes in the vicinity of steady states. 
For example if we are interested only in the properties and behaviour of the subsystem of a more complex system, 
the evolution of such system in general depends on all degrees of freedom of the composite system, which are basically intractable. 
However in some cases we are able to find an effective Markovian evolution for such subsystem, so they can be treated in the same manner as above mentioned systems with few external parameters. 
The basic questions are in which systems there exists an effective Markovian description of the time evolution and how to properly construct the description from the description of more complex systems?

In this chapter we will focus on these questions in a particular class of composite systems with so called \emph{separation of timescales}.\index{separation of timescales}
We speak about the separation of timescales, when we can sort the degrees of freedom of a complex system into two or more distinct sets characterized by typical relaxation times.
Moreover we demand that the typical relaxation times $\tau_i$ of each set greatly differs, $\tau_1 \gg \tau_2 \gg \dots \gg \tau_k$.
We will restrict ourselves further on to the case when all degrees of freedom can be sorted only to two of such sets. 
The degrees of freedom associated with the long (short) typical relaxation time we call ``slow'' (``fast'').
A typical example of such a system is a heavy particle in the bath of light ones, where the degrees of freedom associated with \emph{heavy particle} are considered to be \emph{``slow''}, 
i.e. it takes much more time to relax the heavy particle after a perturbation of the environment than to relax light particles.
The dynamical parameters, which we are typically able to control in the experiment, describe the average density of light particles and their local temperature, 
however it is difficult to control the feedback between the slow and fast particles as well as local fluctuation of density.

In the first section we will focus on the dynamics of the ``fast'' degrees of freedom on their respective timescale, 
while in the second section we will investigate the dynamics of the ``slow'' degrees of freedom again on their respective timescale. 
In the context of our example the first section focuses on the dynamics of light particles under the influence of quasistatically moving slow particle,
while in the second section we will describe the movement of the heavy particle in the continuum of the light ones, which keeps nearby a steady state.

Through this chapter we will be again using the framework of Kolmogorov generators firstly introduced in chapter \ref{chapter:models} section \ref{sec:Markov_processes}, 
where we describe our full system by probabilistic density $\mu_t(x_S,x_F)$, 
where by $x_S$ ($x_F$) we explicitly denote the dependence on the ``slow'' (``fast'') degrees of freedom. 
We also assume that the time evolution of the full system is Markovian and given by \eqref{equ:Markov_stochastic_time_evolution} 
\[
\partial_t \mu_t(x_S,x_F) = \gen^*_\epsilon \left[ \mu_t \right] (x_S,x_F) 
\]
where the total forward Kolmogorov generator $\gen^*_\epsilon$ acts on the overall system and does not explicitly depend on time,
and $\epsilon$ denotes the ratio of the ``fast'' and ``slow'' degrees of freedom's relaxation times ($\tau_F$ and $\tau_S$), $\epsilon = \tau_F/\tau_S$. 
The formal solution of such evolution is given by
\begin{equation}
\mu_t (x_F,x_S) = \rme^{ \left( t-t_0 \right) \gen^*_\epsilon } \left[ \mu_{t_0} \right] (x_S,x_F). 
\label{equ:full_solution}
\end{equation}

\section{Autonomous dynamics for fast degrees of freedom}
\label{sec:dynamics_for_fast}
The first situation, which we want to analyse, is the behaviour of the system on the timescale which is comparable to the typical relaxation times of the ``fast'' degrees of freedom. 
In that case we would expect to see almost no time evolution of the ``slow'' degrees of freedom. 
Hence we presumably expect to obtain a situation very similar to the quasistatic limit, which is a central topic of this thesis, where the ``slow'' degrees of freedom represent the external parameters, although in this case they don't have a prescribed trajectory, but rather evolve on their own.

To show more precisely in what sense the proposition holds true, 
we start with the decomposition of the total forward Kolmogorov generator to the part evolving just the ``slow'' degrees of freedom $\gen^*_S$ and the part responsible for the time evolution of ``fast'' degrees of freedom $\gen^*_F$. 
To have a clear distinction between what are ``fast'' and what are ``slow'' degrees of freedom, we assume that they are entangled only by the interaction potential, 
i.e. in the forward Kolmogorov generator for ``fast'' degrees of freedom the ``slow'' degrees of freedom occurs only as parameters and vice versa. 
To simplify the analysis we introduce the parameter $\epsilon$ which explicitly characterizes the timescale separation. 
Hence the time evolution of the joint probability distribution \eqref{equ:Markov_stochastic_time_evolution} is given by
\[
\partial_t \mu_t (x_S,x_F) = \epsilon \gen^*_S [\mu_t] (x_S,x_F) + \gen^*_F[\mu_t](x_S,x_F) ,
\]
where the only epsilon dependence lies in the pre-factor.

Our main aim is to obtain the time evolution equation for the marginal probability distribution on fast degrees of freedom $\widetilde\mu(x_F)$. 
We start with the decomposition of the joint probability distribution to the conditional probability distribution of the ``fast'' degrees of freedom conditioned to the particular configuration of the ``slow'' degrees of freedom $\nu_t(x_F|x_S)$ and the marginal probability distribution of ``slow'' degrees of freedom $\sigma_t(x_S)$, 
\[
\mu_t(x_S,x_F) = \nu_t (x_F | x_S) \, \sigma_{\epsilon t} (x_S) ,
\]
where we have also assumed that the appropriate timescale of the ``slow'' degrees of freedom is $\epsilon t$, i.e. the marginal distribution of ``slow'' degrees of freedom $\sigma$ is function in $\epsilon t$.
The time evolution equation is then given by
\begin{multline*}
\partial_t \nu_t (x_F|x_S) \, \sigma_{\epsilon t}(x_S) + \epsilon \, \nu_t(x_F|x_S) \, \left. \partial_s \sigma_s(x_S) \right|_{s=\epsilon t} = \\
= \sigma_{\epsilon t}(x_S) \, \gen^*_F[\nu_t](x_F|x_S) + \epsilon \, \gen^*_S\left[\nu_t \, \sigma_{\epsilon t}\right](x_S,x_F) ,
\end{multline*}
from which we can obtain a time evolution equation for marginal distribution of the ``slow'' degrees of freedom $\sigma(x_S)$ by integration over the ``fast'' degrees of freedom
\begin{equation}
\epsilon \left. \partial_s \sigma_s (x_S) \right|_{s = \epsilon t} = \epsilon \int \rmd \Gamma(x_F) \; \gen^*_S\left[ \nu_t \, \sigma_{\epsilon t} \right](x_S,x_F) ,
\label{equ:time_evolution_slow}
\end{equation}
where we also assumed that the normalization condition \eqref{equ:normalization_condition} is valid also for the forward Kolmogorov generator responsible for time evolution of ``fast'' degrees of freedom $\gen^*_F$. 
Notice that we have assumed that the dependence of the forward Kolmogorov generator for ``slow'' degrees freedom $\gen^*_S$ on fast degrees of freedom $x_F$ lies only in the interaction potential, 
hence the time evolution equation \eqref{equ:time_evolution_slow} depends on the averaged potential $\langle V \rangle_{\nu} (x_S)$. 
Similarly we can obtain the time evolution equation for marginal distribution on ``fast'' degrees of freedom $\widetilde{\mu}(x_F)$ 
\[
\partial_t \widetilde{\mu}_t(x_F) 
= \left\langle \gen^*_F[\nu_t](x_F|x_S) \right\rangle_{\sigma_{\epsilon t}} + \epsilon \int \rmd \Gamma(x_S) \; \gen^*_S\left[ \nu_t \, \sigma_{\epsilon t} \right](x_S,x_F) ,
\]
where we can see two contributions, the first term is the time evolution of the ``fast'' degrees of freedom, where the ``slow'' degrees of freedom occurs only as parameters. 
The second term is correction corresponding to the time evolution of ``slow'' degrees of freedom.

If we assume that the feedback of the action of ``slow'' degrees of freedom is small, i.e. the interaction potential in \eqref{equ:time_evolution_slow} is at least of first order in $\epsilon$, 
then up to the zeroth order in epsilon the time evolution of marginal distribution of ``slow'' degrees of freedom is independent of the distribution of ``fast'' degrees of freedom,
i.e. we have autonomous dynamics for ``slow'' degrees of freedom in the leading order 
\[
\left. \partial_s \sigma_s (x_S) \right|_{s = \epsilon t} = \gen^*_S\left[ \sigma_s \right](x_S) + \err{\epsilon} .
\]
Hence the time evolution of the marginal distribution on fast degrees of freedom can also be considered autonomous, 
in the manner that we consider the ``slow'' degrees of freedom as time dependent parameters of the dynamics.

If the dynamics of the ``slow'' degrees of freedom is deterministic
\[
\sigma_t (x_S) = \delta(x_S - x_S(t)) \equiv \delta_{x_S(t)}(x_S)
\] 
then the time evolution equation of the marginal distribution of ``fast'' degrees of freedom simplifies to 
\[
\partial_t \widetilde{\mu}_t(x_F) 
= \left\langle \gen^*_F[\nu_t](x_F|x_S) \right\rangle_{\sigma_{\epsilon t}} + \err[2]{\epsilon},
\]
where we have also used that in case the $\gen^*_S$ is independent of $x_F$ in the leading order,  
hence it has to obey the normalization condition \eqref{equ:normalization_condition} as being the forward Kolmogorov generator for autonomous time evolution up to the same order
and so forth the first order term on the right hand side is zero. 
The left hand side can be rewritten as 
\begin{multline*}
\partial_t \widetilde{\mu}_t(x_F) 
= \partial_t \int \rmd \Gamma(x_S) \nu_t(x_F|x_S) \, \delta(x_S - x_S(\epsilon t)) = \\
= \int \rmd \Gamma(x_S) \left[ \partial_t \nu_t(x_F|x_S) \, \delta(x_S - x_S(\epsilon t)) + \nu_t(x_F|x_S) \, \partial_t \delta(x_S - x_S(\epsilon t)) \right] = \\
= \partial_t \nu_t(x_F|x_S(\epsilon t)) + \epsilon \, \dot{x}_S(\epsilon t) \cdot \left. \nabla_{x_S} \nu_t(x_F|x_S) \right|_{x_S = x_S(\epsilon t)} ,
\end{multline*}
which provides a more precise formulation of our claim in introduction.

We have seen that up to the zeroth order there is autonomous dynamics for ``slow'' degrees of freedom under the assumption that the time evolution of the ``slow'' degrees of freedom depends on the ``fast'' ones only through interaction potential and the feedback is weak, i.e. of order $\epsilon$. 
We have also seen that under the same assumption the time evolution of ``fast'' degrees of freedom depends only on ``slow'' degrees of freedom parametrically. 
Moreover if we also assume that the dynamics of ``slow'' degrees of freedom is deterministic, 
we obtain the quasistatic expansion of the conditional distribution of ``fast'' degrees of freedom on ``slow'' ones up to the first order, as in chapter \ref{chapter:non-equilibrium_thermodynamics} section \ref{sec:quasistatic_processes_noneq},
where we can identify the ``slow'' degrees of freedom with external parameters, whose time dependence is in this given implicitly be evolution equation \eqref{equ:time_evolution_slow}.

\section{Autonomous dynamics for slow degrees of freedom}
\label{sec:dynamics_for_slow} 
%TODO
In this section we describe the time-evolution of slow degrees of freedom on their appropriate timescale.
Although Berglund and Gentz \cite{Berglund2006} provided a rigorous and thorough analysis of slow-fast coupling in case the dynamics is deterministic,
the extension of these ideas to stochastic dynamics isn't treated in such a rigorous manner, 
see for example \cite{Elimination_of_fast_stochastic_variables-Thomas}. 
We provide an alternative more exact derivation of the effective Markovian dynamics for ``slow'' degrees of freedom. 
Our method enables us to systematically quantify the corrections due to the slow-fast coupling,
and also to estimate their propagation in time.

We start by using Dyson-like expansion to provide zeroth order contribution of the effective Kolmogorov generator for slow degrees of freedom, see subsection \ref{ssec:slow-fast_zeroth_order}. 
Furthermore by expanding the Dyson-like series in the parameter characterizing the separation of timescales $\epsilon$ we show that the effective dynamics is Markovian up to the first order with effective generator being \eqref{equ:effective_time_evolution_slow}. 
At the end of the section we illustrate the results on examples.

In order to simplify the investigation we consider a specific form of $\epsilon$ dependency, namely 
\[
\gen^*_\epsilon = \gen^*_S + \frac{1}{\epsilon} \gen^*_F,
\]
where $\gen^*_S$ is the evolution of slow degrees of freedom given the fast ones fixed and $\gen^*_F$ is the opposite, i.e. the evolution of fast degrees of freedom given the slow ones. 
This particular decomposition corresponds to independent dynamics for fast and slow degrees of freedom apart, 
hence the interaction between the fast and slow degrees of freedom is provided only by mutual dependence of some parameters, usually in the form of interaction potential. 
Also the $\epsilon$ dependency corresponds to the situation when the fast degrees of freedom are more active,
e.g. in case of jump processes with continuous time we scale the symmetric part of jump rates or in case of diffusion we scale the \emph{mobility}\index{mobility}.

We further assume that there exists a unique stationary distribution on fast degrees of freedom $\rho_F(x_S,x_F)$ for arbitrary yet fixed configuration of slow degrees of freedom $x_S$ 
\[
\gen^*_F [ \rho_F ] (x_S,x_F) = 0 ,
\]
and that the evolution up to the infinite time of fast degrees of freedom 
is equivalent to the projection $\proj^*_0$ of the fast degrees of freedom to the steady state $\rho_F$ 
\begin{equation}
\lim_{t \rightarrow \infty} \rme^{ t \gen^*_F } [ \mu ] = \proj^*_0 [ \mu ] = \rho_F(x_S,x_F) \int \rmd y \; \mu(x_S,y) .
\label{equ:projection}
\end{equation}

Our starting point is the formal solution \eqref{equ:full_solution}. 
To find an appropriate timescale at which there exists an autonomous dynamics for slow degrees of freedom, 
we slice the total time interval $t$ to time steps of length $\Delta t$ 
\[
\rme^{ t \left( \gen^*_S + \frac{1}{\epsilon} \gen^*_F \right) } [\mu_0] = \left( \rme^{ \Delta t \left( \gen^*_S + \frac{1}{\epsilon} \gen^*_F \right) } \right)^{\frac{t}{\Delta t}} [\mu_0]
\]
and we choose the time step $\Delta t \ll \tau_S $ in a suitable way that we can simultaneously expand in both $\Delta t$ and $\epsilon$. 
To separate the contributions of fast and slow degrees of freedom we use the Dyson-like expansion 
\begin{equation}
\rme^{\Delta t \left( \gen^*_S + \frac{1}{\epsilon} \gen^*_F \right) } 
= \rme^{ \frac{\Delta t}{\epsilon} \gen^*_F } 
+ \int\limits_0^{\Delta t} \rmd s \; \rme^{ \frac{ \left( \Delta t - s \right) }{\epsilon} \gen^*_F} \gen^*_S \rme^{ s \left( \gen^*_S + \frac{1}{\epsilon} \gen^*_F \right) }
\label{equ:dyson_expansion_noniterated}
\end{equation}
recursively. 
So for the evolution over the single time step $\Delta t$ we obtain a series 
\begin{multline}
\rme^{\Delta t \left( \gen^*_S + \frac{1}{\epsilon} \gen^*_F \right) } 
= \rme^{ \frac{\Delta t}{\epsilon} \gen^*_F } + \\
+ \sum\limits_{n=1}^\infty \quad \idotsint\limits_{\Delta t \ge t_1 \ge \dots \ge t_n \ge t_{n+1} \equiv 0 } \rmd t_1 \dots \rmd t_n \; 
\rme^{ \frac{ \left( \Delta t - t_1 \right) }{\epsilon} \gen^*_F} \prod\limits_{k=1}^n \left[ \gen^*_S \rme^{ \left( t_k - t_{k+1} \right) \frac{1}{\epsilon} \gen^*_F } \right],
\label{equ:dyson_expansion}
\end{multline}
where exponential terms have only the evolution of the fast dynamics as the argument.

\subsection{Infinite timescale separation}
\label{ssec:slow-fast_zeroth_order}
In order to describe the time evolution of slow degrees of freedom even in the zeroth order by effective Markovian dynamics, we need to observe the system on an appropriate timescale. 
Which corresponds to the $\Delta t$ being much larger than the characteristic time of fast degrees of freedom $\Delta t \gg \tau_F$ 
and doesn't scale with fast degrees of freedom, i.e.  
\begin{equation}
\lim_{\epsilon \to 0^+} \frac{\epsilon}{\Delta t} = 0 . 
\label{equ:lower_bound}
\end{equation}
%Otherwise we are observing our system at the corresponding timescale of the fast degrees of freedom as in section \ref{sec:dynamics_for_fast}. 

Obtaining the zeroth order contribution is quite straightforward, we take the limit $\epsilon \rightarrow 0^+$ of the Dyson-like expansion \eqref{equ:dyson_expansion} 
\begin{equation}
\lim_{\epsilon \rightarrow 0} \rme^{\Delta t \left( \gen^*_S + \frac{1}{\epsilon} \gen^*_F \right) } 
= \proj^*_0 + \sum\limits_{n=1}^\infty \frac{\left(\Delta t\right)^n}{n!} \proj^*_0 \left[ \gen^*_S \proj^*_0 \right]^n
= \rme^{\Delta t \proj^*_0 \gen^*_S \proj^*_0} \proj^*_0 , 
\label{equ:zeroth_order_limit_slow}
\end{equation}
where we have used the asymptotic of the time evolution on the fast degrees of freedom \eqref{equ:projection} along with the assumption \eqref{equ:lower_bound}.
By introducing the marginal distribution of the slow degrees of freedom $\nu_t$ 
\[
\nu_t (x_S) = \int \rmd x_F \; \mu_t(x_S,x_F) 
\]
we can indeed show that the time evolution is Markovian.
We show that the marginal distribution after the time step $\Delta t$ depends only on the marginal distribution in the initial time 
using the limit \eqref{equ:zeroth_order_limit_slow} 
\[
\nu_{t+\Delta t} (x_S) = \int \rmd x_F \; \rme^{\Delta t \proj^*_0 \gen^*_S \proj^*_0} \proj^*_0 \left[ \mu_t \right] \left( x_S , x_F \right) 
= \rme^{\Delta t \left\langle \gen^*_S \right\rangle_{\rho_F} } \left[ \nu_t \right] \left( x_S \right) .
\]
And hence by taking the limit $\Delta t \rightarrow 0^+$ we obtain a Markovian time evolution equation 
\begin{equation}
\partial_t \nu_t (x_S) = \left\langle \gen^*_S \right\rangle_{\rho_F} \left[ \nu_t \right] (x_S) .
\label{equ:zeroth_order_slow}
\end{equation}
One can see that the leading order describes the situation where the fast degrees of freedom can already be considered relaxed to the steady state 
given by fixed values of the external parameters but also by the fixed values slow degrees of freedom. 
However the assumption \eqref{equ:lower_bound} alone is not sufficient to obtain higher order contributions, where we would likely to see some feedback from the fast degrees of freedom.

\subsection{First-order correction}
In order to find corrections up to higher order of epsilon let us assume that the generator of fast degrees of freedom has a spectral decomposition  
\begin{equation}
\gen^*_F = - \sum_{i>0} \lambda_i P_i^*, 
\label{equ:spectral_projection}
\end{equation}
where eigenvalues has positive real part $\Re \lambda_i > 0$. 
Similarly the time evolution of the fast degrees of freedom alone can also be written as the decomposition to the same eigenstates 
\[
\rme^{\frac{s}{\epsilon} \gen^*_F} = \proj^*_0 + \sum_{i>0} \rme^{- \frac{s \lambda_i}{\epsilon}} \proj^*_i.
\]
Using this decomposition in the Dyson-like series \eqref{equ:dyson_expansion} we can again collect all the zeroth order terms in $\epsilon$, 
hence all the remaining terms are at least of order $\epsilon$. 
While we don't expect an existence of an effective Markovian dynamics behind the first order corrections in $\epsilon$ we include only those.
Finally we can neglect all the terms containing the exponential term $ \rme^{ - \frac{\Delta t}{\epsilon} \lambda_i }$, which are much smaller because of the condition \eqref{equ:lower_bound}.
All these considerations yield to  
\begin{multline}
\rme^{\Delta t \left( \gen^*_S + \frac{1}{\epsilon} \gen^*_F \right) }
= \rme^{ \Delta t \, \proj^*_0 \gen^*_S \proj^*_0} \proj^*_0 
- \epsilon \sum\limits_{n=0}^{\infty} \frac{\left( \Delta t \right)^n}{n!} \sum\limits_{k=0}^{n+1} \left( \proj^*_0 \gen^*_S \right)^k \frac{1}{\gen^*_F} \left( \gen^*_S \proj^*_0 \right)^{n+1-k} + \\  
+ \err{ \rme^{- \frac{\Delta t}{\epsilon} \min\limits_{i>0} \lambda_i}, \epsilon^2} ,
\label{equ:dyson_first_order}
\end{multline}
where $1/\gen^*_F$ denotes the pseudoinverse\index{pseudoinverse} 
\begin{equation}
\frac{1}{\gen^*_F} = \int\limits_0^\infty \rmd s \; \left[ \proj^*_0 - \rme^{s \gen^*_F} \right] = - \sum\limits_{i>0} \frac{1}{\lambda_i} \proj^*_i , 
\label{equ:pseudoinverse_fast} 
\end{equation}
introduced in chapter \ref{chapter:non-equilibrium_thermodynamics} as \eqref{def:forward_pseudoinverse},  
which emerges here as the consequence of the asymptotic behaviour of the time evolution in the memory kernel  
\begin{multline*}
\int\limits_0^t \rmd s \; \rme^{\frac{s}{\epsilon} \gen^*_F} = t \proj^*_0 + \epsilon \int\limits_0^{\frac{t}{\epsilon}} \rmd u \; \left( \rme^{u \gen^*_F} - \proj^*_0 \right) \longrightarrow \\
\longrightarrow t \proj^*_0 + \epsilon \int\limits_0^\infty \rmd u \; \left( \rme^{u \gen^*_F} - \proj^*_0 \right) + \err[2]{\epsilon}= t \proj^*_0 - \epsilon \frac{1}{\gen^*_F} + \err[2]{\epsilon}
\end{multline*}
By rearranging all the terms in \eqref{equ:dyson_first_order} we obtain more compact expression for the expansion of the Dyson-like series up to first order in $\epsilon$  
\begin{multline}
\rme^{\Delta t \left( \gen^*_S + \frac{1}{\epsilon} \gen^*_F \right) }
= \exp \left[ \Delta t \, \proj^*_0 \left( \gen^*_S - \epsilon \gen^*_S \frac{1}{\gen^*_F} \gen^*_S \right) \proj^*_0 \right] \proj^*_0 - \\
- \epsilon \left[ \frac{1}{\gen^*_F} \gen^*_S \proj^*_0 \rme^{ \Delta t \, \proj^*_0 \gen^*_S \proj^*_0} + \rme^{ \Delta t \, \proj^*_0 \gen^*_S \proj^*_0} \proj^*_0 \gen^*_S \frac{1}{\gen^*_F} \right]  
+ \err{ \rme^{- \frac{\Delta t}{\epsilon} \min\limits_{i>0} \lambda_i}, \epsilon^2}.  
\label{equ:dyson_first_order_reordered}
\end{multline}

One can see that up to this point we haven't assume anything about the total probability density. 
The basic idea of the separation of timescale is that in every time instance the system can be considered to be nearby the state with fast degrees of freedom sampled from a stationary distribution 
\begin{equation}
\mu_t (x_S,x_F ) = \rho_F(x_S,x_F) \nu_t (x_S) + \Delta \mu_t (x_S,x_F),
\label{equ:prob_density_structure}
\end{equation}
where $\Delta \mu_t$ is expected to be small 
and by $\nu_t$ we denote the marginal distribution of the slow degrees of freedom hence 
\[
\int \rmd x_F \; \Delta \mu_s (x_S,x_F) = 0 .
\]
Furthermore we assume that the correction is order $\epsilon$.  
To check if the proposed structure of the total probability density is conserved during the time evolution, 
we evolve the total probability density in the form \eqref{equ:prob_density_structure} using \eqref{equ:dyson_first_order_reordered} and expanding it only up to the first order
\begin{multline*}
\mu_{t+\Delta t} 
=\rme^{\Delta t \left( \gen^*_S + \frac{1}{\epsilon} \gen^*_F \right) } \left[ \mu_t \right] 
= \rho_F \exp \left\{ \Delta t \, \left\langle \left( \gen^*_S - \epsilon \gen^*_S \frac{1}{\gen^*_F} \gen^*_S \right) \right\rangle_{\rho_F} \right\} \left[ \nu_t \right] - \\
- \epsilon \frac{1}{\gen^*_F} \gen^*_S \left[ \rho_F \exp \left\{ \Delta t \, \left\langle \gen^*_S \right\rangle_{\rho_F} \right\} \left[ \nu_t \right] \right]
+ \err{ \rme^{- \frac{\Delta t}{\epsilon} \min\limits_{i>0} \lambda_i}, \epsilon^2}.  
\end{multline*}
We can see that the structure \eqref{equ:prob_density_structure} is really preserved by having 
\[
\nu_{t+\Delta t} (x_S) = \exp \left\{ \Delta t \, \left\langle \left( \gen^*_S - \epsilon \gen^*_S \frac{1}{\gen^*_F} \gen^*_S \right) \right\rangle_{\rho_F} \right\} \left[ \nu_t \right]
\] 
and 
\begin{equation}
\Delta \mu_{t+\Delta t} = - \epsilon \frac{1}{\gen^*_F} \gen^*_S \left[ \rho_F \exp \left\{ \Delta t \, \left\langle \gen^*_S \right\rangle_{\rho_F} \right\} \left[ \nu_t \right] \right],
\label{equ:difference_exact_vs_markovian}
\end{equation}
where one can see that the correction is of the first order in $\epsilon$ and also does not depend on the previous correction $\Delta \mu_t$ at that specific order.

The assumption on the correction being small is equivalent to be in a state close to steady on the fast degrees of freedom and also that the magnitude of the correction is bounded.
Next we prove that the correction remain small in this sense during the time evolution. 
To prove this proposition we need to assume that the dynamics is converging to its steady state $\rho$ on the arbitrary large timescale
\begin{equation}
\left\| \left( \rme^{t \left( \gen^*_S + \frac{1}{\epsilon} \gen^*_F \right) } - \proj^* \right) [ \mu ] \right\|_1 \le \left\| \mu \right\|_1 , 
\label{equ:contraction} 
\end{equation}
where $\|\cdot\|_1$ is the $L^1$ norm, $\proj^*$ is the projection to the full-system steady state 
\[
\proj^* = \rho \iint \rmd x_S \, \rmd x_F . 
\]
Then the correction after a single time-step 
\[
\left\| \rme^{\Delta t \left( \gen^*_S + \frac{1}{\epsilon} \gen^*_F \right) } [ \Delta \mu ] \right\|_1
\le \left\| \rme^{\frac{\Delta t}{\epsilon} \gen^*_F} [ \Delta \mu ] \right\|_1 
+ \left\| \int\limits_0^{\Delta t} \rmd s \; \rme^{\left(\Delta t - s \right) \left( \gen^*_S + \frac{1}{\epsilon} \gen^*_F \right) } \gen^*_S \rme^{\frac{s}{\epsilon} \gen^*_F } [ \Delta \mu ] \right\|_1, 
\]
where we've already used the Dyson-like expansion \eqref{equ:dyson_expansion_noniterated} and the triangle inequality.  
Using the spectral decomposition for evolution of the fast degrees of freedom~\eqref{equ:spectral_projection} and estimating the integral by its norm we obtain 
\[
\left\| \rme^{\Delta t \left( \gen^*_S + \frac{1}{\epsilon} \gen^*_F \right) } [ \Delta \mu ] \right\|_1
\le \left\| \sum\limits_{i>0} \rme^{- \frac{\Delta t}{\epsilon} \lambda_i} \proj^*_i [ \Delta \mu ] \right\|_1 
+ \int\limits_0^{\Delta t} \rmd s \; \left\| \rme^{\left(\Delta t - s \right) \left( \gen^*_S + \frac{1}{\epsilon} \gen^*_F \right) } \gen^*_S \rme^{\frac{s}{\epsilon} \gen^*_F } [ \Delta \mu ] \right\|_1, 
\]
using the assumption \eqref{equ:contraction} together with the fact that the evolution on slow degrees of freedom conserves the probability $\int \rmd x_S \; \gen^*_S$ we get
\[
\left\| \rme^{\Delta t \left( \gen^*_S + \frac{1}{\epsilon} \gen^*_F \right) } [ \Delta \mu ] \right\|_1
\le \rme^{- \frac{\Delta t}{\epsilon} \lambda_g } \left\| \Delta \mu \right\|_1 
+ \int\limits_0^{\Delta t} \rmd s \; \left\| \gen^*_S \rme^{\frac{s}{\epsilon} \gen^*_F } [ \Delta \mu ] \right\|_1, 
\]
where we also estimated the first term in terms of spectral gap 
\[
\lambda_g = \min_{i > 0 } \Re \lambda_i . 
\]
Using the supremal norm of $\gen^*_S$ and the estimate on the exponential we have 
\[
\left\| \rme^{\Delta t \left( \gen^*_S + \frac{1}{\epsilon} \gen^*_F \right) } [ \Delta \mu ] \right\|_1
\le \rme^{- \frac{\Delta t}{\epsilon} \lambda_g } \left\| \Delta \mu \right\|_1 
+ \left\| \gen^*_S \right\|_\infty \int\limits_0^{\Delta t} \rmd s \; \rme^{- \frac{s}{\epsilon} \lambda_g } \left\| \Delta \mu \right\|_1, 
\] 
and by integration and using the fact $\frac{\epsilon}{\lambda_g}$ is of order $\tau_F$ we obtain
\begin{equation}
\left\| \rme^{\Delta t \left( \gen^*_S + \frac{1}{\epsilon} \gen^*_F \right) } [ \Delta \mu ] \right\|_1
\le \rme^{- \frac{\Delta t}{\tau_F} } \left( 1 - \tau_F \left\| \gen^*_S \right\|_\infty \right) \left\| \Delta \mu \right\|_1 
+ \tau_F \left\| \gen^*_S \right\|_\infty \left\| \Delta \mu \right\|_1 .  
\label{equ:difference_time_evolution}
\end{equation}
From this we can see that the assumption $\Delta t \gg \tau_F$ makes the first term negligible and the second term is of order of $\epsilon$, 
as the consequence of the norm $\left\| \gen^*_S \right\|_\infty$ being of order $1/\tau_S$ and also $\tau_F/\tau_S$ being of order $\epsilon$, 
so after one iteration the correction is already of order epsilon.

Putting all the results together,
we have verified that at the timescale much longer than characteristic time of the fast degrees of freedom $\Delta t \gg \tau_F$ we have an autonomous dynamics for slow degrees of freedom,
which for small time steps $\tau_F \ll \Delta t \ll \tau_S$ is effectively governed by 
\begin{equation}
\partial_t \nu_t \approx \frac{ \nu_{t+\Delta t} - \nu_t }{\Delta t} = \left\langle \gen^*_S - \epsilon \gen^*_S \frac{1}{\gen^*_F} \gen^*_S \right\rangle_{\rho_F} [ \nu_t ] .
\label{equ:effective_time_evolution_slow}
\end{equation}
The correction to the full probability density is of order $\epsilon$ and is determined only by the distribution on the slow degrees of freedom.
Up to the first order, if the evolution on slow degrees of freedom is smooth, the memory term
\[
\Delta \mu_t = - \epsilon \frac{1}{\gen^*_F} \gen^*_S \left[ \rho_F \lim_{s \rightarrow t^-} \nu_s \right] 
\]
is not present. 
Furthermore we can see that the correction diminishes with time.

%Coupled particles as first example
\subsection{Example I: Diffusive particles entangled by harmonic potential}
To demonstrate our theory we introduce a simple model of two diffusive particles coupled by harmonic interaction potential with the fast particle trapped in an optical trap. 
Both slow $(R,P)$ and fast $(x,p)$ particles has mass $m$ and diffuse through the same environment with inverse temperature $\beta$ and friction $\gamma$. 
The only coupling between them is by interaction harmonic potential $\frac{1}{2} \omega^2 (x-R)^2$, moreover the fast variable is trapped by the optical trap with effective potential $\frac{1}{2} k (x-X_0)^2$, where $X_0$ is the center of optical trap and $k$ is the strength of the trap. 
The time evolution is then introduced by Kolmogorov generators for the fast and slow particles 
\begin{align*}
\gen^*_F [ \mu ] 
&= - \frac{p}{m} \partial_x \mu + \left[ \omega^2 (x-R) + k (x-X_0) \right] \partial_p \mu + \frac{\gamma}{m} \partial_p \left( p \mu \right) + \frac{\gamma}{\beta} \partial_p^2 \mu, \\
\gen^*_S [ \mu ] &= - \frac{P}{m} \partial_R \mu + \omega^2 (R-x) \partial_P \mu + \frac{\gamma}{m} \partial_P \left( P \mu \right) + \frac{\gamma}{\beta} \partial_P^2 \mu. 
\end{align*}
One can easily find the stationary distribution of the fast particle conditioned on the slow one fixed
\[
\rho_F(x,p) = \frac{1}{Z_R} \exp \left[ - \beta \left( \frac{p^2}{2m} + \frac{1}{2} \omega^2 (x-R)^2 + \frac{1}{2} k \left(x-X_0\right)^2 \right) \right], 
\]
where $Z_R$ is the partition function
\begin{multline*}
Z_R = \iint \rmd x \, \rmd p \; \exp \left[ - \beta \left( \frac{p^2}{2m} + \frac{1}{2} \omega^2 (x-R)^2 + \frac{1}{2} k \left(x-X_0\right)^2 \right) \right] = \\
= \frac{2 \pi}{\beta} \sqrt{\frac{m}{\omega^2+k}} \exp \left[ \frac{1}{2} k \omega^2 (R-X_0)^2 \right]. 
\end{multline*}

The correction term in the effective generator \eqref{equ:effective_time_evolution_slow} has the form $ \gen^*_S (1/\gen^*_F) \gen^*_S [ \rho_F \nu ]$ so we start by applying $\gen^*_S$  
\[
\gen^*_S [ \rho_F \nu ] = \rho_F \gen^*_S [ \nu ] - \frac{\beta \omega^2 P}{m} \left( x - \frac{\omega^2 R + k X_0}{\omega^2 + k} \right) \rho_F \nu. 
\]
The computation of pseudoinverse is in general difficult, however at least for this particular example the pseudoinverse can be computed explicitly, see appendix \ref{sec:pseudoinverse_underdamped} for further details, we obtain
\[
\frac{1}{\gen^*_F} \gen^*_S [ \rho_F \nu ] = - \frac{\omega^2}{\omega^2 + k} \left( \partial_P \nu + \frac{\beta P}{m} \nu \right) 
\left[ p - \gamma \left( x - \frac{\omega^2 R + k X_0}{\omega^2 + k} \right) \right] \rho_F . 
\] 
The last step is to apply the $\gen^*_S$ and integrate over fast degrees of freedoms $x$ and $p$ 
\[
\iint \rmd x \, \rmd p \; 
\gen^*_S \frac{1}{\gen^*_F} \gen^*_S [ \rho_F \nu ] = - \frac{\omega^4}{\left(\omega^2 + k\right)^2} \left[ \frac{\gamma}{m} \partial_P \left( P \nu \right) + \frac{\gamma}{\beta} \partial_P^2 \nu \right]. 
\]
Therefore we obtain the effective generator \eqref{equ:effective_time_evolution_slow} for slow degrees of freedom 
\[
\gen^*_{\eff} [\nu] = - \frac{P}{m} \partial_R \nu + \frac{ k \omega^2 }{\omega^2 + k} \left(R - X_0\right) \partial_P \nu + \left( 1 + \epsilon \frac{\omega^4}{\left(\omega^2 + k\right)^2} \right) \left[ \frac{\gamma}{m} \partial_P \left(P \nu \right) + \frac{\gamma}{\beta} \partial_P^2 \nu \right],
\]
where the slow particle is again diffusive in effective potential and the first order correction presents itself as the correction of the friction 
\[
\gamma_{\eff} = \gamma \left( 1 + \epsilon \frac{\omega^4}{\left(\omega^2 + k \right)^2} \right) .
\]

%Ovedamped diffusion as an another example of scaling
\subsection{Example II: Overdamped diffusion} 
Overdamped regime can be seen as another example of separation of fast and slow degrees freedom. 
In this case the fast variable is the momentum $p$, the slow is the position $x$ and the small parameter is $\epsilon = 1/\gamma$.
We will further assume that the diffusion occurs in non-homogeneous environment with spatially dependent inverse temperature $\beta(x)$.
Based on these considerations, we split the generator \eqref{equ:forward_generator_underdamped} for underdamped diffusion into two parts 
\begin{align*}
\gen^*_S [\mu] &= - \frac{p}{m} \partial_x \mu - F \partial_p \mu, \\
\gen^*_F [\mu] &= \frac{1}{m} \partial_p \left( p \mu \right) + \frac{1}{\beta} \partial_p^2 \mu, 
\end{align*}
where the second one describes the relaxation dynamics of momentum.  
The stationary distribution for the fast degree of freedom conditioned on the slow one is 
\[
\rho(p,x)=\frac{1}{Z(x)} \rme^{- \beta(x) \frac{p^2}{2m}}. 
\] 
We will proceed with the same protocol to determine the effective generator as in the previous example. 
First we use the stationary distribution of fast degrees of freedom to determine the structure the first term in the correction
\[
\gen^*_S [ \nu \rho ] = - \left[ \partial_x \nu + \left( \frac{\partial_x \beta}{2 \beta} - \beta F \right) \nu \right] \frac{p}{m} \rho + \left( \partial_x \beta \, \nu \right) \frac{p^3}{2 m^2} \rho,
\]
upon which we apply the pseudoinverse, 
see appendix \ref{sec:pseudoinverse_overdamped} in particular \eqref{equ:pseudoinverse2_const} to \eqref{equ:pseudoinverse2_p3} for further details 
\[
\frac{1}{\gen^*_F} \gen^*_S [ \nu \rho] = - \frac{p^3}{6m} \left( \partial_x \beta \, \nu \right) \rho + \left[ \partial_x \nu - \left( \frac{\partial_x \beta}{2 \beta} + \beta F \right) \nu \right] p \rho
\]
to obtain the total correction
\[
\int \rmd p \; \gen^*_S \frac{1}{\gen^*_F} \gen^*_S [ \nu \rho ] = \partial_x \left[ \left( F - k_B \partial_x T \right) \nu - \frac{1}{\beta} \partial_x \nu \right]. 
\]
While the average of the $\gen^*_S$ is zero, the only contribution to the effective generator is the first order correction
\[
\gen^*_{\eff} [\nu] = - \frac{1}{\gamma} \partial_x \left[ \left( F - k_B \partial_x T \right)  \nu - k_B T \partial_x \nu \right],
\]
which is in agreement with Smoluchowski equation \eqref{equ:time_evolution_overdamped} for diffusion in inhomogeneous environment \cite{Inhomogeneous_diffusion1-vanKampen,Inhomogeneous_diffusion2-vanKampen}. 
In case the force is conservative with potential $V(x)$, the stationary distribution can be easily determined 
\[
\nu_{st} (x) = \frac{1}{Z} \beta(x) \exp\left[ - \int\limits \rmd x \; \beta(x) V'(x) \right], 
\]
which corresponds to the zero steady current $j=0$. 
In case the boundaries are not placed in infinity but rather restrict the diffusion to the finite interval $[x_0,x_1]$, we obtain another stationary solution this time with a non-zero steady current $j$
\[
\nu_{st} (x) = j \beta(x) \int\limits_{x_0}^x \rmd y \; \exp\left[ - \int\limits_{y}^x \rmd z \; \beta(z) V'(z) \right] .
\]
Notice that the steady current $j$ represents here the normalization condition on the probability distribution and hence is fully determined by the potential $V(x)$ and the temperature profile $\beta(x)$.

We can see that the temperature gradient acts here as an additional force pushing the particles to the colder regions. 
We can better understand this effect in the underdamped case, where it is caused by the thermalization of the kinetic energy.
As a consequence of the equipartition theorem in the hot spot the average kinetic energy of the particle is larger than in its surroundings and so the typical momentum.
This means that also the pressure is larger there, which induce the flux of the particles out of the hot spot. 
While the physical reality does not depend on the level of our description, hence in the overdamped diffusion, which is a time coarse grained description of the underdamped diffusion, such effect has to be present too. 
Moreover if we interpret the whole force as the gradient of the total energy
\[
E(x) = V(x) + \frac{1}{2} k_B T(x) ,
\] 
where we can see that the kinetic term is represented by the mean value of the kinetic energy according to \emph{the equipartition theorem}. 
This means that this is the correct total energy of the particle undergoing the overdamped diffusion, which does not neglect the coarse grained kinetic energy.
However in the homogeneous system the term is in most cases the additional term can be neglected, because it does not depend on position thus only shifts the total energy. 
It only manifests itself as the constant contribution to the heat capacity.

\section{Conclusions}
In the first section we have briefly discussed the possibility of the autonomous dynamics for ``fast'' degrees of freedom on their appropriate timescale.  
The main result of that particular section is the observation 
that the autonomous dynamics on the characteristic timescale of ``fast'' degrees of freedom can be in some cases associated with quasistatic process, where the autonomously time evolved ``slow'' degrees of freedom  act as external parameters.

In the second section we have discussed the possibility of Markovian autonomous dynamics on the level of ``slow'' degrees of freedom on their proper timescale. 
We have shown that up to the first order in the parameter characterizing the timescale separation the dynamics is Markovian and can be effectively described by the effective forward Kolmogorov generator \eqref{equ:effective_time_evolution_slow}, 
where the time derivative $\lim_{\Delta t \to 0^+} \Delta f_t/\Delta t$ has to be taken in the sense that $\tau_S \gg \Delta t \gg \tau_F$. 
What is new is that we have also obtained an exact expression how to determine the difference between the effective Markovian solution and the actual solution \eqref{equ:difference_exact_vs_markovian}
and an estimate how it behaves under the time evolution \eqref{equ:difference_time_evolution}.

At the end we have illustrated the results on several models, 
most notably we have obtained a Smoluchovski equation describing the overdamped diffusion in the inhomogeneous medium as the first order correction. 
We can also see that the overdamped diffusion is the limiting case of the underdamped diffusion in case of fast thermalization of the momentum and kinetic energy in comparison with the slow relaxation of positions. 
Moreover notice that the Smoluchovski equation is in fact the first order correction. 

\chapter{Conclusions}
\label{chapter:conclusion}
In this thesis we have studied the quasistatic processes of small driven systems described by stochastic Markovian dynamics. 
Specifically, we have considered (1) Markov jump processes with continuous time, which in physical reality can describe ratchets, various semi-classical models of molecular motors, etc.,
(2) diffusion, both underdamped or overdamped, as describing e.g. colloid particle moving in a rotationally driven medium. 
In both cases the time evolution of such systems can be described by Kolmogorov generators, which provide unified framework, see chapter \ref{chapter:models}. 
After a brief recollection of well known facts for these systems in equilibrium, see chapter \ref{chapter:equilibrium_stat_phys}, 
we focused on the analysis of the heat and work for thermodynamic processes in the quasistatic limit, which is the main topic of this theses.

\section{Quasistatic limit for non-equilibrium systems}
\subsection{Non-potential force driven systems}
The first class of systems described mainly in chapter \ref{chapter:non-equilibrium_thermodynamics} are systems driven out of equilibrium by the action of non-potential force, or by the action of multiple thermal or particle baths. 
For these systems we provide an analytical formalism for the quasistatic expansion of the mean values of work and heat.  
We have shown that the mean heat and work in the quasistatic limit can be naturally decomposed into what we call the ``housekeeping'' and ``reversible'' components and we have obtained explicit formulas for their computation, 
see equations \eqref{def:housekeeping_work}, \eqref{def:reversible_work}, \eqref{def:reversible_heat}. 
The diverging ``housekeeping'' component of the heat and work is related to the steady dissipation of the system out of equilibrium and as such it is symmetric with respect to protocol reversal $\Theta \alpha$,  
while the finite ``reversible'' component is antisymmetric with respect to protocol reversal and also independent of the protocol parametrization and thus geometric. 
In particular, in equilibrium the ``reversible'' component coincides with the total heat or work.

We have also formulated the energy balance equation \eqref{equ:first_law_noneq} stating that the change of the internal energy represented as the mean value of energy of the system is given by the total ``reversible'' heat and work to the system,
i.e. the ``housekeeping'' components do not contribute to the energy balance.  
In this sense the ``reversible'' work and heat are a natural extension of the equilibrium reversible work and heat.

Further, we have introduced a generalized heat capacity as the ``reversible'' heat produced along the quasistatic change of the temperature of the thermal bath \eqref{def:generalized_heat_capacity}. 
As demonstrated on various models the generalized heat capacity has a particularly rich behaviour out of equilibrium. 
The most surprising yet not fully understood is the possibility of the generalized heat capacity to be negative as seen in subsections \ref{ssec:two_level_nonp}, \ref{ssec:three_level_nonp}, \ref{ssec:diffusion_on_plane}. 
Although the negative generalized heat capacity occurs in two- and three-level systems close to the population inversion, the actual effect appears to be much more subtle, cf. subsections \ref{ssec:two_level_nonp} and \ref{ssec:three_level_nonp}.  
In case of the diffusion the negative heat capacity typically occurs when the increase of temperature pushes the system towards smaller dissipation, cf. \ref{ssec:diffusion_on_plane}.

\subsection{Periodically driven systems}
In chapter \ref{chapter:periodically_driven_systems} we have developed the extension of the formalism of quasistatic processes to periodically driven systems, which was not previously discussed in literature. 
Our starting point was the application of the Floquet theory to remove the explicit time dependence from the forward Kolmogorov generator by using the Fourier picture.
That helped us to cast the problem into similar framework as in chapter \ref{chapter:non-equilibrium_thermodynamics} and to identify the generalized
``reversible'' and ``housekeeping'' components of the mean value of the total heat and work, see equations \eqref{equ:generalized_reversible_work}, \eqref{equ:generalized_housekeeping_work}, \eqref{equ:generalized_housekeeping_heat} and \eqref{equ:generalized_reversible_heat}.
These components have similar properties as those discussed in the previous subsection such as the ``reversible'' component being again independent of parametrization of the protocol $\alpha$ or the ``housekeeping'' component being generically divergent in the quasistatic limit.
However there also arise new fundamental interpretational problems as the generalized ``housekeeping'' component contains terms of the comparable magnitude as the ``reversible'' component, reflecting the initial and final state within the time period. 
These terms can heavily oscillate in the quasistatic limit and hence they can introduce an additional uncertainty to the experimental accessibility of  the ``reversible'' components.

We have also studied the generalized heat capacity on particular models as the means to analyse the behaviour of the ``reversible'' component of the heat. 
In these models we have seen the generalized heat capacity may become negative in the intermediate regime where the relaxation time is comparable to the period of driving, while it is  generally positive in either singular or slow driving limit.

\subsection{Generalized Clausius relation}
We have also discussed the possibility of generalization of the Clausius relation \eqref{equ:Clausius_relation} to non-equilibrium systems. 
It has been argued that the Clausius relation or some straightforward generalization of that is not valid in general out of equilibrium.
Nevertheless if the steady state is characterized by the Boltzmann-like distribution in the McLennan form \eqref{equ:generalized_Boltzmann} then the generalized Clausius relation is verified; the system presented in subsection \ref{ssec:diffusion_quadratic_potential} is exactly such an example.
What physical conditions cause the steady state to be represented by such a distribution is still an open question as well as whether there exists any weaker condition. 
The only case where the McLennan structure of the stationary distribution is generally verified is the close-to-equilibrium regime, see section \ref{sec:first_order_expansion}.

\section{Markovianness in time-scale separation} 
The chapter \ref{chapter:slow-fast_coupling} was dedicated to provide a link between the concept of quasistatic changes of some external parameters and the time-scale separation, 
as well as to show how one can obtain an effective Markovian dynamics for slow degrees of freedom on their specific time-scale by projecting out the fast degrees of freedom, in a way which is more precise that standard arguments in literature.
In the first section we provided a heuristic analysis when we can consider the system to be quasistatically driven.  
After giving some heuristic arguments on the connection to previously discussed quasistatic processes, 
we have studied in detail the Markovian structure of the effective dynamics for large time-scale separation. 
In particular, the Markovianness has been proven to hold true up to the first order in the expansion around infinite separation limit.

\section{Open problems}
We finish by providing a number of open question that have emerged during the work and which have not yet been mentioned before.

\subsection{Generalization of Nernst theorem}
A natural question is whether the Nernst theorem representing the (equilibrium) Third law allows for an extension beyond equilibrium.
In our systems it might correspond to the conjecture that the heat capacity with the temperature of all thermal baths acting on the system going to zero also tends to zero. 
We have seen on several examples that there are systems in which such a generalization of the Nernst theorem indeed holds true, cf. subsections \ref{ssec:two_level_nonp}, \ref{ssec:three_level_nonp}, as well as other examples where it does not, cf. subsections \ref{ssec:diffusion_on_plane}, \ref{ssec:merry-go-around}, \ref{ssec:diffusion_on_the_ring}. 
While there is no surprise that the conjectured generalization of the Nernst theorem is not valid for diffusion, its invalidity even for system with finite number of state might sound surprising and it needs to be further understood, cf. subsection \ref{ssec:merry-go-around}.

For driven diffusion on the ring we have seen that the zero-temperature transition from the limit fixed point phase to the limit cycle phase is accompanied with the transition from the equilibrium-like to a new and highly nontrivial heat capacity asymptotics, cf. subsection \ref{ssec:diffusion_on_the_ring}. 
Even a richer collection of different low-temperature patterns have been obtained for the driven 2D diffusion, cf. subsection \ref{ssec:diffusion_on_plane}. 
It remains to be seen on a more general basis in what precise sense the emergence of diverging or negative low-temperature heat capacity patterns characterize the low-temperature steady states.
There are still no definite answers why in some finite systems the Nernst theorem holds true whether in others does not or what are the exact conditions for the zero temperature phase transition to occurs.

\subsection{Generalized quasistatic response functions}
The heat capacity has been introduced in order to quantify the reversible component of heat along the specific quasistatic process with temperature as the only time-dependent parameter. 
Analogously, we can analyze other quasistatic processes in terms of generalized (quasistatic) response functions like the compressibility, thermal expansion coefficient etc. 
Beyond obvious questions about their possible anomalous behavior in far-from-equilibrium regimes, still a more fundamental issue arises: 
Are those mutually related in a way similar to the equilibrium Mayer of Maxwell relations? 
Clearly, this question has much to do with generalizations of the Clausius relation and the existence of non-equilibrium entropy.
Within the heat-renormalization scheme adopted in this work, these question remain largely open and will require further research.

\appendix
\renewcommand{\chaptername}{Appendix}

\chapter{Kolmogorov generators for Markov jump processes} 
\label{ap:jump_processes_generators}
\index{jump processes}
\index{Kolmogorov generator!forward}
In this appendix we derive the forward and backward Kolmogorov generator from the conditional path measure for Markov jump processes. 
Our starting point is the definition of the forward Kolmogorov generator \eqref{def:forward_generator}, first introduced in chapter \ref{chapter:models} section \ref{sec:Markov_processes},
\[
\gen^*_t [\mu_t] (x) = 
\lim_{\Delta t \rightarrow 0^+} \frac{1}{\Delta t} \left( \int \rmd y \; \mu_{t-\Delta t}(y) \int \cprob[{(t-\Delta t,t]}]{\omega}{X_{t-\Delta t}=y} \; \delta_{X_t}(x) - \mu_{t-\Delta t}(x) \right) ,
\]
in terms of the conditional probability measure $\rmd \mathbb P$, in this case, for Markov jump processes.  
The conditional path probability measure for Markov jump processes \eqref{def:jump_process_path_probability}, 
was first introduced in chapter \ref{chapter:models} section \ref{sec:jump_processes}, and was defined as  
\[
\cprob[{(0,T]}]{\omega}{X_0=x_0}
= \exp \left[ - \int\limits_0^T \rmd s \; \lambda_s(x_s) \right] 
\prod_{i=1}^{k} \rate[t]{x_{t_i^-}}{x_{t_i}} \; \rmd t_i ,
\]
where $\rate[t]{x}{y}$ denotes transition rate from $x$ to $y$ at time $t$, $x_{s^-}$ the state right before the jump at time $s$, 
$\lambda_t(x)$ is the escape rate from the state $x$ at time $t$ and $t_i$ denotes jump times. 
We start by expanding the first term in the definition of the forward Kolmogorov generator \eqref{def:forward_generator} in terms of $\Delta t$ 
\begin{multline*}
\int \cprob[{(t-\Delta t,t]}]{\omega}{X_{t-\Delta t}=y} \; \delta_{X_t}(x) = 
\delta_{xy} \exp\biggl[- \int\limits_{t-\Delta t}^{t} \rmd s \; \lambda_s(y) \biggr] + \\
+ (1 - \delta_{xy} ) \int\limits_{t-\Delta t}^t \rmd s \; \rate[s]{y}{x} 
\exp\biggl[ - \int\limits_{t-\Delta t}^s \rmd u \; \lambda_u(y) - \int\limits_s^t \rmd u \; \lambda_u(x) \biggr] + \err[2]{\Delta t} ,
\end{multline*}
where we have used the fact that the only contributions up to the linear order in $\Delta t$ are from trajectories without any jump (the first term)
or with one jump (the second term). 
Combining this result with the definition of the forward Kolmogorov generator above and taking the limit we obtain 
\begin{align*}
\gen^*_t [\mu_t] (x) 
&=  \sum\limits_{y \neq x} \mu_{t}(y) \rate[t]{y}{x} - \mu_t(x) \lambda_t(x) \\
&=  \sum\limits_{y \neq x} \left[ \mu_{t}(y) \rate[t]{y}{x} - \mu_t(x) \rate[t]{x}{y} \right] 
\end{align*}
where we have also used the definition of the escape rate \eqref{def:escape_rate}. 

The backward Kolmogorov generator\index{Kolmogorov generator!backward} \eqref{def:backward_generator} is then defined by the forward Kolmogorov generator as  
\[
 \int \rmd x \; \gen_t[A](x) \, \mu_t(x) = \int \rmd x \; A(x) \, \gen^*_t [\mu_t] (x) .
\]
Inserting the forward Kolmogorov generator into this definition we obtain 
\begin{align*}
\sum\limits_x \gen_t[A](x) \, \mu_t(x) 
&= \sum\limits_x A(x) \sum\limits_{y \neq x} \left[ \mu_t(y) \rate[t]{y}{x} - \mu_t(x) \rate[t]{x}{y} \right], \\
&= \sum\limits_x \mu_t(x) \sum\limits_{y \neq x} \left[ A(y) \rate[t]{x}{y} - A(x) \rate[t]{x}{y} \right] . 
\end{align*}
Hence by comparison we obtain the backward Kolmogorov generator
\[
\gen_t[A](x) = \sum\limits_{y \neq x} \rate[t]{x}{y} \left[ A(y) - A(x) \right] .
\]

\chapter{Stochastic calculus} 
\label{ap:stochastic_calculus}
In this appendix we sketch some of the proofs of statements given in chapter \ref{chapter:models} subsection \ref{ssec:stochastic_calculus}. 
We will mostly follow the text \cite{Evans2001}, where one can find an additional details.

\section{Mean value of It\^{o} integral}
\label{sec:mean_value_Ito}
We have stated that the mean value of the Ito stochastic integral is zero \eqref{equ:Ito_zero_mean_value}. 
We start the proof with the definition of the integral \eqref{def:Ito_integral} 
and distribute the mean value throughout the sum 
\[
\left\langle \int\limits_0^T \vec{f}(\vec{W}_t,t) \cdot \rmd \vec{W}_t \right\rangle_{\mu_0} 
= \lim\limits_{N \to \infty} \sum\limits_{i=0}^{N-1} 
\left\langle \vec{f}(\vec{W}_{t_i},t_i) \cdot \left( \vec{W}_{t_{i+1}} - \vec{W}_{t_i} \right) \right\rangle_{\mu_0} ,
\] 
where $\vec{W}_t$ is the Wiener process, 
$N$ is the number of points in the partition of the time interval $[0,T]$ 
and $t_i$ denotes the respective times of the points of the partitioning. 
We use the definition of the conditional mean value to extract the vector field $\vec{f}(\vec{W}_{t_i},t_i)$ from the mean value
\begin{multline*}
\left\langle \vec{f}(\vec{W}_{t_i},t_i) \cdot \left( \vec{W}_{t_{i+1}} - \vec{W}_{t_i} \right) \right\rangle_{\mu_0} = \\
= \int\limits_{\reals^d} \rmd^d \vec{x} \; \vec{f}(\vec{x},t_i) \cdot \left\langle \vec{W}_{t_{i+1}} - \vec{W}_{t_i} \middle| \vec{W}_{t_i} = \vec{x} \right\rangle_{\mu_0} \, \cprob{ \vec{W}_{t_i} = \vec{x} }{\mu_0} . 
\end{multline*}
Now we use the Markov property of the Wiener process, 
i.e. the future displacement of the position does not depend on the history up to this time,
to express the conditional expectation by the expectation over all realizations of the Wiener process with the new initial condition $\delta_{\vec{x}}$ at time $t_i$ 
\[
\left\langle \vec{W}_{t_{i+1}} - \vec{W}_{t_i} \middle| \vec{W}_{t_i} = \vec{x} \right\rangle_{\mu_0} 
= \left\langle \vec{W}_{t_{i+1}} - \vec{W}_{t_i} \right\rangle_{\delta_\vec{x}} .
\]
Using the fact that the mean value of the displacement of position along the Wiener process is zero \eqref{equ:Wiener_properties} concludes the statement.

\section{Covariance of It\^{o} integrals} 
\label{sec:covariance_Ito}
The second statement connected the covariance of the It\^{o} stochastic integrals \eqref{equ:Ito_covariance} with the time integral of the covariance. 
In this case we start by dividing it to three terms 
\begin{multline*}
\left\langle \int\limits_0^T \vec{f}(\vec{W}_t,t) \cdot \rmd \vec{W}_t \int\limits_0^T \vec{g}(\vec{W}_t,t) 
\cdot \rmd \vec{W}_t \right\rangle_{\mu_0} = \\
= \underbrace{ \sum_{i=0}^{N-1} \sum_{j=i+1}^{N-1} \left\langle \left( \vec{W}_{t_{i+1}} - \vec{W}_{t_i} \right) 
\cdot \vec{f}(\vec{W}_{t_i},t_i) \, \vec{g}(\vec{W}_{t_j},t_j) 
\cdot \left( \vec{W}_{t_{j+1}} - \vec{W}_{t_j} \right) \right\rangle_{\mu_0} }_\text{A} + \\ 
+ \underbrace{ \sum_{i=0}^{N-1} \sum_{j=0}^{i-1} \left\langle \left( \vec{W}_{t_{i+1}} - \vec{W}_{t_i} \right) 
\cdot \vec{f}(\vec{W}_{t_i},t_i) \, \vec{g}(\vec{W}_{t_j},t_j) 
\cdot \left( \vec{W}_{t_{j+1}} - \vec{W}_{t_j} \right) \right\rangle_{\mu_0} }_\text{B} + \\
+ \underbrace{ \sum_{i=0}^{N-1} \left\langle \left( \vec{W}_{t_{i+1}} - \vec{W}_{t_i} \right) 
\cdot \vec{f}(\vec{W}_{t_i},t_i) \, \vec{g}(\vec{W}_{t_i},t_i) 
\cdot \left( \vec{W}_{t_{i+1}} - \vec{W}_{t_i} \right) \right\rangle_{\mu_0} }_\text{C} 
\end{multline*}
taking the limit $N \to \infty$ afterwards. 
By following the same line of thoughts as in the previous case see that parts A and B are zero. 
The remaining term C can be rearranged in the similar fashion using the conditional mean value 
\begin{multline*}
\left\langle \left( \vec{W}_{t_{i+1}} - \vec{W}_{t_i} \right) 
\cdot \vec{f}(\vec{W}_{t_i},t_i) \, \vec{g}(\vec{W}_{t_i},t_i) 
\cdot \left( \vec{W}_{t_{i+1}} - \vec{W}_{t_i} \right) \right\rangle_{\mu_0} = \\
\begin{aligned}
= \int\limits_{\reals^d} \rmd^d \vec{x} \; & \cprob{\vec{W}_{t_i} = \vec{x}}{\mu_0} \times \\
& \times \vec{f}(\vec{x},t_i) 
\cdot \left\langle \left( \vec{W}_{t_{i+1}} - \vec{W}_{t_i} \right) \left( \vec{W}_{t_{i+1}} - \vec{W}_{t_i} \right) \middle| \vec{W}_{t_i} = \vec{x} \right\rangle_{\mu_0} 
\cdot \vec{g}(\vec{x},t_i) 
\end{aligned}
\end{multline*} 
and using the fact that the variance is proportional to the time interval while the mean value is zero \eqref{equ:Wiener_properties} we obtain 
\begin{multline*}
\left\langle \left( \vec{W}_{t_{i+1}} - \vec{W}_{t_i} \right) 
\cdot \vec{f}(\vec{W}_{t_i},t_i) \, \vec{g}(\vec{W}_{t_i},t_i) 
\cdot \left( \vec{W}_{t_{i+1}} - \vec{W}_{t_i} \right) \right\rangle_{\mu_0} = \\
= \int\limits_{\reals^d} \rmd^d \vec{x} \; \vec{f}(\vec{x},t_i) 
\cdot \vec{g}(\vec{x},t_i) \left( t_{i+1} - t_i \right) 
\cprob{\vec{W}_{t_i} = \vec{x}}{\mu_0} = \\
= \left\langle \vec{f}(\vec{W}_{t_i},t_i) \cdot \vec{g}(\vec{W}_{t_i},t_i) \right\rangle_{\mu_0} \left( t_{i+1} - t_i \right) ,
\end{multline*} 
which concludes the proof.

\section{Riemann sum of square of displacement}
\label{sec:Riemann_sum}
Another statement connected the sum of square of displacements with the length of the time interval. 
Here we sketch the proof of the more general version \eqref{equ:pre_Ito_lemma}. 
We will show that the variance of the difference between the Riemann sum representing the integral and the sum of square displacement which represents the mean value tends to vanish with $N \to \infty$, 
\[
\sigma^2 = \left\langle \left( \sum\limits_{i=0}^{N-1} f(\vec{W}_{t_i},t_i) \left[ \left( \vec{W}_{t_{i+1}} - \vec{W}_{t_i} \right)^2 - \left( t_{i+1} - t_i \right) \right] \right)^2 \right\rangle_{\mu_0} .
\] 
We start again by dividing the product of the sum into three terms, 
where by following the same pattern the terms corresponding to $i \neq j$ are equal to zero,
which leave us with 
\[
\sigma^2 
= \sum\limits_{i=0}^{N-1} \left\langle f^2(\vec{W}_{t_i},t_i) \left[ \left( \vec{W}_{t_{i+1}} - \vec{W}_{t_i} \right)^2 - \left( t_{i+1} - t_i \right) \right]^2 \right\rangle_{\mu_0} ,
\]
which can be rewritten using the conditional probability to  
\begin{multline*}
\left\langle f^2(\vec{W}_{t_i},t_i) \left[ \left( \vec{W}_{t_{i+1}} - \vec{W}_{t_i} \right)^2 - \left( t_{i+1} - t_i \right) \right]^2 \right\rangle_{\mu_0} = \\
\begin{aligned}
=& \int\limits_{\reals^d} \rmd^d \vec{x} \; \cprob{\vec{W}_{t_i} = \vec{x}}{\mu_0} \, f^2(\vec{x},t_i) \times \\
& \times \left[ \left\langle \left( \vec{W}_{t_{i+1}} - \vec{W}_{t_i} \right)^4 - 2 \left( \vec{W}_{t_{i+1}} - \vec{W}_{t_i} \right)^2 \left( t_{i+1} - t_i \right) \middle| \vec{W}_{t_i} = \vec{x} \right\rangle_{\mu_0} + \left( t_{i+1} - t_i \right)^2 \right] .
\end{aligned}
\end{multline*} 
By using the Gaussian properties of the Wiener process \eqref{equ:Wiener_distribution} we conclude the proof.

\section{It\^{o} lemma}
\label{sec:Ito_lemma}
To prove the It\^{o} lemma \eqref{equ:total_differential_Ito} we need to associate the difference in the result $Y(\vec{W}_T,T) - Y(\vec{W}_0,0)$ with its integral representation. 
Lets take an arbitrary partition $t_i$ of the time interval $[0,T]$ then it is valid 
\[
Y(\vec{W}_T,T) - Y(\vec{W}_0,0) = \sum\limits_{i=0}^{N-1} \left[ Y(\vec{W}_{t_{i+1}},t_{i+1}) - Y(\vec{W}_{t_i},t_i) \right] . 
\]
To obtain terms similar to Riemann sum we expand the particular differences in both arguments to Taylor series we obtain 
\begin{multline*}
Y(\vec{W}_T,T) - Y(\vec{W}_0,0) 
= \sum\limits_{i=0}^{N-1} \Biggl[ \left. \partial_t Y(\vec{W}_{t_i},t) \right|_{t=t_i} ( t_{i+1} - t_i ) + \\
+ \left. \nabla_\vec{x} Y(\vec{x},t_i) \right|_{\vec{x} = \vec{W}_{t_i}} \cdot ( \vec{W}_{t_{i+1}} - \vec{W}_{t_i} ) + \\
+ \frac{1}{2} ( \vec{W}_{t_{i+1}} - \vec{W}_{t_i} ) \cdot \left. \nabla^2_\vec{x} Y(\vec{x},t_i) \right|_{\vec{x} = \vec{W}_{t_i}} \cdot ( \vec{W}_{t_{i+1}} - \vec{W}_{t_i} ) 
+ \dots \Biggr] .
\end{multline*}
In the limit $N \to \infty$ one can see the first two terms correspond to Riemann sums for the Riemann integral and It\^{o} integral \eqref{def:Ito_integral}. 
The third term in the sum under the same limit $N \to \infty$ according to \eqref{equ:pre_Ito_lemma} converges almost surely to
\[
\lim_{N \to \infty} \sum\limits_{i=0}^{N-1} ( \vec{W}_{t_{i+1}} - \vec{W}_{t_i} ) \cdot \left. \nabla^2_\vec{x} Y(\vec{x},t_i) \right|_{\vec{x} = \vec{W}_{t_i}} \cdot ( \vec{W}_{t_{i+1}} - \vec{W}_{t_i} ) 
= \int\limits_0^T \rmd t \; \left. \Delta_\vec{x} Y(\vec{x},t) \right|_{\vec{x} = \vec{W}_t} ,
\]
while other terms in the sum converge almost surely to zero by a similar reasoning as in section \ref{sec:Riemann_sum}.

As an example we will show that the term corresponding to the mixed derivative tends to go to zero by proving that the mean value as well as the variance goes to zero in the limit $N \to \infty$. 
To obtain the mean value we distribute the sum at first and then use that the displacement of the Wiener process is independent on the history \eqref{equ:Wiener_combination} and has zero mean value \eqref{equ:Wiener_properties}
\begin{multline*}
\left\langle
\sum\limits_{i=0}^{N-1} ( \vec{W}_{t_{i+1}} - \vec{W}_{t_i} ) \cdot \left. \vec{\nabla}_\vec{x} \partial_t Y(\vec{x},t) \right|_{t=t_i,\vec{x} = \vec{W}_{t_i}} ( t_{i+1} - t_i ) \right\rangle_{\mu_0} = \\
= \sum\limits_{i=0}^{N-1} \left\langle ( \vec{W}_{t_{i+1}} - \vec{W}_{t_i} ) \cdot \left. \vec{\nabla}_\vec{x} \partial_t Y(\vec{x},t) \right|_{t=t_i,\vec{x} = \vec{W}_{t_i}} \right\rangle_{\mu_0} ( t_{i+1} - t_i ) = \\ 
= \sum\limits_{i=0}^{N-1} ( t_{i+1} - t_i ) \int\limits_{\reals^d} \rmd^d \vec{x} \; \left\langle \vec{W}_{t_{i+1}} - \vec{W}_{t_i} \right\rangle_{\delta_\vec{x}} \cdot \left. \vec{\nabla}_\vec{x} \partial_t Y(\vec{x},t) \right|_{t=t_i} \cprob{\vec{W}_{t_i}=\vec{x}}{\mu_0} = 0.
\end{multline*}
By similar reasons as in section \ref{sec:covariance_Ito} in case of variance the only contributing terms are quadratic,
then we proceed in the similar fashion as for the mean value 
\begin{multline*}
\Biggl\langle 
\sum\limits_{j=0}^{N-1} ( \vec{W}_{t_{j+1}} - \vec{W}_{t_j} ) \cdot \left. \vec{\nabla}_\vec{x} \partial_t Y(\vec{x},t) \right|_{t=t_j,\vec{x} = \vec{W}_{t_j}} ( t_{j+1} - t_j ) \times \\
\times \sum\limits_{i=0}^{N-1} ( \vec{W}_{t_{i+1}} - \vec{W}_{t_i} ) \cdot \left. \vec{\nabla}_\vec{x} \partial_t Y(\vec{x},t) \right|_{t=t_i,\vec{x} = \vec{W}_{t_i}} ( t_{i+1} - t_i ) \Biggr\rangle_{\mu_0} = \\
= \sum\limits_{i=0}^{N-1} \left\langle \left[ ( \vec{W}_{t_{i+1}} - \vec{W}_{t_i} ) \cdot \left. \vec{\nabla}_\vec{x} \partial_t Y(\vec{x},t) \right|_{t=t_i,\vec{x} = \vec{W}_{t_i}} \right]^2 \right\rangle_{\mu_0} ( t_{i+1} - t_i )^2 = \\ 
\begin{split} 
= \sum\limits_{i=0}^{N-1} ( t_{i+1} - t_i )^2 \int\limits_{\reals^d} \rmd^d \vec{x} \; \left. \vec{\nabla}_\vec{x} \partial_t Y(\vec{x},t) \right|_{t=t_i} \cdot \left\langle \left( \vec{W}_{t_{i+1}} - \vec{W}_{t_i} \right) \left( \vec{W}_{t_{i+1}} - \vec{W}_{t_i} \right) \right\rangle_{\delta_\vec{x}} \cdot \quad \\ 
\cdot \left. \vec{\nabla}_\vec{x} \partial_t Y(\vec{x},t) \right|_{t=t_i} \cprob{\vec{W}_{t_i}=\vec{x}}{\mu_0} = \end{split} \\
= \sum\limits_{i=0}^{N-1} ( t_{i+1} - t_i )^3 \left\langle \left\| \left. \vec{\nabla}_\vec{x} \partial_t Y(\vec{x},t) \right|_{t=t_i,\vec{x}=\vec{W}_{t_i}} \right\|^2 \right\rangle_{\mu_0} ,
\end{multline*}
where $\|\cdot\|$ denotes the Euclid norm. 
If we assume that the time step is uniformly bounded 
\[
\forall i : \qquad t_{i+1} - t_i < \frac{C T}{N} 
\]
and that the mean value in the Riemann sum is on the interval $[0,T]$ bounded 
then we can make estimate for variance 
\begin{multline*}
\Biggl\langle 
\sum\limits_{j=0}^{N-1} ( \vec{W}_{t_{j+1}} - \vec{W}_{t_j} ) \cdot \left. \vec{\nabla}_\vec{x} \partial_t Y(\vec{x},t) \right|_{t=t_j,\vec{x} = \vec{W}_{t_j}} ( t_{j+1} - t_j ) \times \\
\times \sum\limits_{i=0}^{N-1} ( \vec{W}_{t_{i+1}} - \vec{W}_{t_i} ) \cdot \left. \vec{\nabla}_\vec{x} \partial_t Y(\vec{x},t) \right|_{t=t_i,\vec{x} = \vec{W}_{t_i}} ( t_{i+1} - t_i ) \Biggr\rangle_{\mu_0} < \\
< \frac{C^3 T^3}{N^2} \max_{t_i} \left\langle \left\| \left. \vec{\nabla}_\vec{x} \partial_t Y(\vec{x},t) \right|_{t=t_i,\vec{x}=\vec{W}_{t_i}} \right\|^2 \right\rangle_{\mu_0} \xrightarrow{N \to \infty} 0 , 
\end{multline*}
which concludes the statement that higher order contributions vanish.

Putting all these partial results altogether we obtain that the difference can be represented as 
\[
Y(\vec{W}_T,T) - Y(\vec{W}_0,0) = \int\limits_0^T \rmd t \; \left[ \partial_t Y(\vec{W}_{t_i},t) + \frac{1}{2} \left. \Delta_\vec{x} Y(\vec{x},t) \right|_{\vec{x}=\vec{W}_t} \right] + \int\limits_0^T \rmd \vec{W}_t \cdot \left. \nabla_\vec{x} Y(\vec{x},t) \right|_{\vec{x}=\vec{W}_t} , 
\]
which together with \eqref{def:stochastic_differential_equation} concludes the proof.

\section{Comparison of It\^{o} and Stratonovich integral} 
\label{sec:Ito_vs_Stratonovich}
We have provided the relation between the Stratonovich and It\^{o} integral \eqref{equ:relation_Ito_Stratonovich}. 
To prove it we start with the expansion of the argument of the Stratonovich integral to the Taylor series up to the first order in the position in a similar fashion to the previous section, while higher orders are almost surely zero  
\begin{multline*}
\int\limits_0^T \vec{f}(\vec{W}_t,t) \circ \rmd \vec{W}_t = \lim\limits_{N \to \infty} \sum\limits_{i=0}^{N-1} \vec{f}(\vec{W}_{\tau_i},\tau_i) \cdot \left[\vec{W}_{t_{i+1}} - \vec{W}_{t_i}\right] \\
= \lim\limits_{N \to \infty} \sum\limits_{i=0}^{N-1} \left[ \vec{f}(\vec{W}_{t_i},t_i) + \frac{1}{2} \left( \vec{W}_{t_{i+1}} - \vec{W}_{t_i} \right) \cdot \left. \vec{\nabla}_\vec{x} \vec{f}(\vec{x},t_i) \right|_{\vec{x}=\vec{W}_{t_i}} \right] \cdot \left[\vec{W}_{t_{i+1}} - \vec{W}_{t_i}\right] . 
\end{multline*} 
Using again the \eqref{equ:pre_Ito_lemma} gives us the desired relation.

\chapter{Generator pseudoinverse}
\label{ap:pseudoinverse}

\section{Alternative definitions}
The definition of forward \eqref{def:forward_pseudoinverse} and backward \eqref{def:backward_pseudoinverse} pseudoinverse is not suitable for practical computations. 
In this section we will provide other two equivalent more convenient expressions.

The first alternative definition of forward Kolmogorov generator is based on the spectral decomposition of the forward Kolmogorov generator (on the assumption it exists)
\[
\gen^* [\mu] (x) = \sum_{i>0} \lambda_i \, {\mathcal P}^*_i [\mu](x) ,
\]
where ${\mathcal P}^*_i$ is the projection to the eigenstate with the eigenvalue of $\lambda_i$. 
Because we also assume that the system starting from an arbitrary state $\mu(x)$ will reach the stationary state as $t \to \infty$, the zero eigenvalue has to be non-degenerate and all the nonzero eigenvalues has to have negative real part $\Re \lambda_i < 0$. 
By default we denote by the index $0$ the projection to the zero eigenvalue, i.e. the projection to the stationary state $\rho(x)$, defined as 
\begin{equation}
{\mathcal P}^*_0 [\mu](x) = \rho(x) \int \rmd \Gamma(y) \; \mu(y) .
\label{def:projector}
\end{equation}
From the definition it can be directly verified that $\mathcal P^*_0$ is indeed a projection to the stationary state 
\begin{gather*}
{\mathcal P}^*_0 [ \rho ] (x) = \rho(x) , \\
\left( {\mathcal P}^*_0 \right)^2 = {\mathcal P}^*_0 .
\end{gather*}
Using the linearity of the forward Kolmogorov generator and the normalization condition \eqref{equ:normalization_condition} we also obtain
\[
\gen^* {\mathcal P}^*_0 = {\mathcal P}^*_0 \gen^* = 0 .
\]
Using the orthogonality of projections, ${\mathcal P}^*_i {\mathcal P}^*_j = 0 \quad \forall i \neq j$, we obtain the first alternative expression for the forward pseudoinverse 
\begin{equation}
\frac{1}{\gen^*}[\mu] (x) 
= - \int\limits_0^\infty \rmd t \; \sum_{i>0} \rme^{\lambda_i t} {\mathcal P}^*_i [\mu] (x) 
= \sum_{i>0} \frac{1}{\lambda_i} {\mathcal P}^*_0 [\mu](x) .
\label{equ:forward_pseudoinverse_spectral}
\end{equation}
This expression has proved to be particularly convenient for the case of models with quadratic potential, where the eigenvectors for low powers in position and momentum can be explicitly found, e.g. see sections \ref{sec:pseudoinverse_underdamped} and \ref{sec:pseudoinverse_overdamped} .

The second alternative expression for the forward pseudoinverse is by including the projection $\mathcal P^*_0$ into the forward Kolmogorov generator making thus the total operator invertible and subtracting the projection afterwards 
\[
\left[ \gen^* + {\mathcal P}^*_0 \right]^{-1} - {\mathcal P}^*_0 = \left[ \frac{1}{1} {\mathcal P}^*_0 + \sum_{i>0} \frac{1}{\lambda_i} {\mathcal P}^*_i \right] - {\mathcal P}^*_0 = \frac{1}{\gen^*} ,
\]
which on the other hand proved to be useful in numerical calculations in case of models describing Markov jump processes. 

These expressions are also valid for backward pseudoinverse 
\begin{align*}
\frac{1}{\gen}[A](x) &= \sum_{i>0} \frac{1}{\lambda_i} {\mathcal P}_i [A] (x), \\ 
\frac{1}{\gen}[A](x) &= \frac{1}{\gen + {\mathcal P}_0} [A] (x) - {\mathcal P}_0 [A] (x) ,
\end{align*} 
where projections $\mathcal P_i$ are related to the projections $\mathcal P^*_i$ by 
\[
\int \rmd \Gamma(x) \; A(x) \, {\mathcal P}^*_i [ \mu ] (x) = \left\langle {\mathcal P}_i [A] \right\rangle_\mu ,
\]
from where we conclude that the projection to the zero eigenvalue $\mathcal P_0$ corresponds to taking the stationary mean value of the observable $A$ 
\[
{\mathcal P}_0 [A] (x) = \left\langle A \right\rangle_\rho .
\]

\section{Identities for the forward pseudoinverse}
In this section we will show some of the identities for the forward pseudoinverse and also show that the forward pseudoinverse is in fact the Drazin pseudoinverse of the forward Kolmogorov generator. 
The first identity states that the forward pseudoinverse applied to the stationary state $\rho$ is zero, which can be obtained directly from the definition 
\begin{equation}
\frac{1}{\gen^*}[\rho](x) = \int\limits_0^\infty \rmd t \; \left[ \rho(x) - \rme^{t \gen^*}[\rho](x) \right] = 0 . 
\label{equ:pseudoinverse_stationary_state}
\end{equation}
Because the forward Kolmogorov generator and the integral are linear operators, the forward pseudoinverse is also a linear operator
\begin{multline*}
\frac{1}{\gen^*}[\mu+\nu](x) 
= \int\limits_0^\infty \rmd t \; \left[ \rho(x) \int \rmd \Gamma(y) \; \left( \mu(y) + \nu(y) \right) + \rme^{t \gen^*} [ \mu + \nu ] (x) \right] = \\
= \int\limits_0^\infty \rmd t \; \left[ \rho(x) \int \rmd \Gamma(y) \; \mu(y) + \rme^{t \gen^*} [ \mu ] (x) \right] 
+ \int\limits_0^\infty \rmd t \; \left[ \rho(x) \int \rmd \Gamma(y) \; \nu(y) + \rme^{t \gen^*} [ \nu ] (x) \right] = \\
= \frac{1}{\gen^*}[\mu](x) + \frac{1}{\gen^*}[\nu](x) .
\end{multline*}
If we apply the forward Kolmogorov generator to the forward pseudoinverse and use the linearity we obtain 
\[
\gen^* \frac{1}{\gen^*} [\mu] (x) = - \int\limits_0^\infty \rmd t \; \partial_t \rme^{t \gen^*} [\mu] (x) = - \left[ \mu_t(x) \right]_{t=0}^\infty = \mu(x) - \rho(x).  
\]

The same result can also be obtained by application of the forward pseudoinverse to the forward Kolmogorov generator, 
however in this case it is the consequence of normalization of the probability distribution \eqref{equ:normalization_condition}
\[
\frac{1}{\gen^*} \gen^* [\mu] (x) = - \int\limits_0^\infty \rmd t \; \rme^{t \gen^*} \gen^* [\mu] (x) = \mu(x) - \rho(x) . 
\]
Putting all these facts together we can see that it is also valid 
\begin{equation}
\begin{gathered}
\frac{1}{\gen^*} \gen^* [\mu] (x) = \gen^* \frac{1}{\gen^*} [\mu] (x), \\
\left(\gen^*\right)^n \frac{1}{\gen^*} [\mu] (x) = \left( \gen^* \right)^{n-1} [\mu] (x) , \qquad n > 2 \\ 
\left(\frac{1}{\gen^*}\right)^n \gen^* [\mu] (x) = \left( \frac{1}{\gen^*} \right)^{n-1} [\mu] (x) . \qquad n > 2 
\end{gathered} 
\label{equ:forward_pseudoinverse_Drazin}
\end{equation}
These properties also uniquely characterizes the Drazin pseudoinverse\index{Drazin pseudoinverse}, for further details see \cite{Drazin1958}.

\section{Identities for backward pseudoinverse}
To show that the backward pseudoinverse is also a Drazin pseudoinverse, we need to show for that same set of identities \eqref{equ:forward_pseudoinverse_Drazin} is valid.

By putting the definition of the backward Kolmogorov generator \eqref{def:backward_generator} together with the normalization condition \eqref{equ:normalization_condition} we see that for every state $\mu$ and every constant $c$,
\[
\left\langle \gen [c] \right\rangle_\mu = c \int \rmd \Gamma(x) \; \gen^* [ \mu ] (x) = 0 ,
\] 
which leads to the conclusion that the constant function $c(x)\equiv c$ is invariant under time evolution   
\[
\gen[c](x) = 0.
\]
Hence constant functions in case of the backward Kolmogorov plays the role of stationary state in case the forward Kolmogorov generator. 
From there immediately follows that also the backward pseudoinverse applied to constant function $c$ is zero
\[
\frac{1}{\gen}[c](x) = \int\limits_0^\infty \rmd t \; \left[ c - \rme^{t \gen}[c](x) \right] = 0 .
\]
Because the mean value does not depend on configuration $x$ it is also valid that 
\[
\gen \frac{1}{\gen} [A] (x) 
= - \int\limits_0^\infty \rmd t \; \partial_t \rme^{t \gen} [A] (x)  
= - \left[ \left\langle \rme^{t \gen} [A] \right\rangle_{\delta_x} \right]_{t=0}^\infty 
= A(x) - \left\langle A \right\rangle_\rho .
\]

The stationary mean value of the observable to which the backward Kolmogorov pseudoinverse is applied is also zero 
\begin{equation}
\left\langle \gen[A] \right\rangle_\rho = \int \rmd \Gamma(x) \; A(x) \, \gen^*[\rho] (x) = 0 
\label{equ:generator_stationary_mean_value}
\end{equation}
In a similar fashion from \eqref{equ:pseudoinverse_stationary_state} follows 
\begin{equation}
\left\langle \frac{1}{\gen} [A] \right\rangle_\rho = \int \rmd \Gamma(x) \; A(x) \, \frac{1}{\gen^*}[\rho](x) = 0 .
\label{equ:backward_pseudoinverse_zero_mean}
\end{equation}
By using \eqref{equ:generator_stationary_mean_value} we can conclude 
\begin{equation}
\frac{1}{\gen} \gen [A] (x)  
= - \int\limits_0^\infty \rmd t \; \rme^{t \gen} \gen [A] (x) 
= A(x) - \left\langle A \right\rangle_\rho ,
\label{equ:backward_pseudoinverse_identity} 
\end{equation}
hence the backward Kolmogorov generator and the backward pseudoinverse are again commutating. 

By putting all these facts together we obtain the identities 
\begin{gather*}
\frac{1}{\gen} \gen [\mu] (x) = \gen \frac{1}{\gen} [\mu] (x), \\
\left(\gen\right)^n \frac{1}{\gen} [\mu] (x) = \left( \gen \right)^{n-1} [\mu] (x), \qquad n > 2 \\ 
\left(\frac{1}{\gen}\right)^n \gen [\mu] (x) = \left( \frac{1}{\gen} \right)^{n-1} [\mu] (x), \qquad n > 2 
\end{gather*} 
we can again see that the backward pseudoinverse is the Drazin pseudoinverse.

%Underdamped case
\section{Underdamped diffusion with harmonic potential}
\label{sec:pseudoinverse_underdamped}
The forward Kolmogorov generator for underdamped diffusion in harmonic potential has the form  
\[
\gen^* [ \mu ] = - \frac{p}{m} \partial_x \mu + \omega^2 (x-x_0) \partial_p \mu + \frac{\gamma}{m} \partial_p \left( p \mu \right) + \frac{\gamma}{\beta} \partial_p^2 \mu, 
\]
where $x$ and $p$ are the degrees corresponding degrees of freedom, $x_0$ is the center of the harmonic potential and $\omega^2$ is its strength. 
The friction $\gamma$ and inverse temperature $\beta$ of the environment are constant in the whole volume. 
One can easily find the stationary state 
\[
\rho = \frac{1}{Z} \exp \left[ - \beta \left( \frac{p^2}{2m} + \frac{1}{2} \omega^2 \left(x-x_0\right)^2 \right) \right], 
\]
which also corresponds to the eigenvector to the eigenvalue zero. 
The task to find all the others eigenvalues and eigenvectors is hard,
however for the purpose of computing the pseudoinverse of the linear correction by \eqref{equ:forward_pseudoinverse_spectral} it is not necessary to have them all. 
The quadratic nature of the generator implies that any application of it does not increase the order of the polynomial in $p$ neither in $x$. 
From these considerations we expect the eigenvector to be a linear combination of $x-x_0$ and $p$.
We are looking for the solution of the equation for eigenvectors 
\[
\gen^*\left[ \left( C (x-x_0) + D p \right) \rho \right] = \lambda \left( C (x-x_0) + D p \right) \rho ,
\]
which can be expressed as a set of algebraic equations
\begin{align*}
\lambda D &= - \left( \frac{C}{m} + \frac{\gamma D}{m} \right), \\
\lambda C &= \omega^2 D .
\end{align*}
There are two independent solutions with eigenvalues 
\[
\lambda_{1,2} = \frac{ - \gamma \pm \sqrt{ \gamma^2 - 4 m \omega^2 }}{2 m} 
\]
and corresponding coefficients
\begin{align*}
C_{1,2} &= \frac{\omega^2}{\sqrt{\omega^4 + \lambda_{1,2}^2}}, &
D_{1,2} &= \frac{\lambda_{1,2}}{\sqrt{\omega^4 + \lambda_{1,2}^2}}.
\end{align*}
Using these results, we compute the pseudoinverse up to the linear order corrections
\begin{subequations} 
\begin{align}
\frac{1}{\gen^*} \left[ \rho \right] &= 0, \label{equ:pseudoinverse_const} \\
\frac{1}{\gen^*} \left[ (x-x_0) \rho \right] &= - \frac{\gamma}{\omega^2} (x-x_0) \rho + \frac{1}{\omega^2} p \rho, \label{equ:pseudoinverse_x} \\
\frac{1}{\gen^*} \left[ p \rho \right] &= - m (x-x_0) \rho. \label{equ:pseudoinverse_p}
\end{align}
\label{equ:pseudoinverse_underdamped} 
\end{subequations}

%Overdamped case 
\section{From underdamped to overdamped diffusion}
\label{sec:pseudoinverse_overdamped}
To obtain the overdamped diffusion we assume that the velocity and in our case the momentum $p$ is the fast variable, which is predominantly governed by 
\[
\gen^*[\mu] = \frac{1}{m} \partial_p \left( p \mu \right) + \frac{1}{\beta(x)} \partial_p^2 \mu . 
\]
The stationary distribution of the momentum $p$ with respect to fixed position $x$ is the Maxwell distribution corresponding to the inverse temperature $\beta(x)$ 
\[
\rho(x,p) = \frac{1}{Z(x)} \rme^{- \beta(x) \frac{p^2}{2m} }. 
\]
In general it is valid that   
\begin{align*}
\gen^*[ p^k \rho ] &= \frac{k}{\beta(x)} \partial_p \left( p^{k-1} \rho \right) \\
&= - \frac{k}{m} p^k \rho + \frac{k (k-1)}{\beta(x)} p^{k-2} \rho , 
\end{align*}
which suggests that eigenvectors will be polynomials times stationary density $\rho$. 
For our purpose we need the first four, which are listed bellow. 
\begin{align*}
\gen^*[ \rho ] &= 0  \\
\gen^*[p \rho] &= - \frac{1}{m} p \rho \\
\gen^*\left[\left( p^2 - \frac{m}{\beta(x)} \right) \rho \right] &= - \frac{2}{m} \left( p^2 - \frac{m}{\beta(x)} \right) \rho \\
\gen^*\left[\left( p^3 - \frac{3m}{\beta(x)} p \right) \rho \right] &= - \frac{3}{m} \left( p^3 - \frac{3m}{\beta(x)} p \right) \rho 
\end{align*}
Using them we obtain 
\begin{subequations} 
\begin{align}
\frac{1}{\gen^*} \left[ \rho \right] &= 0 , \label{equ:pseudoinverse2_const} \\ 
\frac{1}{\gen^*} \left[ p \rho \right] &= - m p \rho , \label{equ:pseudoinverse2_p} \\ 
\frac{1}{\gen^*} \left[ p^2 \rho \right] &= - \frac{m}{2} \left( p^2 - \frac{m}{\beta(x)} \right) \rho , \label{equ:pseudoinverse2_p2}\\ 
\frac{1}{\gen^*} \left[ p^3 \rho \right] &= - \frac{m}{3} \left( p^3 + \frac{6m}{\beta(x)} p \right) \rho . \label{equ:pseudoinverse2_p3}
\end{align}
\label{equ:pseudoinverse_overdamped}
\end{subequations}

\cleardoublepage
\setcounter{page}{1}
\pagenumbering{roman}

\bibliographystyle{alphaurl}
\bibliography{main.lib}

\cleardoublepage

\listoffigures
\listoftables

\cleardoublepage

\printindex

\end{document}